\documentclass[11pt]{book}  

\usepackage{amsfonts, amsmath, amsthm, amssymb, color, mathrsfs, mdsymbol}   
\allowdisplaybreaks
\usepackage{times}
\usepackage[usenames, dvipsnames ]{xcolor}  
\usepackage[all]{xy} 
\usepackage{multirow, makecell}
\usepackage{siunitx}
\usepackage{booktabs}
\usepackage{adjustbox}
\usepackage{pdflscape}
\usepackage[figuresright]{rotating}
\usepackage{subcaption}
\usepackage{caption}
\usepackage{url}
\usepackage[atend]{bookmark}
\usepackage{seqsplit}
\usepackage[skip = 0pt, indent = 2em]{parskip}
\usepackage{indentfirst}
\usepackage{changepage}

\bookmarksetup{
  open,
  openlevel=2,
  numbered,
  addtohook={%
    \ifnum\bookmarkget{level}>1 %
      \DisableBookmarkNumbering
    \fi
  },
}
\makeatletter
\newcommand*{\DisableBookmarkNumbering}{%
  \let\numberline\@gobble
}
\makeatother

\usepackage[english]{babel}
\usepackage{csquotes}
\usepackage[style=apa, backend=biber]{biblatex}

\usepackage{hyperref}
\hypersetup{pdfborder=0 0 0}

\usepackage{graphicx}
\DeclareGraphicsExtensions{.pdf,.png,.jpg}

\theoremstyle{definition}
\newtheorem{assump}{Assumption}[section]

\newcommand\dif{\mathop{}\!\mathrm{d}}

\begin{document}
\pagestyle{empty}


\begin{center}

\vspace{1cm}


{\Huge  Bayesian inference on average treatment effects in the PreventS trial data in the presence of unmeasured confounding}

\vspace{55mm}


{\Large       Jinghong Zeng}

	\vspace{1ex}

Department of Statistics

The University of Auckland

	\vspace{5ex}

 
Supervisor:             Alain C. Vandal

	\vspace*{60mm}

A thesis submitted in partial fulfilment of the requirements for the degree of MSc in Statistics, The University of Auckland, 2022.

\end{center}

  \newpage


\chapter*{Acknowledgements}  
\setcounter{page}{1}
\pagestyle{headings}

\addcontentsline{toc}{chapter}{Acknowledgements}

My supervisor, Alain, provided me with great advice and professional instructions during my research. Discussions with him inspired me to create many original ideas. His strong support and encouragement is greatly appreciated.

Online help obtained from the Stan Forum and Stack Overflow is also much appreciated.

\chapter*{Abstract}       
\pagestyle{headings}

\addcontentsline{toc}{chapter}{Abstract}

Using the PreventS trial data, our objective is to estimate the average effect of a Health Wellness Coaching intervention on improvement of cardiovascular health at 9 months post randomization and in each of three consecutive 3-month periods over 9 months post randomization. Conventional approaches, including instrumental variable models, are not applicable in the presence of multiple correlated multivalued exposures and unmeasured confounding. We propose a general causal framework to identify and estimate the average effects of one or multiple multivalued exposures on one outcome in the presence of unmeasured confounding, noncompliance and missing data, in a two-arm randomized trial. We implement our causal framework through Bayesian models. We also propose general estimation methods of unmeasured confounders, where the exposure and outcome distributions are conditional on unmeasured confounders and then unmeasured confounders are imputed as completely missing variables. Model non-identifiability is a major problem in estimation of unmeasured confounders. Several types of model non-identifiability and possible solutions are described. There is a risk that estimation methods of unmeasured confounders can fail when multiple posterior solutions are produced and they are contradictory. The random intercept outcome models that only adjust for unmeasured confounding in the outcome distribution are proposed as a good surrogate causal model in the presence of unmeasured confounding. No multiple posterior solutions are found from the random intercept outcome models, but the random intercept outcome models need further development.

There is evidence that the Health Wellness Coaching intervention is beneficial to cardiovascular health at 9 months post randomization. On average, completing one Health Wellness Coaching session improves the Life's Simple Seven total score by 0.16 (0.09, 0.22) and reduces systolic blood pressure by 0.54 (0.19, 0.90) mm Hg, but due to sensitivity of Bayesian models, these changes may become smaller or statistically insignificant. On average, completing one Health Wellness Coaching session increases the 5-year cardiovascular disease risk score recalculated with PREDICT by 0.00 (-0.03, 0.02), with little sensitivity of Bayesian models, which indicates completing one Health Wellness Coaching session does not change the 5-year cardiovascular disease risk score recalculated with PREDICT on average. There is also evidence from a random intercept model that based on the Life's Simple Seven total score, the Health Wellness Coaching intervention has a larger beneficial effect on cardiovascular health during 3 months post randomization than in either of two consecutive 3-month periods between 3 and 9 months post randomization. During 3 months post randomization, attending one HWC session reduces the LS7 outcome by 0.10 (-0.14, 0.35) on average. Between 3 and 6 months post randomization, attending one HWC session reduces the LS7 outcome by 0.02 (-0.36, 0.40) on average. Between 6 and 9 months post randomization, attending one HWC session reduces the LS7 outcome by 0.04 (-0.37, 0.45) on average. However, the average effect of one HWC session on the LS7 outcome in each of these three periods is not statistically significant, while the Health Wellness Coaching intervention has a slightly more statistically significant beneficial effect on cardiovascular health during 9 months post randomization than in each of three consecutive 3-month periods over 9 months post randomization. There is no clear evidence that the Health Wellness Coaching intervention benefits or harms mental health based on the Patient Health Questionnaire-9 score, but there is a risk that the Health Wellness Coaching intervention may harm mental health due to sensitivity of Bayesian models. On average, completing one Health Wellness Coaching session increases the Patient Health Questionnaire-9 score by 0.08 (-0.01, 0.16). More evidence supports this change is not statistically significant, but due to sensitivity of Bayesian models, it may become statistically significant. Strong compliance to the Health Wellness Coaching intervention is recommended to see a potential obvious improvement in cardiovascular health.


\setcounter{secnumdepth}{3} 
\setcounter{tocdepth}{3}    
\tableofcontents            


	
\chapter{Introduction}

Stroke is the second leading cause of death in the world \autocite{mahon_primary_2018}. It is necessary for people to control the stroke risk in order to main good health conditions and life quality. Stroke has modifiable risk factors and non-modified risk factors, which indicates we can reduce the stroke risk through modifiable risk factors, such as smoking and high blood pressure \autocite{mahon_primary_2018}. Modifiable risk factors are associated with behaviors and lifestyles. The Health Wellness Coaching is a multidimensional psychological intervention \autocite{mahon_primary_2018}. It can raise awareness of health in people through coaching and motivate people to change their lifestyles, so that modifiable risk factors may be reduced and health may be improved. The Health Wellness Coaching is proposed as a primary prevention approach to stroke by the PreventS trial. The PreventS trial, where ``PreventS'' stands for primary prevention of stroke and cardiovascular disease in the community, is ``a parallel, prospective, randomized, open-treatment, single-blinded end-point trial of participants who have a five-year CVD risk of $\geq$10\%'' (Mahon et al., 2018, p. 3), ``calculated using the PREDICT web-based clinical tool'' (Mahon et al., 2018, p. 1). Participants are randomized to either the Health Wellness Coaching intervention group or the usual care group (Mahon et al., 2018). This trial is developed to understand the stroke risk among people in New Zealand and test effectiveness of the Health Wellness Coaching on reduction in the cardiovascular disease and stroke risk among people in New Zealand. The primary results of this trial have not been published.

The Health Wellness Coaching intervention in the PreventS trial consists of 15 individual Health Wellness Coaching sessions with trained coaches, which will be spread over 9 months post randomization \autocite{mahon_primary_2018}. We aim to know how effective one Health Wellness Coaching session will be on one outcome during 9 months post randomization and whether the average effect of one Health Wellness Coaching session on one outcome will differ among different periods during 9 months post randomization. In order to understand effectiveness of the Health Wellness Coaching intervention on improving cardiovascular health, three outcomes are selected for analysis: the change in the 5-year cardiovascular disease risk score recalculated with PREDICT at 9 months post randomization, the change in the Life's Simple Seven total score at 9 months post randomization, the change in systolic blood pressure at 9 months post randomization \autocite{mahon_primary_2018}. Ideally, the Health Wellness Coaching intervention should not harm mental health when it improves cardiovascular health. In order to understand how the Health Wellness Coaching intervention affects mood, the change in the Patient Health Questionnaire-9 score is selected for analysis. Understanding effectiveness of the Health Wellness Coaching intervention on cardiovascular and mental health may not only benefit people in New Zealand, but also help revise or improve the design of the Health Wellness Coaching intervention. First, if we can find statistical evidence to support that the Health Wellness Coaching intervention is effective to cardiovascular health, then people in New Zealand can have access to the Health Wellness Coaching intervention as a primary prevention of stroke and cardiovascular diseases, which reduces the risk of stroke and cardiovascular diseases and improves the health conditions of these people in their future lives. Second, currently the Health Wellness Coaching intervention lasts for at least 9 months. This duration may be too long for some people and project management. If we can find statistical evidence that the Health Wellness Coaching intervention is more effective in the earlier periods than in the later periods, then we may consider shortening the Health Wellness Coaching intervention, which may be more time-saving and cost-effective. Other comparisons of the effectiveness of Health Wellness Coaching intervention among different periods may lead to different decisions on the intervention improvement. Hence, it is necessary to investigate the effectiveness of Health Wellness Coaching intervention both over 9 months post randomization and in different periods during 9 months post randomization.

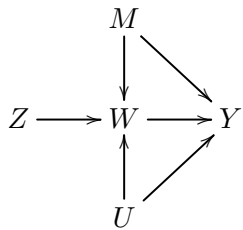
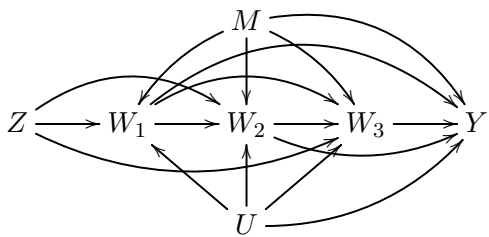
\begin{figure}[htbp]
     \centering
     \begin{subfigure}{0.4\textwidth}
\xymatrix{
&  & M \ar[d] \ar[dr] & \\
&Z \ar[r] & W \ar[r] & Y \\
& & U  \ar[u] \ar[ur]  & 
}       
\caption{First question.}
     \end{subfigure}

     \begin{subfigure}{0.4\textwidth}
\xymatrix{
&    & M \ar@/_0.5pc/[dl] \ar[d] \ar@/^0.5pc/[dr] \ar@/^1.5pc/[drr]&& \\
Z \ar[r] \ar@/^1.5pc/[rr] \ar@/_1.5pc/[rrr] & W_1 \ar[r] \ar@/^1.5pc/[rr] \ar@/^2.5pc/[rrr] & W_2 \ar[r] \ar@/_1pc/[rr] & W_3 \ar[r] & Y \\
&    & U \ar[ul] \ar[u] \ar[ur] \ar@/_1pc/[urr]&& 
}      
\caption{Second question.}
     \end{subfigure}
        \caption{Possible causal relationships among variables in our questions. $Z$ is the assignment or randomization. $W$ is an exposure. $W_1$, $W_2$, $W_3$ are three sequential exposures. $Y$ is an outcome. $M$ is measured confounders. $U$ is unmeasured confounders.}
        \label{fig: intro-cg}
\end{figure}

Our first question is, how can we estimate the average effect of one Health Wellness Coaching session on each of four outcomes during 9 months post randomization? Our second question is, how can we estimate the average effect of one Health Wellness Coaching session on the Life's Simple Seven total score at 9 months post randomization, in each of three consecutive 3-month periods during 9 months post randomization? More specifically in our second question, there are three average effects during three periods, and three periods are during 3 months post randomization, between 3 and 6 months post randomization, between 6 and 9 months post randomization. For our first question, the exposure is the number of Health Wellness Coaching sessions attended at 9 months post randomization. For our second questions, three exposures are the numbers of Health Wellness Coaching sessions attended in three periods during 9 months post randomization. Figure \ref{fig: intro-cg} shows possible causal relationships among variables with abstract symbols in our questions.

In the two questions mentioned above, we take unmeasured confounding into account and we aim to estimate causal effects unbiasedly. We first conduct a literature review in causal inference and pay particular attention to unmeasured confounding. 


To draw causal inference, Rubin's model is most commonly used \autocite{rubin_estimating_1974, rubin_bayesian_1975, rubin_assignment_1977, rubin_bayesian_1978, holland_causal_1980, holland_lords_1983, holland_statistics_1986, holland_causal_1988, angrist_identification_1996, durlauf_rubin_2010}. It defines a causal effect and focuses on what causation is. Rubin's model started with the simplest case where a binary exposure is considered \autocite{rubin_estimating_1974}. Suppose there is a binary exposure denoted by $W$, an outcome denoted by $Y$ and participants with a sample size of $N$. $W=1$ means the experimental intervention or the treatment under research. $W=0$ means the control intervention. For a particular participant, if we let $Y(W=0)$ be the outcome when the control intervention is received and let $Y(W=1)$ be the outcome when the experimental intervention is received, then $Y(W=1)-Y(W=0)$ is the causal effect of the experimental intervention versus the control intervention on the outcome \autocite{rubin_estimating_1974, holland_statistics_1986, angrist_identification_1996}. $Y(W=0)$ and $Y(W=1)$ are potential outcomes, because they are what would happen in potential scenarios and in reality only one of them can be observed \autocite{angrist_identification_1996}. The average causal effect of the experimental intervention versus the control intervention on the outcome is then an average of all the individual causal effects, that is, $\frac{1}{N}\sum_{i=1}^N{\left[Y_i(W=1)-Y_i(W=0)\right]}$ or $E\left(Y(W=1)-Y(W=0)\right)$ \autocite{rubin_estimating_1974}. Rubin then extended the definition of a causal effect from a binary exposure to a discrete multivalued exposure \autocite{rubin_bayesian_1975, rubin_bayesian_1978}. Suppose the exposure $W$ is now multivalued and takes values from a discrete set $\Omega = \{0, 1, 2,\ldots, J\}$, for $J \geq 3$, where $W=0$ means the control intervention and the other exposure levels mean the experimental intervention. For a particular participant, the causal effect of the exposure level $t \in \Omega$ versus the exposure level $c \in \Omega$ ($t \neq c$) is $Y(W=t)-Y(W=c)$. Causal effects from all combinations of two different exposure levels can be calculated separately. Usually we want $c=0$ so as to compare the control intervention with the experimental intervention. The average causal effect of the exposure level $t \in \Omega$ versus the exposure level $c \in \Omega$ ($t \neq c$) is still an average of all the individual causal effects. Rubin's definition of causal effects for a multivalued exposure is clear, but the practical calculation can be a long process when the exposure has too many levels. It can be more feasible if we find an approach to avoid calculating causal effects from all combinations of two different exposure levels, but Rubin did not mention clearly how to estimate the average causal effect over all the exposure levels. 

Rubin further examined the importance of randomization in experiments to estimation of causal effects and adjusted for a covariate in estimation of causal effects through experiment design \autocite{rubin_bayesian_1975, rubin_assignment_1977, rubin_bayesian_1978, basu_randomization_1980}. Holland and Rubin \autocite*{holland_causal_1980, holland_causal_1988} discussed causal inference through Rubin's model in prospective and retrospective observational studies, and they proposed the definitions and estimation methods of population-level causal effects, intermediate-level causal effects and unit-level causal effects, which are causal effects in different groups of people. Holland and Rubin \autocite*{holland_lords_1983} applied Rubin’s model in several Lord's examples including causal and descriptive studies to tell the differences between causal and descriptive inference. Their applications gave a very clear explanation on derivations of causal effect estimands through Rubin's model, but they did not mention clearly what statistical models should be used in estimation. Rubin's model is a definition on causation and not a modelling technique. To implement estimation of causal effects defined by Rubin's model in the presence of different biases, specific statistical models should be developed and used.


One fundamental statistical model to estimate causal effects defined by Rubin's model in the presence of unmeasured confounding is instrumental variable models \autocite{imbens_identification_1994, bound_problems_1995, angrist_identification_1996, imbens_estimating_1997, imbens_instrumental_2014, imbens_causal_2015}. Instrumental variable models usually require at least one valid instrument and certain assumptions such as the exclusion restriction assumption. Angrist and Imbens \autocite*{angrist_identification_1991, imbens_identification_1994} proposed local average treatment effect of a binary exposure in 1991 through instrumental variable models with a binary instrument, and they defined local average treatment effect as ``the average treatment effect for individuals whose treatment status is influenced by changing an exogenous regressor that satisfies an exclusion restriction'' (1994, p. 467) in 1994. Simply, local average treatment effect is the average causal effect in some population whose treatment status can be changed by changing the exposure. We use the article \autocite{imbens_identification_1994} for further reference. 
Angrist, Imbens and Rubin \autocite*{angrist_identification_1996} mentioned noncompliance, first proposed by Balke and Pearl \autocite{balke_nonparametric_1993}, to further clarify what subpopulation local average treatment effect defined in 1994 is estimated from. Noncompliance is a feature of any experiment that some participants do not comply with the intervention assigned to them. It is a special case of principal stratification \autocite{frangakis_principal_2002}. Four compliance behaviors are proposed: compliers, never-takers, always-takers, defiers; the last three compliance behaviors are noncompliers \autocite{angrist_identification_1996}. Local average treatment effect of a binary exposure on an outcome is the average treatment effect of this exposure on the outcome in compliers, when instrumental variable models assume there is no defier \autocite{imbens_identification_1994, angrist_two-stage_1995, angrist_identification_1996, imbens_causal_2015}.

Angrist and Imbens \autocite*{angrist_two-stage_1995} started with two-stage least squares (TSLS) with a binary instrument, a widely used instrumental variable method, and figured out the local average treatment effect for a multivalued exposure that can take at least three possible values, such as ``drug dosage'' (p. 431) and ``years of schooling'' (p. 431). From Rubin's model, there is a local average treatment effect of each exposure level versus the control intervention, for a particular unit. They \autocite{angrist_two-stage_1995} found that the local average treatment effect for a multivalued exposure through TSLS is a weighted average over local average treatment effects of all exposure levels versus the control intervention. Little progress on causal effects of multivalued exposures following this work \autocite{angrist_two-stage_1995} has been found.

Instrumental variable models can fail when necessary assumptions are violated or there are complex causal relationships such as causal relationships among multiple exposures. Researchers have been trying hard to extend instrumental variable models in more difficult problems. Examples are generalized instrumental variable models that extend the form and use of instruments such as conditional instruments and instruments in generalized linear models \autocite{brito_generalized_2002, johnston_use_2008, chalak_viewpoint_2011, van_der_zander_efficiently_2015, salois_generalized_2015, chesher_generalized_2017, shingaki_selection_2021, wong_calculus_2021, bollen_model-implied_2014}, estimation with invalid instruments \autocite{kolesar_identification_2015, bowden_mendelian_2015, kang_simple_2015, kang_instrumental_2016, guo_confidence_2018, windmeijer_use_2019}, and other applications \autocite{angrist_non-parametric_1995, imbens_estimating_1997}.

Another important statistical method to estimate causal effects defined by Rubin's model in the presence of unmeasured confounding is Bayesian causal inference. Instrument variable models can be implemented through both frequentist and Bayesian approaches. But, Bayesian causal inference mentioned here is a different inferential approach from Bayesian instrument variable models, and specifically refers to the Bayesian method proposed by Imbens and Rubin \autocite*{imbens_bayesian_1997, imbens_causal_2015}. The Bayesian ideas in the joint work of Imbens and Rubin \autocite*{imbens_bayesian_1997, imbens_causal_2015} have similarity to those in Rubin's early work \autocite*{rubin_bayesian_1975, rubin_bayesian_1978}, such as they both proposed fitting distributions of the potential outcomes, and they are also mentioned in Rubin's later paper \autocite*{rubin_causal_2005}.

It is very likely that an experiment involves noncompliance. Some compliance behaviors such as never-takers may not be able to provide information about causal effects of exposures. To obtain more precise causal effect estimates, Imbens and Rubin \autocite*{imbens_bayesian_1997} proposed fitting distributions of the outcome for each compliance behavior separately given assumptions on parameter dependence, which allows the causal effect of the exposure to be estimated in each subpopulation if possible, and they defined two local average treatment effects of a binary exposure in the presence of a binary instrument. The two local average treatment effects are complier average causal effect (CACE) in compliers and defier average causal effect (DACE) in defiers \autocite{imbens_bayesian_1997}. Their approaches \autocite{imbens_bayesian_1997} can be summarized in several steps: define a compliance behavior variable in causal frameworks, use the potential outcome framework to construct exposure and outcome distributions for potential or counterfactual variables, transform distributions of counterfactual variables to distributions of observed variables, compute posterior distributions of causal estimands. Imbens and Rubin \autocite*{imbens_causal_2015} clearly described the process of deriving posterior distributions with transformation between distributions of potential variables and distributions of observed variables.

Bayesian causal inference from Imbens and Rubin \autocite*{imbens_bayesian_1997, imbens_causal_2015} is more flexible on identification of causal effects, but it does not explicitly take unmeasured confounding into account. Further, Rubin's definition of causal effects of multivalued exposures \autocite{rubin_bayesian_1975, rubin_bayesian_1978} and the instrumental variable formula of causal effects of multivalued exposures \autocite{angrist_two-stage_1995} may not be feasible in Bayesian causal inference \autocite{imbens_bayesian_1997}, because of different modelling procedures. 

There is much similar methodological research and interdisciplinary applications in Bayesian causal inference through Rubin's model. Yau and Little \autocite*{yau_inference_2001} developed a Bayesian model to estimate causal effects of a binary exposure in longitudinal data with inclusion of baseline covariates and handling of missing data. 
Karabatsos and Walker \autocite*{karabatsos_bayesian_2012} proposed a Bayesian nonparametric causal model to estimate causal effects of a binary exposure in observational data. Mattei et al. \autocite*{torelli_exploiting_2013} described Bayesian inference procedures to estimate causal effects of a binary exposure on multiple outcomes. Alaa et al. \autocite*{alaa_bayesian_2018} developed Bayesian nonparametric learning algorithms to estimate causal effects of a binary exposure in observational studies. Ning et al. \autocite*{ning_bayesian_2019} developed a Bayesian method to estimate causal effects of a binary exposure in time series data in the presence of spatial correlation. Park and Kellis \autocite*{park_cocoa-diff_2021} proposed a causal inference framework for single-cell gene expression analysis and implemented it through Bayesian modelling. Pang et al. \autocite*{pang_bayesian_2022} proposed a Bayesian alternative to the synthetic control method for comparative case studies to estimate causal effects of a binary exposure.

Particular attention is given to Bayesian causal inference with multivalued or multiple exposures. Hoshino \autocite*{hoshino_bayesian_2008} proposed a Bayesian propensity score method to estimate causal effects of a multivalued exposure in observational data through Rubin's model. Hill \autocite*{hill_bayesian_2011} proposed Bayesian nonparametric modeling approaches to estimate causal effects of a binary, multivalued or continuous exposure. Graham et al. \autocite*{graham_approximate_2016} proposed a Bayesian bootstrap approach that can produce doubly robust estimation of causal quantities, where the exposure can be binary, multi-valued or continuous but non-randomized. Alaa and van der Schaar \autocite*{alaa_bayesian_2017} proposed a Bayesian multi-task learning framework to estimate individual treatment effects of a multivalued exposure in observational data. Gutman and Rubin \autocite*{gutman_robust_2013, gutman_estimation_2015, gutman_estimation_2017} proposed multiply imputing missing or unobservable potential outcomes through Bayesian modelling to estimate causal effects of a binary exposure. Then Silva and Gutman \autocite*{silva_multiple_2022} proposed new multiple imputation procedures to estimate causal effects of multiple binary exposures through Rubin's model and implemented these procedures through Bayesian approaches. Their methods \autocite{silva_multiple_2022} can also be used to estimate causal effects of a single multivalued exposure. The studies mentioned above indicate that development of different specific Bayesian causal inference methods seems necessary for different causal questions in different data.


Another interesting statistical method to estimate causal effects in the presence of unmeasured confounding that has been applied in practice is two-stage residual inclusion estimation \autocite{hausman_specification_1978, nagelkerke_estimating_2000, have_causal_2003, terza_two-stage_2008, baiocchi_tutorial_2014, geraci_testing_2014, palmer_correcting_2017, terza_two-stage_2018, koladjo_instrumental_2018, ying_twostage_2019}. This method first estimates unmeasured confounders as the residuals from the model of the exposure against the instrument and then adjusts for estimated unmeasured confounders in the model of the outcome against the exposure. Two-stage residual inclusion estimation can be applied in both linear and nonlinear models. Linear models guarantee that two-stage residual inclusion estimators are equal to two-stage least squares estimators \autocite{terza_two-stage_2008, ateba_effect_2021}. In nonlinear models, two-stage residual inclusion estimators are not explicitly proved to be defined by Rubin's model. 

There are also other methods through Rubin's model to estimate causal effects. Rubin \autocite*{rubin_assignment_1977} mentioned some causal effect estimators through Rubin's model, including model fitting, blocking and matching. These estimators may be applied in both frequentist and Bayesian models, but they are different from instrumental variable models and Bayesian causal inference. Uysal \autocite*{uysal_doubly_2015} proposed a multivalued treatment framework through Rubin's model and implemented it through doubly robust estimation methods, where pairwise treatment effects were defined as the difference between causal effects of two different treatment levels, same as Rubin's definition \autocite{rubin_bayesian_1975, rubin_bayesian_1978, uysal_doubly_2015}. Different from average treatment effects, Xu et al. \autocite*{xu_bayesian_2018} proposed a Bayesian nonparametric framework to estimate quantile treatment effects, and Xu et al. \autocite*{xu_bayesian_2022} proposed a Bayesian semiparametric framework to estimate quantile treatment effects. 

There are other methods or frameworks for causal inference, outside or beyond Rubin's model. In observational studies for causal effects of a binary or multivalued exposure, the propensity score is commonly used in covariate adjustment such as confounding adjustment to reduce bias \autocite{rosenbaum_central_1983, imbens_role_2000, hirano_propensity_2004, imai_causal_2004, hoshino_bayesian_2008, austin_introduction_2011, feng_generalized_2012}. Robins \autocite*{robins_new_1986, robins_addendum_1987} proposed a new approach to causal inference in longitudinal studies, where generalized treatments as a sequence of exposures over time and G-causal parameters associated with two generalized treatments were mathematically defined. This definition of exposure \autocite{robins_new_1986, robins_addendum_1987} makes it possible to estimate causal effects of time-varying or continuous exposures. Then structural nested models were proposed through a counterfactual approach to estimate causal effects of a wider class of randomized or non-randomized exposures including time-varying and continuous exposures in longitudinal studies, under certain assumptions \autocite{robins_correcting_1994, bickel_causal_1997, cooper_testing_1999, ostrow_analytic_2002}. As an alternative, marginal structural models were proposed to estimate causal effects of a wider class of randomized or non-randomized exposures including time-varying and continuous exposures in longitudinal studies, possibly with adjustment for time-varying confounding \autocite{robins_marginal_1998, robins_marginal_2000, miller_marginal_2000, hernan_marginal_2001, oba_how_2011, ewald_marginal_2019, shinozaki_understanding_2020}. 
Wong \autocite*{wong_calculus_2021, wong_integral_2021} proposed the definition and estimation of causal effects of binary, multidimensional, discrete and continuous exposures in a larger class of models including nonlinear models through the structural equation framework. Wong's definition of a causal effect \autocite*{wong_calculus_2021, wong_integral_2021} is based on derivatives rather than potential outcomes.

Biases include selection bias, unmeasured confounding, measurement error. Among them, unmeasured confounding is one significant bias source. To obtain unbiased average treatment effect estimates, unmeasured confounding should be adjusted for. Unmeasured confounders include unmeasured known confounders and unknown, unmeasureable confounders. Since unmeasured confounders are unmeasured in reality, we need to estimate them as precisely as possible.

More generally, analysis related to any bias is bias analysis. Bias analysis aims to assess the impact of one or multiple biases on causal effect estimates in observational studies or imperfect experimental designs. We use unmeasured confounding and one bias for descriptions about bias analysis. Usually a bias is associated with a bias variable, denoted by $S$, and some bias parameters related to $S$, denoted by $\xi$. In our descriptions, this bias variable is an unmeasured confounder. To do bias analysis, we first need to find the bias and then use methods to correct it. There are two main approaches to adjust for unmeasured confounders in causal effect estimates. The first approach is to bias-correct the causal effect estimate \autocite{rosenbaum_assessing_1983, lin_assessing_1998, steenland_monte_2004, greenland_multiple-bias_2005, vanderweele_bias_2011, arah_bias_2017, groenwold_adjustment_2018}. The steps are: (1) construct the models for the exposure and the outcome without $S$ to produce the unadjusted effect, (2) also construct the models for the exposure and the outcome with $S$ to produce the adjusted or true effect, (3) derive the mathematical relationship between the true effect, the unadjusted effect and the bias as the formula for the bias-corrected effect estimate. The bias usually can be written as a function of $S$, $\xi$ and possibly some other coefficients in the models.  The second approach is to adjust for imputed unmeasured confounders in statistical models \autocite{lash_semi-automated_2003,mccandless_bayesian_2007, greenland_bayesian_2009, faries_evaluating_2013}. The steps are: (1) construct the model for $S$ conditional on the exposure and/or the outcome, (2) impute $S$ possibly through resampling methods, Bayesian models or multiple imputation, (3) adjust for imputed $S$ in the model for the outcome to produce the adjusted causal effect estimate. $S$ can be completely missing, but limited data about $S$ may also be available from external sources such as validation studies. Some methods have demonstrated application potential in many types of bias, and they can apply to both one bias and multiple biases.

Unmeasured confounders are not measured in reality, and thus statistical models that adjust for unmeasured confounders are usually non-identified. For example, unmeasured confounders or some associated parameters are non-identified. Model non-identifiability related to unmeasured confounders as a bias source has been treated by sensitivity analysis of this bias, in order to obtain interpretable causal effect estimates such as the ranges of causal effects and further quantify the impact of the bias on causal effect estimates. Sensitivity analysis includes fixed sensitivity analysis \autocite{rosenbaum_assessing_1983, lin_assessing_1998, vanderweele_bias_2011} and probabilistic sensitivity analysis, including Monte Carlo sensitivity analysis (MCSA) and Bayesian sensitivity analysis (BSA) \autocite{steenland_monte_2004, greenland_multiple-bias_2005, mccandless_bayesian_2007, greenland_bayesian_2009, dorie_flexible_2016, zheng_bayesian_2021, zheng_copula-based_2021, zheng_sensitivity_2022}. In fixed sensitivity analysis, we give fixed values to the bias variable, bias parameters and possibly other parameters to obtain a fixed bias, then may repeat this process many times to obtain many fixed biases and see how the causal effect estimate varies over these biases. In probabilistic sensitivity analysis, we give prior distributions to the bias variable, bias parameters and possibly other parameters to obtain a range of the bias variable or the bias and also the range of the causal effect estimate. MCSA only uses the information from the priors to generate samples about the bias variable or the bias, while BSA uses the information from both the prior and data. Fixed sensitivity analysis usually fails to well account for variability in the bias-corrected causal effect estimate, and thus probabilistic sensitivity analysis is preferred, where BSA is proven better than MCSA because data is used and interval estimates can be more reliable \autocite{mccandless_comparison_2017}. This whole process that adjusts for unmeasured confounders and uses sensitivity analysis is useful to understand the range of causal effects, but it may not be used to draw inference on causal effects because causal effects may be non-identifiable. In order to identify causal effects, model non-identifiability should been resolved.

Wang and Blei \autocite*{wang_blessings_2018, wang_multiple_2019} proposed the deconfounder algorithm that uses proxy variables to adjust for unmeasured confounders in the presence of multiple exposures. They also attempted to establish conditions on model identifiability in the presence of unmeasured confounding, but their approaches have been questioned by counterexamples \autocite{damour_multi-cause_2019, ogburn_counterexamples_2020}. However, their attempts to establish conditions on model identifiability is a good step, though their conditions on model identifiability are mathematical and seem too absolute. Earlier, Pearl \autocite{pearl_parameter_2001} also proposed mathematical definitions on parameter and model identifiability. Mathematical concepts of model identifiability are useful if they can be used in practice such as derivations of posterior distributions in Bayesian models, otherwise establishment of model non-identifiability criteria regarding practical modelling procedures such as model diagnostics would be more helpful. 

Some discovered sources of model non-identifiability include relaxation of the exclusion restriction assumption and inappropriate supports of exposures \autocite{hirano_assessing_2000,jo_estimation_2002,forcina_causal_2006}. Weakly identified models have been proposed to estimate causal effects when model identifiability cannot be guaranteed \autocite{hirano_assessing_2000,jo_estimation_2002}. Weakly identified models have proper posterior distributions, but they do not have unique maximum likelihood estimates \autocite{jo_estimation_2002}. To use weakly identified models, additional auxiliary assumptions usually have to be made \autocite{hirano_assessing_2000,jo_estimation_2002}. These results do not directly focus on unmeasured confounding, and thus it is also necessary to explore model non-identifiability associated with unmeasured confounding. Above all, no existing causal framework that considers multiple correlated multivalued exposures and unmeasured confounding simultaneously has been found.


Instrumental variables models \autocite{angrist_identification_1991, imbens_identification_1994, angrist_two-stage_1995, angrist_identification_1996} can apply to our first question. For our second question, instrumental variables models cannot apply in the presence of multiple exposures that are associated with each other because there is no valid instrument. For example, in figure  \ref{fig: intro-cg} (b), to estimate the direct effect of one exposure on the outcome, the other two exposures have to be conditioned on, which violates the exclusion restriction assumption in instrumental variables models. In addition, statistical models that do not adjust for unmeasured confounding may produce biased estimates of the average effects of multiple exposures. We are thinking, can we adjust for unmeasured confounding directly in order to obtain unbiased estimates of the average effects of multiple exposures?

We find one existing causal framework that can estimate the average effect of one multivalued exposure on one outcome in the presence of unmeasured confounding is the two-stage least squares method \autocite{angrist_two-stage_1995}, which is one instrumental variable method. Hence, this causal framework cannot be extended to multiple exposures. We find Bayesian models seem to be able to deal with multiple multivalued exposures, missing data and noncompliance simultaneously. Inspired by the Bayesian causal framework developed for binary exposures in the presence of noncompliance \autocite{imbens_bayesian_1997}, we propose a general causal framework to identify and estimate the average effects of one or multiple multivalued exposures on one outcome in the presence of unmeasured confounding and noncompliance. We implement our causal framework through Bayesian models. Our causal framework identifies the unmeasured confounding effects, and then we propose estimation methods of unmeasured confounders, where the exposure and outcome distributions are conditional on unmeasured confounders and unmeasured confounders are imputed as completely missing variables. Model non-identifiability is a major problem in estimation of unmeasured confounders. We find out several types of model non-identifiability and propose possible solutions for each of them. However, there is a risk that estimation methods of unmeasured confounders can fail when multiple posterior solutions are produced and they are contradictory. In this case, we propose using the random intercept outcome models that only adjust for unmeasured confounding in the outcome distribution, though the random intercept outcome models are insufficiently developed in our thesis.

In development of our causal framework and estimation methods of unmeasured confounders, we start with one exposure and then extend to multiple exposures. Our methods also show a potential applicability in frequentist models, and cohort studies in addition to two-arm randomized trials. Then our methods are applied in analysis of the PreventS trial data with the exposures and outcomes mentioned above. 

An outline of our thesis is given. Chapter \ref{sec:develop} describes methodological development in Bayesian causal inference. It consists of five parts: development of causal frameworks with one multivalued exposure in section \ref{sec:framework-1}, estimation methods of unmeasured confounders in section \ref{sec:u-1}, extension of causal frameworks and estimation methods of unmeasured confounders in multiple multivalued exposures that have common or different unmeasured confounders in section \ref{sec:ext}, simulation study on application of causal frameworks and estimation of unmeasured confounders in section \ref{sec:sim}, establishment of a Bayesian modelling procedure through our methods in section \ref{sec:model-procedure}. Section \ref{sec:framework-1} starts with variable notations and an overview of assumptions, then focuses on development of causal frameworks in subsection \ref{sec:framework-2}. Subsection \ref{sec:framework-2} consists establishment of causal frameworks for potential outcomes or counterfactual variables, identification of the average treatment effect, variations in causal frameworks for counterfactual variables, establishment of causal frameworks for observed variables, extension to cohort studies and a different definition of the control intervention effect. Section \ref{sec:u-1} first discusses estimation methods of one unmeasured confounder in subsection \ref{sec:u-2} and then extends these estimation methods in multiple unmeasured confounders in subsection \ref{sec:u-5}. Subsection \ref{sec:u-2} describes model non-identifiability and reparameterization as one important tool to reduce model non-identifiability. Two subsections \ref{sec:ri-1} and \ref{sec:ri-2} in simulation study describe the modelling performance of the random intercept outcome models.

Chapter \ref{sec:analysis-1} focuses on analysis of the PreventS trial data through our methods and Bayesian models. It starts with definitions of the exposures, the outcomes and measured confounders, where descriptive statistics and missing data statistics are shown. Then derivation of complete-data models and calculation of sample standard deviations, which will be used in Bayesian models, are described. Two major parts are analysis with one exposure to estimate the average effect of one Health Wellness Coaching session during 9 months post randomization in section \ref{sec:analysis-2} and analysis with three exposures to estimate the average effect of one Health Wellness Coaching session on the Life's Simple Seven total score at 9 months post randomization in three periods during 9 months post randomization in section \ref{sec:analysis-3}. Section \ref{sec:analysis-2} presents results from four outcomes and compares unmeasured confounding among four outcomes. Our thesis ends with discussions in chapter \ref{sec:dis} and conclusions in chapter \ref{sec:con}. All the work is done by the student. The supervisor provides comments and advice through regular supervisory meetings to help the student improve the methods and models.

\chapter{Developments in Bayesian causal inference}
\label{sec:develop}

We focus on developing a general causal framework that can identify and estimate the average effects of one or multiple multivalued exposures on one outcome in the presence of unmeasured confounding, noncompliance and missing data. We choose to implement our causal framework through Bayesian models. 

Our approach to develop a causal framework is original by our knowledge. We establish causal relationships between counterfactual variables through the potential outcome framework and identify necessary assumptions that are different from assumptions in instrumental variable models. Then we construct structural equations for counterfactual variables, in order to identify the average treatment effects. The assumptions in our causal framework enable us to estimate the average treatment effects rather than local average treatment effect. We use the Imbens-Rubin method \autocite{imbens_bayesian_1997} to adjust for noncompliance, which is to define a compliance behavior variable in our causal framework and condition on this variable in structural equations. We also use Imbens-Rubin method \autocite{imbens_causal_2015} to transform the distributions from counterfactual variables to observed variables, but we use different mathematical relationships between counterfactual variables and observed variables and develop different complete-data models to suit estimation of the average treatment effects. Except that adjustment for noncompliance and distribution transformation between counterfactual variables and observed variables are the main ideas that we learn from prior works, settings including assumptions, variable relationships, identification of the average treatment effect and incorporation of the effect of study design in our causal framework and organization of the structure of our causal framework are our own ideas.

To adjust for unmeasured confounding, we propose estimation methods of unmeasured confounders ,where the exposure and outcome distributions are conditional on unmeasured confounders and unmeasured confounders are imputed as completely missing variables. Model non-identifiability is a major problem in estimation of unmeasured confounders. Several types of model non-identifiability have been discovered and possible solutions are proposed, in order to make Bayesian models identified.

\section{Causal framework with one multivalued exposure}
\label{sec:framework-1}

We start with one multivalued exposure to introduce our Bayesian causal framework and estimation methods of unmeasured confounders, before extending the methods into multiple multivalued exposures. For the methods with one multivalued exposure, variable definitions and notations are first described in detail. All the basic information also applies to multiple multivalued exposures, unless otherwise stated.

\subsection{Variable definitions and notations}




Our Bayesian causal framework is based on the potential outcome framework. Relevant variable definitions and notations are first introduced.

For our Bayesian causal framework, the target design is a two-arm randomized trial, involving a control arm and a treatment arm, a random assignment, one multivalued exposure, one outcome, pre-exposure covariates including measured and unmeasured confounders, a compliance behavior, the sample size of $N$. We want to estimate the average treatment effect in this target randomized trial. We use an uppercase letter to represent a random vector, an uppercase letter with subscript $i$ to represent a random variable for the $i$-th participant, and a lowercase letter with subscript $i$ to represent a realization of a random variable for the $i$-th participant.

An exposure represents an intervention received by participants in the randomized trial. It includes both the control intervention and the treatment of research interest. The average exposure effect is not used, because it also includes the control intervention effect. To avoid ambiguity, when we talk about the average treatment effect, we mean how the treatment of research interest can affect the outcome, compared to the control intervention. We only focus on the average treatment effect. A multivalued exposure means this exposure as a random variable has at least three possible values, one value for the control arm and at least two values for the treatment arm. The multivalued exposure is denoted by $W=(W_1, \ldots, W_i, \ldots, W_N)$, for $i = 1, 2, \ldots, N$. If the multivalued exposure is a discrete variable, then the domain of $W_i$ is a discrete set that contains at least three distinct values. If the multivalued exposure is a continuous variable, then the domain of $W_i$ is a continuous interval. For the PreventS trial data, the exposures we will use are discrete variables and we use a set $\{0,1,\ldots,J\}$ as the domain for $W_i$, with $J$ finite and measured, where zero represents the control intervention. When introducing our methods, we do not define whether the exposure is discrete or not because all methods apply to both discrete and continuous variables. In some situations where derivatives are used, we treat discrete variables as if they are continuous to apply differentiation. The outcome is denoted by $Y=(Y_1, \ldots, Y_i, \ldots, Y_N)$, for $i = 1, 2, \ldots, N$. It can also be either discrete or continuous, and it can be either binary or multivalued. The notations for the exposure and the outcome, $W$ and $Y$, are general at this stage. When we further describe counterfactual variables, multiple counterfactual exposures and outcomes are involved. Each counterfactual exposure or outcome shares the same symbol, $W$ or $Y$. We further use superscripts and subscripts to distinguish exposures and outcomes of different meanings.

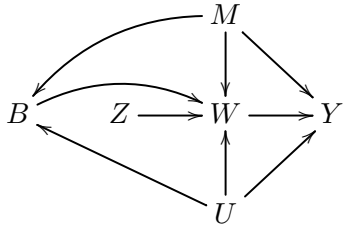
\begin{figure}[!h]
    \centering
\xymatrix{
&&&&&  & M \ar[d] \ar[dr] \ar@/_1pc/[dll] &   \\
&&&& B \ar@/^1pc/[rr]& Z \ar[r]& W \ar[r]& Y \\
&&&&&   & U \ar[u] \ar[ur] \ar[ull]&  
}

    \caption{A basic causal mechanism for the introduced variables.}
    \label{fig:method-setup}
\end{figure}

The assignment is denoted by $Z=(Z_1, \ldots, Z_i, \ldots, Z_N)$, and  

\begin{equation}
Z_i =
    \begin{cases}
      0, & \text{if assigned to the control arm}\\
      1, & \text{if assigned to the treatment arm}
    \end{cases}       
\end{equation}
for $i = 1, 2, \ldots, N$. The $p$ pre-exposure measured confounders are denoted by $M=(M_1,$  $\ldots, M_j, \ldots, M_p)$, for $j = 1, 2, \ldots, p$. Each measured confounder is denoted by $M_j = (M_{j1}, $. $\ldots, M_{ji}, \ldots, M_{jN})$, for $i = 1, 2, \ldots, N$. A pre-exposure unmeasured confounder is denoted by $U=(U_1, \ldots, U_i, \ldots, U_N)$, for $i = 1, 2, \ldots, N$. We do not fix the number of unmeasured confounders here. When multiple unmeasured confounders are mentioned, their notations will be given but each denotation has the same structure as $U$. The compliance behavior is denoted by $B=(B_1, \ldots, B_i, \ldots, B_N)$, $i = 1, 2, \ldots, N$. This variable is about whether a participant complies with the assigned exposure. More details about the compliance behavior will be introduced soon. Generally, we assume the compliance behavior affects the exposure and itself is affected by pre-exposure covariates. The compliance behavior may or may not affect the outcome directly. All unknown parameters separate from these variables are denoted by $\theta$ as a parameter vector. Figure \ref{fig:method-setup} shows a basic causal mechanism among these variables. We will use similar causal graphs to explain methods. 

Then all the notations are applied in the potential outcome framework. Potential outcomes are what would happen in hypothetical situations. In the potential outcome framework, every variable is counterfactual and we call all variables ``counterfactual variables''. If one hypothetical situation comes true, all associated variables become observable and we call them ``observed variables'', where the realizations of counterfactual variables are equal to the realizations of observed variables. For simplicity, counterfactual variables are in hypothetical situations while observed variables are in reality.


We assume the counterfactual assignment, the counterfactual pre-exposure covariates and the counterfactual compliance behavior are identical in two counterfactual arms. Since the counterfactual assignment has two possible values that lead to two counterfactual arms including a counterfactual control arm and a counterfactual treatment arm, the exposure and the outcome have two potential outcomes from two counterfactual arms, while the other variables including the assignment, the covariates and the compliance behavior are assumed to have one potential outcome. For the assignment, the covariates and the compliance behavior, the realizations of the counterfactual variables are equal to the realizations of the observed variables. In the following notations, the subscript $i$ is hidden for simplicity, so all the equations are assumed to be in a vector or matrix form. To avoid ambiguity between counterfactual and observed variables, the notations mentioned above without any superscript are used to represent counterfactual variables, and we add a superscript ``$^{obs}$'' to these notations to represent observed variables.

The counterfactual assignment is denoted by $Z$, where $Z \in \{0,1\}$. $Z=0$ means a participant is assigned to the counterfactual control arm. We imagine this participant is assigned to the control arm, whether he is assigned to the control arm in reality or not. Similarly, $Z=1$ means a participant is assigned to the counterfactual treatment arm. In reality, each participant is assigned to one arm and we only have one observed assignment variable. The observed assignment is denoted by $Z^{obs}$. Actual observed assignment information in the PreventS trial data is a realization of the observed assignment. 

Two counterfactual exposures are denoted by $W(Z)$, where $Z \in \{0,1\}$. We use $W(Z) \in \{0,1,\ldots,J\}$. For simplicity, we write $W(0)=W(Z=0)$ and $W(1)=W(Z=1)$. $W(0)$ is the counterfactual exposure received by a participant if assigned to the counterfactual control arm. $W(1)$ is the counterfactual exposure received by a participant if assigned to the counterfactual treatment arm. Sometimes, we also name a counterfactual exposure as the counterfactual control intervention if it represents the control intervention or as the counterfactual treatment if it represents the treatment. This alternative name will be used if ambiguity is present. When we directly mention ``two counterfactual exposures'', this is to mean both the counterfactual control intervention and the counterfactual treatment. In addition, given our chosen domain for $W(Z)$, $W(0)=0$ and $W(1) > 0$ hold in the PreventS trial data. In reality, we only have one observed exposure variable and one observed value from this observed exposure variable for each participant. The observed exposure is denoted by $W^{obs}(Z^{obs})$. The realizations of the observed exposure are equal to the realizations from one of two counterfactual exposures. The realizations from the other counterfactual exposure are thus missing.


Two counterfactual outcomes are denoted by $Y(Z, W(Z))$, where $Z \in \{0,1\}$ and $W(Z) \in \{0,1,\ldots,J\}$. For simplicity, we write $Y(0, W(0)) = Y(Z=0, W(Z=0))$ and $Y(1, W(1)) = Y(Z=1, W(Z=1))$. $Y(0, W(0))$ is the counterfactual outcome from a participant if he is assigned to the counterfactual control arm. $Y(1, W(1))$ is the counterfactual outcome from a participant if he is assigned to the counterfactual treatment arm. In reality, we only have one observed outcome variable and one observed value from this observed outcome variable for each participant. The observed outcome is denoted by $W^{obs}(Z^{obs})$. The realizations of the observed outcome are equal to the realizations from one of two counterfactual outcomes. The realizations from the other counterfactual outcome are thus missing.

$M$ denotes counterfactual measured confounders. $M^{obs}$ denotes observed measured confounders. $M$ and $M^{obs}$ should be random matrices. They can contain multiple confounders, where each confounder contains $N$ random variables. Similarly, $U$ denotes counterfactual unmeasured confounders and $U^{obs}$ denotes observed unmeasured confounders. The observed unmeasured confounders just mean variables of unmeasured confounders in reality, but since they are unmeasured, they cannot be actually observed and thus their realizations are missing.

In a randomized trial, participants either comply or do not comply with the assigned exposure. This is associated with the latent compliance behavior of each participant. We can define four compliance behaviors: complier, never-taker, always-taker and defier \autocite{imbens_bayesian_1997}. The last three compliance behaviors are noncompliers. Noncompliance happens when there are noncompliers in a randomized trial. A complier always complies with the assigned exposure in either arm. A never-taker always takes the control intervention in either arm. An always-taker always takes the treatment in either arm. A defier always complies with the exposure in the other arm where he is not assigned. For each never-taker or each always-taker, two counterfactual exposures are equal and two counterfactual outcomes are equal. Usually, we assume there is no defier, which is related to the monotonicity assumption \ref{assump:mono} in subsection \ref{sec:assumptions}. The PreventS trial data meets this assumption, since participants in the control arm cannot take the treatment. In all the following content, we assume there is no defier. This assumption is not mandatory, as explained in the next section, but will simplify descriptions of our methods. 

Noncompliance can be one-sided or two-sided. One-sided noncompliance involves compliers and never-takers, where participants all take the control intervention in the control arm while some participants take the control intervention in the treatment arm. Two-sided noncompliance involves compliers, never-takers and always-takers, where some participants take the treatment in the control arm while some participants take the control intervention in the treatment arm. The counterfactual compliance behavior is denoted by $B_1$ if in the presence of one-sided noncompliance and $B_2$ if in the presence of two-sided noncompliance, defined as
\begin{equation}
B_1 =
    \begin{cases}
      co, & \text{if the participant is a complier,}\\
      nt, & \text{if the participant is a never-taker.}
    \end{cases}  
    \label{eqn:compliance-B1}
\end{equation}

\begin{equation}
B_2 =
    \begin{cases}
      co, & \text{if the participant is a complier,}\\
      nt, & \text{if the participant is a never-taker,} \\
      at, & \text{if the participant is an always-taker.}
    \end{cases}  
    \label{eqn:compliance-B2}
\end{equation}

$B_1^{obs}$ and $B_2^{obs}$ denote the observed compliance behavior. Table \ref{tab:compliance-one} shows how compliance behaviors are distributed in the presence of one-sided noncompliance, while table \ref{tab:compliance-two} shows how compliance behaviors are distributed in the presence of two-sided noncompliance. 

\begin{table}[htbp]
\centering
\begin{tabular}{ccr}
  \hline
$Z^{obs}$ & $W^{obs}$ & $B_1^{obs}$ \\ 
  \hline
\begin{minipage}[t]{0.2\textwidth} \centering
0
\end{minipage} & \begin{minipage}[t]{0.2\textwidth} \centering $W(0)$ \end{minipage}& \begin{minipage}[t]{0.2\textwidth}  \raggedleft $co$, $nt$ \end{minipage}\\ 
1 & $W(0)$ & $nt$ \\ 
1 & $W(1)$ & $co$ \\ 
   \hline
\end{tabular}
\caption{Compliance behaviors in the presence of one-sided noncompliance} 
\label{tab:compliance-one}
\end{table}

\begin{table}[htbp]
\centering
\begin{tabular}{ccr}
  \hline
$Z^{obs}$ & $W^{obs}$ & $B_2^{obs}$ \\ 
  \hline
\begin{minipage}[t]{0.2\textwidth} \centering
0
\end{minipage} & \begin{minipage}[t]{0.2\textwidth} \centering $W(0)$ \end{minipage}& \begin{minipage}[t]{0.2\textwidth}  \raggedleft $co$, $nt$ \end{minipage}\\ 
0 & $W(1)$ & $at$ \\ 
1 & $W(0)$ & $nt$ \\ 
1 & $W(1)$ & $co$, $at$ \\ 
   \hline
\end{tabular}
\caption{Compliance behaviors in the presence of two-sided noncompliance} 
\label{tab:compliance-two}
\end{table}

The relationships between observed variables and counterfactual variables are shown in equations \ref{eqn:wobs} and \ref{eqn:yobs}, based on the observed compliance behavior. Equality between observed variables and counterfactual variables means observed variables and counterfactual variables have the same realizations, which implies that observed variables and counterfactual variables are identically distributed in a broader sense.

\begin{equation}
W^{obs} =
    \begin{cases}
      W(Z = Z^{obs}), & \text{if $B_1^{obs} = co$ or $B_2^{obs} = co$,}\\
      W(0), & \text{if $B_1^{obs} = nt$ or $B_2^{obs} = nt$,} \\
      W(1), & \text{if $B_2^{obs} = at$.}
    \end{cases}  
    \label{eqn:wobs}
\end{equation}

\begin{equation}
Y^{obs} =
    \begin{cases}
      Y(Z = Z^{obs}, W(Z = Z^{obs})), & \text{if $B_1^{obs} = co$ or $B_2^{obs} = co$,}\\
      Y(0, W(0)), & \text{if $B_1^{obs} = nt$ or $B_2^{obs} = nt$,} \\
      Y(1, W(1)), & \text{if $B_2^{obs} = at$.}
    \end{cases}  
    \label{eqn:yobs}
\end{equation}

The PreventS trial data has one-sided noncompliance. The methods, analyses and discussions in our thesis all focus on one-sided noncompliance. They can be easily generalized to two-sided noncompliance, and where necessary, two-sided noncompliance will be mentioned for comparison with one-sided noncompliance directly in text.

\subsection{Overview of assumptions}
\label{sec:assumptions}


Assumptions are made to facilitate statistical analysis through our proposed Bayesian causal framework. Assumptions in the Bayesian causal framework come from but are usually weaker than assumptions in instrumental variable methods \autocite{angrist_identification_1996, yau_inference_2001, imbens_causal_2015}. Hence, we give an overview of assumptions, including almost all the assumptions in instrumental variable methods and additional assumptions that suit Bayesian causal inference, as a reference source. We also justify some assumptions for the PreventS trial data, in order to demonstrate more details rooted in our Bayesian models. 

Identification and estimation of local average treatment effect requires assumptions \ref{assump:sutv} to \ref{assump:mono} through instrumental variable methods.

\begin{assump}[Stable unit treatment value] 
\label{assump:sutv}
The subjects are independent of each other. Hence, their outcomes will not be affected by each other. 
\end{assump}

\begin{assump}[Randomization] 
\label{assump:random}
\seqsplit{$P(Z=1~|~\theta)=P(Z=1~|~W(0), W(1),Y(0, W(0)),Y(1, W(1)),\theta)$}, where $\theta$ are unknown parameters. Simply, \seqsplit{$Z \upvDash W(0), W(1),Y(0, W(0)),Y(1, W(1)) ~|~ \theta$}. The assignment is random, whether in the counterfactual scenario or in reality, which means unconfoundedness between the assignment and all counterfactual variables. If $X$ are pre-randomization covariates, we assume $Z \upvDash X$. Further, the assignment is conditionally random if $P(Z=1~|~X,\theta)=P(Z=1~|~W(0),W(1),Y(0, W(0)),Y(1, $ $W(1)), X,\theta)$, or simply \seqsplit{$Z \upvDash W(0),W(1),Y(0, W(0)), Y(1, W(1)) ~|~ X, \theta$}, where $X$ are pre-exposure but post-randomization covariates. Equations also hold if $Z$ is substituted with $Z^{obs}$.  
\end{assump}

\begin{assump}[Exclusion restriction for never-takers] 
\label{assump:exclu-nt}
For never-takers, $Y(0,0)=Y(1,0)$. 
\end{assump}

\begin{assump}[Exclusion restriction for always-takers] 
\label{assump:exclu-at}
For always-takers, $Y(0,W(1))=Y(1,W(1))$ where $W(1) \neq 0$.
\end{assump}

\begin{assump}[Stochastic exclusion restriction for never-takers or always-takers] 
\label{assump:exclu-nt-sto}
For never-takers or always-takers, $Z \upvDash Y(Z,W(Z))$. This means that in the superpopulation, the distributions of the counterfactual outcomes for never-takers or always-takers do not depend on the assignment. Weakly, for never-takers or always-takers, $Z \upvDash Y(Z,W(Z)) ~|~ X$, where $X$ are pre-exposure covariates.
\end{assump}

\begin{assump}[Exclusion restriction for compliers] 
\label{assump:exclu-co}
For compliers, $Y(0,W)=Y(1,W)$, where $W$ represents a fixed value.
\end{assump}

\begin{assump}[Monotonicity] 
\label{assump:mono}
$W(Z=1) \geq W(Z=0)$, for each compliance type. Alternatively, defiers do not exist.
\end{assump}

Assumption \ref{assump:sutv} is likely to be met in the PreventS trial data, because each participant undergoes their own exposure independently. In the treatment arm, each participant has a coach in each HWC session while this coach can have many participants at a time. Hence, there is a possibility that these participants have some correlation due to having the same coach. But each coach follows strict coaching rules and is not allowed to provide participants with direct advice on how to improve their health and wellness. In addition, coaching is a personalized process. A coach coaches a participant based on their own conditions and progress. Other participants' conditions do not affect the way the coach treats this participant. Hence, it is plausible to assume independence between participants.

Assumption \ref{assump:random} is met by design in the PreventS trial data because the PreventS trial is a randomized trial. Randomization is expected to produce exchangeability \autocite{hernan_causal_2020}. Exchangeability means people in either arm will have the same counterfactual variables if they are assigned to the alternative arm. In other words, the counterfactual variables of the exposure and the outcome are independent of the assignment.

Assumptions \ref{assump:exclu-nt} to \ref{assump:exclu-co} are the exclusion restriction assumption. Assumption \ref{assump:exclu-nt-sto} is a weaker version than assumption \ref{assump:exclu-nt}. 


Imbens and Rubin \autocite*{imbens_bayesian_1997} found that Bayesian modelling does not require exclusion restriction and monotonicity assumptions. Even if the two assumptions are violated, the distributions for compliers, never-takers, always-takers or defiers can be modelled separately in two arms. This indicates that the Bayesian causal framework can have different assumptions from instrumental variable methods.

Modelling assumptions, \ref{assump:compliance} to \ref{assump:mar}, are also necessary in our methods and analyses of the PreventS trial data. They validate our Bayesian models, but do not represent general assumptions in Bayesian causal inference. Modelling assumptions vary by data and model, and they can be relaxed for generalization or more complex situations. 

\begin{assump}[Time-independent compliance behavior] 
\label{assump:compliance}
The compliance behavior does not change over time. 
\end{assump}

Assumption \ref{assump:compliance} is used to avoid possible time-varying compliance behaviors due to noncompletion. Noncompletion of the treatment includes withdrawal and loss to follow up from the participants. A participant in the treatment arm who at first takes the treatment but withdraws from the trial later is a complier during the period under treatment but has unmeasured compliance behavior after withdrawal. We assume the participant is still a complier after their withdrawal and impute their missing data after withdrawal as if still a complier. This means each participant has only one compliance behavior during the entire trial, and any noncompletion status will not affect the compliance behavior. Our treatment of the compliance behavior is crude. In fact, the compliance behavior can change over time. It is very likely that the compliance behavior has changed when a participant leaves the trial or stops the treatment halfway. There should be some factors to explain the association between the compliance behavior and noncompletion. However, we do not investigate the issue of time-varying compliance behaviors further in this work.

\begin{assump}[Time-independent confounders] 
\label{assump:confounders}
Measured and unmeasured confounders used in Bayesian models do not change over time. They can be baseline or pre-exposure covariates.
\end{assump}

There are two types of confounders: time-independent and time-varying. Assumption \ref{assump:confounders} is used to ignore possible time-varying confounders in Bayesian models. This does not mean time-varying confounders do not exist, but simplifies the causal relationships between the variables. Methods to accommodate time-varying confounders need development but we do not focus on them. For one multivalued exposure, this assumption does not matter much, since the exposure level and the outcome level are measured at a single time point. For multiple multivalued exposures taken sequentially in the same trial, there can be several measurement time points over which some confounders can change. Due to the temporal order among multiple exposures, time-varying confounders may play an important role in estimation of the average treatment effect. Hence, this assumption may overlook extant time-varying confounders and cause bias in causal effect estimates. 

\begin{assump}[Independent confounders] 
\label{assump:idconfounders}
All the confounders are independent. This means, measured or unmeasured confounders are independent of each other, and measured confounders are independent of unmeasured confounders. 
\end{assump}

We assume that the confounders are independent through assumption \ref{assump:idconfounders} and we ignore possible association or interaction among the confounders to simplify the modelling process. This assumption can serve as a starting point, and when further relationships between the confounders are needed, we can then include them in Bayesian models.

\begin{assump}[MCAR and MAR] 
\label{assump:mar}
Missing data are missing completely at random (\seqsplit{MCAR}) or missing at random (MAR). For the exposure, the outcome and other endogenous variables, missing data are assumed to be MAR. For exogenous variables, missing data are assumed to be MCAR. 
\end{assump}

Assumption \ref{assump:mar} stipulates that missingnesss in the PreventS trial data is completely at random or at random. It is not related to any missing realization of counterfactual variables, but missing realizations of observed variables. Missing data is imputed in Bayesian models in the presence of all observed variables and causal relationships. Unlike predictive imputation that builds predictive Bayesian models to estimate missing values based on correlation or association, missing data in our methods and Bayesian models is imputed based on causal relationships among variables. We prefer causal imputation to predictive imputation, because causal imputation makes the entire causal framework more complete and convincing.

Causal imputation of missing data in exogenous observed variables may not be as precise as predictive imputation where data from endogenous observed variables related to these exogenous observed variables are used to impute missing data in these exogenous observed variables. For example, variability in imputations based on causality may be greater than that based on correlation. However, variability in imputation is expected because it represents uncertainty in missing data. Hence, we think causal imputation is plausible. Exogenous variables should be determined among causal relationships. In the PreventS trial data, we think some variables are exogenous because no other variables seem to affect them, such as baseline age, when we do not take other human-related bias into account. 

To sum up, these assumptions are not comprehensive, but some of them are needed in Bayesian causal inference and data analysis. Additional assumptions will be listed alongside the methodological descriptions or analysis results, because we think this can maintain the completeness of the methods or results.

\subsection{Causal framework}
\label{sec:framework-2}

Existing causal frameworks resolve the problem of noncompliance well \autocite{imbens_bayesian_1997}. We can use the well-established methods to adjust for noncompliance. On the other hand, unmeasured confounders can bias causal effect estimates. To obtain less biased average treatment effect estimates, a causal framework incorporating unmeasured confounding needs developing before data analysis is carried out.



A causal framework rests in part on assumptions and their plausibility. Assumptions make a causal framework viable under certain conditions. They should be plausible in the data context and they should be justified where applicable. However, some assumptions are untestable, which makes it hard to justify a causal framework completely, but this does not preclude usefulness of a causal framework. We should expect a causal framework to help us solve data-related questions to some extent. From the last section, two assumptions necessary to the causal framework for counterfactual variables and observed variables are: (1) stable unit treatment value, (2) randomization. These ``necessary'' assumptions can be relaxed, provided the causal framework is modified and analytical approaches are adjusted. These assumptions are called ``necessary'', because they represent a general situation and thus can fit in generalization. We will discuss how some assumptions can be removed in order to understand how the causal framework can be applied in different situations.

Our Bayesian causal framework is based on the potential outcome framework. So far we have not yet seen any research on causal relationships between counterfactual variables. We argue that causal relationships among counterfactual variables should also be investigated, in order to understand the role of unmeasured confounders in these causal relationships and create a causal framework that effectively adjusts for unmeasured confounding. Hence, causal graphs for both counterfactual variables and observed variables are drawn to describe methods. The causal relationships in causal graphs need some assumptions, such as assuming that there exists, or not, a causal path between some variables, when we do not have actual information or knowledge about this relationship.

\subsubsection{Simplest framework for counterfactual variables}
\label{sec:framework-3}


Innovatively, we propose investigating the causal relationships among counterfactual variables within and between two counterfactual arms, before investigating the causal relationships among observed variables, in order to understand how causal relationships can be used in Bayesian models and how unmeasured confounders affect estimation of the average treatment effect. We start with one multivalued exposure, and then we will explain how a Bayesian causal framework can be applied with multiple multivalued exposures. Bayesian causal frameworks are developed for a single outcome.

To facilitate understanding of the Bayesian causal framework when more methods are introduced, the denotation of the compliance behavior is now changed to $G$ from $B_1$ and $B_2$. $G$ represents the counterfactual variable while $G^{obs}$ represents the observed variable. Other symbols, $nt$, $co$ and $at$, are the same as in $B_1$ and $B_2$. 

For one-sided noncompliance,

\begin{eqnarray}
G &=& 
    \begin{cases}
      0, & \text{if $g = nt$} \\
      1, & \text{if $g = co$}  \nonumber 
    \end{cases} 
\label{equ:g-1-s}
\end{eqnarray}

For two-sided noncompliance,

\begin{eqnarray}
G &=& 
    \begin{cases}
      0, & \text{if $g = nt$} \\
      1, & \text{if $g \in \{co, at\}$}  \nonumber 
    \end{cases} 
\label{equ:g-2-s}
\end{eqnarray}

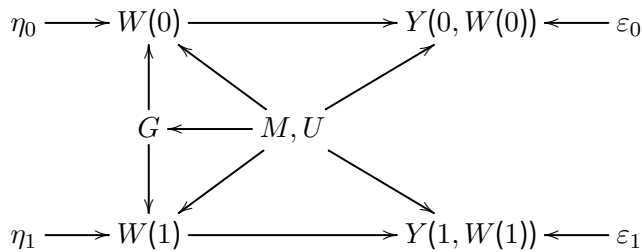
\begin{figure}[htbp]
    \centering
\xymatrix{
&&\eta_0 \ar[r]  & W(0) \ar[rr] && Y(0, W(0)) & \varepsilon_0 \ar[l]\\
&&& G \ar[u] \ar[d] & M, U \ar[l] \ar[dr] \ar[ur] \ar[ul] \ar[dl] &&  \\
&&\eta_1 \ar[r] & W(1) \ar[rr] && Y(1, W(1)) & \varepsilon_1 \ar[l]
}  
    \caption{Causal directed acyclic graph for counterfactual variables. $W(0)$ represents the counterfactual exposure in the counterfactual control arm. $W(1)$ represents the counterfactual exposure in the counterfactual treatment arm. $Y(0, W(0))$ represents the counterfactual outcome in the counterfactual control arm. $Y(1, W(1))$ represents the counterfactual outcome in the counterfactual treatment arm. $G$ is the counterfactual compliance behavior. $Z$ is the counterfactual assignment. $M$ represents counterfactual measured confounders and $U$ represents counterfactual unmeasured confounders. $\eta_0$, $\eta_1$, $\varepsilon_0$ and $\varepsilon_1$ are the error terms corresponding to counterfactual exposures and outcomes. The error terms of the other covariates are not shown, if any. $G$ is assumed to have no direct effect on the counterfactual outcomes, though this assumption can be relaxed later. 
    The effect of $W(0)$ on $Y(0, W(0))$ and the effect of $W(1)$ on $Y(1, W(1))$ are not affected by $G$, conditional on $M$ and $U$. It is implausible to assume $Y(0, W(0))$ and $Y(1, W(1))$ are independent without conditioning on $W, M, U$.}
    \label{fig:cdag-potential-noz}
\end{figure}

At first, we create two identical counterfactual scenarios, the counterfactual control arm and the counterfactual treatment arm. They have identical conditions, except the name. Identical conditions include identical exposure, identical outcome, identical compliance behavior, identical confounders and identical causal relationships. Figure \ref{fig:cdag-potential-noz} shows a simplest causal mechanism in two identical counterfactual arms. This mechanism is simplest partly because no interaction among counterfactual variables is considered and the compliance behavior does not affect the outcome directly. More complex mechanisms will be considered later when we talk about variations of the causal framework.

We do not constrain the exposure in either arm; for example, we do not fix the exposure in the counterfactual control arm to be the control intervention. Hence, the assignment has no effect on any variable and interacts with no variable. Under identical conditions, all the variables should be identical for the same person in two counterfactual arms. If there is any interaction between some counterfactual variables in one counterfactual arm, there should be the same interaction in the other counterfactual arm. This means when the variables that causally affect the exposure are fixed at the same levels, two counterfactual exposures are identical. Since time-varying exposures and confounders are not considered, assumed is a constant treatment effect on the counterfactual outcomes, constant confounding effects on the counterfactual exposures and the counterfactual outcomes, a constant effect of the compliance behavior on the counterfactual exposures, for all participants in a two-arm randomized trial.

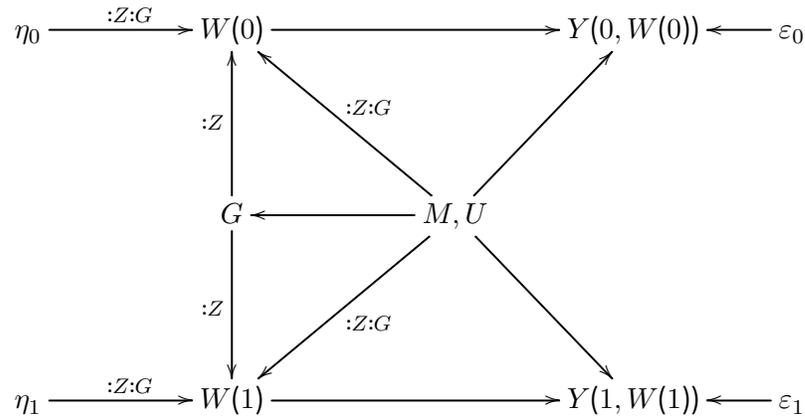
\begin{figure}[htbp]
    \centering
\xymatrix{
&&\eta_0 \ar^{:Z:G}[rr] & & W(0) \ar[rrr] &&& Y(0, W(0)) & \varepsilon_0 \ar[l]\\
&&&&  &  && & \\
&&&& G \ar^{:Z}[uu] \ar_{:Z}[dd] && M, U \ar[ll] \ar[ddr] \ar[uur] \ar_{:Z:G}[uull] \ar^{:Z:G}[ddll] &&  \\
&&&& &   &&&  \\
&&\eta_1 \ar^{:Z:G}[rr] &  & W(1) \ar[rrr] &&& Y(1, W(1)) & \varepsilon_1 \ar[l]
}  
    \caption{Causal directed acyclic graph for counterfactual variables. $W(0)$ represents the counterfactual exposure in the counterfactual control arm. $W(1)$ represents the counterfactual exposure in the counterfactual treatment arm. $Y(0, W(0))$ represents the counterfactual outcome in the counterfactual control arm. $Y(1, W(1))$ represents the counterfactual outcome in the counterfactual treatment arm. $G$ is the compliance behavior. $Z$ is the counterfactual assignment. $M$ represents counterfactual measured confounders and $U$ represents counterfactual unmeasured confounders. $\eta_0$, $\eta_1$, $\varepsilon_0$ and $\varepsilon_1$ are the error terms corresponding to counterfactual exposures and outcomes. The error terms of the other covariates are not shown, if any. $G$ is assumed to have no direct effect on the counterfactual outcomes, though this assumption can be relaxed later. 
    ``$:$'' is the interaction symbol used in \texttt{R} software. $:Z$ means the effect where it is located is interacted with $Z$. $:Z:G$ means the effect where it is located is interacted with both $Z$ and $G$. How the causal effects in this graph are defined, interacted and identified are described in the content.}
    \label{fig:cdag-potential-all}
\end{figure}

In the presence of one-sided noncompliance, $W(0)=0$ holds by design for all participants, representing the control intervention. In the presence of two-sided noncompliance, $W(0)=0$ holds by design for compliers. Hence, in the presence of either noncompliance, with any other covariate or confounder fixed, $W(0)$ is not always identical to $W(1)$. Then we want to explain how $W(0)$ can be different from $W(1)$ with the variables that causally affect the exposure fixed, through the causal relationships in figure \ref{fig:cdag-potential-noz}. One plausible explanation is that the effects of the counterfactual compliance behavior and the counterfactual confounders on the counterfactual exposures are different in two counterfactual arms. When all the variables except the assignment are considered in figure \ref{fig:cdag-potential-noz}, differences in the effects on the counterfactual exposures are very likely to be caused by the counterfactual assignment. We propose a causal mechanism where the counterfactual assignment modifies the effects of the counterfactual compliance behavior and the counterfactual confounders on the counterfactual exposures, rather than has a direct effect on the counterfactual exposures. Figure \ref{fig:cdag-potential-all} shows a simplest causal mechanism for our proposal. With inclusion of the counterfactual assignment, the causal mechanism becomes a causal framework that can be used in data analysis.

The counterfactual assignment does not modify any initial conditions related to the counterfactual outcomes. Hence, the treatment effect and the confounding effects on the counterfactual outcomes are still identical in two counterfactual arms. This is a strong assumption that the conditions for two counterfactual outcomes are identical, which guarantees two counterfactual outcomes are identical with the exposure and any other parent covariate fixed. This assumption is named as the identical condition assumption, shown in assumption \ref{assump:idcond}.

\begin{assump}[Identical condition] 
\label{assump:idcond}
$Y(0, w) = Y(1, w) ~|~ W(0) = W(1) = w, M = m, U = u$. Or, $P(Y(0, w) = Y(1, w) ~|~ W(0) = W(1) = w, M = m, U = u) = 1$, even if this probability may not be identified because in reality $W(0)$ and $W(1)$ may not have all possible values in common. The idea is that, for each participant, two counterfactual outcomes are identical, with any counterfactual parent variable fixed identically in two counterfactual arms. This implies that two counterfactual outcomes are identically distributed with the same exposure level and confounders.
\end{assump}

Assumption \ref{assump:idcond} corresponds to the exclusion assumptions in the last section. It is untestable, because two counterfactual outcomes cannot have counterparts in reality simultaneously and thus we do not have data related to two counterfactual outcomes simultaneously. We never know whether this assumption is plausible or not, but this assumption is necessary to establish a Bayesian causal framework. With no prior practical knowledge about this assumption, we prefer holding it.

So far we have explicitly used ``exposure'' to represent both the control intervention and the treatment of research interest and separated the treatment effect from the exposure effect and the control intervention effect. The treatment effect here is solely regarded as the causal effect of the treatment on the outcome. Hence, $W(0)$ and $W(1)$ themselves are neither the control intervention nor the treatment. They represent the same exposure. When their values are zero, we define their exposure status as the control intervention. When their values are non-zero, we define their exposure status as the treatment of research interest. If $W(0)$ is fixed at zero by design, the effect of $W(0)$ on $Y(0, W(0))$ is the control intervention effect. If we imagine $W(0)$ is the treatment whether it is fixed at zero by design or not, the effect of $W(0)$ on $Y(0, W(0))$ becomes the treatment effect that is equal to the treatment effect from $W(1)$ on $Y(1, W(1))$. In practice, the control intervention can be a placebo or standard care. A placebo may have no effect on the outcome, while the standard care may have a known effect on the outcome. For a placebo control, we can treat the control intervention as the natural zero level of the treatment and thus the control intervention is identically distributed as the treatment. For the standard care with known efficacy, we may have to redefine the treatment as the treatment level compared to the standard care and thus the treatment effect becomes the effect of the treatment of research interest with the standard care as the reference level. If we only use participants who take the treatment to build models, then we may treat the treatment effect as the effect of the treatment of research interest, with no reference level. Hence, the treatment effect needs a clear definition. This definition does not affect validity of the causal framework, but affects the way we interpret results in data analysis.



The causal framework should be easily applied in Bayesian modelling, which means we should have handy structural equations from the causal framework to turn into code. Structural equations are also more useful to find the estimands of interest mathematically. We assume the control intervention is not the natural zero level of the treatment. In two identical counterfactual arms, we define models for the counterfactual exposures with an additive part and a multiplicative term. The additive part is a linear combination of all the confounders that affect the counterfactual exposures. We assume the confounding effects are additive. The compliance behavior is used as the multiplicative term. We assume the effect of the compliance behavior is multiplicative. For both one-sided and two-sided noncompliance,
\begin{eqnarray}
W(0) &=& G ~ (\alpha_0 + \alpha_1 M + \alpha_2 U + \eta_0),  \nonumber \\
W(1) &=& G ~ (\alpha_0 + \alpha_1 M + \alpha_2 U + \eta_1),
\label{equ:mechanism-expomodel-id}
\end{eqnarray}
where $\alpha_0$ is the intercept, $\alpha_1$ and $\alpha_2$ are the coefficients for the confounders. $\eta_0$ and $\eta_1$ are two independent error terms. $E(\eta_0) = E(\eta_1) = 0$, $Var(\eta_0) = Var(\eta_1) = \sigma^2_{\eta}$. We can assume $\eta_0=\eta_1$ holds, so that $W(0) = W(1)$. We also assume $\eta_0$ and $\eta_1$ are Normally distributed. These model formula can distinguish compliers and/or always-takers from never-takers. When $G \neq 0$, the right-hand side in equations \ref{equ:mechanism-expomodel-id} will not be fixed at zero and the confounding effects may be estimated from equations \ref{equ:mechanism-expomodel-id}. In this case, we say that the confounding effects are identifiable.

The effect of the counterfactual assignment is introduced into the causal framework through an interaction function between the counterfactual compliance behavior $G$ and the counterfactual assignment $Z$, denoted by $h(Z, G)$. Additionally, this function can be viewed as the effect of $G$ modified by $Z$, the effect of $Z$ modified by $G$, or the interaction effect between $G$ and $Z$. For one-sided noncompliance,
\begin{eqnarray}
h(Z, G) &=& 
    \begin{cases}
      0, & \text{if $Z=0$ and $g \in \{nt, co\}$} \\
      0, & \text{if $Z=1$ and $g = nt$} \\
      1, & \text{if $Z=1$ and $g = co$}
    \end{cases} 
\label{equ:fzg-1-s}
\end{eqnarray}

For two-sided noncompliance,
\begin{eqnarray}
h(Z, G) &=& 
    \begin{cases}
      0, & \text{if $Z=0$ and $g \in \{nt, co\}$} \\
      1, & \text{if $Z=0$ and $g = at$} \\
      0, & \text{if $Z=1$ and $g = nt$} \\
      1, & \text{if $Z=1$ and $g \in \{co, at\}$}
    \end{cases} 
\label{equ:fzg-2-s}
\end{eqnarray}

We then define models for the counterfactual exposures in the causal framework by replacing $G$ in equation \ref{equ:mechanism-expomodel-id} with the interaction function $h(Z, G)$. For both one-sided and two-sided noncompliance,
\begin{eqnarray}
W(0) &=& h(Z=0, G) ~ (\alpha_0 + \alpha_1 M + \alpha_2 U + \eta_0), \nonumber \\
W(1) &=& h(Z=1, G) ~ (\alpha_0 + \alpha_1 M + \alpha_2 U + \eta_1),
\label{equ:mechanism-expomodel}
\end{eqnarray}
where all the coefficients and error terms have the same definitions as in equation \ref{equ:mechanism-expomodel-id}.

In addition to equations \ref{equ:mechanism-expomodel-id} and \ref{equ:mechanism-expomodel}, other structural equations for the counterfactual exposures may also be able to explain the causal relationships in causal graphs \ref{fig:cdag-potential-noz} and \ref{fig:cdag-potential-all}. Structural equations we proposed are just one plausible possibility, under our assumptions. Then we focus on explaining our structural equations.

$h(Z, G)$ represents an approach to incorporate noncompliance in the models for counterfactual exposures. It is also a mechanism about how we assign the ``control'' to compliers in the counterfactual control arm. Hence, these model formula not only can distinguish compliers and/or always-takers from never-takers, but also can distinguish the treatment from the contorl. More generally, we can define various interaction functions $h(Z, G)$ in different study designs to incorporate our understanding or assumptions in the causal framework. For example, we may define $h(Z=1, G) = 1$ to indicate that the counterfactual treatment arm only has compliers, or from another perspective, never-takers also take the treatment under certain conditions.

In equations \ref{equ:mechanism-expomodel}, the measured and unmeasured confounding effects and the error effect exist identically in two counterfactual arms, but these effects are modified by the interaction function $h(Z, G)$ differently in two counterfactual arms. To be more specific, the interaction function $h(Z, G)$ controls identifiability of all the effects on the counterfactual exposures. It allows these effects to appear under some conditions and disappear under other conditions, but it does not change these effects when it allows them to appear. When $h(Z, G) = 1$, the effects on the counterfactual exposures can be identified from $W(1)$, and also from $W(0)$ if in the presence of two-sided noncompliance. This may suggest that information about the treatment from always-takers is also useful. Similarly to equations \ref{equ:mechanism-expomodel-id}, when $h(Z, G) \neq 0$, we say that the confounding effects are identifiable in equations \ref{equ:mechanism-expomodel}. On the other hand, there is a strong assumption that when $W(0)$ and $W(1)$ both represent the treatment, they are identical with the confounders fixed. We name this assumption the identical conditional assumption for counterfactual exposures, shown in assumption \ref{assump:idcond-w}.
\begin{assump}[Identical condition] 
\label{assump:idcond-w}
$W(0) = W(1) = w ~|~ M = m, U = u$, for $w \geq 0$. Or $P(W(0) = W(1) = w ~|~ w \geq 0, M = m, U = u) = 1$. For each participant, when two counterfactual exposures both represent the treatment, they are identical, with any counterfactual parent variable except the counterfactual assignment fixed identically in two counterfactual arms. This implies that two counterfactual exposures are identically distributed with the same confounders when they both represent the treatment.
\end{assump}

Assumption \ref{assump:idcond-w} is also untestable, because two counterfactual exposures cannot be the treatment simultaneously except always-takers and two counterfactual exposures cannot have counterparts in reality simultaneously. Similarly to assumption \ref{assump:idcond}, we hold assumption \ref{assump:idcond-w} to establish a Bayesian causal framework. When the identical condition assumption is further mentioned, it represents both assumptions \ref{assump:idcond} and \ref{assump:idcond-w}.

There may also be various causal mechanisms that can explain the difference between $W(0)$ and $W(1)$ as the proposed causal framework does, for a given $h(Z, G)$. The proposed causal framework assumes when the confounding effects are identifiable, they are additive and linear. However, multiplicative and nonlinear confounding effects are also possible. Choices of the functional form of the confounders should be made in the data context.

In addition, we can show that $f(W(0)~|~G, M, U, \alpha) = f(W(0)~|~Z, G, M, U, \alpha)$, where $f$ is the conditional density function of $W(0)$ and $\alpha$ is a vector containing all the unknown parameters in the model of $W(0)$, because $f(W(0)~|~Z=0, G, M, U, \alpha)=f(W(0)~|~Z=1, G, M, U, \alpha)=f(W(0)~|~G, M, U, \alpha)$. Hence, $Z \upvDash W(0) ~|~G, M, U, \alpha$. Similarly, $Z \upvDash W(1)~|~G, M, U, \alpha$. Intuitively, the models of two counterfactual exposures have a fixed value of $Z$, indicating the counterfactual assignment is fixed for two counterfactual exposures. Hence, the counterfactual assignment does not affect the models of two counterfactual exposures. This means that equations \ref{equ:mechanism-expomodel} do not violate the randomization assumption for the counterfactual exposures.

Since the counterfactual assignment does not modify any initially identical conditions related to two counterfactual outcomes based on our assumptions, the models of two counterfactual outcomes in the causal framework should be identical. For both one-sided and two-sided noncompliance,
\begin{eqnarray}
Y(0, W(0)) &=& \beta_0 + \beta_1 W(0) + \beta_2 M + \beta_3 U + \varepsilon_0,  \nonumber \\
Y(1, W(1)) &=& \beta_0 + \beta_1 W(1) + \beta_2 M + \beta_3 U + \varepsilon_1,
\label{equ:mechanism-outcomes}
\end{eqnarray}
where $\beta_0$ is the intercept, $\beta_2$ and $\beta_3$ are the coefficients for the confounders, $\varepsilon_0$ and $\varepsilon_1$ are two independent error terms. $E(\varepsilon_0) = E(\varepsilon_1) = 0$, $Var(\varepsilon_0) = Var(\varepsilon_1) = \sigma^2_{\varepsilon}$. $\varepsilon_0=\varepsilon_1$ from our assumptions such that $Y(0, W(0)) = Y(1, W(1))$. We can but we do not relax the strict equality between two counterfactual outcomes by assuming $\varepsilon_0$ and $\varepsilon_1$ are independently identically distributed. Then we assume $\varepsilon_0$ and $\varepsilon_1$ are Normally distributed.

The meaning of $\beta_1$ needs clarification. Before clarification, we give a more clear explanation about the control intervention effect and the treatment effect from two counterfactual exposures. At first, in equations \ref{equ:mechanism-outcomes}, we assume the control intervention effect and the treatment effect is constant for all the participants. 
In fact, the treatment may not have a constant effect on all the participants. Some participants may find the treatment is extremely beneficial, while other participants may find the treatment is slightly better than the standard care they have taken before the treatment. Hence, the individual treatment effect can vary. When we assume the treatment effect is constant among all the participants, what we can estimate from equations \ref{equ:mechanism-outcomes} is the average treatment effect (ATE) among all the participants. Similarly, the control intervention effect should also be the average control effect among all the participants. We use ``average'' effect in the subsequent content for more precise descriptions.

In the presence of one-sided noncompliance, $W(0)$ does not have a treatment effect on $Y(0, W(0))$ in a randomized trial because $W(0)$ is not the treatment. However, this does not mean the average effect of $W(0)$ is fixed at zero. We can imagine in the counterfactual scenario, if $W(0)$ becomes the treatment, there should be an average effect of $W(0)$ on $Y(0, W(0))$. Fixing $W(0)$ at zero as the control intervention only makes the average effect of $W(0)$ non-identifiable by letting the average control effect of $W(0)$ appear, but does not eliminate the average treatment effect. In the presence of two-sided noncompliance, the average effect of $W(0)$ on $Y(0, W(0))$ exists in always-takers. Since never-takers and compliers both exist in the counterfactual treatment arm, $W(1)=0$ holds for never-takers, where $W(1)$ has the average control effect, while $W(1) \neq 0$ holds for compliers, where $W(1)$ has the average treatment effect. Hence, which effect a counterfactual exposure has depends on what intervention this counterfactual exposure is.


Some never-takers may be forced to take the treatment under certain conditions in a randomized trial and thus there may be a significant treatment effect in them. We ignore ethicality of the trial here. As mentioned before, we can adjust the definition of $h(Z, G)$ to incorporate this information in the models of the counterfactual exposures. Hence, for never-takers who take the treatment, $h(Z, G)=1$ holds regardless of the value of $Z$. Then, identifiability of the average treatment effect in these never-takers is opened by $h(Z, G)$. Through the control of $h(Z, G)$, we obtain a flexibility in determining the effect type of the counterfactual exposures.

Now we can clarify the meaning of $\beta_1$ intuitively. In the presence of one-sided noncompliance, $W(0)$ is the control intervention. If the control intervention has no efficacy, $W(0)$ does not affect $Y(0, W(0))$, but if the control intervention has efficacy, $W(0)$ has an average control effect on $Y(0, W(0))$. This means $W(0)$ has the average control effect which may not be constantly zero. But $W(0) = 0$ holds, whether the control intervention has efficacy or not. For a participant whose $W(0)$ is the control intervention and $W(1)$ is the treatment, the difference between $Y(0, W(0))$ and $Y(1, W(1))$ is caused by the difference between $W(0)$ and $W(1)$. When $W(0)$ has efficacy, the difference between $Y(0, W(0))$ and $Y(1, W(1))$ is counted from some efficacy level. Analogously to a factor in a regression model whose coefficients should be interpreted with regard to the reference level, the average treatment effect $\beta_1$ should be interpreted with the reference level that is the control intervention effect. This means, $\beta_1$ is the average treatment effect when the control intervention has no efficacy and is the average treatment effect compared to the control intervention when the control intervention has efficacy. The meaning of $\beta_1$ in the presence of two-sided noncompliance is identical to that in the presence of one-sided noncompliance, through similar comparisons. 

There may also be various causal mechanisms for the counterfactual outcomes, such as involving multiplicative and nonlinear effects of the confounders. Choices of the functional form is also flexible in the models for the counterfactual outcomes. Similarly, the randomization assumption holds for the counterfactual outcomes.

Finally, we focus on unmeasured confounders in the models for the counterfactual exposures and outcomes. There is a chance to estimate unmeasured confounders if they exist in the models for both the counterfactual exposures and outcomes. This condition is sufficient because of the definition that a confounder affects both the exposure and outcome. Identifiability of unmeasured confounders in the models for the counterfactual outcomes is always open, since the counterfactual outcomes are not fixed at a certain value by design. However, identifiability of unmeasured confounders in the models for the counterfactual exposures is not always open, but controlled by the interaction function $h(Z, G)$. 
Accurate estimates about $\beta_1$ can be obtained only if unmeasured confounders are first well estimated and then well adjusted for in the models for the counterfactual outcomes. 

\subsubsection{Local average treatment effect and average treatment effect}
\label{sec:framework-4}

For a multivalued exposure, what is the causal effect of a treatment on an outcome? Before estimating this treatment effect, we have to define it in our Bayesian causal framework. Below we show the process of how we develop our definition of the treatment effect for a multivalued exposure. 


From the perspective of Imbens and Angrist \autocite*{imbens_identification_1994}, local average treatment effect is ``the average treatment effect for individuals whose treatment status is influenced by changing an exogenous regressor that satisfies an exclusion restriction'' (p. 467). Imbens and Angrist \autocite*{imbens_identification_1994} used a binary exposure, whose value is 1 if the treatment is received by a participant and zero if the control intervention is received by a participant. We still denote the exposure by $W$. Local average treatment effect can then be expressed mathematically as $E(Y(1, W(1)) - Y(0, W(0)) ~|~ W(1) \neq W(0))$ \autocite{imbens_identification_1994}. 

When Imbens and Rubin \autocite*{imbens_bayesian_1997} extended identification and estimation of local average treatment effect from instrumental variable methods to Bayesian methods, they gave a more clear explanation on the meaning of local average treatment effect with regard to the mathematical formula. Two local average treatment effects were proposed, complier average causal effect (CACE) and defier average causal effect (DACE) \autocite{imbens_bayesian_1997}. Complier average causal effect is equal to $E(Y(1, W(1)) - Y(0, W(0)) ~|~ W(1) - W(0)=1)$, while defier average causal effect is equal to $-E(Y(1, W(1)) - Y(0, W(0)) ~|~ W(1) - W(0)=-1)$. For compliers and defiers, one counterfactual exposure is the treatment, while the other counterfactual exposure is the control intervention. Hence, the unit-level treatment effect can always be obtained from $Y(1, W(1)) - Y(0, W(0))$ and the average treatment effect can be obtained by averaging all the unit-level treatment effects. If two counterfactual exposures are identical, $Y(1, W(1)) - Y(0, W(0))$, if non-zero, should not be the treatment effect at the unit level. Usually, the monotonicity assumption holds in a randomized trial, which means defiers do not exist. Hence, local average treatment effect is sometimes regarded as complier average causal effect.

Instrumental variable methods \autocite{imbens_identification_1994} and Bayesian methods \autocite{imbens_bayesian_1997} were proposed to estimate local average treatment effect. Initially, they were developed from a binary exposure. Angrist and Imbens \autocite*{angrist_two-stage_1995} started with two-stage least squares (TSLS), a widely used instrumental variable method, and figured out local average treatment effect for a multivalued exposure that can take at least three possible values, such as ``drug dosage'' (p. 431) and ``years of schooling'' (p. 431). We want to develop Bayesian methods to estimate average effects for one or multiple multivalued exposures, but the definition of local average treatment effect from this work \autocite{angrist_two-stage_1995} cannot be directly applied to our methods, because (1) the two-stage least squares method to estimate local average treatment effect for one multivalued exposure \autocite{angrist_two-stage_1995} cannot be easily extended to multiple multivalued exposures, and (2) instrumental variable methods can fail when there are causal relationships among multiple exposures.

However, the two-stage least squares method to estimate local average treatment effect of one multivalued exposure \autocite{angrist_two-stage_1995} shows an interesting idea that local average treatment effect is estimated from participants whose exposure status can be affected by the random assignment, either significantly or slightly. Hence, for each individual, $W(1)-W(0) \geq 1$ holds for a multivalued exposure. There may be a local average treatment effect like $E(Y(1, W(1)) - Y(0, W(0)) ~|~ W(1) - W(0)=1)$, and there may be another local average treatment effect like $E(Y(1, W(1)) - Y(0, W(0)) ~|~ W(1) - W(0)=2)$. If the domain of $W$ is $\{0,1,2,\ldots,J\}$ where $J \geq 2$, then $W(1)-W(0) \in \{1, \ldots, J\}$ and local average treatment effects include $E(Y(1, W(1)) - Y(0, W(0)) ~|~ W(1) - W(0)=j)$, for $j \in \{1, \ldots, J\}$. Angrist and Imbens \autocite*{angrist_two-stage_1995} called local average treatment effect of a multivalued exposure from two-stage least squares ``average causal response'' (p. 435), because the result from two-stage least squares is a weighted average among all local average treatment effects $E(Y(1, W(1)) - Y(0, W(0)) ~|~ W(1) - W(0)=j)$ for $j \in {1, \ldots, J}$. Then we think, is it possible to develop a regression model like $E(Y(1, W(1)) - Y(0, W(0)) ~|~ W(1) - W(0))$ such that we can directly estimate the average treatment effect from a regressor coefficient, rather than use a weighted average? 

Wong \autocite*{wong_calculus_2021, wong_integral_2021} presented how to estimate the causal effect of an exposure on an outcome through instrumental variables in various cases where binary, multidimensional, discrete and continuous exposures and instruments were considered. Wong's approach \autocite{wong_calculus_2021, wong_integral_2021} focused on nonlinear, nonparametric instrumental variable methods, 
but our primary interest in Wong's work \autocite{wong_calculus_2021, wong_integral_2021} is a definition of the causal effect that ``the causal effect of $X$ on $Y$ at $X=x$ is $\theta(x)=E(\frac{\partial Y}{\partial x})$'' (p. 6) given that ``$Y = f(X, U)$'' (p. 6) and ``$f$ is a smooth function of $x$'' (p. 6). Hence, ``the causal effect of $X$ on $Y$ is the expected change of $Y$ induced by changing $X$, \ldots, while keeping $U$ fixed'' (Wong, 2021, p. 3). $X$ is continuous and multidimensional \autocite{wong_calculus_2021}, but we can extend this definition in discrete, one-dimentional, multivalued exposures. If we can find a structure like $E(Y(1, W(1)) - Y(0, W(0)) ~|~ W(1) - W(0)) = \beta~(W(1) - W(0))$, then $\beta$ should be the causal effect of $W(1) - W(0)$ on $Y(1, W(1)) - Y(0, W(0))$, because
\begin{eqnarray}
\beta_1 &=& \frac{\partial E(Y(1, W(1))-Y(0, W(0)))}{\partial (W(1) - W(0))} \nonumber \\
 &=& \frac{\partial \int{(y(1, w(1))-y(0, w(0))) f(y(1, w(1)), y(0, w(0))) }\dif{y(1, w(1))}\dif{y(0, w(0))}}{\partial (W(1) - W(0))} \nonumber \\
 &=& \int {\frac{\partial (y(1, w(1))-y(0, w(0)))}{\partial (W(1) - W(0))} f(y(1, w(1)), y(0, w(0)))} \dif{y(1, w(1))}\dif{y(0, w(0))} \nonumber \\
 &=& E\left(\frac{\partial (Y(1, W(1))-Y(0, W(0)))}{\partial (W(1) - W(0))}\right ),
\label{equ:mechanism-beta}
\end{eqnarray}
when some regularity conditions are met and we assume so. Inspired by prior works \autocite{angrist_two-stage_1995, wong_calculus_2021}, we further investigate how to define the average treatment effect from structural equations developed in our Bayesian causal framework.

At first, we start with $\beta_1$. Through subtraction between the models for two counterfactual outcomes in equations \ref{equ:mechanism-outcomes}, we can obtain two equations where $W$ is a multivalued exposure, as
\begin{eqnarray}
Y(1, W(1))-Y(0, W(0)) &=& \beta_1 (W(1) - W(0)) + (\varepsilon_1 - \varepsilon_0) \label{equ:mechanism-late-reg} \\
E(Y(1, W(1))-Y(0, W(0))~|~W(1) - W(0)) &=& \beta_1 (W(1) - W(0)) 
\label{equ:mechanism-late-mean}
\end{eqnarray}

Equation \ref{equ:mechanism-late-reg} is a regression model, because $\varepsilon_1$ and $\varepsilon_0$ are independent of $W(1)$ and $W(0)$, thus making $\varepsilon_1 - \varepsilon_0$ independent of $W(1) - W(0)$. $\beta_1$ is the average treatment effect that we want to estimate, because 
\begin{eqnarray}
\beta_1 = E\left(\frac{\partial (Y(1, W(1))-Y(0, W(0)))}{\partial (W(1) - W(0))}\right ) \label{eqn:late-math-def}
\end{eqnarray}

We mentioned in the former sections that even under the assumption that the treatment effect is constant, the individual treatment effect can actually vary when we want to estimate an average treatment effect. This point can be understood through equation \ref{eqn:late-math-def}: the average treatment effect $\beta_1$ is an average over all the individual treatment effects, whose estimation is not affected by differences in the individual treatment effects.

Equation \ref{equ:mechanism-late-mean} is the mean model of equation \ref{equ:mechanism-late-reg}, from which $\beta_1$ is the causal effect of $W(1) - W(0)$ on $Y(1, W(1))-Y(0, W(0))$, as discussed before. Through equation \ref{equ:mechanism-late-mean}, we can use a regression perspective to interpret the meaning of $\beta_1$ as the average difference in $Y(1, W(1))-Y(0, W(0))$ per one unit change in $(W(1) - W(0))$. 

Equation \ref{equ:mechanism-late-mean} can help us understand why the control intervention effect affects the meaning of $\beta_1$ through compliers whose $W(0)$ is the control intervention and $W(1)$ is the treatment. When the control intervention has no efficacy, then $W(0)=0$ purely means there is no treatment of interest and thus $W(1) - W(0) = W(1)$, where $W(1)$ is the treatment of research interest. In this case, $\beta_1$ is simply the average treatment effect. But when the control intervention has known efficacy, for example, the control intervention is as half effective as the treatment, then $W(0)=0$ does not mean there is no treatment of interest but means the reference level is set at some known efficacy. $W(1)$ is not the treatment of research interest, but now becomes the treatment taken above the control intervention. $E(Y(1, W(1))-Y(0, W(0))~|~W(1) - W(0))$ becomes the average difference in the outcome from the reference level. In this case, $\beta_1$ is the average treatment effect, compared to the control intervention. $\beta_1>0$ may mean the treatment of research interest is better than the control intervention, but may not mean the treatment of research interest is actually effective. The change in the meaning of $\beta_1$ in fact is resulted from the change in the definition of the treatment used in the causal framework. We suggest fixing the definition of the treatment used in the causal framework before data analysis based on the control intervention and the treatment of research interest, rather than changing the definition of the treatment during any analysis. 

$\beta_1$ is identifiable from equation \ref{equ:mechanism-late-mean} only if $(W(1) - W(0)) \neq 0$. In the presence of one-sided and two-sided noncompliance, $W(0) = W(1)$ holds in never-takers and/or always-takers, while $(W(1) - W(0)) \neq 0$ only holds in compliers whose $W(1)$ is the treatment. Hence, $\beta_1$ is identifiable from compliers through equation \ref{equ:mechanism-late-mean}. However, that $\beta_1$ is non-identifiable from never-takers and always-takers through equation \ref{equ:mechanism-late-mean} does not mean that $\beta_1$ is solely the average treatment effect in compliers. $\beta_1$ is defined as the average treatment effect among all the participants based on our former assumptions. For never-takers, if they take the treatment, their average treatment effect is also assumed to be $\beta_1$. For always-takers who always take the treatment, both of their counterfactual exposures have the average treatment effect as $\beta_1$. 

Equation \ref{equ:mechanism-late-mean} is the conventional definition of the average causal effect, applicable to both binary and multivalued exposures. On the other hand, we recognize difficulty in applying this equation in Bayesian modelling, because (1) two counterfactual exposures and outcomes do not have counterparts in reality simultaneously, and thus missing realizations of one counterfactual exposure and one counterfactual outcome need imputation \autocite{imbens_bayesian_1997}, which can involve a large amount of unknown parameters, and (2) the unknown compliance behavior needs imputation to find out compliers who take the control intervention, which involves more computational complexity. Then we think, can we use simpler equations or models to estimate $\beta_1$? We find, in equations \ref{equ:mechanism-outcomes}, $\beta_1$ is the coefficient of the counterfactual exposure in both the model for $Y(0, W(0))$ and the model for $Y(1, W(1))$. If our interest is to estimate $\beta_1$, we can also use the models for two counterfactual outcomes independently through equations \ref{equ:mechanism-outcomes}, without subtraction between the two models. This means that for each participant, we can use either the model for $Y(0, W(0))$ or the model for $Y(1, W(1))$, rather than both models in equation \ref{equ:mechanism-late-mean}.

In the presence of one-sided noncompliance, $W(0)=0$ and $\beta_1$ disappears in the model for $Y(0, W(0))$. Hence, the model for $Y(0, W(0))$ is not helpful in estimation of $\beta_1$ but possibly in estimation of the other coefficients. $\beta_1$ also disappears in the model for $Y(1, W(1))$ in never-takers. Hence, $\beta_1$ is identifiable from the model for $Y(1, W(1))$ in compliers in the treatment arm. Identifiability of $\beta_1$ from the model for $Y(1, W(1))$ in compliers in the treatment arm requires that $W(1)$ takes at least two possible values except the control intervention, which means $\beta_1$ is estimated in this model by comparing different levels of the treatment. This also means our estimation approach of the average treatment effect is not applicable to binary exposures.

The way we use the model for $Y(1, W(1))$ in compliers in the treatment arm is that we directly build this model as a regression model. For the other participants, their models for either $Y(0, W(0))$ or $Y(1, W(1))$ can be used to estimate the other coefficients except $\beta_1$, because based on the causal framework, these coefficients except $\beta_1$ are identical for all the participants. This may improve estimation accuracy of the coefficients except $\beta_1$ compared to only using compliers in the treatment arm, which may in turn improve estimation accuracy of $\beta_1$. Hence, we think the models from the other participants except compliers in the treatment arm are still useful, alongside with a condition that we have to find a way to estimate unmeasured confounders $U$ in these models. 

In the presence of two-sided noncompliance, $\beta_1$ is identifiable from two sources, only if the treatment used in the causal framework can take at least two possible values except the control intervention. The definition of the treatment is identical to that mentioned above. One source is the model for $Y(1, W(1))$ in compliers in the treatment arm, similarly to the case in the presence of one-sided noncompliance. The other source is the models for two counterfactual outcomes in always-takers. For always-takers, $\beta_1$ can be estimated from either the model for $Y(0, W(0))$ or the model for $Y(1, W(1))$, because $W(0)$ and $W(1)$ are both the treatment in always-takers. The models of two counterfactual outcomes for never-takers and the model for $Y(0, W(0))$ in compliers in the control arm can be used to estimate the other coefficients except $\beta_1$. $\beta_1$ is identifiable from both compliers and always-takers, but it is not defined as the average treatment effect in compliers and always-takers. From our causal framework, it is still defined as the average treatment effect among all the participants. 

Since only one of two counterfactual exposures and one of two counterfactual outcomes have corresponding observed variables in reality, we transform either the model for $Y(1, W(1))$ or the model for $Y(0, W(0))$ into the models for observed outcomes, in order to estimate $\beta_1$, rather than conduct subtraction between the models for $Y(1, W(1))$ and $Y(0, W(0))$. We compare our approach with the Bayesian method proposed by Imbens and Rubin \autocite{imbens_bayesian_1997}. The practical Bayesian modelling process for the Bayesian method proposed by Imbens and Rubin \autocite{imbens_bayesian_1997} involves four steps: (1) figure out which counterfactual exposure/outcome the observed exposures/outcomes correspond to, (2) build models for observed exposures/outcomes and obtain the models for two counterfactual exposures/outcomes from the corresponding observed exposures/outcomes, (3) impute missing realizations of one counterfactual exposure and one counterfactual outcome using the model for this counterfactual exposure/outcome from step (2), and (4) after obtaining two counterfactual exposures and outcomes for each participant, use equation \ref{equ:mechanism-late-mean} to estimate $\beta_1$. Our approach involves three steps: (1) figure out which counterfactual exposure/outcome the observed exposures/outcomes correspond to, (2) build models for observed exposures/outcomes and obtain the models for two counterfactual exposures/outcomes from the corresponding observed exposures/outcomes, and (3) use equations \ref{equ:mechanism-outcomes} to estimate $\beta_1$. Hence, our approach can greatly simplify the Bayesian modelling process by avoiding imputation on missing realizations of one counterfactual exposure and one counterfactual outcome. 

It is plausible to imagine that, if we assume the average treatment effect is different between compliers and never-takers, or more understandably we assume the average treatment effect is different between compliers and always-takers, there can be an average treatment effect in compliers and there can also be an average treatment effect in always-takers or never-takers. These treatment effects are local average treatment effects in original sense, because they come from particular subpopulations. Instead of estimating the average treatment effect for all the participants as $\beta_1$, we may want to estimate the average treatment effect in one particular subpopulation. This is related to subgroup analysis. And this means that in one causal framework, we may obtain various average treatment effects, or more specifically various local average treatment effects. 

Our causal framework mentioned above is the simplest, and we also mentioned there can be many variations. Methods and ideas can be transferred to more complex causal frameworks. A causal framework should be established from our knowledge and assumptions about causal relationships in the data context. Different data involves different causal relationships among variables. Hence, a causal framework should change with data, just like a regression model changes with data. It is also plausible to imagine that, if we change the causal framework and thus structural equations change, $\beta_1$ may not be simply defined as the average treatment effect for all the participants but has some more complex meaning based on the causal relationships. This means that we may obtain different average treatment effects from different causal frameworks. However, we can try to estimate the average treatment effect for all the participants in every causal framework.

Hence, we do not intend to define local average treatment effect as from some subpopulations. And we do not deliberately focus on some subpopulations to estimate the causal effect of a treatment on an outcome. Instead, we focus on the average treatment effect among all the participants and how to estimate this average treatment effect from the causal framework in use. Further, we only use ``average treatment effect'' in subsequent descriptions and analyses. ``Local average treatment effect'' is still an interesting name. We think, ``local '' may mean being in a specific causal framework with certain assumptions, though this does not affect data analysis any more. 

Imbens \autocite*{imbens_instrumental_2014} mentioned confusion about local average treatment effect from many researchers. Pearl wondered why the average treatment effect from some subpopulation is needed (Imbens, 2014, p. 342). We think Pearl's question is interesting: is local average treatment effect needed because of convenience, necessity or research interest \autocite{imbens_instrumental_2014}? The answer to this question is straightforward. Local average treatment effect is needed because we want to know the causal effect of a treatment on an outcome as precisely as possible. From Imbens' works \autocite{imbens_identification_1994, imbens_bayesian_1997}, when we want to know the causal effect of a treatment on an outcome, this causal treatment effect is ``credibly'' (Imbens, 2014, p. 342) identified from compliers. Hence, local average treatment effect is simply a type of causal effect.

Our goal is to estimate the average causal effect of a treatment and the average causal effects of three sequential treatments on an outcome in the PreventS trial data. We developed a Bayesian causal framework and structural equations to understand the average treatment effect in the causal relationships among variables. With the help of a Bayesian causal framework, we do not have to restrict the average treatment effect in compliers. We think, average treatment effect may be easier to understand than local average treatment effect, because it does not focus on any subpopulation and can distinguish itself from subgroup analysis. One possible controversy is the untestable assumption that the average treatment effect for never-takers exists and can be equal to the average treatment effect for compliers. Whether we assume the average treatment effect for never-takers is equal to the average treatment effect for compliers or not, the average treatment effect is always non-identifiable from never-takers. The primary function of the models for counterfactual outcomes from never-takers is to estimate the coefficients except the average treatment effect. Hence, the causal relationships between the confounders and the outcome are more important than the causal relationship between the exposure and the outcome in the models for counterfactual outcomes from never-takers. We should instead focus on the assumptions related to the causal relationships between the confounders and the outcome, when we use the models for counterfactual outcomes from never-takers. 




From the above descriptions, we have three issues remaining: (1) What are more complex Bayesian causal frameworks and how can the assumptions change? Will the average treatment effect be affected? (2) How are the models for observed variables built and connected to the models for counterfactual variables? (3) How can unmeasured confounders be estimated in a Bayesian causal framework?  These issues will be treated in the following sections.

\subsubsection{Complex frameworks for counterfactual variables}
\label{sec:framework-5}

The simplest framework for counterfactual variables can be extended to more complex settings. This is analogous to adding more complex associations or functional forms in regression models, such as additional covariates, interaction and polynomials. Complex frameworks for counterfactual variables discussed in this section mainly focus on more complex causal relationships among counterfactual variables instead of functional forms. Different functional forms can be considered in both the simplest and complex frameworks.

Three necessary assumptions for complex frameworks are identical to those for the simplest framework. They are stable unit treatment value, randomization, identical condition. Other assumptions related to causal relationships among counterfactual variables vary so that different complex framework are born. We name a complex framework ``variation'' from the simplest framework, for simplicity. Causal graphs are not drawn for complex frameworks, but structural equations are given instead. Structural equations should be interpreted as in the simplest framework, where each coefficient has a meaning related to a variable. We do not elaborate on parameter definitions in complex frameworks, unless confusion occurs.


The purpose of variations is to suit different data. The simplest framework may be suitable for some data, but not for all types of data, because causal relationships among variables differ among various data. 
We present some but not all variations to show how a causal framework can change from the simplest one. The first variation is that the counterfactual compliance behavior is assumed to have a direct additive effect on the counterfactual outcomes. The structural equations for the counterfactual outcomes now become:
\begin{eqnarray}
Y(0, W(0)) &=& \beta_0 + \beta_1 W(0) + \beta_2 M + \beta_3 U + \beta_4 G + \varepsilon_0,  \label{equ:mechanism-variation-1-0} \\
Y(1, W(1)) &=& \beta_0 + \beta_1 W(1) + \beta_2 M + \beta_3 U + \beta_4 G + \varepsilon_1.
\label{equ:mechanism-variation-1-1}
\end{eqnarray}

The counterfactual compliance behavior does not affect the treatment effect, but without adjusting for the counterfactual compliance behavior as an additional confounder in the models for counterfactual outcomes, the average treatment effect may be estimated biasedly. 

The second variation is that the counterfactual compliance behavior is assumed to modify the treatment effect on the counterfactual outcomes. The structural equations for the counterfactual outcomes now become:
\begin{eqnarray}
Y(0, W(0)) &=& \beta_0 + \beta_1 W(0) + \beta_2 M + \beta_3 U + \beta_4 W(0)\cdot G + \varepsilon_0,  \label{equ:mechanism-variation-2-0} \\
Y(1, W(1)) &=& \beta_0 + \beta_1 W(1) + \beta_2 M + \beta_3 U + \beta_4 W(1)\cdot G + \varepsilon_1,
\label{equ:mechanism-variation-2-1}
\end{eqnarray}
which means the compliance behavior is interacted with the counterfactual exposures. In the presence of one-sided noncompliance, since the treatment effect from never-takers is always non-identifiable, the effect modification from the compliance behavior on the counterfactual exposures does not make a difference. In the presence of two-sided noncompliance, if we separate compliers from always-takers in the definition of $G$, then the effect modification from the compliance behavior on the counterfactual exposures is related to subgroups analysis and we may obtain the average treatment effect separately for compliers and always-takers.

A more complex variation is that the counterfactual compliance behavior is assumed to modify all the effects except the error effect on the counterfactual outcomes. The structural equations for the counterfactual outcomes now become:
\begin{eqnarray}
Y(0, W(0)) &=& \beta_0 + \beta_1 W(0) + \beta_2 M + \beta_3 U + G~(\beta_4 + \beta_5 W(0) + \beta_6 M + \beta_7 U) \nonumber \\
&& + ~ \varepsilon_0, \label{equ:mechanism-variation-3-0} \\
Y(1, W(1)) &=& \beta_0 + \beta_1 W(1) + \beta_2 M + \beta_3 U + G~(\beta_4 + \beta_5 W(1) + \beta_6 M + \beta_7 U) \nonumber \\
&&  +~ \varepsilon_1, \label{equ:mechanism-variation-3-1}
\end{eqnarray}
which indicates the treatment and confounding effects on the counterfactual outcomes differ by the compliance behavior. To adjust for the compliance behavior, the models of the counterfactual outcomes can be built separately for compliers, never-takers, and/or always-takers. If the interaction between the compliance behavior and the confounders is not adjusted for, there may be residual confounding.

Another variation is that the counterfactual confounders are assumed to modify the treatment effect. The structural equations for the counterfactual outcomes now become:
\begin{eqnarray}
Y(0, W(0)) &=& \beta_0 + \beta_1 W(0) + \beta_2 M + \beta_3 U + W(0)~(\beta_4 M + \beta_5 U) +  \varepsilon_0, \label{equ:mechanism-variation-4-0} \\
Y(1, W(1)) &=& \beta_0 + \beta_1 W(1) + \beta_2 M + \beta_3 U + W(1)~(\beta_4 M + \beta_5 U) + \varepsilon_1. \label{equ:mechanism-variation-4-1}
\end{eqnarray}

Essentially, any modification on the treatment effect is related to subgroup analysis. We can exclude the interaction related to the counterfactual exposures out of the models for the counterfactual outcomes, in order to estimate the average treatment effect among all the participants. This can be accomplished by using the simplest framework instead of this variation.


Multiplicative and nonlinear effects are not mentioned in variations but can be considered in a causal framework similarly to additive and linear effects. The fundamental idea is that structural equations are created from assumed causal relationships among counterfactual variables. We can even assume one covariate affects only one counterfactual outcome but not the other. However, we have to think about plausibility of causal relationships among counterfactual variables when we make assumptions.


\subsubsection{Frameworks for observed variables}
\label{sec:framework-6}

The counterfactual exposure or outcome and its counterpart in reality as an observed variable should have some relationship that can be expressed mathematically. Through this mathematical relationship, their models can be connected. Hence, we can construct the model for the observed variable from the model for the counterfactual variable and obtain parameter estimates of the model for the counterfactual variable through building the model for the observed variable. Before understanding how the models for counterfactual variables and the models for observed variables are connected, we first investigate causal relationships among observed variables and try to compare these causal relationships with causal relationships among counterfactual variables.

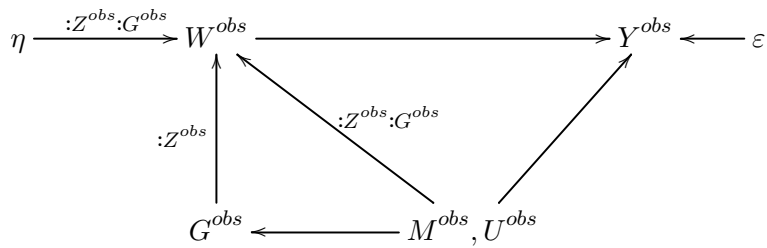
\begin{figure}[htbp]
    \centering
\xymatrix{
&& \eta \ar^{:Z^{obs}:G^{obs}}[rr] &  & W^{obs} \ar[rrr] &&& Y^{obs} & \varepsilon \ar[l]\\
&&&& & && & \\
&&&& G^{obs} \ar^{:Z^{obs}}[uu] && M^{obs}, U^{obs} \ar[ll] \ar[uur] \ar_{:Z^{obs}:G^{obs}}[uull] && 
}  
    \caption{Causal directed acyclic graph for observed variables.}
    \label{fig:cdag-observed-all}
\end{figure}

In the counterfactual scenario, each participant has two counterfactual exposures and outcomes with regard to two counterfactual arms. But in reality, each participant is only assigned to one arm. Hence, all observed variables are associated with a single arm. Figure \ref{fig:cdag-observed-all} shows a simplest causal mechanism among observed variables. Causal relationships among observed variables are similar to those among counterfactual variables shown in figure \ref{fig:cdag-potential-all}. Figure \ref{fig:cdag-observed-all} looks like one arm in figure \ref{fig:cdag-potential-all}.

As mentioned above, the counterfactual assignment, compliance behavior, confounders are equal to the observed assignment, compliance behavior, confounders in the sense that their realizations and distributions are identical. Simply, $Z = Z^{obs}$, $G = G^{obs}$, $M = M^{obs}$, $U = U^{obs}$. We regard $W(Z)$ and $Y(Z, W(Z))$ as functions of $Z$. Then the relationship between counterfactual exposures/outcomes and observed exposure/outcome can be expressed mathematically, where structural equations for the observed exposure and outcome can be easily obtained.

In the presence of one-sided or two-sided noncompliance, the relationship between the counterfactual exposures and the observed exposure and the structural equation for the observed exposure are

\begin{eqnarray}
W^{obs} &=& W(Z = Z^{obs}) \nonumber \\
&=& h(Z=Z^{obs}, G^{obs}) ~ (\alpha_0 + \alpha_1 M^{obs} + \alpha_2 U^{obs} + \eta_1)  
\label{equ:structural-wobs}
\end{eqnarray}

Similarly, the relationship between the counterfactual outcomes and the observed outcome and the structural equation for the observed outcome are

\begin{eqnarray}
Y^{obs} &=& Y(Z = Z^{obs}, W(Z) = W^{obs}) 
\label{equ:structural-yobs}
\end{eqnarray}


The Bayesian modelling process in RStan \autocite{stan_development_team_rstan_2022} involves three steps: (1) choose a suitable causal framework and construct models for counterfactual exposures and outcomes, (2) establish relationships between counterfactual variables and observed variables and convert the models for counterfactual exposures and outcomes into the models for the observed exposure and outcome in which the parameter for the average treatment effect is transferred, (3) build the models for the observed exposure and outcome and obtain parameter estimates including the estimate of the average treatment effect. To understand why this modelling process works, we use a theoretical perspective. Bayesian inference is to obtain the posterior distributions of all the unknown parameters including the average treatment effect. To obtain the posterior distributions of all the unknown parameters, we need: (1) the complete data model, or the likelihood function, (2) prior distributions of all the unknown parameters. We suppose the complete data model is $f(y^{obs}, w^{obs}|Z^{obs}, G^{obs}, M^{obs}, U^{obs}, \theta)$ and prior distributions of all the unknown parameters are $f(u^{obs}, \theta)$. Then we can obtain the posterior distributions of all the unknown parameters as
\begin{eqnarray}
&& f(u^{obs}, \theta~|~Y^{obs}, W^{obs},Z^{obs}, G^{obs}, M^{obs}, U^{obs}) \nonumber \\
&=& \frac{f(y^{obs}, w^{obs}|Z^{obs}, G^{obs}, M^{obs}, U^{obs}, \theta)~f(u^{obs}, \theta)}{\int f(y^{obs}, w^{obs}|Z^{obs}, G^{obs}, M^{obs}, U^{obs}, \theta)~f(u^{obs}, \theta)~\dif{u^{obs}}~\dif{\theta}}
\label{equ:posterior-bayes}
\end{eqnarray}

In equation \ref{equ:posterior-bayes}, we assume there is no missing data in the observed variables. The following process shows the construction of the complete data model $f(y^{obs}, w^{obs}|Z^{obs}, G^{obs},$ $M^{obs}, U^{obs}, \theta)$ with regard to counterfactual variables.
\begin{eqnarray}
&& f(y^{obs}, w^{obs}~|~Z^{obs}, G^{obs}, M^{obs}, U^{obs}, \theta) \nonumber \\
&=& \prod_i^N~f_i(y^{obs}, w^{obs}~|~Z^{obs}, G^{obs}, M^{obs}, U^{obs}, \theta) \nonumber \\
&& (\text{SUTV assumption}) \nonumber \\
&=& \prod_i^N~ \{ {\rm I}_i(Z^{obs}=0)~f_i(y^{obs}, w^{obs}~|~Z^{obs}=0, G^{obs}, M^{obs}, U^{obs}, \theta)  \nonumber \\
&& + ~ {\rm I}_i(Z^{obs}=1)~f_i(y^{obs}, w^{obs}~|~Z^{obs}=1, G^{obs}, M^{obs}, U^{obs}, \theta)\} \nonumber \\
&=& \prod_i^N~ \{ {\rm I}_i(Z^{obs}=0)~f_i(y(0,w(0)), w(0)~|~Z^{obs}=0, G^{obs}, M^{obs}, U^{obs}, \theta)  \nonumber \\
&& + ~ {\rm I}_i(Z^{obs}=1)~f_i(y(1,w(1)), w(1)~|~Z^{obs}=1, G^{obs}, M^{obs}, U^{obs}, \theta)\} \nonumber \\
&& (\text{Equations \ref{equ:structural-wobs}, \ref{equ:structural-yobs}}) \nonumber \\
&=& \prod_i^N~ \{ {\rm I}_i(Z^{obs}=0)~f_i(y(0,w(0)), w(0)~|~G^{obs}, M^{obs}, U^{obs}, \theta)  \nonumber \\
&& + ~ {\rm I}_i(Z^{obs}=1)~f_i(y(1,w(1)), w(1)~|~G^{obs}, M^{obs}, U^{obs}, \theta)\} \nonumber \\
&& (\text{Randomization assumption}) \nonumber \\
&=& \prod_i^N~ \{ {\rm I}_i(Z^{obs}=0)~f_i(y(0,w(0))~|~W(0), G^{obs}, M^{obs}, U^{obs}, \theta)  \nonumber \\
&& \cdot~f_i(w(0)~|~G^{obs}, M^{obs}, U^{obs}, \theta) \nonumber \\
&& + ~ {\rm I}_i(Z^{obs}=1)~f_i(y(1,w(1))~|~W(1), G^{obs}, M^{obs}, U^{obs}, \theta)  \nonumber \\
&& \cdot~f_i(w(1)~|~G^{obs}, M^{obs}, U^{obs}, \theta) \}
\label{equ:posterior-completedata}
\end{eqnarray}

Equation \ref{equ:posterior-completedata} can be further expanded into compliers, never-takers and/or always-takers. In equation \ref{equ:posterior-completedata}, $f_i(y(0,w(0))~|~w(0), G^{obs}, M^{obs}, U^{obs}, \theta)$ and $f_i(y(1,w(1))~|~w(1), G^{obs}, M^{obs}, $ $U^{obs}, \theta)$ are the models for two counterfactual outcomes, while
$f_i(w(0)~|~G^{obs}, M^{obs}, U^{obs}, \theta)$ and $f_i(w(1)~|~G^{obs}, M^{obs}, U^{obs}, \theta)$ are the models for two counterfactual exposures. These models are constructed when a Bayesian causal framework is chosen and thus are available for use. Basically, equation \ref{equ:posterior-completedata} means we use the models for $W(0)$ and $Y(0, W(0))$ in the actual control arm and the models for $W(1)$ and $Y(1, W(1))$ in the actual treatment arm. On the other hand, for each participant, only the models for one counterfactual exposure and outcome are used, where the average treatment effect can be estimated from a regressor coefficient from compliers and/or always-takers, as mentioned before. In addition, the simplest framework for observed variables can be extended in more complex frameworks. Since we construct the causal framework for observed variables from the causal framework for counterfactual variables, we can choose complex causal frameworks for counterfactual variables, such as variations mentioned before, to develop complex frameworks for observed variables. Framework change affects the model specification for counterfactual variables and in turn the model specification for observed variables, but never affects the estimation method and process of the average treatment effect as shown in equations \ref{equ:posterior-bayes} and \ref{equ:posterior-completedata}. 

In reality there can be missing data in the observed variables. $W^{obs}_{mis}$, $Y^{obs}_{mis}$, $G^{obs}_{mis}$ and $M^{obs}_{mis}$ denote in order missing data in the observed exposure, outcome, compliance behavior and measured confounders. Then the complete data model becomes $f(y^{obs}, y^{obs}_{mis}, w^{obs}, w^{obs}_{mis}~|~Z^{obs},$ $G^{obs}, G^{obs}_{mis}, M^{obs}, M^{obs}_{mis}, U^{obs}, \theta)$. The priors become $f(g^{obs}_{mis})$, $f(m^{obs}_{mis})$, $f(u^{obs})$, $f(\theta)$, And the joint posterior distribution of unknown parameters becomes $f(y^{obs}_{mis}, w^{obs}_{mis}, g^{obs}_{mis}, c^{obs}_{mis}, u^{obs}, \theta~|$ $Y^{obs}, W^{obs}, Z^{obs}, G^{obs}, M^{obs})$. We can still obtain the posterior distribution of the average treatment effect, no matter how complex the complete data model is. All the methods mentioned are transferable to complex complete data models.

Similarly to equation \ref{equ:posterior-completedata}, 
the complete data model with missing data can be constructed with regard to counterfactual variables, as shown in equation \ref{eqn:likelihood-with-missing}.
\begin{eqnarray}
&& f(y^{obs}, y^{obs}_{mis}, w^{obs}, w^{obs}_{mis}~|~Z^{obs}, G^{obs}, G^{obs}_{mis}, M^{obs}, M^{obs}_{mis}, U^{obs}, \theta) \nonumber \\
&=& \prod_i^N~ \{ {\rm I}_i(Z^{obs}=0)~f_i(y(0,w(0))^{obs}, y(0,w(0))^{mis}~|~W(0)^{obs}, W(0)^{mis}, G^{obs}, G^{obs}_{mis}, M^{obs},   \nonumber \\
&& ~M^{obs}_{mis}, U^{obs}, \theta)~f_i(w(0)^{obs}, w(0)^{mis}~|~G^{obs}, G^{obs}_{mis}, M^{obs}, M^{obs}_{mis}, U^{obs}, \theta) \nonumber \\
&& + ~ {\rm I}_i(Z^{obs}=1)~f_i(y(1,w(1))^{obs}, y(1,w(1))^{mis}~|~W(1)^{obs}, W(1)^{mis}, G^{obs}, G^{obs}_{mis}, M^{obs},  \nonumber \\
&& ~M^{obs}_{mis}, U^{obs}, \theta)~f_i(w(1)^{obs}, w(1)^{mis}~|~G^{obs}, G^{obs}_{mis}, M^{obs}, M^{obs}_{mis}, U^{obs}, \theta) \}
\label{eqn:likelihood-with-missing}
\end{eqnarray}
where the subscript ``$^{obs}$'' indicates the counterfactual variables correspond to the observed variables with complete cases, and the subscript ```$^{mis}$'' indicates the counterfactual variables correspond to the observed variables with missing data.



In the Bayesian causal frameworks for both counterfactual and observed variables, the assignment and the compliance behavior do not have an additive effect on the exposure. This does not conflict with two stage least squares from instrumental variable methods, where we usually build linear models on the observed exposure and outcome conditional on the observed assignment. This issue is irrelevant to our Bayesian causal inference methods, but some more details may be helpful to understand our causal framework and it is plausible that a new causal framework should not conflict with instrumental variable methods when instrumental variable methods are proven valid.

Local average treatment effect from two stage least squares \autocite{imbens_causal_2015} is based on intention-to-treat (ITT) effects of the assignment on the observed exposure and the observed outcome. When the exposure is binary, local average treatment effect is the ratio of $E(Y^{obs} ~|~ Z^{obs}=1)-E(Y^{obs} ~|~ Z^{obs}=0)$ to $E(W^{obs} ~|~ Z^{obs}=1)-E(W^{obs} ~|~ Z^{obs}=0)$ under certain assumptions. In this ratio, the numerator is the coefficient of $Z^{obs}$ in the regression model $Y^{obs}\sim Z^{obs}$, and the denominator is the coefficient of $Z^{obs}$ in the regression model $W^{obs}\sim Z^{obs}$. We can use linear models to estimate the two coefficients of $Z^{obs}$, but this does not necessarily indicate the assignment $Z^{obs}$ has an additive effect on both the observed exposure and the observed outcome.  When the exposure is multivalued, two stage least squares yields an average causal response \autocite{angrist_two-stage_1995}, without assuming the assignment has an additive effect on the observed exposure and the observed outcome. Two stage least squares, and any other instrumental variable method, do not require any specific functional form of the effect of the assignment on the observed exposure and the observed outcome. Hence, our causal frameworks for counterfactual and observed variables do not conflict with instrumental variable methods.

\subsubsection{Removal of the randomization assumption}
\label{sec:framework-7}


In the following discussions, we assume the counterfactual compliance behavior has a direct additive effect on the counterfactual outcomes. If there are any variables not mentioned, their relationships with the counterfactual outcomes can be analyzed similarly.

The randomization assumption guarantees that 

(1) The counterfactual assignment does not affect the counterfactual confounders and compliance behavior directly;

(2) The counterfactual assignment does not modify the effect of the counterfactual confounders and compliance behavior on the counterfactual outcomes;

(3) The counterfactual assignment does not modify the treatment effect;

(4) The counterfactual assignment does not affect the counterfactual outcomes directly. 

The random errors on the counterfactual outcomes are assumed to be independent of the counterfactual assignment all the time. Hence, we do not discuss the random errors in the randomization assumption. Then we consider what would happen if each of the four conditions mentioned above is not met.

If the counterfactual assignment does affect the counterfactual confounders and compliance behavior, the models for counterfactual outcomes will not be changed, because after conditioning on the counterfactual exposures, confounders and compliance behavior, the causal paths between the counterfactual outcomes and the counterfactual assignment are blocked. The counterfactual assignment may become conditional randomization.

If the counterfactual assignment modifies the effect of the counterfactual confounders and compliance behavior on the counterfactual outcomes, which means these effects on two counterfactual outcomes are different, the models for two counterfactual outcomes should not share the same coefficients for the confounders and compliance behavior, otherwise the average treatment effect may be estimated badly. In this case, the models for two counterfactual outcomes can have the same coefficients for the intercept and the average treatment effect but different coefficients for the confounders and compliance behavior. 

If the counterfactual assignment modifies the treatment effect, then the models for two counterfactual outcomes should not share the same coefficient for the average treatment effect. We may still want to estimate the average treatment effect for all the participants. 

If the counterfactual assignment affects the counterfactual outcomes directly, there can be many ways to explain the effect of the counterfactual assignment on the counterfactual outcomes. If we assume the effect of the counterfactual assignment is additive, then the models for two counterfactual outcomes should not share the same intercept. In this case, the models for two counterfactual outcomes can have different intercepts but same coefficients for the other variables. How to build the models for two counterfactual outcomes depends on the way we assume the counterfactual assignment affects the counterfactual outcomes.

These discussions indicate that the randomization assumption is not necessary. We can remove it from the causal framework in use. It can simplify causal relationships among counterfactual variables, but without it we can take more complex causal relationships among counterfactual variables into account and gain a flexibility in building more complex models. When we do not use the randomization assumption, equation \ref{equ:posterior-completedata} will not have the last two lines and we will instead use the models for $f_i(w(0)~|~ Z^{obs}=0, G^{obs}, M^{obs}, U^{obs}, \theta)$, $f_i(w(1)~|~ Z^{obs}=1, G^{obs}, M^{obs}, U^{obs}, \theta)$, $f_i(y(0,w(0))~|~ w(0), Z^{obs}=0, G^{obs}, M^{obs}, U^{obs}, \theta)$, $f_i(y(1,w(1))~|~ w(1), Z^{obs}=1, G^{obs}, M^{obs}, U^{obs}, \theta)$.

The randomization assumption usually holds in a two-arm randomized trial. Hence, our goal is not to remove the randomization assumption from a two-arm randomized trial discussed before, but to mention another two practical considerations. The first consideration is that in the PreventS trial data measured confounders are pre-exposure but post-randomization. The baseline assessment is carried out post randomization. Hence, there is a possibility that randomization can affect baseline covariates, with regard to the temporal order. This situation is similar to the first condition we discussed above on the randomization assumption. After conditioning on the confounders, we can obtain conditional randomization and block possible causal paths from randomization to the outcome through measured and unmeasured confounders. Hence, we can use the randomization assumption in the PreventS trial data and do not have to assume unmeasured confounders are pre-randomization.


The second consideration is that we can apply our methods in observational studies by removing the randomization assumption. Observational studies do not have the (random) assignment. Hence, we can simply remove the assignment from the causal graphs for counterfactual and observed variables. We describe in detail how to extend our methods in cohort studies with a multivalued exposure including the control intervention and the treatment of research interest.

The control intervention effect affects the definitions of the treatment and the average treatment effect, as mentioned in a two-arm randomized trial. Hence, we consider a cohort study where the control intervention means no treatment of research interest is taken. The control intervention is neither the placebo nor some standard care, but should be the natural zero level in the domain of the treatment of research interest. For example, if the treatment is drug dosage and can be taken at 10 mg, 100 mg and so on, then the control intervention should be 0 mg, with no other form of exposure taken. In a cohort study, participants themselves choose the exposure to take. Hence, there is no noncompliance and the compliance behavior is not necessary. The potential outcome framework is still applicable, which means that there are still two counterfactual exposures and outcomes. Two counterfactual exposures represent the control intervention and the treatment, while two counterfactual outcomes represent the outcome from the counterfactual control intervention and the outcome from the counterfactual treatment. We do not have two arms in a cohort study, but we have two groups, the control group and the treatment group. We define an indicator variable for group membership of participants with regard to the exposure taken, denoted by $Z$ in equation \ref{eqn:z-cohort}. $Z$ represents the counterfactual scenarios with counterfactual exposures, while $Z^{obs}$ represents the actual scenarios with observed exposures. Since $Z$ solely indicates group membership, it has no effect on any variable.
\begin{equation}
Z =
    \begin{cases}
      0, & \text{if the control intervention is taken,}\\
      1, & \text{if the treatment is taken.}
    \end{cases}  
    \label{eqn:z-cohort}
\end{equation}

Counterfactual exposures and outcomes are also defined more clearly, denoted by $W(Z)$ and $Y(Z, W(Z))$ as in equations \ref{eqn:w-cohort} and \ref{eqn:y-cohort}. Similarly, $W^{obs}(Z)$ and $Y^{obs}(Z, W(Z))$ represent the observed variables.
\begin{equation}
W(Z) =
    \begin{cases}
      \text{counterfactual control intervention}, & \text{if $Z=0$,}\\
      \text{counterfactual treatment}, & \text{if $Z=1$.}
    \end{cases}  
    \label{eqn:w-cohort}
\end{equation}

\begin{equation}
Y(Z, W(Z)) =
    \begin{cases}
      \text{counterfactual outcome from counterfactual control intervention}, & \text{if $Z=0$,}\\
      \text{counterfactual outcome from counterfactual treatment}, & \text{if $Z=1$.}
    \end{cases}  
    \label{eqn:y-cohort}
\end{equation}

To establish a Bayesian causal framework for a cohort study, we first make the stable unit treatment value and identical condition assumptions. The stable unit treatment value assumption is identical to that in a two-arm randomized trial. After some modifications, the identical condition assumption is given in assumption \ref{assump:idcond-cohort}. 

\begin{assump}[Identical condition] 
\label{assump:idcond-cohort}
(i) $W(0) = W(1) ~|~ M=m, U=u$. Or $P(W(0) = W(1) ~|~ M=m, U=u)=1$. For each participant, two counterfactual exposures are identical, with any counterfactual parent variable fixed. (ii) $Y(0, W(0)) = Y(1, W(1)) ~|~ W(0) = W(1)=w, M=m, U=u$. Or $P(Y(0, W(0)) = Y(1, W(1)) ~|~ W(0) = W(1)=w, M=m, U=u)=1$. For each participant, two counterfactual outcomes are identical, with any counterfactual parent variable fixed. 
\end{assump}

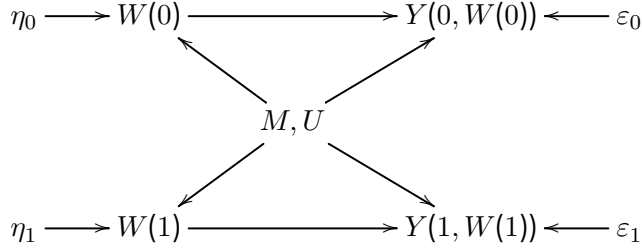
\begin{figure}[!h]
    \centering
\xymatrix{
&&\eta_0 \ar[r]  & W(0) \ar[rr] && Y(0, W(0)) & \varepsilon_0 \ar[l]\\
&&& & M, U \ar[dr] \ar[ur] \ar[ul] \ar[dl] &&  \\
&&\eta_1 \ar[r] & W(1) \ar[rr] && Y(1, W(1)) & \varepsilon_1 \ar[l]
}
    \caption{Causal directed acyclic graph for counterfactual variables.}
    \label{fig:cdag-cohort-counterfactual}
\end{figure}

We want to show how to establish the simplest frameworks for counterfactual and observed variables in a cohort study. More complex frameworks can be established similarly as variations mentioned before. Figure \ref{fig:cdag-cohort-counterfactual} shows the simplest causal mechanism for counterfactual variables. Figure \ref{fig:cdag-cohort-observed} shows the simplest causal mechanism for observed variables. 

\begin{figure}[!h]
    \centering
\xymatrix{
&&\eta \ar[r] & W^{obs} \ar[rr] && Y^{obs} & \varepsilon \ar[l]\\
&& & & M^{obs}, U^{obs}\ar[ul] \ar[ur] &&
}  
    \caption{Causal directed acyclic graph for observed variables.}
    \label{fig:cdag-cohort-observed}
\end{figure}

The we can develop structural equations for the counterfactual exposures as
\begin{eqnarray}
W(0) &=& \alpha_0 + \alpha_1 M + \alpha_2 U + \eta = 0,  \nonumber \\
W(1) &=& \alpha_0 + \alpha_1 M + \alpha_2 U + \eta,
\label{equ:strucmodel-w-cohort}
\end{eqnarray}

and structural equations for the counterfactual outcomes as
\begin{eqnarray}
Y(0, W(0)) &=& \beta_0 + \beta_1 W(0) + \beta_2 M + \beta_3 U + \varepsilon,  \nonumber \\
Y(1, W(1)) &=& \beta_0 + \beta_1 W(1) + \beta_2 M + \beta_3 U + \varepsilon.
\label{equ:strucmodel-y-cohort}
\end{eqnarray}

$W(0)=0$ holds by definition. This does not conflict with assumption \ref{assump:idcond-cohort}, because the values of the confounders can be different for $W(0)=0$ and $W(0) \neq 0$. Assumption \ref{assump:idcond-cohort} imagines the case where $W(0)$ can take many possible values and means the causal mechanism for two counterfactual exposures are identical. The meanings of $W(Z)$ and $Y(Z, W(Z))$ are different between two-arm randomized trials and cohort studies. We use $W(Z)$ for explanations. In cohort studies, $W(Z)$ is a deterministic function of the exposure and represents either the control intervention or the treatment. In two-arm randomized trials, $W(Z)$ is a function of the assignment and represents the exposure taken in one arm with the exposure unnecessarily defined. For example, $W(1)$ is the treatment in cohort studies, while in two-arm randomized trials it is the exposure taken in the counterfactual treatment arm where it is the control intervention for never-takers and the treatment for compliers and/or always-takers. Two-arm randomized trials and cohort studies are two different mechanisms. Hence, causal graphs and structural equations should be carefully interpreted with regard to the meanings of $W(Z)$ and $Y(Z, W(Z))$.

In equation \ref{equ:strucmodel-w-cohort}, since the control intervention is the natural zero level of the treatment, it fits into the model for the treatment of research interest. In reality, this means that some factors make some participants decide not to take the treatment. In equation \ref{equ:strucmodel-y-cohort}, we still assume a constant treatment effect. Hence, $\beta_1$ is the average treatment effect.

The relationship between the counterfactual exposures and the observed exposures and the relationship between the counterfactual outcomes and the observed outcomes are shown in equation \ref{equ:structural-obs-cohort}, similar to the relationships in a two-arm randomized trial. 
\begin{eqnarray}
W^{obs} &=& W(Z = Z^{obs}) \nonumber \\
&=& \alpha_0 + \alpha_1 M + \alpha_2 U + \eta \nonumber \\
Y^{obs} &=& Y(Z = Z^{obs}, W(Z) = W^{obs})  \nonumber \\ 
&=& \beta_0 + \beta_1 W^{obs} + \beta_2 M + \beta_3 U + \varepsilon
\label{equ:structural-obs-cohort}
\end{eqnarray}

The complete data model $f(y^{obs}, w^{obs}~|~ M^{obs}, U^{obs}, \theta)$ now becomes
\begin{eqnarray}
&& f(y^{obs}, w^{obs}~|~ M^{obs}, U^{obs}, \theta) \nonumber \\
&=& \prod_i^N~f_i(y^{obs}, w^{obs}~|~M^{obs}, U^{obs}, \theta) \nonumber \\
&& (\text{SUTV assumption}) \nonumber \\
&=& \prod_i^N~f_i(y^{obs}~|~W^{obs}, M^{obs}, U^{obs}, \theta) f_i(w^{obs}~|~M^{obs}, U^{obs}, \theta) 
\label{equ:posterior-completedata-cohort}
\end{eqnarray}

From equations \ref{equ:structural-obs-cohort} and \ref{equ:posterior-completedata-cohort}, we can obtain posterior distributions of all the unknown parameters. The modelling process for a cohort study is simpler, because equation \ref{equ:structural-obs-cohort} are basically two regression models and indicates that we can use all participants to build models. Two another issues that should be treated in the modelling process are unmeasured confounders and missing data.

Compared to a cohort study, one limitation from the structural equations for the counterfactual exposures for a two-arm randomized trial where we assume the control intervention is not the natural zero level of the treatment, as shown in equation \ref{equ:mechanism-expomodel}, is revealed that the natural zero level of the treatment is always suppressed by the assignment through $h(Z, G)=0$. This means that we cannot build the same model for participants who take the control intervention as for participants who take the treatment of research interest, even if the control intervention is truly the natural zero level of the treatment.

However, researchers may want to assume the control intervention represents the natural zero level of the treatment. There is a way to incorporate the natural zero level of the treatment in never-takers, when the control intervention can be regarded as the natural zero level of the treatment. For both one-sided and two-sided noncompliance, we define a new compliance behavior and a new function $h(Z, G)$ as
\begin{eqnarray}
G &=& 
    \begin{cases}
      0, & \text{if $g = co$} \\
      1, & \text{otherwise}  \nonumber 
    \end{cases} \\
h(Z, G) &=& 
    \begin{cases}
      0, & \text{if $Z = 0$ and $G = 0$} \\
      1, & \text{otherwise}  \nonumber 
    \end{cases} 
\end{eqnarray}

The simplest structural equations for the counterfactual exposures and outcomes now become
\begin{eqnarray}
W(0) &=& h(Z=0, G)~(\alpha_0 + \alpha_1 G + \alpha_2 M + \alpha_3 U + \eta), \nonumber \\
W(1) &=& h(Z=1, G)~(\alpha_0 + \alpha_1 G + \alpha_2 M + \alpha_3 U + \eta), \nonumber \\
Y(0, W(0)) &=& \beta_0 + \beta_1 W(0) + \beta_2 M + \beta_3 U + \varepsilon,  \nonumber \\
Y(1, W(1)) &=& \beta_0 + \beta_1 W(1) + \beta_2 M + \beta_3 U + \varepsilon,
\label{equ:modif-trial-counterfactual}
\end{eqnarray}
where now  we include the compliance behavior as a covariate in the model for counterfactual exposures, because we think possibly being a complier indicates a higher treatment level.

In equations \ref{equ:modif-trial-counterfactual}, $h(Z, G)=0$ only holds in compliers in the counterfactual control arm. For never-takers who have $W(0) = 0$, $\alpha_0 + \alpha_1 G + \alpha_2 M + \alpha_3 U + \eta = 0$ should hold. Since $\alpha_0 + \alpha_1 G + \alpha_2 M + \alpha_3 U + \eta$ is the causal model for the treatment of research interest, the control intervention taken by never-takers should represent the natural zero level of the treatment. Hence, for never-takers, their exposures now have a causal model that can be used to estimate unmeasured confounders. This can improve the estimation accuracy of $\beta_0$, $\beta_2$, $\beta_3$ in the model for the counterfactual outcomes and thus in turn improve the estimation accuracy of $\beta_1$.

In equation \ref{equ:modif-trial-counterfactual}, we explicitly assume if a participant achieves the natural zero level of the treatment, then he is a never-taker. A complier takes the control intervention in the control arm because he is assigned with the control intervention. This is not because a complier decides not to take the treatment, otherwise he will become a never-taker. If we introduce the confounders of compliers in the control arm into the model for the counterfactual exposure from compliers in the treatment arm, we may obtain a non-zero exposure level in compliers in the control arm. Hence, it is impossible yet unnecessary to incorporate the natural zero level of the treatment in compliers in the control arm.

On the other hand, our structural equations with the assumption that the control intervention is not the natural zero level of the treatment and the structural equations above that incorporate the natural zero level of the treatment in never-takers, which both are developed for a two-arm randomized trial, are also applicable to single-arm randomized trials where participants are first randomized to two arms but only participants in the treatment arm actually take the exposure while the participants randomized to the control arm do not participate in the trial further. This is because the structural equations do not require the average treatment effect to be estimated from the control arm. Single-arm randomized trials seem to cost less and may be helpful in some situations.

A final note about assumptions is that, in either a two-arm randomization trial or a cohort study, the stable unit treatment value and identical condition assumptions are always necessary for a Bayesian causal framework. The identical condition assumption is related to the fundamental problem in causal inference: only one of two counterfactual arm can come true and not all counterfactual variables have counterparts in reality. We may not test these two assumptions, but we may be able to justify the stable unit treatment value assumption in the data context. Untestable assumptions should not be the reason to reject a causal framework. Every causal framework needs assumptions. We should be aware of the strengths and weaknesses of a causal framework, with regard to the assumptions used.

Another note about the causal framework is that all the causal frameworks discussed above are applicable to both Bayesian and frequentist inference. We choose the Bayesian approach because it is straightforward to implement a causal framework with estimation of unmeasured confounders shown in next section. Frequentist implementation of a causal framework has not yet been investigated, but is not impossible. Further research on this issue may be necessary.



\section{Estimation of unmeasured confounders}
\label{sec:u-1}

Our method to estimate unmeasured confounders is original by our knowledge. It is inspired by the definition that a confounder affects the exposure and the outcome simultaneously. Hence, can we use unmeasured confounders as covariates in both the exposure and outcome distributions, such that they can be estimated? Since we do not measure unmeasured confounders, their observed values are all missing. A natural idea is to treat unmeasured confounders as completely missing variables and then impute them in Bayesian models. We assume unmeasured confounders are Missing Completely at Random (MCAR) and we name this idea ``the missing variable model'' for following descriptions.

Bias analysis from literature recognizes that the bias, the bias variable, relevant parameters or the bias models are non-identifiable \autocite{rosenbaum_assessing_1983, lin_assessing_1998, steenland_monte_2004, greenland_multiple-bias_2005, vanderweele_bias_2011, arah_bias_2017, groenwold_adjustment_2018, lash_semi-automated_2003,mccandless_bayesian_2007, greenland_bayesian_2009, faries_evaluating_2013, dorie_flexible_2016, zheng_bayesian_2021, zheng_copula-based_2021, zheng_sensitivity_2022}. Our missing variable model also uses the missing variable method. But different from these prior works, especially works related to bias imputation, our approach aims to make the bias variable and relevant parameters identified. In other words, our estimation methods of unmeasured confounders are not sensitivity analysis, but they are developed to estimate unmeasured confounders and associated unknown parameters directly. Hence, we first study how non-identifiability can exist in Bayesian models and then propose methods to resolve model non-identifiability, in order to obtain identified models from which every part is identified. We choose Bayesian models, because not only priors can be used and tested, but also more evidential information from data is included, which may yield better estimates of the bias. Our methods can be transferrable to different types of bias, but we focus on unmeasured confounding and only discuss this type of bias here.

\subsection{One unmeasured confounder}
\label{sec:u-2}

We start with the simplest scenario by assuming there is one unmeasured confounder. We use the observed variable denotation $U^{obs}$ for the unmeasured confounder. Implementation of the missing variable model is to directly declare the unmeasured confounder as a parameter vector in RStan \autocite{stan_development_team_rstan_2022} and put this parameter vector as a covariate in the models for the exposure and the outcome. The length of the parameter vector or the number of parameters in this vector is equal to the number of participants from whom we can identify the unmeasured confounding effect, rather than the total number of participants. Participants from whom we can identify the unmeasured confounding effects are those whose observed exposure distributions are not constantly zero but conditional on the confounders. Each parameter corresponds to a missing value in this unmeasured confounder for a participant. 

The missing variable model does not involve complicated theories, but is more like an intuitive understanding of the unmeasured confounding effect. Its underlying theory is simple: when the complete data model from the observed variables is denoted by $f(y^{obs}, w^{obs}|Z^{obs},$ $G^{obs}, M^{obs}, U^{obs}, \theta)$, it can be decomposed into the product of the model for the outcome and the model for the exposure, as shown in equation \ref{equ:u-start-decomp}.

\begin{eqnarray}
&& f(Y^{obs}, W^{obs}~|~Z^{obs}, G^{obs}, M^{obs}, U^{obs}, \theta) \nonumber \\
&=& f(Y^{obs}~|~W^{obs}, Z^{obs}, G^{obs}, M^{obs}, U^{obs}, \theta) \nonumber \\
&& \cdot~ f(W^{obs}~|~Z^{obs}, G^{obs}, M^{obs}, U^{obs}, \theta)
\label{equ:u-start-decomp}
\end{eqnarray}

In equation \ref{equ:u-start-decomp}, $f(Y^{obs}~|~W^{obs}, Z^{obs}, G^{obs}, M^{obs}, U^{obs}, \theta)$ is the model for the outcome and $f(W^{obs}~|~Z^{obs}, G^{obs}, M^{obs}, U^{obs}, \theta)$ is the model for the exposure. The model for the outcome and the model for the exposure are both conditional on the unmeasured confounder $U^{obs}$. Hence, we use the unmeasured confounder $U^{obs}$ as a covariate in the exposure and outcome distributions. With the complete data model and priors for unknown parameters, we can obtain the posterior distributions of the unmeasured confounder. 

We give a clear description on how the unmeasured confounder is modelled in RStan. For participants whose exposure distribution is not fixed at zero but conditional on the confounders, such as compliers in the treatment arm, the unmeasured confounder is adjusted for in both the exposure and outcome distributions. For participants whose exposure distribution is fixed at zero or a point distribution, such as compliers in the control arm, the unmeasured confounder is adjusted for only in the outcome distribution.

Adjusting for unmeasured confounding with both the exposure and outcome distributions reflects the definition of a confounder and we expect this approach can estimate the unmeasured confounder well. On the other hand, in simulation study, adjusting for unmeasured confounding with only the outcome distribution is also effective, though slightly less effective than adjusting for unmeasured confounding with both the exposure and outcome distributions. Hence, we think adjusting for unmeasured confounding with only the outcome distribution can be a good alternative when the exposure distribution is not available to adjust for unmeasured confounding. Since no theoretical proof has been developed and comprehensive comparisons have not been conducted, we suggest adjusting for unmeasured confounding with only the outcome distribution should be a topic for future research. More details are given in simulation study.


The missing variable model is straightforward in RStan, but it is not free of problems. Next we use the simplest causal framework and the models for compliers in the treatment arm to further elaborate on the missing variable model, because their model for the exposure and their model for the outcome both contain the unmeasured confounder. These models are shown in equations \ref{equ:unmeasured-examp-w} and \ref{equ:unmeasured-examp-y}.
\begin{eqnarray}
W^{obs} &=& \alpha_0 + \alpha_1 M^{obs} + \alpha_2 U^{obs} + \eta, \label{equ:unmeasured-examp-w} \\
Y^{obs} &=& \beta_0 + \beta_1 W^{obs} + \beta_2 M^{obs} + \beta_3 U^{obs} + \varepsilon,
\label{equ:unmeasured-examp-y}
\end{eqnarray}
where $E(\eta) = E(\varepsilon) = 0$, $Var(\eta) = \sigma^2_\eta$, $Var(\varepsilon) = \sigma^2_\varepsilon$, $\alpha = (\alpha_0, \alpha_1, \alpha_2)$, $\beta = (\beta_0, \beta_1, \beta_2, \beta_3)$, $\sigma = (\sigma_\eta, \sigma_\varepsilon)$.

\subsubsection{Model non-identifiability}
\label{sec:u-3}

The missing variable model is related to model non-identifiability. A model is non-identifiable if various sets of parameters can produce the same model. For non-identifiable Bayesian models, McCandless et al. \autocite*{mccandless_bayesian_2007} wrote ``Depending on the prior distribution, the posterior mean may be asymptotically biased and credible intervals will not have nominal large sample coverage probability. Hence without identifiability, standard large-sample theory does not guarantee good performance of interval estimators regardless of the choice of prior distributions.''(McCandless et al., 2007, p. 2332). Bias in the posterior mean may bias the conclusions. Although we do not require 95\% posterior intervals from Bayesian models have nominal frequentist large sample coverage probability, it does not hurt to have this property. Further, even if the average treatment effect is identifiable in some non-identifiable models, we do not know how the priors of the non-identifiable parameters would affect the posterior distribution of the average treatment effect and identifiability of the average treatment effect, and thus we cannot guarantee that model non-identifiability does not affect identifiability of the average treatment effect. Hence, we prefer identifiable Bayesian models. To make sure our Bayesian models identified, our approach is to find out reasons that cause model non-identifiability and then develop methods to prevent these reasons. We have identified four types of model non-identifiability in the missing variable model with one unmeasured confounder and proposed methods to resolve them.

The first type of model non-identifiability is reversion of the sign of $U^{obs}$. This means if we multiply $U^{obs}$ and its coefficient by -1 simultaneously, the models will not change. For example, in equation \ref{equ:unmeasured-examp-w},
\begin{eqnarray}
W^{obs} &=& \alpha_0 + \alpha_1 M^{obs} + \alpha_2 U^{obs} + \eta \nonumber \\
&=& \alpha_0 + \alpha_1 M^{obs} + (-\alpha_2)~(- U^{obs}) + \eta. \nonumber 
\end{eqnarray}

This type of model non-identifiability can be resolved by holding the coefficient of $U^{obs}$, $\alpha_2$, to be positive. Through fixing the sign of $\alpha_2$, the sign of $U^{obs}$ is fixed. To estimate $U^{obs}$, we need to fit the model for the exposure and the model for the outcome simultaneously. Hence, there are two coefficients of $U^{obs}$ from two models. In practice, we only have to hold one coefficient of $U^{obs}$ to be positive in either model. This is because after fixing the sign of one coefficient in one model, the sign of $U^{obs}$ is fixed and consequently the sign of the other coefficient is fixed. If we know the signs of two coefficients, we can incorporate this information in the models, but in reality we usually do not know the signs. Hence, a more viable way is to fix one sign of two coefficients.

The second type of model non-identifiability is shifting and scaling of $U^{obs}$. For example, in equation \ref{equ:unmeasured-examp-w}, if we shift or scale $U^{obs}$ by a non-zero constant $\Delta$, the exposure distribution will not change, such as
\begin{eqnarray}
W^{obs} &=& (\alpha_0 + \alpha_2~\Delta) + \alpha_1 M^{obs} + \alpha_2 (U^{obs} - \Delta) + \eta \nonumber \\
&=& \alpha_0  + \alpha_1 M^{obs} + (\alpha_2~\Delta)~\frac{U^{obs}}{\Delta} + \eta, \nonumber 
\end{eqnarray}
where the coefficient of $U^{obs}$ and the intercept in the exposure distribution may change to keep the entire model unchanged. This means, except the coefficient of $U^{obs}$ and the intercept in the exposure distribution, the other parameters including $U^{obs}$ do not change. Similar results can come from the model for the outcome. Hence, the models with the unmeasured confounder directly as a missing variable are non-identifiable. The underlying idea is that shifting and scaling $U$ does not change the complete data model or the likelihood function, and thus for the other parameters except the coefficient of $U^{obs}$ and the intercept, when their priors remain unchanged, their posterior distributions remain unchanged or are changed little.

Shifting and scaling of $U^{obs}$ can be implemented in RStan in two ways: (1) as indicated above, add the constant $\Delta$ in the models for the exposure and the outcome, with the priors for the coefficient of $U^{obs}$ and the intercept varying and the priors for the other parameters fixed, or (2) treat the shifted and scaled unmeasured confounder as a new unmeasured confounder $U'$, replace $U$ by $U'$ in the exposure and outcome distributions, then shift and scale the prior of $U'$. Way (2) is equivalent to shifting and scaling the prior of $U$ directly, but we distinguish the original unmeasured confounder from the shifted and scaled one to avoid confusion in following discussions. In way (2), shifting or scaling the prior of $U'$ is equivalent to shifting or scaling $U'$ because the prior location and scale dominates the posterior location and scale to a large extent. When $U'$ is shifted or scaled, the magnitude in the change of its posterior estimates depends on the magnitude in the change of its prior. For the coefficient of $U'$ and the intercept, their posterior distributions are also affected by their priors, but even if their priors remain unchanged, their posterior distributions can also change to match the changes in the posterior distribution of $U'$, in order to make the entire model unchanged or changed little. In this case, the models are still non-identifiable, but the other parameters except $U'$, its coefficient and the intercepts may be estimated well.

For illustration, we fit the outcome model on complete cases from compliers in the treatment arm in the PreventS trial data, using the simplest causal framework. The exposure is the total number of coaching sessions attended by the 9-month assessment and the outcome is the change in the LS7 total score between the baseline and 9-month assessments. We want to see how posterior estimates of the intercept, $\beta_0$, the coefficient of $U'$, $\beta_3$, and the coefficient of the exposure as the average treatment effect, $\beta_1$, will change with shifting or scaling of $U'$. Table \ref{tab:nonid-u-priors} shows the prior of $U'$, the prior and posterior distributions of three parameters of interest, with the priors for the other parameters fixed.
\begin{table}[htbp]
\centering
\footnotesize
\begin{tabular}{llrrr}
  \hline
  & & \multicolumn{3}{c} {Posterior mean (95\% interval)} \\ \cline{3-5}
$U$ change & Priors & $\hat \beta_0$ & $\hat \beta_3$ & $\hat \beta_1$ \\ 
  \hline
$U$ (original) &  
\begin{minipage}[t]{0.2\textwidth} \raggedright $U \sim N(1,1)$ \end{minipage}
& -1.88 (-3.77, -0.04) & 1.68 (1.22, 2.18) & 0.00 (-0.14, 0.15) \\ 
& \begin{minipage}[t]{0.2\textwidth} \raggedright $\beta_0 \sim N(-2, 1)$  \end{minipage} &&& \\
& \begin{minipage}[t]{0.2\textwidth} \raggedright $\beta_3 \sim N(2, 1)$* \end{minipage} &&& \\
$U' = U + 9$ &  
\begin{minipage}[t]{0.2\textwidth} \raggedright $U' \sim N(10,1)$ \end{minipage}
& -19.77 (-21.67, -17.87) & 1.82 (1.45, 2.24) & 0.10 (-0.19, 0.37) \\ 
& \begin{minipage}[t]{0.2\textwidth} \raggedright $\beta_0 \sim N(-20, 1)$  \end{minipage} &&& \\
& \begin{minipage}[t]{0.2\textwidth} \raggedright $\beta_3 \sim N(2, 1)$* \end{minipage} &&& \\
$U' = 2U$ &  
\begin{minipage}[t]{0.2\textwidth} \raggedright $U' \sim N(2,4)$ \end{minipage}
& -1.86 (-3.75, 0.04) & 0.45 (0.33, 0.60) & 0.06 (-0.09, 0.20) \\ 
& \begin{minipage}[t]{0.2\textwidth} \raggedright $\beta_0 \sim N(-2, 1)$  \end{minipage} &&& \\
& \begin{minipage}[t]{0.2\textwidth} \raggedright $\beta_3 \sim N(1, 0.25)$* \end{minipage} &&& \\
   \hline
\multicolumn{5}{l} {* indicates this distribution is left-truncated at 0.}
\end{tabular}
\caption{Posterior estimates under shifting or scaling of the unmeasured confounder} 
\label{tab:nonid-u-priors}
\end{table}

In two decimal places, when the unmeasured confounder is increased by 9 or multiplied by 2, the changes in the posterior estimates of the intercept and the coefficient of the unmeasured confounder are as expected. The average treatment effect does not change materially, which does not affect the conclusions about the average treatment effect. In nature, model non-identifiability from shifting and scaling of $U'$ comes from the fact that $U'$ is a set of parameters. Parameters interact with each other in the Bayesian sampling process. Hence, to resolve model non-identifiability from shifting and scaling of $U'$, we need to integrate $U'$ and other relevant parameters into a whole, in order to reduce parameter interaction. We propose reparameterization as a way to resolve this issue. Reparameterization is elaborated after other types of model non-identifiability are introduced.

The third type of model non-identifiability is degeneration of the causal model. This means that the unmeasured confounder is not successfully adjusted for in the models for the exposure and the outcome, and the seemingly causal model becomes an association model that does not adjust for unmeasured confounding and produces biased estimates of the average treatment effect. We use equations \ref{equ:unmeasured-examp-w} and \ref{equ:unmeasured-examp-y} with $U^{obs}$ for descriptions and assume there is no shifting and scaling of the unmeasured confounder $U^{obs}$.

Model degeneration can occur when at least one coefficient of the unmeasured confounder in the exposure and outcome distributions is estimated to be zero, or when the unmeasured confounder itself is estimated to be zero. Specific situations include:

(1) In equation \ref{equ:unmeasured-examp-w}, $\alpha_2$ is estimated to be zero;

(2) In equation \ref{equ:unmeasured-examp-y}, $\beta_3$ is estimated to be zero;

(3) In equations \ref{equ:unmeasured-examp-w} and \ref{equ:unmeasured-examp-y}, $U^{obs}$ is estimated to be zero.

In situation (1), the models for the exposure and the outcome become
\begin{eqnarray}
W^{obs} &=& \alpha_0 + \alpha_1 M^{obs} + \eta, \nonumber \\
Y^{obs} &=& \beta_0 + \beta_1 W^{obs} + \beta_2 M^{obs} + \beta_3 U^{obs} + \varepsilon, \nonumber
\end{eqnarray}
where $U^{obs}$ is no longer defined as a confounder but is purely a random variable. There is no connection between the model for the exposure and the model for the outcome, and unmeasured confounding may not be completely adjusted for in the model for the outcome. Situation (2) is similar to situation (1). In situation (3), the models for the exposure and the outcome become
\begin{eqnarray}
W^{obs} &=& \alpha_0 + \alpha_1 M^{obs} + \eta, \nonumber \\
Y^{obs} &=& \beta_0 + \beta_1 W^{obs} + \beta_2 M^{obs} + \varepsilon, \nonumber
\end{eqnarray}
where unmeasured confounding disappears and two models act independently. The model for the outcome is an association model that only adjusts for measured confounding, where the average treatment effect estimates are likely to be biased. 

\begin{figure}[htbp]
     \centering
     \includegraphics[width=\textwidth]{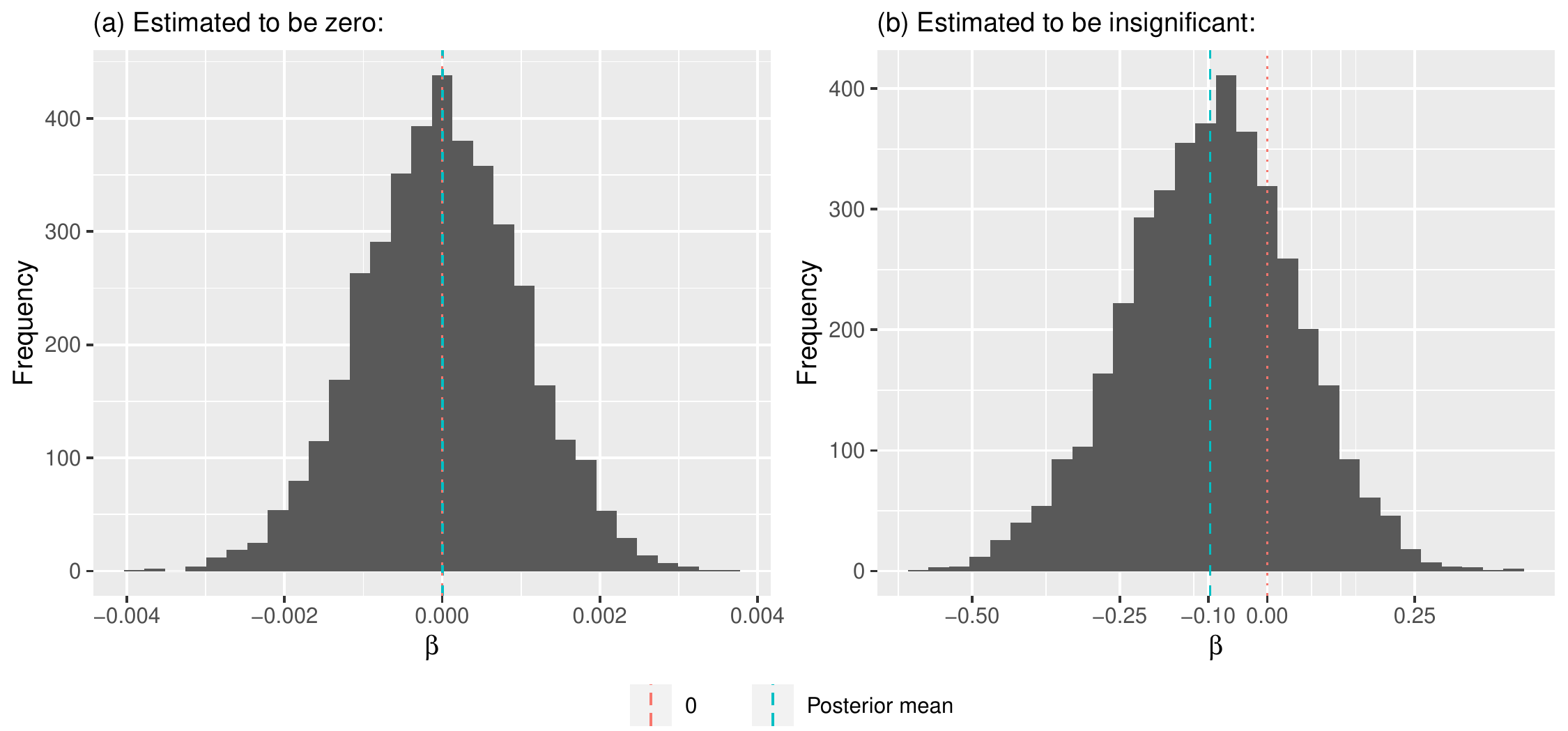}
        \caption{Example posterior distributions for parameter $\beta$ estimated to be zero or insignificant.}
        \label{fig:post-est}
\end{figure}

A parameter estimated to be zero is different from a parameter estimated to be insignificant. By our definition, one parameter is said to be estimated to be zero, if its posterior mean is exactly or almost zero. Its posterior standard deviation can be very small. On the other hand, one parameter is said to be estimated to be insignificant, if its 95\% posterior interval covers zero. Its posterior mean can be very close to zero, but usually should not be exactly zero. Figure \ref{fig:post-est} shows example posterior distributions of a parameter, $\beta$, estimated to be zero or insignificant.

The posterior distributions of the two types of parameter may or may not look like a Normal distribution. For example, if a parameter is positive, then its posterior distribution is a truncated distribution, and when it is estimated to be zero, its posterior distribution may look like a point distribution at zero. Sometimes, these two types of parameter are hard to distinguish, when their posterior distributions are quite similar. We may directly build a model that only adjusts for measured confounding to obtain the association estimates, and compare the association estimates with the causal effect estimates from the model that generates this parameter. If this parameter is estimated to be zero, then the association estimates should be similar to the causal effect estimates. This may help us identify the type of parameter.

When the unmeasured confounder is actually non-zero or at least has a non-zero mean, and it has a non-zero effect on the exposure and the outcome, it is natural that the unmeasured confounder and its effect will be estimated to be significant. However, in reality, the unmeasured confounder itself can be very small or varies around zero, or the unmeasured confounding effect is negligible after many possible measured confounders are considered. In these cases, we can use shifting and scaling to transform the unmeasured confounder. We first shift and scale the underlying true unmeasured confounder to a significant level that will not be estimated to be zero. For simplicity, we call the shifted and scaled unmeasured confounder ``the working unmeasured confounder'' in Bayesian models. Shifting and scaling can be achieved by assigning the working unmeasured confounder with a prior that has a non-zero mean, then the coefficient of the working unmeasured confounder and/or the intercept will change accordingly such that the entire model is unchanged with regard to the original unmeasured confounder.


After the working unmeasured confounder becomes significant, if the original unmeasured confounder is small or the unmeasured confounding effect is small, at least one coefficient of the working unmeasured confounder in the models for the exposure and the outcome should be estimated to be insignificant. This means that we can control the behaviors of the posterior distributions of two coefficients of the working unmeasured confounder in order to prevent degeneration of the causal model, by using a significant working unmeasured confounder in modelling and adjusting the priors of two coefficients, whether the original true unmeasured confounder is significant or not and whether the unmeasured confounding effect is significant or not. This also means that we now only have to prevent two coefficients of the unmeasured confounder from being estimated to be zero.

One more issue worth mentioning is disappearance of a random error. This means that one random error from either the model for the exposure or the model for the outcome has a posterior standard deviation estimated to be zero. In other words, the standard deviation of the exposure or the standard deviation of the outcome is estimated to be zero. Since the expectations of two random errors are zero, the random error with a posterior standard deviation estimated to be zero is also estimated to be zero. For example, if $\varepsilon$ is estimated to be zero, then the models for the exposure and the outcome now become
\begin{eqnarray}
W^{obs} &=& \alpha_0 + \alpha_1 M^{obs} + \alpha_2 U^{obs} + \eta,  \\
Y^{obs} &=& \beta_0 + \beta_1 W^{obs} + \beta_2 M^{obs} + \beta_3 U^{obs},
\end{eqnarray}
where the new outcome distribution is a mathematical equation with no uncertainty. In reality, data seldom satisfies the strict mathematical relationships in the new outcome distribution. Hence, Bayesian models are usually problematic. For example, it can be divergent, or the Monte Carlo samples are strongly autocorrelated. This phenomenon is not directly related to model non-identifiability, because we cannot directly derive alternative solutions from the models. But it is usually occurs with the third type of model non-identifiability. The coefficient of the working unmeasured confounder in the model where the random error disappears is usually estimated to be zero, or at least one coefficient of the working unmeasured confounder is usually estimated to be zero. We find disappearance of a random error from Bayesian models with Normally distributed exposures and outcomes. We have not yet investigated other distributions, but we suspect this phenomenon may occur for any distribution where the mean and the variance are independent.

Both standard deviations of the exposure and the outcome have to be estimated to be non-zero or significant. Sometimes, prior adjustment can solve the problem that standard deviations are estimated to be zero but it is not a general approach from our experience. In some Bayesian models, the priors for standard deviations have to be strongly informative with a very small prior standard deviation, but it is hard to determine the locations of these priors. Hence, we use sample standard deviations of the exposure and the outcome in Bayesian models to estimate standard deviations of the exposure and the outcome. The exposure and the outcome we have considered are Normally distributed. We can even fix standard deviations to be sample standard deviations in the models for the exposure and the outcome. We can also use sample standard deviations of the exposure and the outcome as the prior means of standard deviations of the exposure and the outcome, then construct strongly informative priors to make posterior estimates of standard deviations nearly equal to sample standard deviations. Use of sample standard deviations has several advantages: (1) unlike coefficients that we do not know for sure, sample standard deviations are valid, useful estimates for true standard deviations, (2) use of sample standard deviations can reduce the number of parameters in Bayesian models and ease the modelling process, and (3) standard deviations can be prevented from being estimated to be zero in Bayesian models. In simulation study, we will compare different approaches of using sample standard deviations in Bayesian models. 

The fourth type of model non-identifiability is correlation between exposure parameters and outcome parameters of the same meaning. This means that two parameters of the same meaning from the exposure and outcome distributions, such as two intercepts or two coefficients for the same confounder or two standard deviations, have posterior correlation. We assume model non-identifiability from the other types does not exist. In the outcome distribution from equation \ref{equ:unmeasured-examp-y}, if we decrease the coefficient of the exposure by a non-zero constant $\Delta$, then the outcome distribution becomes
\begin{eqnarray}
Y^{obs} &=& \beta_0 + \beta_1 W^{obs} + \beta_2 M^{obs} + \beta_3 U^{obs} + \varepsilon \nonumber \\ 
&=& \beta_0 + (\beta_1-\Delta) W^{obs} + \Delta W^{obs} + \beta_2 M^{obs} + \beta_3 U^{obs} + \varepsilon \nonumber \\ 
&=& \beta_0 + (\beta_1-\Delta) W^{obs} + \Delta~(\alpha_0 + \alpha_1 M^{obs} + \alpha_2 U^{obs} + \eta) + \beta_2 M^{obs} + \beta_3 U^{obs} + \varepsilon \nonumber \\ 
&& (\text{Equation \ref{equ:unmeasured-examp-w}}) \nonumber \\ 
&=& (\beta_0 + \Delta~\alpha_0) + (\beta_1-\Delta) W^{obs} + (\beta_2+\Delta~\alpha_1) M^{obs} + (\beta_3+\Delta~\alpha_2) U^{obs} + (\varepsilon+\Delta~\eta) \nonumber \\ 
&\doteq& \beta_0' + \beta_1' W^{obs} + \beta_2' M^{obs} + \beta_3' U^{obs} + \varepsilon'
\end{eqnarray}
where $\beta_0' = \beta_0 + \Delta~\alpha_0$, $\beta_1' = \beta_1-\Delta$, $\beta_2' = \beta_2+\Delta~\alpha_1$, $\beta_3' = \beta_3+\Delta~\alpha_2$, $\varepsilon' = \varepsilon+\Delta~\eta$. We build the outcome distribution using $Y^{obs} = \beta_0' + \beta_1' W^{obs} + \beta_2' M^{obs} + \beta_3' U^{obs} + \varepsilon'$, which has the same structure as $Y^{obs} = \beta_0 + \beta_1 W^{obs} + \beta_2 M^{obs} + \beta_3 U^{obs} + \varepsilon$ but is ignorant of the effect of $\Delta$. Then the intercept and the coefficients of the confounders from the outcome distribution are correlated to the same parameters from the exposure distribution. For example, $\beta_0'$ is correlated with $\alpha_0$, $\beta_2'$ is correlated with $\alpha_1$, $\beta_3'$ is correlated with $\alpha_2$, for $\Delta \neq 0$. Further, for $\Delta \neq 0$, the variance of the outcome denoted by $\sigma_y^2$ is also associated with the variance of the exposure denoted by $\sigma_w^2$, through below equations
\begin{eqnarray}
\sigma_y^2 &=& Var(\varepsilon+\Delta~\eta) \nonumber \\ 
&=& \sigma_\varepsilon^2+\Delta^2~\sigma_\eta^2 \nonumber \\ 
&& (\text{$\varepsilon$ and $\eta$ are independent}) \nonumber \\ 
\sigma_w^2 &=& \sigma_\eta^2 \nonumber \\ 
\Longrightarrow ~~~ \sigma_y &=& \sqrt{\sigma_\varepsilon^2+\Delta^2~\sigma_w^2} \nonumber \\
&\approx& \sigma_\varepsilon+|\Delta|~\sigma_w \nonumber
\end{eqnarray}

Hence, there will appear to be some positive correlation between the standard deviation of the outcome and the standard deviation of the exposure in their posterior distributions, when two standard deviations are actually independent. When this type of model-nonidentifiability occurs, the average treatment effect estimates change. Hence, to obtain unbiased average treatment effect estimates, we want $\Delta = 0$. We can check if this type of model-nonidentifiability occurs by investigating the posterior relationships between two parameters of the same meaning from the exposure and outcome distributions. If posterior correlation is found, it is very likely that $\Delta \neq 0$. We can eliminate posterior correlation between standard deviations of the exposure and the outcome such that $\Delta$ is fixed at zero and thus model-nonidentifiability is eliminated, possibly by use of sample standard deviations of the exposure and the outcome, as mentioned before.

The features of problems in each type of model non-identifiability may not be comprehensively concluded, because even under the same type of model non-identifiability, different data can yield different posterior distributions and model problems can be different. We simulated many data sets and tried to understand each type of model non-identifiability as more as possible, but still found ourselves unable to summarize all the features. 
Our solutions are a good starting point in problem solving. They are helpful in our investigations, but we believe there are other solutions. We suggest model problems should be solved in the data context and types of model non-identifiability mentioned above should be used to help identify what problems are and further what type model non-identifiability is, in order to construct a useful solution to problems.

\subsubsection{Reparameterization}
\label{sec:u-4}

In this section, we focus on how to resolve model non-identifiability from shifting and scaling of one unmeasured confounder through reparameterization and how reparameterization is related to reversion of the sign of the unmeasured confounder and correlation between exposure and outcome parameters of the same meaning can be resolved by use of sample standard deviations. Hence, they are not mentioned in this section.

Reparameterization is to combine parts in one model from the exposure and outcome distributions to define a new unmeasured confounder and transform the original unmeasured confounder into the new unmeasured confounder in the other model. It is carried out from the outcome distribution to the exposure distribution, because the average treatment effect is estimated from the outcome distribution and thus the outcome distribution is of our interest. However, there is actually no order in reparameterization but simply preference of order. We can also first reparameterize the exposure distribution and then extend to the outcome distribution. In some simulated data, we found reparameterization from the exposure distribution to the outcome distribution is more likely to lead to degeneration of the causal model.

The first reparameterization is to define a new unmeasured confounder $U'$ in the outcome distribution such that $U' = \beta_0 + \beta_3 U^{obs}$. Then the exposure and outcome distributions become
\begin{eqnarray}
Y^{obs} &=& \beta_1 W^{obs} + \beta_2 M^{obs} + U' + \varepsilon \nonumber \\
\Longrightarrow ~~~ W^{obs} &=& \alpha_0 + \alpha_1 M^{obs} + \alpha_2 ~\left (\frac{U' - \beta_0}{\beta_3}\right ) + \eta  \nonumber \\ 
&=& (\alpha_0 - \frac{\alpha_2}{\beta_3}~\beta_0) + \alpha_1 M^{obs} + \frac{\alpha_2}{\beta_3} U' + \eta  \nonumber \\
&\doteq& \alpha_0' + \alpha_1 M^{obs} + \alpha_2' U' + \eta  \nonumber
\end{eqnarray}

In the reparameterized outcome distribution, $U'$ cannot be shifted or scaled by the other parameters. $U'$ also looks like a random intercept in the reparameterized outcome distribution. Hence, we name this reparameterization ``random intercept reparameterization'', for further reference.

The random intercept reparameterization can resolve model non-identifiability from reversion of the sign of the original unmeasured confounder. This is because $U' = \beta_0 + \beta_3 U^{obs} = \beta_0 + (-\beta_3) (-U^{obs})$. When the sign of $U^{obs}$ reverses, the signs of $\alpha_2$ and $\beta_3$ reverse simultaneously such that the sign of $\alpha_2'$ does not change. Hence, there is no further constraint on the signs of parameters.

In addition, the posterior distribution of $\alpha_2'$ usually exists. When the posterior mean of $\alpha_2'$ does not approach $\infty$ and we assume $\alpha_2$ is finite, this implies that $\beta_3$ is not estimated to be zero. When $\alpha_2'$ is not estimated to be zero and we assume $\beta_3$ is finite, this implies that $\alpha_2$ is not estimated to be zero. We do not consider other more complex mathematical relationships between $\alpha_2$ and $\beta_3$, such as $\alpha_2 \rightarrow \infty$ and $\beta_3 \rightarrow \infty$, because degeneration of the causal model only involves $\alpha_2 =0$ and $\beta_3 =0$. Hence, the random intercept reparameterization can prevent model degeneration from zero estimated coefficients by obtaining a proper posterior distribution of $\alpha_2'$, where being proper means the posterior mean of $\alpha_2'$ is finite and $\alpha_2'$ is not estimated to be zero. However, the random intercept reparameterization cannot prevent model degeneration from zero estimated $U^{obs}$. When $U^{obs}$ is estimated to be zero, $U' = \beta_0$, that is, $U'$ is estimated to be a constant. Hence, the posterior distribution of $u_i'$, for $i = 1, 2, \ldots, N$, would look similar around a common value. Since $U^{obs}$ and $\beta_0$ are not directly involved in the modelling process, there is no way to adjust their priors. Hence, when we find suspicious posterior distributions of $U'$ related to zero estimated $U^{obs}$, we may try adjusting the prior of $U'$ or may have to try another reparameterization.

The second reparameterization is to define a new unmeasured confounder $U'$ in the outcome distribution such that $\beta~U' = \beta_0 + \beta_3 U^{obs}$ where $\beta$ is a new parameter and the first entry in $U'$ is 1. Then the exposure and outcome distributions become
\begin{eqnarray}
Y^{obs} &=& \beta_1 W^{obs} + \beta_2 M^{obs} + \beta~U' + \varepsilon \nonumber \\
\Longrightarrow ~~~ W^{obs} &=& \alpha_0 + \alpha_1 M^{obs} + \alpha_2 ~\left (\frac{\beta~U' - \beta_0}{\beta_3}\right ) + \eta  \nonumber \\ 
&=& \left(\alpha_0 - \frac{\alpha_2}{\beta_3}~\beta_0\right) + \alpha_1 M^{obs} + \frac{\alpha_2~\beta}{\beta_3} U' + \eta  \nonumber \\
&\doteq& \alpha_0' + \alpha_1 M^{obs} + \alpha_2' U' + \eta  \nonumber
\end{eqnarray}

We use the vector form to understand the relationship between $U^{obs}$ and $U'$. If we write $U^{obs} = (u_1, u_2, \ldots, u_N)$ and $U' = (1, u_2', \ldots, u_N')$ , then $\beta~U' = \beta_0 + \beta_3 U^{obs}$ implies that 
\begin{eqnarray}
 \beta~\left[ \begin{array}{l}
1 \\ u_2' \\ \vdots \\ u_n' \end{array} \right] &=& \beta_0 + \beta_3 ~\left[ \begin{array}{l}
u_1 \\ u_2 \\ \vdots \\ u_n \end{array} \right] \nonumber \\
\Longrightarrow \quad \quad \quad ~~~ \beta &=& \beta_0 + \beta_3 ~ u_1 \nonumber \\
\beta~u_i' &=& \beta_0 + \beta_3 ~ u_i, ~ \text{for}~ i=2, 3, \ldots, N \nonumber \\
\Longrightarrow \quad \quad \quad ~~ u_i' &=& \frac{\beta_0 + \beta_3 ~ u_i}{\beta_0 + \beta_3 ~ u_1} \nonumber 
\end{eqnarray}

If we denote the reparameterized unmeasured confounder in the random intercept reparameterization by $U'' = (u_1'', u_2'', \ldots, u_N'')$, then $u_i' = \frac{u_i''}{u_1''}$, for $i=2, 3, \ldots, N$. Hence, each entry in $U'$ is the ratio of the same entry from $U''$ to the first entry from $U''$. Hence, we name this reparameterization ``ratio reparameterization'', for further reference. On the other hand, the ratio reparameterization is a reparameterization on the reparameterized unmeasured confounder in the random intercept reparameterization, where $\beta~U' = U''$ holds. However big or small the entries in $U''$ are, they should have some comparability as they come from the same distribution, if we ignore extreme cases, such that we can expect entries in $U'$ would be moderate. In addition, since $\beta=\beta_0 + \beta_3 ~ u_1$ holds, $\beta$ will not scale $U'$, and since ratios among entries from $U''$ are fixed, $U'$ will not be able to shift. Hence, there is no shifting or scaling of $U'$ by other parameters.

Usually, we want the posterior distributions of $\beta$ and $\alpha_2'$ to be proper, such that $\beta$ and $\alpha_2'$ are not estimated to be zero and their posterior means are finite. The first entry in $U'$ is fixed to be 1, but any entry in $U'$ at any place can be fixed to be 1. This means, if $\beta$ is estimated to be zero, we may then fix another value in $U'$ to be 1 except the first entry, until the posterior distribution of $\beta$ is proper. When the posterior distribution of $\beta$ is proper and we assume $\alpha_2$ is finite, if the posterior mean of $\alpha_2'$ does not approach infinity, then $\beta_3$ is not estimated to be zero. When the posterior distribution of $\beta$ is proper and we assume $\beta_3$ is finite, if $\alpha_2'$ is not estimated to be zero, then $\alpha_2$ is not estimated to be zero. Further, if $U^{obs}$ is estimated to be zero, then $u_i' = 1$, for $i=2, 3, \ldots, N$. Hence, the posterior distribution of $u_i'$ for $i=2, 3, \ldots, N$ would be similarly center on 1, and if this phenomenon occurs, we can then adjust the prior of $U'$ to pull the posterior distribution of $U'$ out of 1. Hence, the ratio reparameterization can prevent degeneration of the causal model.

However, from our practice, we find the ratio reparameterization cannot prevent model non-identifiability from reversion of the sign of the original unmeasured confounder. The magnitude of $U'$ can be estimated well, but the model can produce a bimodal posterior distribution symmetric about the $y$-axis for $U'$. In other words, the model may regard $\beta U' = (-\beta)~(-U')$. Sometimes, fixing the sign of $\beta$ to be positive can help, but may still be prone to other model problems. If bimodal posterior distributions occurs, we may first check the models, especially to see if $\beta$ is estimated to be zero, but may further have to try another reparameterization.

We find the random intercept reparameterization and the ratio reparameterization are very useful in practice. There are many other reparameterizations in addition to those mentioned above. One reparameterization cannot resolve model-nonidentifiability in all data nor all types of model-nonidentifiability. Reparameterization is a good approach to resolve model-nonidentifiability from shifting or scaling of the unmeasured confounder, but after this type of model-nonidentifiability is prevented, reversion of the sign of the unmeasured confounder and degeneration of the causal model can still occur. We thus suggest choosing or developing a reparameterization that can resolve all types of model-nonidentifiability in the data context. The principle in developing a reparameterization is to reduce the number of parameters used in the models and thus reduce model flexibility.

\subsection{Multiple unmeasured confounders}
\label{sec:u-5}

Now we consider a more complex situation by assuming there are two unmeasured confounders. Their counterfactual variables are not mentioned here but are similarly used as in former sections. Their observed variables are denoted by $U_1^{obs}$ and $U_2^{obs}$. We assume two unmeasured confounders are independent. the exposure and outcome distributions then become
\begin{eqnarray}
W^{obs} &=& \alpha_0 + \alpha_1 M^{obs} + \alpha_2 U_1^{obs} + \alpha_3 U_2^{obs} + \eta, 
\label{equ:unmeasured-examp-w-2} 
\\
Y^{obs} &=& \beta_0 + \beta_1 W^{obs} + \beta_2 M^{obs} + \beta_3 U_1^{obs} + \beta_4 U_2^{obs} + \varepsilon.
\label{equ:unmeasured-examp-y-2}
\end{eqnarray}

We want to estimate two unmeasured confounders in equations \ref{equ:unmeasured-examp-w-2} and \ref{equ:unmeasured-examp-y-2} through the missing variable model, in order to obtain unbiased average treatment effect estimates. The first method is an implementation of equations \ref{equ:unmeasured-examp-w-2} and \ref{equ:unmeasured-examp-y-2}, that is, to directly declare two unmeasured confounders $U_1^{obs}$ and $U_2^{obs}$ as two parameter vectors in RStan \autocite{stan_development_team_rstan_2022} and put two parameter vectors as two covariates in the models for the exposure and the outcome. Four types of model non-identifiability with one unmeasured confounder also apply to Bayesian models with two unmeasured confounders and reparameterization can also apply to each unmeasured confounder. This approach is valid but involves much more parameters.

To simplify Bayesian models, we proposed the second method that two unmeasured confounders are combined into a single new unmeasured confounder. This method is a reparameterization, though not developed to solve model non-identifiability. We define a new unmeasured confounder $U'$ in the outcome distribution such that $U' = \beta_0+\beta_3~U_1^{obs}+\beta_4~U_2^{obs}$. Then the exposure and outcome distributions become
\begin{eqnarray}
Y^{obs} &=& \beta_1 W^{obs} + \beta_2 M^{obs} + U' + \varepsilon \nonumber \\
\Longrightarrow ~~~ W^{obs} &=& \alpha_0 + \alpha_1 M^{obs} + \frac{\alpha_2}{\beta_3}~\beta_3~U_1^{obs} + \frac{\alpha_3}{\beta_4}~\beta_4~U_2^{obs} + \eta  \nonumber \\ 
&=& \left\{\alpha_0-\beta_0~(\frac{\alpha_2}{\beta_3}+\frac{\alpha_3}{\beta_4})\right\} + \alpha_1 M^{obs} + \frac{\alpha_2}{\beta_3}~(\beta_0 + \beta_3~U_1^{obs}+\beta_4~U_2^{obs}) \nonumber \\
&& +\frac{\alpha_3}{\beta_4}~(\beta_0 + \beta_3~U_1^{obs}+\beta_4~U_2^{obs}) -\left(\frac{\alpha_2}{\beta_3}~\beta_4~U_2^{obs} + \frac{\alpha_3}{\beta_4}~\beta_3~U_1^{obs}\right) + \eta  \nonumber \\
&=& \left\{\alpha_0-\beta_0~\left(\frac{\alpha_2}{\beta_3}+\frac{\alpha_3}{\beta_4}\right)\right\} + \alpha_1 M^{obs} + \left(\frac{\alpha_2}{\beta_3}+\frac{\alpha_3}{\beta_4}\right)~U' \nonumber \\
&& -\left(\frac{\alpha_2}{\beta_3}~\beta_4~U_2^{obs} + \frac{\alpha_3}{\beta_4}~\beta_3~U_1^{obs}\right) + \eta \nonumber \\
&\doteq& \alpha_0' + \alpha_1 M^{obs} + \alpha_2' U' + U'' + \eta,
\label{eqn:models-2ukc-combined}
\end{eqnarray}
where $U''$ is the residual unmeasured confounder term after reparameterization.

When the condition $\frac{\alpha_2}{\beta_3} = \frac{\alpha_3}{\beta_4}$ holds, which means for each unmeasured confounder its unmeasured confounding effect is identically proportionally distributed to the exposure and the outcome, then the exposure distribution in equation \ref{eqn:models-2ukc-combined} can be written as
\begin{eqnarray}
W^{obs} &=& \alpha_0 + \alpha_1 M^{obs} + \frac{\alpha_2}{\beta_3}~\beta_3~U_1^{obs} + \frac{\alpha_3}{\beta_4}~\beta_4~U_2^{obs} + \eta  \nonumber \\ 
&=& \left\{\alpha_0-\beta_0~\frac{\alpha_2}{\beta_3}\right\} + \alpha_1 M^{obs} + \frac{\alpha_2}{\beta_3}~(\beta_0 + \beta_3~U_1^{obs}+\beta_4~U_2^{obs}) + \eta \nonumber \\
&=& \left\{\alpha_0-\beta_0~\frac{\alpha_2}{\beta_3}\right\} + \alpha_1 M^{obs} + \frac{\alpha_2}{\beta_3}~U' + \eta \nonumber \\
&\doteq& \alpha_0' + \alpha_1 M^{obs} + \alpha_2' U' + \eta,
\label{eqn:models-2ukc-combined-w-withcond}
\end{eqnarray}
where there is no residual unmeasured confounder term $U''$. When $\frac{\alpha_2}{\beta_3} \neq \frac{\alpha_3}{\beta_4}$, which seems to be the usual situation in reality, and we use equation \ref{eqn:models-2ukc-combined-w-withcond} as the exposure distribution, then the residual unmeasured confounder term $U''$ will not be adjusted for in the exposure distribution and there should be residual confounding. 

To further reduce residual confounding, we may treat $U''$ as another completely missing variable and build the models with $U'$ and $U''$ together. The implementation can be using a random intercept as $\alpha_0' + U''$ in the exposure distribution in addition to $U'$. However, the analytical relationships between $U'$ and $U''$ cannot be incorporated in the exposure distribution. On the other hand, this approach is similar to directly using two unmeasured confounders in the first method, which means we can use the first method instead. When we want to simplify the models, we may drop $U''$ out of Bayesian models, use one reparameterized unmeasured confounder only and let residual confounding be in Bayesian models. Then the strength of residual confounding may be associated with the amount of bias in the average treatment effect estimates.

Usually we assume two unmeasured confounders are Normally distributed, as 
\begin{eqnarray}
U_1^{obs} &\sim& N(\mu_1,~\sigma_1), \nonumber \\
U_2^{obs} &\sim& N(\mu_2,~\sigma_2).
\end{eqnarray}

The true distributions of two unmeasured confounders may not be Normal distributions, but their asymptotic distributions in large samples will be Normal distributions. Hence, we can use normal approximation on the distributions of unmeasured confounders. Then $U'$, the linear transformation on two unmeasured confounders, is also Normally distributed, as 
\begin{eqnarray}
U' &\sim& N(\beta_0+\beta_3~\mu_1+\beta_4~\mu_2,~\sqrt{\beta_3^2~\sigma_1^2+\beta_4^2~\sigma_2^2}) \nonumber \\
&\doteq& N(\mu, \sigma), \nonumber 
\end{eqnarray}
which means we can use a Normal prior for $U'$.

In reality, we do not know how many there are unmeasured confounders. It is plausible that there can be more than two unmeasured confounders. Hence, we assume there are more than two unmeasured confounders. The number of unmeasured confounders is unmeasured but is assumed to be finite. This number is denoted by $K$, where $K \in \mathbb{Z}^+$, $K > 2$. The unmeasured confounders are denoted by $U_1^{obs}, \ldots, U_K^{obs}$. the exposure and outcome distributions then become
\begin{eqnarray}
W^{obs} &=& \alpha_0 + \alpha_1 M^{obs} + \alpha_2 U_1^{obs} + \ldots + \alpha_{K+1} U_K^{obs} + \eta, 
\label{equ:unmeasured-examp-w-more} 
\\
Y^{obs} &=& \beta_0 + \beta_1 W^{obs} + \beta_2 M^{obs} + \beta_3 U_1^{obs} + \ldots +  \beta_{K+2} U_K^{obs} + \varepsilon.
\label{equ:unmeasured-examp-y-more}
\end{eqnarray}

In this case, it is not feasible to build the missing variable model by directly using more than two unmeasured confounders as covariates, because the number of unmeasured confounders used in the models is hard to determine, and even if we choose a reasonable number, too many completely missing variables in the models can lead to greater modelling complexity and difficulty. One plausible approach is to combine all the unmeasured confounders into a single new unmeasured confounder. This approach is also a reparameterization. This means, we define a new unmeasured confounder $U'$ in the outcome distribution such that $U' = \beta_0+\beta_3 U_1^{obs} + \ldots +  \beta_{K+2} U_K^{obs}$. Then the exposure can also be written as 
\begin{eqnarray}
W^{obs} &\doteq& \alpha_0' + \alpha_1 M^{obs} + \alpha_2' U' + U'' + \eta,
\label{eqn:expo-more-ukc}
\end{eqnarray}
where $U''$ is still the residual unmeasured confounder term after reparameterization.

To further reduce residual confounding, we can then use a random intercept as $\alpha_0' + U''$ in equation \ref{eqn:expo-more-ukc}. If we ignore $U''$, the models have the same formula as the models with one reparameterized unmeasured confounder. Hence, we may regard the models with one reparameterized unmeasured confounder as the models that truly have one unmeasured confounder, or the models that truly have multiple unmeasured confounders and also residual confounding due to partial adjustment for unmeasured confounders.

The models with one reparameterized unmeasured confounder have several advantages. First, they are the simplest models that adjust for unmeasured confounding and modelling can be simpler. Second, they may be enough to adjust for unmeasured confounding. In practice, after we adjust for many measured confounders, there should not be too many unmeasured confounders. Hence, it may be plausible to assume there is one unmeasured confounder. Third, they can represent the models with multiple unmeasured confounders. Although they are still prone to residual confounding, residual confounding may be negligible, especially when many measured confounders have been adjusted for.

Then we discuss how model non-identifiability is associated with the models with one reparameterized unmeasured confounder in the presence of two or multiple unmeasured confounders. Four types of model non-identifiability with one measured confounder can be prevented by the models with one reparameterized unmeasured confounder in the presence of two or multiple unmeasured confounders, through similar reasoning in the presence of one unmeasured confounder. However, we found another type of model non-identifiability in the presence of two or multiple unmeasured confounders from simulation study. It is multiple posterior solutions due to residual confounding in the presence of multiple unmeasured confounders. 

Under this new type of model non-identifiability, some parameters have multi-modal posterior distributions. For example, the coefficient of the reparameterized unmeasured confounder may have a bimodal posterior distribution. In each posterior solution, Bayesian models are identified, which means they do not have the other four types of model non-identifiability. Posterior estimates of the reparameterized unmeasured confounder can be different across different posterior solutions. In some posterior solutions, the reparameterized unmeasured confounder may adjust for the combined unmeasured confounding effects better. Multiple posterior solutions can include unbiased solutions and biased solutions. Hence, there is a risk that we cannot identify unbiased solutions. 

One possible reason to explain existence of this new type of model non-identifiability is that estimation of the reparameterized unmeasured confounder may be affected by or interacted with the unadjusted residual unmeasured confounder term. Hence, we think it is residual confounding that introduces extra flexibility in estimation of the reparameterized unmeasured confounder and thus leads to multiple posterior solutions. However, our opinion has not been proved theoretically. And the models with two reparameterized unmeasured confounders also show multiple posterior solutions. Hence, there may be some other reasons. We have not yet developed any method to eliminate this new type of model non-identifiability. We suggest in data analysis reporting all posterior solutions where Bayesian models should be identified, if any. More details are given in simulation study.

\section{Extension to multiple multivalued exposures}
\label{sec:ext}

All the methods including the Bayesian causal framework and estimation of unmeasured confounders that have been described so far are for one multivalued exposure and one outcome. We extend these methods into the settings with multiple multivalued exposures and one outcome. Multiple exposures and the outcome are assumed to be Normally distributed. We give a demonstration of three multivalued exposures through simplest causal frameworks, because our data analysis will use three exposures and more complex frameworks can be easily extended from simplest ones. We assume the past exposures have a direct effect on the future exposures. Hence, there is a temporal order among three exposures. Measured and unmeasured confounders are still assumed to be time-independent. We assume three exposures have common measured confounders, but we consider two scenarios where three exposures have common or different unmeasured confounders. It is straightforward to understand that three exposures have common unmeasured confounders, but different unmeasured confounders may be able to account for temporal differences between three exposures. For example, one unmeasured confounder related to the first exposure can be different from another unmeasured confounder related to the second exposure that happens before the second exposure but after the first exposure. In addition, if known confounders are different for three exposures but only common known confounders are measured, then different unmeasured confounders can make up for the unmeasured known confounders. 

Three multivalued exposures are denoted by $W^1$, $W^2$ and $W^3$ in temporal order. For example, $W^1$ is the first exposure. $W^h=(W^h_1, \ldots, W^h_i, \ldots, W^h_N)$, for $i = 1, 2, \ldots, N$ and $h = 1, 2, 3$. We use a finite discrete set $\{0,1,\ldots,J\}$ as the domain for $W^h_i$, with $J$ finite and measured. Three exposures have the same natural zero level as 0.

Three exposures all have a direct effect on the outcome. Hence, the denotation of the counterfactual outcome is modified to $Y(Z, W^1(Z), W^2(Z), W^3(Z))$, for $Z \in \{1, 2\}$. The other counterfactual and observed variables related to three multivalued exposures will have a superscript ``$^h$'' to indicate the exposure to which they correspond. For example, $W^1(Z)$ is the counterfactual first exposure and $W^{1, obs}$ is the observed first exposure. More notations may be needed and will be given in text.

\subsection{Three exposures with common unmeasured confounders}

We assume there is one unmeasured confounder and three exposures have this unmeasured confounder in common. This unmeasured confounder is still denoted by $U$.

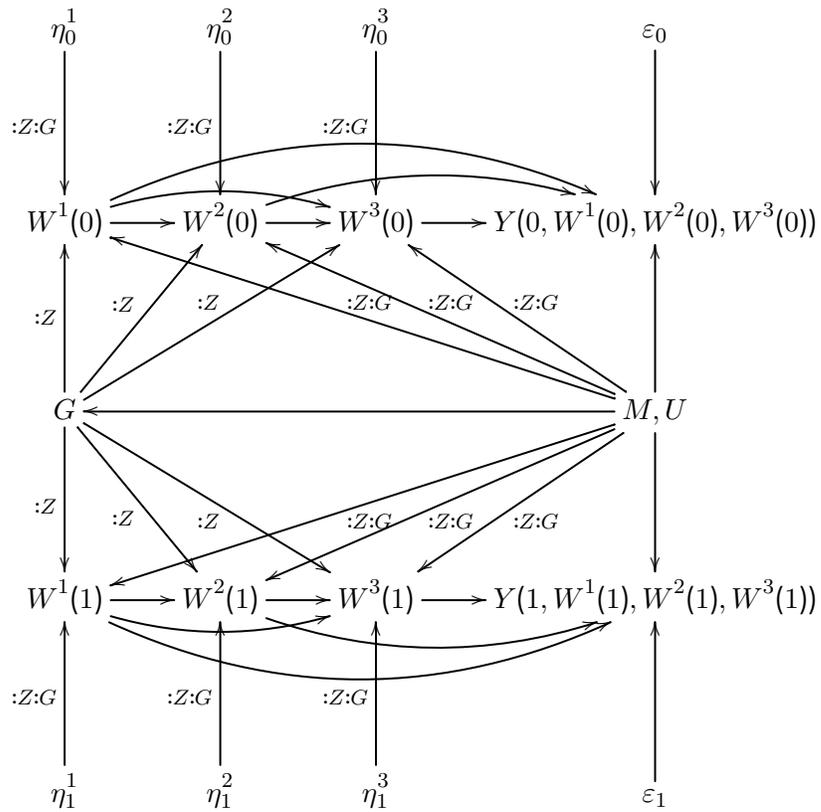
\begin{figure}[htbp]
    \centering
\xymatrix{
&&\eta_0^1 \ar_{:Z:G}[dd]& \eta_0^2 \ar_{:Z:G}[dd]& \eta_0^3 \ar_{:Z:G}[dd]& \varepsilon_0 \ar[dd] \\
&&&&& \\ 
&&W^1(0) \ar[r] \ar@/^1pc/[rr] \ar@/^2.5pc/[rrr] & W^2(0) \ar[r] \ar@/^1.5pc/[rr] & W^3(0) \ar[r] & Y(0, W^1(0), W^2(0), W^3(0)) \\
&&&&& \\ 
&&G \ar_{:Z}[dd] \ar_{:Z}[ddr] \ar_{:Z}[ddrr] \ar^{:Z}[uu] \ar^{:Z}[uur] \ar^{:Z}[uurr] & & & M, U \ar[dd] \ar^{:Z:G}[ddl] \ar^{:Z:G}[ddll] \ar^{:Z:G}[ddlll] \ar[uu] \ar_{:Z:G}[uul] \ar_{:Z:G}[uull] \ar_{:Z:G}[uulll] \ar[lll]\\
&&&&& \\ 
&&W^1(1) \ar[r] \ar@/_1pc/[rr] \ar@/_2.5pc/[rrr] & W^2(1) \ar[r] \ar@/_1.5pc/[rr] & W^3(1) \ar[r] & Y(1, W^1(1), W^2(1), W^3(1)) \\
&&&&& \\ 
&&\eta_1^1 \ar^{:Z:G}[uu]& \eta_1^2 \ar^{:Z:G}[uu]& \eta_1^3 \ar^{:Z:G}[uu]& \varepsilon_1 \ar[uu] 
}  
    \caption{Causal directed acyclic graph for counterfactual variables. The effect of $G$ on counterfactual exposures is modified by $Z$. The effects of $M$ and $U$ on counterfactual exposures are modified by $Z$ and $G$. The effect of $\eta_i^h$ is modified by $Z$ and $G$, for $i = 1, 2$, $h=1, 2, 3$.}
    \label{fig:cdag-3expo-1u}
\end{figure}

Figure \ref{fig:cdag-3expo-1u} shows the simplest causal framework for counterfactual variables in the three-exposure setting that uses the same assumptions as the simplest causal framework for counterfactual variables in the one-exposure setting.

With identical condition assumptions \ref{assump:idcond} and \ref{assump:idcond-w}, $\eta \doteq \eta_0^h = \eta_1^h$, $\varepsilon \doteq \varepsilon_0^h = \varepsilon_1^h$, for $h = 1, 2, 3$. Then structural equations for the counterfactual exposures and outcomes are developed from figure \ref{fig:cdag-3expo-1u}. To develop structural equations, we assume the control intervention is the natural zero level of three exposures. This is because this assumption can hold in the PreventS trial data and we will explain it in the results section. Structural equations without this assumption can be easily developed similarly to those for one exposure. For both one-sided and two-sided noncompliance, we still have
\begin{eqnarray}
G &=& 
    \begin{cases}
      0, & \text{if $g = co$} \\
      1, & \text{otherwise}  \nonumber 
    \end{cases} \\
h(Z, G) &=& 
    \begin{cases}
      0, & \text{if $Z = 0$ and $G = 0$} \\
      1, & \text{otherwise}  \nonumber 
    \end{cases} 
\end{eqnarray}

Structural equations for the counterfactual exposures and outcomes now become
\begin{eqnarray}
W^1(0) &=& h(Z=0, G)~(\alpha_0^1 + \alpha_1^1 G + \alpha_2^1 M + \alpha_3^1 U + \eta), \nonumber \\
W^2(0) &=& h(Z=0, G)~(\alpha_0^2 + \alpha_1^2 G + \alpha_2^2 M + \alpha_3^2 U \nonumber \\
&& + ~\alpha_4^2 W^1(0) + \eta)  \nonumber \\
W^3(0) &=& h(Z=0, G)~(\alpha_0^3 + \alpha_1^3 G + \alpha_2^3 M + \alpha_3^3 U  \nonumber \\
&& +~ \alpha_4^3 W^1(0) + \alpha_5^3 W^2(0) + \eta), \nonumber \\
W^1(1) &=& h(Z=1, G)~(\alpha_0^1 + \alpha_1^1 G + \alpha_2^1 M + \alpha_3^1 U + \eta), \nonumber \\
W^2(1) &=& h(Z=1, G)~(\alpha_0^2 + \alpha_1^2 G + \alpha_2^2 M + \alpha_3^2 U  \nonumber \\
&& + ~\alpha_4^2 W^1(1) + \eta), \nonumber \\
W^3(1) &=& h(Z=1, G)~(\alpha_0^3 + \alpha_1^3 G + \alpha_2^3 M + \alpha_3^3 U \nonumber \\
&& + ~\alpha_4^3 W^1(1) + \alpha_5^3 W^2(1) + \eta),  \nonumber \\
Y(0, W^1(0), W^2(0), W^3(0)) &=& \beta_0 + \beta_1 W^1(0) + \beta_2 W^2(0) + \beta_3 W^3(0)  \nonumber \\
&& +~ \beta_4 M + \beta_5 U + \varepsilon,  \nonumber \\
Y(1, W^1(1), W^2(1), W^3(1)) &=& \beta_0 + \beta_1 W^1(1) + \beta_2 W^2(1) + \beta_3 W^3(1) \nonumber \\
&& +~ \beta_4 M + \beta_5 U + \varepsilon.
\label{equ:struceqn-3expo-1y}
\end{eqnarray}

Then the simplest causal framework and structural equations from counterfactual variables are converted to observed variables. Figure \ref{fig:cdag-3expo-1u-obs} shows the simplest causal framework for observed variables.
\begin{figure}[htbp]
    \centering
\xymatrix{
&&&\eta^1 \ar_{:Z^{obs}:G^{obs}}[dd]& \eta^2 \ar_{:Z^{obs}:G^{obs}}[dd]& \eta^3 \ar_{:Z^{obs}:G^{obs}}[dd]& \varepsilon \ar[dd] \\
&&&&&& \\ 
&&&W^{1, obs} \ar[r] \ar@/^1pc/[rr] \ar@/^2.5pc/[rrr] & W^{2, obs} \ar[r] \ar@/^1.5pc/[rr] & W^{3, obs} \ar[r] & Y^{obs} \\
&&&&&& \\ 
&&&G^{obs} \ar^{:Z^{obs}}[uu] \ar^{:Z^{obs}}[uur] \ar^{:Z^{obs}}[uurr] & & & M^{obs}, U^{obs} \ar[uu] \ar_{:Z^{obs}:G^{obs}}[uul] \ar_{:Z^{obs}:G^{obs}}[uull] \ar_{:Z^{obs}:G^{obs}}[uulll] \ar[lll]
}  
    \caption{Causal directed acyclic graph for observed variables.}
    \label{fig:cdag-3expo-1u-obs}
\end{figure}

Structural equations for three observed exposures and the observed outcome are
\begin{eqnarray}
W^{obs} &=& W(Z = Z^{obs}) \nonumber \\
Y^{obs} &=& Y(Z = Z^{obs}, W^1(1)= W^{1, obs}, W^2(1)= W^{2, obs}, W^3(1)= W^{3, obs}) 
\label{equ:struceqn-3expo-1u-obs}
\end{eqnarray}

We can also find the complete data model $f(y^{obs}, w^{1, obs}, w^{2, obs}, w^{3, obs}|Z^{obs}, G^{obs}, M^{obs},$ $U^{obs}, \theta)$ similarly based on the relationships between counterfactual and observed exposures and outcomes, as
\begin{eqnarray}
&& f(y^{obs}, w^{1, obs}, w^{2, obs}, w^{3, obs}~|~Z^{obs}, G^{obs}, M^{obs}, U^{obs}, \theta) \nonumber \\
&=& \prod_i^N~f_i(y^{obs}, w^{1, obs}, w^{2, obs}, w^{3, obs}~|~Z^{obs}, G^{obs}, M^{obs}, U^{obs}, \theta) \nonumber \\
&& (\text{SUTV assumption}) \nonumber \\
&=& \prod_i^N~ \{ {\rm I}_i(Z^{obs}=0)~f_i(y^{obs}, w^{1, obs}, w^{2, obs}, w^{3, obs}~|~Z^{obs}=0, G^{obs}, M^{obs}, U^{obs}, \theta)  \nonumber \\
&& + ~ {\rm I}_i(Z^{obs}=1)~f_i(y^{obs}, w^{1, obs}, w^{2, obs}, w^{3, obs}~|~Z^{obs}=1, G^{obs}, M^{obs}, U^{obs}, \theta)\} \nonumber \\
&=& \prod_{i \in \rm UC}f_i(y(0,w^1(0), w^2(0), w^3(0)), w^1(0), w^2(0), w^3(0)~|~Z^{obs}=0, G^{obs}, M^{obs},  \nonumber \\
&& U^{obs}, \theta) \cdot \prod_{i \in \rm HWC}f_i(y(1,w^1(1), w^2(1), w^3(1)), w^1(1), w^2(1), w^3(1)~|~Z^{obs}=1,  \nonumber \\
&& G^{obs}, M^{obs}, U^{obs}, \theta) \nonumber \\
&& (\text{Equation \ref{equ:struceqn-3expo-1u-obs}}) \nonumber \\
&=& \prod_{i \in \rm UC}f_i(y(0,w^1(0), w^2(0), w^3(0)), w^1(0), w^2(0), w^3(0)~|~G^{obs}, M^{obs}, U^{obs}, \theta)  \nonumber \\
&& \cdot \prod_{i \in \rm HWC}f_i(y(1,w^1(1), w^2(1), w^3(1)), w^1(1), w^2(1), w^3(1)~|~G^{obs}, M^{obs}, U^{obs}, \theta) \nonumber \\
&& (\text{Randomization assumption}) \nonumber \\
&=& \prod_{i \in \rm UC} \left\{ f_i(y(0,w^1(0), w^2(0), w^3(0))~|~ w^1(0), w^2(0), w^3(0), G^{obs}, M^{obs}, U^{obs}, \theta) \right. \nonumber \\
&&\quad\quad \cdot~ f_i(w^3(0)~|~ w^1(0), w^2(0), G^{obs}, M^{obs}, U^{obs}, \theta) \nonumber \\
&& \quad\quad \cdot~ f_i(w^2(0)~|~ w^1(0), G^{obs}, M^{obs}, U^{obs}, \theta) \nonumber \\
&& \quad\quad \left. \cdot~f_i(w^1(0)~|~G^{obs}, M^{obs}, U^{obs}, \theta) \right\} \nonumber \\
&& \cdot ~ \prod_{i \in \rm HWC} \left\{ f_i(y(1,w^1(1), w^2(1), w^3(1))~|~ w^1(1), w^2(1), w^3(1), G^{obs}, M^{obs}, U^{obs}, \theta) \right. \nonumber \\
&& \quad\quad \cdot~ f_i(w^3(1)~|~w^1(1), w^2(1), G^{obs}, M^{obs}, U^{obs}, \theta) \nonumber \\
&& \quad\quad \cdot~ f_i(w^2(1)~|~ w^1(1), G^{obs}, M^{obs}, U^{obs}, \theta) \nonumber \\
&& \quad\quad \left. \cdot~f_i(w^1(1)~|~G^{obs}, M^{obs}, U^{obs}, \theta) \right\} \nonumber
\end{eqnarray}

From the complete data model, we can then specify the priors for unknown parameters including $U$ and $\theta$, and also include missing data and may derive more complex complete data models for particular questions.

We need to estimate the unmeasured confounder $U^{obs}$ in the simplest causal framework. We notice that equation \ref{equ:struceqn-3expo-1y} is also non-identifiable. Four types of model non-identifiability with one unmeasured confounder and one exposure also exist Bayesian models with multiple exposures. The third type of model non-identifiability needs some modification for three exposures, while the other types of model non-identifiability can directly apply to three exposures. In the third type of model non-identifiability, degeneration of the causal model is defined to occur when the coefficient of the unmeasured confounder in the outcome distribution is estimated to be zero, or when at least one coefficient of the unmeasured confounder in three exposure distributions is estimated to be zero, or when the unmeasured confounder itself is estimated to be zero. If the coefficient of the unmeasured confounder is estimated to be zero in the model of $W^{1, obs}$ but not in the other two exposure distributions, the meaning of structural equations would change, because equation \ref{equ:struceqn-3expo-1y} becomes another system of equations where we adjust for one common unmeasured confounder of $W^{2, obs}$ and $W^{3, obs}$. In this case, some unmeasured confounding effect is still adjusted for, which means the causal model does not degenerate to an association model but another causal model with a different definition of the unmeasured confounder. On the other hand, this ``common'' unmeasured confounder may actually be a confounder between the outcome and some exposures, which means three exposures may not actually have this unmeasured confounder in common. Then we should expect the coefficients of this unmeasured confounder for the exposures that do not have this confounder to be estimated to be insignificant, which will not change the structure of the structural equations \ref{equ:struceqn-3expo-1y} and will also be able to find the correct confounded pairs without human judgement.

Methods to resolve model non-identifiability can apply to the three-exposure setting. For example, reparameterization can be carried out in the same way. We describe how to develop the random intercept reparameterization in three exposures and one outcome. Previously we used the models of compliers in the treatment arm for descriptions, but now we use the models of the counterfactual variables in the treatment arm from equation \ref{equ:struceqn-3expo-1y} for descriptions. We define a new unmeasured confounder $U'$ in the outcome distribution such that $U' = \beta_0 + \beta_5 U$ and then transform $U$ into $U'$ in the exposure distributions. The reparameterized models of the counterfactual variables in the treatment arm become
\begin{eqnarray}
W^1(1) &=& {\alpha_0^1}' + \alpha_1^1 G + \alpha_2^1 M + {\alpha_3^1}' U' + \eta, \nonumber \\
W^2(1) &=& {\alpha_0^2}' + \alpha_1^2 G + \alpha_2^2 M + {\alpha_3^2}' U' + \alpha_4^2 W^1(1) + \eta, \nonumber \\
W^3(1) &=& {\alpha_0^3}' + \alpha_1^3 G + \alpha_2^3 M + {\alpha_3^3}' U' + \alpha_4^3 W^1(1) + \alpha_5^3 W^2(1) + \eta, \nonumber \\
Y(1, W^1(1), W^2(1), W^3(1)) &=& \beta_1 W^1(1) + \beta_2 W^2(1) + \beta_3 W^3(1) + \beta_4 M + U' + \varepsilon, \nonumber
\end{eqnarray}
where the superscript ``$'$'' indicates new coefficients.

We can also assume there are multiple unmeasured confounders that three exposures have in common. For example, we assume there are two common unmeasured confounders denoted by $U_1$ and $U_2$. They can be directly added in equation \ref{equ:struceqn-3expo-1y}. Methods to resolve model non-identifiability still apply. Similarly, after two unmeasured confounders are reparameterized into a single new unmeasured confounder in the outcome distribution, there is a residual unmeasured confounder term in the exposure distributions. If we want to adjust for more residual confounding, we can use a random intercept in each exposure distribution, which greatly increases model complexity. If we want simpler models and do not think there is great residual confounding, then we ignore the residual unmeasured confounder term, which turns the models into the models with one reparameterized unmeasured confounder and residual confounding that we in fact do not know whether is great or not actually. More unmeasured confounders should be used if data can provide enough information and the models can afford them; however, these necessary conditions are hard to judge in practice. In addition, model non-identifiability from multiple posterior solutions due to residual confounding may also occur in the presence of two or multiple unmeasured confounders.

\subsection{Three exposures with different unmeasured confounders and more extensions}

In this section, we consider more complex scenarios in order to give a more complete description of possible applications using our methods. We only mention some possible solutions and aim to stimulate more ideas about our methods, rather than try to solve all the scenarios. Structural equations for the counterfactual variables in the treatment arm instead of more complex causal graphs are given in some scenarios.

In the last section, we assume three exposures have all measured confounders in common. However, in reality, they may have different known confounders and different unknown confounders. To account for these differences, we consider the scenario where three exposures are still assumed to have all measured confounders in common but different unmeasured confounders. Different unmeasured confounders used in the models can represent both different unknown confounders and different unmeasured known confounders.

A simpler scenario is described where we assume each exposure has one different unmeasured confounder and there is no common unmeasured confounder. The unmeasured confounder is denoted by $U^h$, for $h = 1, 2, 3$ to indicate with which exposure it is associated. Structural equations for the counterfactual variables in the treatment arm become
\begin{eqnarray}
W^1(1) &=& \alpha_0^1 + \alpha_1^1 G + \alpha_2^1 M + \alpha_3^1 U^1 + \eta, \nonumber \\
W^2(1) &=& \alpha_0^2 + \alpha_1^2 G + \alpha_2^2 M + \alpha_3^2 U^2 + \alpha_4^2 W^1(1) + \eta, \nonumber \\
W^3(1) &=& \alpha_0^3 + \alpha_1^3 G + \alpha_2^3 M + \alpha_3^3 U^3 + \alpha_4^3 W^1(1) + \alpha_5^3 W^2(1) + \eta, \nonumber \\
Y(1, W^1(1), W^2(1), W^3(1)) &=& \beta_0 + \beta_1 W^1(1) + \beta_2 W^2(1) + \beta_3 W^3(1) + \beta_4 M  \nonumber \\
&& + \beta_5 U^1 + \beta_6 U^2 + \beta_7 U^3 + \varepsilon.
\end{eqnarray}

The random intercept reparameterization starts with each exposure distribution where a new unmeasured confounder is defined as $U^{h, '} = \alpha_0^h + \alpha_3^h U^h$, for $h = 1, 2, 3$ and then the outcome distribution is transformed to have $U^{h, '}$, for $h = 1, 2, 3$.

The assumption of different unmeasured confounders may be unrealistic. When there are many exposures, such as ten exposures over time, the model for one outcome will contain many different unmeasured confounders. If many measured confounders have already been adjusted for in the outcome distribution, then it is implausible that the outcome distribution can afford so many different unmeasured confounders. Hence, it is very likely that some different unmeasured confounders are actually not confounders and their coefficients are estimated to be insignificant in the outcome distribution, or some exposures may have common unmeasured confounders.

If our assumption becomes that some exposures have different unmeasured confounders while some have common unmeasured confounders. Then a problem arises: how can we identify which exposures have different unmeasured confounders from the other exposures? We may have to use many common unmeasured confounders and let the models tell us which coefficients of these unmeasured confounders are estimated to be insignificant. As in the last section, if the coefficient of one unmeasured confounder is estimated to be insignificant for only one exposure, then it is the common unmeasured confounder for the other exposures. However, it is inevitable that we have to include many common unmeasured confounders in the models and the models may not be able to be built with so much flexibility and uncertainty. Reparameterization can help simplify the models, but there can be much residual confounding. 

More complicated scenarios include: (1) each exposure has multiple different unmeasured confounder, (2) in addition to different unmeasured confounder, there are common unmeasured confounders. The missing variable model can also be used: treat every distinct unmeasured confounder as an individual completely missing variable and include all of them in the model. There will be a large amount of unknown parameters. We may have to make simpler assumptions instead to make the models built.

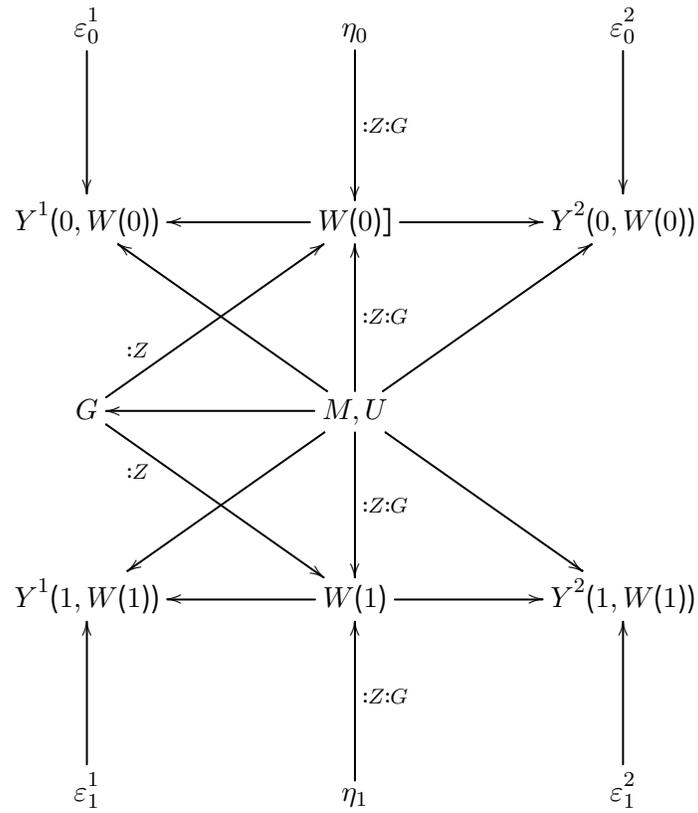
\begin{figure}[htbp]
    \centering
\xymatrix{
&& \varepsilon_0^1 \ar[dd]& & \eta_0 \ar^{:Z:G}[dd]& & \varepsilon_0^2 \ar[dd] \\
&&&&&& \\
&&Y^1(0, W(0)) & & W(0) ]\ar[ll] \ar[rr]& & Y^2(0, W(0)) \\
&&&&&& \\
&&G \ar^<<<<<<<<{:Z}[uurr] \ar_<<<<<<<<{:Z}[ddrr] && M, U \ar[ll] \ar_{:Z:G}[uu] \ar^{:Z:G}[dd] \ar[uull] \ar[uurr] \ar[ddll] \ar[ddrr]&& \\
&&&&&& \\
&&Y^1(1, W(1)) & & W(1) \ar[ll] \ar[rr] & & Y^2(1, W(1)) \\
&&&&&& \\
&&\varepsilon_1^1 \ar[uu]& & \eta_1 \ar_{:Z:G}[uu]& & \varepsilon_1^2 \ar[uu]
}  
    \caption{Causal directed acyclic graph for counterfactual variables. The effect of $G$ on counterfactual exposures is modified by $Z$. The effects of $M$ and $U$ on counterfactual exposures are modified by $Z$ and $G$. The effect of $\eta_i$ is modified by $Z$ and $G$, for $i = 1, 2$.}
    \label{fig:cdag-2y-1expo-1u}
\end{figure}

So far discussions have been around one outcome. We can also build the models for multiple outcomes together. This requires causal relationships among multiple outcomes, one or multiple exposures, measured and unmeasured confounders should be carefully established in order to understand the associations among multiple outcomes and figure out how to model their possible correlations after conditioning on exposures and confounders. There are too many combinations of outcomes, exposures and confounders, and thus we do not discuss them all. We choose a simpler scenario where there are two outcomes, one exposure, common measured confounders and one common unmeasured confounder. Figure \ref{fig:cdag-2y-1expo-1u} shows the causal graph for counterfactual variables in this scenario.

Two outcomes are independent conditional on the exposure and confounders. Hence, we do not need to model their correlation explicitly in the models conditional on the exposure and confounders. If two outcomes are still dependent conditional on the exposure and confounders, we may use a multivariate distribution to build their models where their correlation is included, such as a bivariate normal distribution with a covariance matrix that has non-zero off-diagonal entries.

Then we assume there are multiple outcomes and one exposure. We also assume they have common measured confounders and different unmeasured confounders. And the outcomes are independent conditional on the exposure and confounders. If we build the models for the outcomes together, the exposure distribution will include measured confounders and all different unmeasured confounders. When the number of the outcomes is relatively large, there can be so many different unmeasured confounders in the exposure distribution that it becomes implausible that the exposure can be affected by so many confounders. Hence, one plausible explanation can be that some outcomes have common unmeasured confounders. A similar problem arises: how can we identify which outcomes have different unmeasured confounders from the other outcomes? We may instead choose to build the models for each outcome separately, or build the models for all the outcomes together with simpler assumptions. One simpler assumption is that the outcomes have one common unmeasured confounder. This assumption is plausible for the outcomes that measure the same thing, such as the same disease risk, but seems implausible for the outcomes that measure different things, such as different disease risks. Hence, how to make plausible assumptions is also a challenging task.

A final note is that, application viability of complex scenarios should be determined in the data context.

\section{Simulation study}
\label{sec:sim}

Simulation study is carried out to illustrate if the average treatment effect is well estimated in some scenarios in the presence of unmeasured confounding. It is not comprehensive and we do not intend to carry out massive simulation study to discover every characteristic of every methodological detail. We aim to better understand our methods including the Bayesian causal framework and estimation of unmeasured confounders before applying them in real data. More simulation study may be needed to investigate in detail computational performance of our methods.


Data is simulated from the causal framework for observed variables, without knowing counterfactual variables. We think the causal framework for counterfactual variables helps us interpret the true data generating mechanism but itself may not be the true data generating mechanism, and simulating data from a causal framework for counterfactual variables may be in favor of this framework in inference.

In our simulation study, the sample size $N = 300$. A small sample size may be insufficient to estimate the average treatment effect well. We consider a relatively large sample size, comparable to the sample size in the PreventS trial data. We do not consider the effect of the sample size on estimation of the average treatment effect, though this can be done by comparing results from different sample sizes. No missing data is involved in simulation study.

Notations from former sections apply. The compliance behavior $G^{obs}$ is fully identified in the data simulation process but partially identified in simulated data. We pretend to know nothing about the true compliance behavior in analysis. When a causal framework for counterfactual variables is applied to the data, the observed compliance behavior can only be inferred from the observed assignment $Z^{obs}$ and the observed exposure $W^{obs}$. In the presence of one-sided noncompliance, when $Z^{obs} =1$, the known observed compliance behavior is never-taker if $W^{obs}=0$, otherwise is complier. In the presence of two-sided noncompliance, the known observed compliance behavior is never-taker if $Z^{obs} = 1$ and $W^{obs}=0$, and is always-taker if $Z^{obs} = 0$ and $W^{obs}>0$. The observed exposure  $W^{obs}$ and the observed outcome $Y^{obs}$ we use are continuous. The observed exposure is multivalued. The observed outcome is Normally distributed. 

One important step in Bayesian causal inference is to choose a plausible causal framework. The assumptions in the chosen causal framework should be compatible with the data in use. Hence, we first look at some different causal frameworks. Unmeasured confounding should also be well adjusted for, and then we explore different scenarios about unmeasured confounders to see how our estimation methods would work. We investigate the two aspects separately in order to understand each aspect more: the Bayesian causal framework is studied without unmeasured confounders, and estimation of unmeasured confounders is studied given a plausible causal framework.

Simulation study is carried out through the \texttt{rstan} \texttt{R} package \autocite{stan_development_team_rstan_2022}.
Data simulation models are built using the fitted-parameter (``\texttt{Fixed\_param}'') algorithm. Bayesian models are built using the No-U-Turn sampler (NUTS). In each model, we use 4 Markov chains. Each chain has 2000 iterations, from which 1000 iterations are warmup iterations. Other modelling settings except the seed for random number generation are default. Posterior predictive checks and other model diagnostic procedures are carried out if necessary but the intermediate results are not presented. We mainly report estimates from the final fitted models. Sensitivity analysis of the priors is carried out in some scenarios. Unlike simulation study, we will present more complete results from models for the PreventS trial data.

\subsection{When the compliance behavior has no effect on the outcome}

This scenario has one-sided noncompliance, no confounders, no effect of the compliance behavior on the outcome. It becomes a simple regression problem between the exposure and the outcome, and the simplest framework for counterfactual variables is applicable. 

The distributions used in data simulation are described in equation \ref{eqn:sim-1}. The data generating process looks like a two-arm randomized trial. There is an equal probability to be assigned to either arm. We still use terminology of a two-arm randomized trial for further descriptions. The proportion of never-takers in the population is set to be 10\%. The control intervention is observed to be zero and the treatment is observed to be greater than zero. The control intervention is assumed to not be the natural zero level of the treatment so that never-takers cannot be used to build the exposure distribution. We compare two treatment effect sizes. A big treatment effect size is 2, and a small treatment effect size is 0.1. The treatment effect to estimate is the average treatment effect.
\begin{eqnarray}
Z^{obs} &\sim& Bern(0.5), \nonumber \\
G^{obs} &\sim& Bern(0.9), \nonumber \\
G^{obs} &=& 
    \begin{cases}
      0, & \text{if $g = nt$} \\
      1, & \text{if $g = co$}  \nonumber 
    \end{cases} \\
W^{obs} &\sim& 
    \begin{cases}
      N(5, 1)~{\rm T}[0.5,~], & \text{if $Z^{obs}=1$, $G^{obs}=co$} \\
      0, & \text{otherwise}  \nonumber 
    \end{cases} \\
Y^{obs} &\sim& 
    \begin{cases}
      N(1+2~W^{obs}, 1), & \text{with a big treatment effect} \\
      N(1+0.1~W^{obs}, 1), & \text{with a small treatment effect}
    \end{cases}    
\label{eqn:sim-1}
\end{eqnarray}
where $Bern(p)$ is Bernoulli distribution with the probability of success to be $p$, $N(\mu,\sigma)$ is Normal distribution with mean $\mu$ and standard deviation $\sigma$, and $N(\mu,\sigma)~{\rm T}[a,~]$ is Normal distribution left-truncated at $a$.

The Bayesian model is simply
\begin{eqnarray}
Y^{obs} &\sim& N(\beta_0 + \beta_1~W^{obs}, \sigma).
\label{eqn:sim1-model1}
\end{eqnarray}

The results are shown in table \ref{tab:sim1-result}. All the parameter estimates are very accurate compared to their true values, under both effect sizes. The results indicate that the average treatment effect size is not a problem in modelling when the chosen causal framework is plausible. 

\begin{table}[htbp]
\small
  \begin{subtable}{0.9\textwidth}
   \centering
   \begin{tabular}{ccr}
  \hline
  $Z^{obs}$ & $G^{obs}$ & $N$ \\ 
  \hline
\begin{minipage}[t]{0.2\textwidth} \centering 0 \end{minipage} &  \begin{minipage}[t]{0.2\textwidth} \centering $nt$ \end{minipage} & \begin{minipage}[t]{0.15\textwidth} \raggedleft 15 \end{minipage}  \\ 
0 &   $co$ & 138 \\ 
1 &   $nt$ &  12 \\ 
1 &   $co$ & 135 \\ 
   \hline 
   \end{tabular}
   \caption{Simulated data. $N$ is the sample size in each category.}
   \end{subtable}

\vspace{1em}

  \begin{subtable}{1\textwidth}
   \centering
\begin{tabular}{rrrr}
  \hline
Parameter & True value & Prior & Posterior mean (95\% interval)  \\ 
  \hline
Model 1 &&& \\
$e_{ate}$ & 2.00 & $N(0, 1)$ & 2.00 (1.96, 2.04)   \\ 
$\beta_0$ & 1.00 & $N(0, 1)$ & 0.99 (0.84, 1.13)   \\ 
$\sigma$ & 1.00 & $N(0, 1)~{\rm T}[0,~]$ & 0.99 (0.91, 1.07)  \\ 
Model 2 &&& \\  
$e_{ate}$ & 0.10 & $N(0, 1)$ & 0.10 (0.05, 0.14)   \\ 
$\beta_0$ & 1.00 & $N(0, 1)$ & 1.00 (0.84, 1.15)   \\ 
$\sigma$ & 1.00 & $N(0, 1)~{\rm T}[0,~]$ & 1.02 (0.94, 1.10)   \\ 
   \hline
\end{tabular}  
\caption{Comparison of true values and estimates from Bayesian models.}
   \end{subtable}  
 \caption{Simulation results from the simplest causal framework.}
 \label{tab:sim1-result}
\end{table}

\subsection{When the compliance behavior has a direct additive effect on the outcome}

This scenario has one-sided noncompliance, no confounders, a direct additive effect of the compliance behavior on the outcome. Now the compliance behavior becomes a confounder between the exposure and the outcome. We compare the simplest framework and a variation with the direct additive effect of the compliance behavior on the outcome, given that the average treatment effect is big or small, in order to learn if data is sensitive to the causal framework.

The distributions used in data simulation are described in equation \ref{eqn:sim-2}. They are similar to the first simulation scenario, except the outcome. The intercept in the mean of $Y^{obs}$ is 1.5 for never-takers and 1 for compliers. Two different effect sizes are compared. A big 40\% proportion of never-takers in the population and a small 10\% proportion of never-takers in the population are compared.
\begin{eqnarray}
Z^{obs} &\sim& Bern(0.5), \nonumber \\
G^{obs} &\sim& Bern(0.9)~/~Bern(0.6), \nonumber \\
G^{obs} &=& 
    \begin{cases}
      0, & \text{if $g = nt$} \\
      1, & \text{if $g = co$}  \nonumber 
    \end{cases} \\
W^{obs} &\sim& 
    \begin{cases}
      N(5, 1)~{\rm T}[0.5,~], & \text{if $Z^{obs}=1$, $G^{obs}=co$} \\
      0, & \text{otherwise}  \nonumber 
    \end{cases} \\
Y^{obs} &\sim& 
    \begin{cases}
      N(1.5+2~W^{obs}-0.5~G^{obs}, 1), & \text{with a big treatment effect} \\
      N(1.5+0.1~W^{obs}-0.5~G^{obs}, 1), & \text{with a small treatment effect}
    \end{cases}    
\label{eqn:sim-2}
\end{eqnarray}

The Bayesian model from the simplest framework are also in equation \ref{eqn:sim1-model1}. Using the variation, we need to estimate the unknown compliance behavior in the control arm to distinguish compliers from never-takers. A Bernoulli distribution with the unknown probability for success to be the parameter $p$ with a Uniform prior on $[0, ~1]$ is used as the distribution for the compliance behavior. We treat the unknown compliance behavior in the control arm as missing data $G^{obs}_{mis}$. RStan does not support integer optimization, so it is not possible to impute the unknown behavior compliance directly from the Bernoulli distribution. Instead, marginalization is used to integrate $G^{obs}_{mis}$ out of the complete data model $f_{Y^{obs}, W^{obs}, G^{obs}_{mis}}$, so that we do not have to impute the unknown compliance behavior explicitly. The distributions for counterfactual variables through the variation are given in equation \ref{eqn:sim2-model2-potential}. For simplicity, we use $f_N(x ~|~ \mu, \sigma)$ to represent the probability density function of a marginal or conditional Normal distribution with mean $\mu$ and standard deviation $\sigma$ for a random variable $X$.
\begin{eqnarray}
f(y(0, w(0))~|~W(0), G) &=& f_N(y(0, w(0) ~|~ \beta_0 + \beta_1~W(0) + \beta_2~G, \sigma_y), \nonumber \\
f(y(1, w(1))~|~W(1), G) &=& f_N(y(1, w(1)~|~ \beta_0 + \beta_1~W(1) + \beta_2~G, \sigma_y), \nonumber \\
f(w(0)~|~G) &=& {\rm I}(W(0) = 0), \nonumber \\
f(w(1)~|~G) &=& {\rm I}(G = 0)~{\rm I}(w(1) = 0) + {\rm I}(G = 1)~{\rm I}(w(1) \geq 0.5) \nonumber \\
&&\cdot~f_N(w(1)~|~\alpha, \sigma_w), \nonumber \\
f(g) &=& p~{\rm I}(g = 1) + (1-p)~{\rm I}(g = 0).
\label{eqn:sim2-model2-potential}
\end{eqnarray}

From the randomization assumption, $Z^{obs} \upvDash W(0), W(1), Y(0, W(0)), Y(1, W(1))~|~G$. Hence, the distributions in equation \ref{eqn:sim2-model2-potential} can all condition on $Z^{obs}$ without changing the formulation. Then we represent the relationship between the counterfactual variables and the observed variables in a different way in equation \ref{eqn:sim2-model2-observe}.
\begin{eqnarray}
W^{obs} &=& 
    \begin{cases}
      W(0), & \text{if $Z^{obs}=0$} \\
      W(1), & \text{if $Z^{obs}=1$}  \nonumber 
    \end{cases} \\
Y^{obs} &=& 
    \begin{cases}
      Y(0, W(0)), & \text{if $Z^{obs}=0$} \\
      Y(1, W(1)), & \text{if $Z^{obs}=1$}
    \end{cases}    
\label{eqn:sim2-model2-observe}
\end{eqnarray}

Then, the joint distribution of $Y^{obs}, W^{obs}, G^{obs}, G^{obs}_{mis}$ conditional on $Z^{obs}$ is 
\begin{eqnarray}
&& f(y^{obs}, w^{obs}, g^{obs}, g^{obs}_{mis}~|~Z^{obs}) \nonumber \\
 &=& \prod_i^N \left\{ {\rm I}_i(Z^{obs}=0)~f_i(y^{obs}, w^{obs}, g^{obs}, g^{obs}_{mis}~|~Z^{obs} = 0) \right.  \nonumber \\
&& \left. + ~ {\rm I}_i(Z^{obs}=1)~ f_i(y^{obs}, w^{obs}, g^{obs}, g^{obs}_{mis}~|~Z^{obs} = 1)\right\} \nonumber \\
 &=& \prod_i^N \left\{ {\rm I}_i(Z^{obs}=0)~f_i(y(0, w(0))~|~W^{obs}, G^{obs}, G^{obs}_{mis}) \right.  \nonumber \\
 && \cdot~f_i(w(0)~|~ G^{obs}, G^{obs}_{mis})~f_i(g^{obs}, g^{obs}_{mis})\nonumber \\
&& + ~ {\rm I}_i(Z^{obs}=1)~ f_i(y(1, w(1))~|~W^{obs}, G^{obs}, G^{obs}_{mis})  \nonumber \\
&&\left. \cdot ~f_i(w(1)~|~G^{obs}, G^{obs}_{mis})~ f_i(g^{obs}, g^{obs}_{mis}) \right\}.
\label{eqn:sim2-model2-complete}
\end{eqnarray}

We marginalize equation \ref{eqn:sim2-model2-complete} over $G^{obs}_{mis}$ to obtain the joint distribution of $Y^{obs}, W^{obs}, G^{obs}$ conditional on $Z^{obs}$ through
\begin{eqnarray}
&& f(y^{obs}, w^{obs}, g^{obs}~|~Z^{obs}) \nonumber \\
&=& \sum_{G^{obs}_{mis} = 0}^1 {f(y^{obs}, w^{obs}, g^{obs}, g^{obs}_{mis}~|~Z^{obs})}
\label{eqn:sim2-model2-margi}
\end{eqnarray}

Equation \ref{eqn:sim2-model2-margi} is then coded as log probability in RStan. With equation \ref{eqn:sim2-model2-margi} and the priors for the unknown parameters, the posterior distributions of the unknown parameters can be found. The results from two frameworks are shown in table \ref{tab:sim2-result}. 
\begin{table}[htbp]
\small
  \begin{subtable}{0.9\textwidth}
   \centering
   \begin{tabular}{cccr}
  \hline
$p$ & $Z^{obs}$ & $G$ & $N$ \\ 
  \hline
\begin{minipage}[t]{0.2\textwidth} \centering 0.9 \end{minipage} & \begin{minipage}[t]{0.2\textwidth} \centering 0 \end{minipage} &  \begin{minipage}[t]{0.2\textwidth} \centering $nt$ \end{minipage} & \begin{minipage}[t]{0.15\textwidth} \raggedleft 15 \end{minipage}  \\ 
0.9 & 0 &   $co$ & 138 \\ 
0.9 & 1 &   $nt$ &  12 \\ 
0.9 & 1 &   $co$ & 135 \\ 
0.6 &   0 &   $nt$ &  60 \\ 
0.6 &   0 &   $co$ &  96 \\ 
0.6 &   1 &   $nt$ &  54 \\ 
0.6 &   1 &   $co$ &  90 \\ 
   \hline 
   \end{tabular}
   \caption{Simulated data. $N$ is the sample size in each category.}
   \end{subtable}
   
\vspace{1em}

 \begin{subtable}{1\textwidth}
   \centering
\begin{tabular}{rrrr}
  \hline
Parameter & True value & Prior & Posterior mean (95\% interval)  \\ 
  \hline
Model 1 &&& \\
$p$ &0.90 &  & \\
$e_{ate}$ & 2.00 & $N(0, 1)$ & 1.99 (1.94, 2.03)   \\ 
$\beta_0$ & 1.00/1.50 & $N(0, 1)$ & 1.07 (0.92, 1.21)  \\ 
$\sigma_y$ & 1.00 & $N(0, 1)~{\rm T}[0,~]$ & 0.99 (0.91, 1.07)  \\ 
Model 2 &&& \\
$p$ &0.90& & \\
$e_{ate}$ & 0.10 & $N(0, 1)$ & 0.08 (0.03, 0.13)   \\ 
$\beta_0$ & 1.00/1.50 & $N(0, 1)$ & 1.07 (0.92, 1.23)  \\ 
$\sigma_y$ & 1.00 & $N(0, 1)~{\rm T}[0,~]$ & 1.03 (0.95, 1.12)  \\ Model 3 &&& \\
$p$ &0.60& & \\
$e_{ate}$ & 2.00 & $N(0, 1)$ & 1.96 (1.91, 2.00)   \\ 
$\beta_0$ & 1.00/1.50  & $N(0, 1)$ & 1.15 (1.01, 1.29)   \\ 
$\sigma_y$ & 1.00 & $N(0, 1)~{\rm T}[0,~]$ & 1.02 (0.94, 1.11)  \\ Model 4 &&& \\
$p$ &0.60& & \\
$e_{ate}$ & 0.10 & $N(0, 1)$ & 0.10 (0.05, 0.15)  \\ 
$\beta_0$ & 1.00/1.50  & $N(0, 1)$ & 1.20 (1.06, 1.35)  \\ 
$\sigma_y$ & 1.00 & $N(0, 1)~{\rm T}[0,~]$ & 1.06 (0.98, 1.15) \\  
\hline
\end{tabular}  
   \caption{Simplest framework: comparison of true values and estimates from Bayesian models.}
   \end{subtable}  
\end{table}

\begin{table}[htbp]
\ContinuedFloat
\small
 \begin{subtable}{1\textwidth}
   \centering
\begin{tabular}{rrrr}
  \hline
Parameter & True value & Prior & Posterior mean (95\% interval)  \\ 
  \hline
Model 1 &&& \\
$p$ &0.90 &$U(0, 1)$ & 0.92 (0.88, 0.95)\\
$e_{ate}$ & 2.00 & $N(0, 1)$ & 1.99 (1.95, 2.04)  \\ 
$\beta_0$ & 1.50 & $N(0, 1)$ & 1.30 (0.81, 1.80)   \\ 
$\beta_2$ &-0.50&$N(0, 1)$& -0.28 (-0.85, 0.28)\\
$\sigma_y$ & 1.00 & $N(0, 1)~{\rm T}[0,~]$ & 0.98 (0.91, 1.06)  \\ 
$\alpha$ &5.00&$N(0, 1)$& 4.81 (4.62, 5.00)\\
$\sigma_w$ & 1.00 & $N(0, 1)~{\rm T}[0,~]$ & 1.09 (0.97, 1.24)  \\ 
Model 2 &&& \\
$p$ &0.90&$U(0, 1)$& 0.91 (0.88, 0.94)\\
$e_{ate}$ & 0.10 & $N(0, 1)$ & 0.10 (0.05, 0.15)   \\ 
$\beta_0$ & 1.50 & $N(0, 1)$ &  1.57 (1.01, 2.11)   \\ 
$\beta_2$ &-0.50&$N(0, 1)$& -0.59 (-1.20, 0.05)\\
$\sigma_y$ & 1.00 & $N(0, 1)~{\rm T}[0,~]$ & 1.01 (0.93, 1.10)  \\ 
$\alpha$ &5.00&$N(0, 1)$& 4.81 (4.62, 5.00)\\
$\sigma_w$ & 1.00 & $N(0, 1)~{\rm T}[0,~]$ &1.09 (0.97, 1.23)  \\ 
Model 3 &&& \\
$p$ &0.60 &$U(0, 1)$ & 0.63 (0.57, 0.68)\\
$e_{ate}$ & 2.00 & $N(0, 1)$ & 1.99 (1.92, 2.05)  \\ 
$\beta_0$ & 1.50 & $N(0, 1)$ & 1.30 (1.05, 1.54)   \\ 
$\beta_2$ &-0.50&$N(0, 1)$& -0.33 (-0.76, 0.13)\\
$\sigma_y$ & 1.00 & $N(0, 1)~{\rm T}[0,~]$ &  1.01 (0.93, 1.09)  \\ 
$\alpha$ &5.00&$N(0, 1)$& 4.98 (4.78, 5.18)\\
$\sigma_w$ & 1.00 & $N(0, 1)~{\rm T}[0,~]$ & 0.97 (0.83, 1.12)  \\ 
Model 4 &&& \\
$p$ &0.60&$U(0, 1)$& 0.63 (0.57, 0.68)\\
$e_{ate}$ & 0.10 & $N(0, 1)$ & 0.16 (0.10, 0.23)  \\ 
$\beta_0$ & 1.50 & $N(0, 1)$ & 1.50 (1.24, 1.74)   \\ 
$\beta_2$ &-0.50&$N(0, 1)$& -0.62 (-1.03, -0.20)\\
$\sigma_y$ & 1.00 & $N(0, 1)~{\rm T}[0,~]$ &1.03 (0.94, 1.12)  \\ 
$\alpha$ &5.00&$N(0, 1)$&4.98 (4.77, 5.17) \\
$\sigma_w$ & 1.00 & $N(0, 1)~{\rm T}[0,~]$ & 0.97 (0.84, 1.12)  \\ 
   \hline
\end{tabular}
   \caption{Variation: comparison of true values and estimates from Bayesian models.}
   \end{subtable}  
 \caption{Simulation results from two frameworks.}
 \label{tab:sim2-result}
\end{table}

The intercept estimate from the simplest framework seems like an average between the intercepts of compliers and never-takers. Overall, the average treatment effect estimates are relatively accurate from both frameworks, possibly because residual confounding from the compliance behavior is small in the simplest framework. When the proportion of never-takers in the population is relatively high and the effect size is relatively big, the simplest framework seems to underestimate the average treatment effect. When the proportion of never-takers in the population is relatively high and the effect size is relatively small, the variation seems to overestimate the average treatment effect. The interaction among the sample size of compliers, residual confounding, the relative magnitude of the treatment effect size to the compliance behavior effect size may be able to explain these phenomena. This may suggest that data is sensitive to the chosen framework to some extent.

\subsection{When the compliance behavior modifies the treatment effect}

This scenario has two-sided noncompliance, no confounders, an effect modification of the compliance behavior on the exposure. Hence, the average treatment effect in compliers is different from the average treatment effect in always-takers. The average treatment effect in never-takers is set to be identical to the average treatment effect in compliers, though the average treatment effect in never-takers does not affect the models.

We compare the simplest framework and a variation with the effect modification of the compliance behavior on the exposure, given that the average treatment effect is big or small. We want to estimate the average treatment effect among compliers and always-takers, rather than the average treatment effect for each of the two compliance behaviors.

The distributions used in data simulation are described in equation \ref{eqn:sim-3}. A categorical distribution $Categorical(0.1, 0.1, 0.8)$ is used to simulate the compliance behavior, where the proportion for never-takers or always-takers in the population is 0.1 and the proportion for compliers in the population is 0.8. Two different effect sizes are compared. 
\begin{eqnarray}
Z^{obs} &\sim& Bern(0.5), \nonumber \\
G^{obs} &\sim& Categorical(0.1, 0.1, 0.8),~G \in \{1, 2, 3\} \nonumber \\
G^{obs}  &=& 
    \begin{cases}
      1, & \text{if $g = nt$} \\
      2, & \text{if $g = at$} \\
      3, & \text{if $g = co$} \nonumber
    \end{cases} \\
W^{obs} &\sim& 
    \begin{cases}
      N(5, 1)~{\rm T}[0.5,~], & \text{if $Z^{obs}=1$, $G^{obs} \in \{at, co\}$} \\
      0, & \text{otherwise}  \nonumber 
    \end{cases} \\
Y^{obs}_{nt} &\sim& 
    \begin{cases}
      N(1 + 2.5~W^{obs}, 1), & \text{with a big treatment effect} \\
      N(1 + 0.25~W^{obs}, 1), & \text{with a small treatment effect}   \nonumber 
    \end{cases}  \\
Y^{obs}_{co,~at} &\sim& 
    \begin{cases}
      N(1+W^{obs}+0.5~W^{obs}~G^{obs}, 1), & \text{with a big treatment effect} \\
      N(1+0.1~W^{obs}+0.05~W^{obs}~G^{obs}, 1), & \text{with a small treatment effect}
    \end{cases}    
\label{eqn:sim-3}
\end{eqnarray}

The Bayesian model from the simplest framework is also equation \ref{eqn:sim1-model1}. The Bayesian models from the variation framework are shown in equation \ref{eqn:sim3-model2}. 
\begin{eqnarray}
f(y(0, w(0))~|~W(0), G) &=& f_N(y(0, w(0))~|~\beta_0 + \beta_1~W(0)+\beta_2~G~W(0), \sigma_y), \nonumber \\
f(y(1, w(1))~|~W(1), G) &=& f_N(y(1, w(1))~|~ \beta_0 + \beta_1~W(1)+\beta_2~G~W(1), \sigma_y), \nonumber \\
f(w(0)~|~G) &=&{\rm I}(G \neq 2)~{\rm I}(w(0) = 0) + {\rm I}(G = 2)~{\rm I}(w(0) \geq 0.5) ~f_N(w(0)~|~\alpha, \sigma_w), \nonumber \\
f(w(1)~|~G) &=& {\rm I}(G = 1)~{\rm I}(w(1) = 0) + {\rm I}(G > 1)~{\rm I}(w(1) \geq 0.5) ~f_N(w(1)~|~\alpha, \sigma_w), \nonumber \\
f(g) &=& p_1~{\rm I}(g = 1) + p_2~{\rm I}(g = 2) + (1-p_1-p_2) ~{\rm I}(g = 3)\nonumber \\
f(y^{obs}, w^{obs}~|~Z^{obs}) &=& f_N(y^{obs}~|~ \beta_0, \sigma_y)~{\rm I}(w^{obs} = 0)~p_1 \nonumber \\
&& +~ f_N(y^{obs}~|~ \beta_0 + (\beta_1+2~\beta_2)~w^{obs}, \sigma_y) \nonumber \\
&& \cdot ~{\rm I}(w^{obs} \geq 0.5)~f_N(w^{obs}~|~ \alpha, \sigma_w)~p_2 \nonumber \\
&& +~ f_N(y^{obs}~|~ \beta_0, \sigma_y)~{\rm I}(w^{obs} = 0)~p_3~{\rm I}(Z^{obs} = 0) \nonumber \\
&& +~ f_N(y^{obs}~|~ \beta_0 + (\beta_1+3~\beta_2)~w^{obs}, \sigma_y) \nonumber \\
&& \cdot ~{\rm I}(w^{obs} \geq 0.5)~f_N(w^{obs}~|~ \alpha, \sigma_w)~p_3~{\rm I}(Z^{obs} = 1).
\label{eqn:sim3-model2}
\end{eqnarray}

The results from two frameworks are shown in table \ref{tab:sim3-result}. 

\begin{table}[htbp]
\small
  \begin{subtable}{0.9\textwidth}
   \centering
   \begin{tabular}{ccr}
  \hline
$Z^{obs}$ & $G$ & $N$ \\ 
  \hline
\begin{minipage}[t]{0.2\textwidth} \centering 0 \end{minipage} &  \begin{minipage}[t]{0.2\textwidth} \centering $nt$ \end{minipage} & \begin{minipage}[t]{0.15\textwidth} \raggedleft 18 \end{minipage}  \\ 
0 &   $at$ &  16 \\ 
0 &   $co$ & 115 \\ 
1 &   $nt$ &  13 \\ 
1 &   $at$ &  18 \\ 
1 &   $co$ & 120 \\ 
   \hline 
   \end{tabular}
   \caption{Simulated data. $N$ is the sample size in each category.}
   \end{subtable}
\end{table}

\begin{table}[htbp]
\ContinuedFloat
\small
  \begin{subtable}{1\textwidth}
   \centering
\begin{tabular}{rrrr}
  \hline
Parameter & True value & Prior & Posterior mean (95\% interval)  \\ 
  \hline
Model 1 &&& \\
$e_{ate}$ & 2.00/2.50 & $N(0, 1)$ & 2.42 (2.37, 2.47)  \\ 
$\beta_0$ & 1.00 & $N(0, 1)$ & 1.05 (0.88, 1.22)  \\ 
$\sigma_y$ & 1.00 & $N(0, 1)~{\rm T}[0,~]$ & 1.17 (1.08, 1.26)  \\ 
Model 2 &&& \\
$e_{ate}$ & 0.20/0.25 & $N(0, 1)$ & 0.24 (0.19, 0.29)   \\ 
$\beta_0$ & 1.00 & $N(0, 1)$ & 0.95 (0.79, 1.10)  \\ 
$\sigma_y$ & 1.00 & $N(0, 1)~{\rm T}[0,~]$ & 1.04 (0.96, 1.13)  \\ 
\hline
\end{tabular} 
   \caption{Simplest framework: comparison of true values and estimates from Bayesian models.}
   \end{subtable}  

\vspace{1em}

 \begin{subtable}{1\textwidth}
   \centering
\begin{tabular}{rrrr}
  \hline
Parameter & True value & Prior & Posterior mean (95\% interval)  \\ 
  \hline
Model 1 &&& \\
$p_{nt}$ &0.10 &$Dirichlet(0.5, 0.5, 0.5)$ & 0.09 (0.05, 0.14)\\
$p_{at}$ &0.10 &$Dirichlet(0.5, 0.5, 0.5)$ & 0.03 (0.01, 0.05) \\
$p_{co}$ &0.80 &$Dirichlet(0.5, 0.5, 0.5)$ & 0.88 (0.83, 0.93) \\
$e_{ate}^{co}$ & 2.50 & $N(0, 1)$ & 2.47 (2.42, 2.52) \\ 
$e_{ate}^{at}$ & 2.00 & $N(0, 1)$ & 1.80 (1.64, 1.96)  \\ 
$\beta_0$ & 1.00 & $N(0, 1)$ & 1.03 (0.87, 1.19)  \\ 
$\beta_1$ &1.00&$N(0, 1)$& 0.47 (0.00, 0.91) \\
$\beta_2$ &0.50&$N(0, 1)$& 0.67 (0.52, 0.82) \\
$\sigma_y$ & 1.00 & $N(0, 1)~{\rm T}[0,~]$ & 1.03 (0.94, 1.12)  \\ 
$\alpha$ &5.00&$N(0, 1)$& 4.78 (4.60, 4.96) \\
$\sigma_w$ & 1.00 & $N(0, 1)~{\rm T}[0,~]$ & 1.07 (0.95, 1.21)  \\ 
Model 2 &&& \\
$p_{nt}$ &0.10 &$Dirichlet(0.5, 0.5, 0.5)$ & 0.09 (0.05, 0.14) \\
$p_{at}$ &0.10 &$Dirichlet(0.5, 0.5, 0.5)$ & 0.00 (0.00, 0.01) \\
$p_{co}$ &0.80 &$Dirichlet(0.5, 0.5, 0.5)$ & 0.91 (0.86, 0.95) \\
$e_{ate}^{co}$ & 0.25 & $N(0, 1)$ & 0.24 (0.19, 0.29)  \\ 
$e_{ate}^{at}$ & 0.20  & $N(0, 1)$ & 0.17 (-0.44, 0.75)  \\ 
$\beta_0$ & 1.00 & $N(0, 1)$ & 0.95 (0.80, 1.11)  \\ 
$\beta_1$ &0.10 &$N(0, 1)$& 0.02 (-1.82, 1.78)\\
$\beta_2$ &0.05&$N(0, 1)$& 0.08 (-0.51, 0.69)\\
$\sigma_y$ & 1.00 & $N(0, 1)~{\rm T}[0,~]$ & 1.04 (0.96, 1.13)  \\ 
$\alpha$ &5.00&$N(0, 1)$& 4.78 (4.60, 4.96) \\
$\sigma_w$ & 1.00 & $N(0, 1)~{\rm T}[0,~]$ & 1.07 (0.95, 1.21)  \\ 
   \hline
\end{tabular}
   \caption{Variation: comparison of true values and estimates from Bayesian models.}
   \end{subtable}  
\caption{Simulation results from two frameworks.}
 \label{tab:sim3-result}
\end{table}

The average treatment effect estimate from the simplest framework seems to be an average of the average treatment effects from compliers and always-takers. It is useful that we can obtain a good overall average treatment effect estimate from a simple framework. When we consider separating the compliance behaviors, the variation can estimate the average treatment effects for compliers and always-takers separately. Each average treatment effect is a linear combination of $\beta_1$ and $\beta_2$. $\beta_1$ and $\beta_2$ are not estimated well by the variation. This may be affected by the prior choice. However, the average treatment effect estimates for compliers are relatively accurate and the average treatment effect estimates for always-takers are also plausible yet less accurate. 

The variation is more appropriate to the true data generating mechanism. If we only want an overall estimate, the simplest framework is also appropriate. We should choose an inferential framework based on what estimate we want. A simple framework may be as good as a complex framework.

\subsection{When there is one unmeasured confounder}
\label{sec:ri-1}

This scenario has one-sided noncompliance, one unmeasured confounder, no effect of the compliance behavior on the outcome. We use the simplest framework on the simulated data and the missing variable model, to see if the average treatment effect can be estimated well in the presence of the unmeasured confounding effect.

\begin{figure}[htbp]
     \centering
     \includegraphics[width=\textwidth]{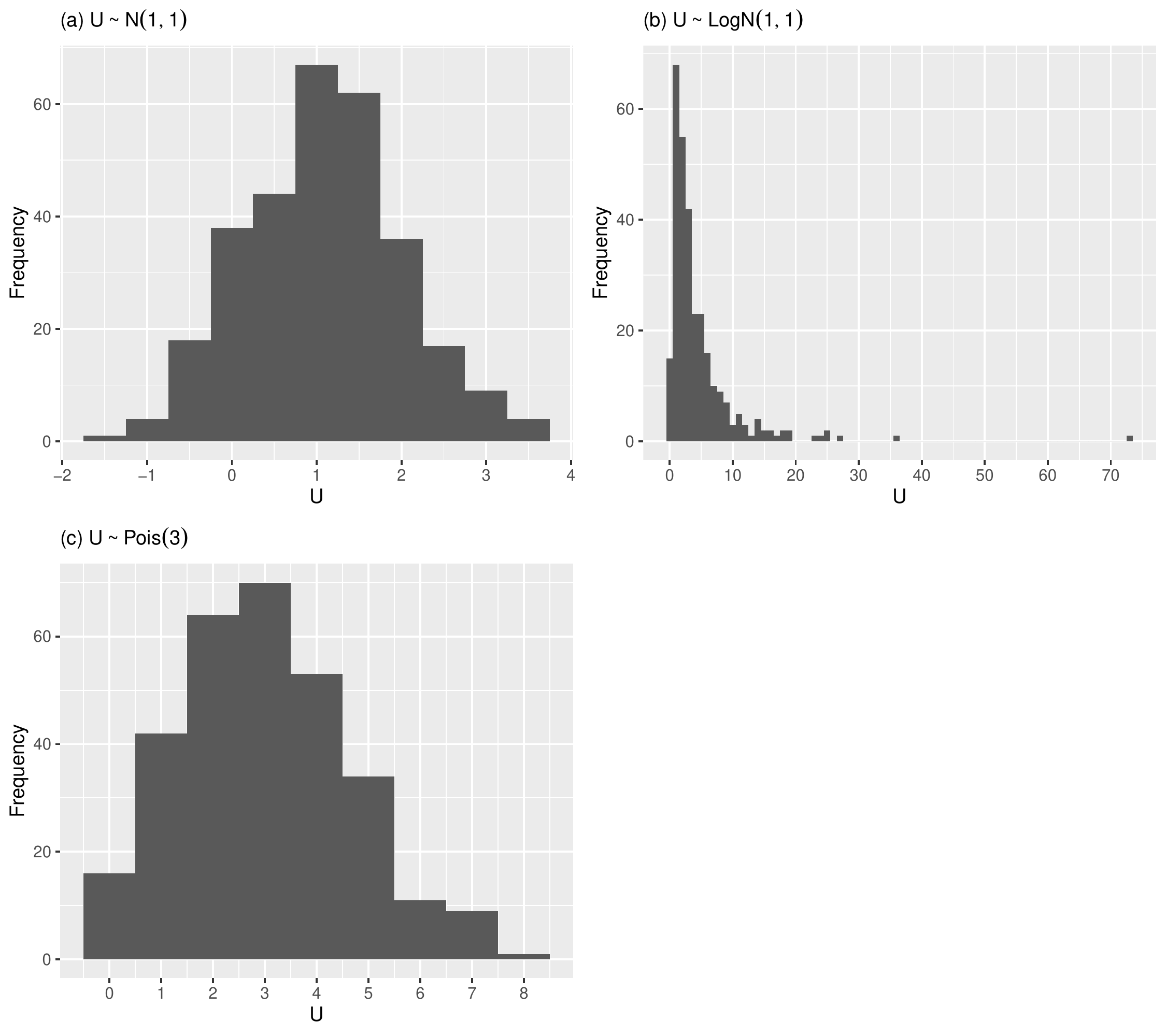}
        \caption{Histograms of three true distributions of the unmeasured confounder.}
        \label{fig:sim-u-scenario1-3true-original}
\end{figure}

The distributions used in data simulation are described in equation \ref{eqn:sim-4}. Some settings are identical to the previous simulation scenarios. The proportion of never-takers in the population is set to be 10\%. The unmeasured confounder is simulated from a Normal, LogNormal or Poisson distribution but will be modelled by a Normal prior, so that we can learn how normal approximation of the distribution of the unmeasured confounder performs in practice. Figure \ref{fig:sim-u-scenario1-3true-original} shows the histograms of the generated unmeasured confounder from three different distributions. The generated unmeasured confounder from a LogNormal distribution is very right-skewed, where normal approximation seems implausible. However, the generated unmeasured confounder from a Normal or Poisson distribution seems more Normally distributed.
\begin{eqnarray}
Z^{obs} &\sim& Bern(0.5), \nonumber \\
G^{obs} &\sim& Bern(0.9), \nonumber \\
U^{obs} &\sim& N(1, 1)~/~LogN(1, 1)~/~Pois(3), \nonumber \\
W^{obs} &\sim& 
    \begin{cases}
      N(3+U^{obs}, 1)~{\rm T}[0.5,~], & \text{if $Z^{obs}=1$, $G^{obs}=co$} \\
      0, & \text{otherwise}  \nonumber 
    \end{cases} \\
Y^{obs} &\sim& N(1+2~W^{obs}-U^{obs}, 1).
\label{eqn:sim-4}
\end{eqnarray}

The Bayesian models use the random intercept reparameterization, as
\begin{eqnarray}
Y^{obs} ~|~ G^{obs}, Z^{obs} &\sim& N(\beta_1~W^{obs} + U', \sigma_y), ~ U' = 1-U^{obs}, \nonumber \\
W^{obs} ~|~ G^{obs} = co, Z^{obs} = 1 &\sim& N(\alpha_0 + \alpha_1~U', \sigma_w).
\label{eqn:sim4-model}
\end{eqnarray}

First, we want to build models with the unmeasured confounder that come from a Normal distribution, and check if models are identified. We use reparameterization to prevent model non-identifiability from reversion of the sign of the unmeasured confounder and shifting and scaling of the unmeasured confounder. We ensure that the coefficients of the reparameterized unmeasured confounder are not estimated to be zero and the reparameterized unmeasured confounder itself is not estimated to be constant, so that model non-identifiability from degeneration of the causal model is prevented. 

We also need to ensure standard deviations of the exposure and the outcome are not estimated to be zero, and check if exposure and outcome parameters in the same meaning are correlated. These two checks require estimation of standard deviations of the exposure and the outcome. In initial models, either or both standard deviations of $W^{obs}$ and $Y^{obs}$ are estimated to be zero. Hence, we try two methods to estimate two standard deviations: (1) fix two standard deviations by sample standard deviations, (2) construct a strongly informative prior with sample standard deviations as the prior mean. In method (1), since the distributions of compliers in the treatment arm and the other people have different means, we pool sample standard deviations of the outcome from these two groups by weighted averaging. We may use the sample standard deviation from the entire sample only if the means between groups are similar. For the exposure, we calculate the sample standard deviation from groups where the exposure distribution is not fixed at zero. In this scenario, the exposure distribution is not fixed in compliers in the treatment arm. After finding (pooled) sample standard deviations, we directly include them in the models for $W^{obs}$ and $Y^{obs}$. In method (2), we use sample standard deviations calculated in the same way as method (1) as the prior means of two standard deviations and construct two strongly informative priors with a relatively small prior standard deviation.

We build and compare models using true standard deviations and these two methods to estimate two standard deviations. Table \ref{tab:sim4-result-3sdmethods} shows the estimates from three estimation methods of two standard deviations.
\begin{table}[htbp]
\small
  \begin{subtable}{0.9\textwidth}
   \centering
   \begin{tabular}{ccr}
  \hline
  $Z^{obs}$ & $G^{obs}$ & $N$ \\ 
  \hline
\begin{minipage}[t]{0.2\textwidth} \centering 0 \end{minipage} &  \begin{minipage}[t]{0.2\textwidth} \centering $nt$ \end{minipage} & \begin{minipage}[t]{0.15\textwidth} \raggedleft 12 \end{minipage}  \\ 
0 &   $co$ & 131 \\ 
1 &   $nt$ &  17 \\ 
1 &   $co$ & 140 \\ 
   \hline 
   \end{tabular}
   \caption{Simulated data. $N$ is the sample size in each category.}
   \end{subtable}
\end{table}

\begin{table}[htbp]
\ContinuedFloat
\small
 \begin{subtable}{1\textwidth}
   \centering
\begin{tabular}{rrrr}
  \hline
Parameter & True value & Prior & Posterior mean (95\% interval)  \\ 
  \hline
Model 1: \text{True standard deviations} &&& \\
$U^{obs}$ & $N(1, 1)$ & & \\
$U'$ & &$N(0, 3)$ & \\
$e_{ate}$ & 2 & $N(0, 1)$ & 2.22 (2.12, 2.33)  \\ 
$\sigma_y$ & 1 & & 1.00 \\
$\alpha_0$ & 4 & $N(0, 1)$& 3.54 (3.20, 3.86) \\
$\alpha_1$ & -1 & $N(0, 1)$& -0.63 (-0.76, -0.50) \\
$\sigma_w$ & 1 &  & 1.00 \\
Model 2: \text{Sample standard deviations} &&& \\
$U^{obs}$ & $N(1, 1)$ & & \\
$U'$ & &$N(0, 3)$ & \\
$e_{ate}$ & 2 & $N(0, 1)$ & 2.07 (1.93, 2.20)   \\ 
$\sigma_y$ & 1 & & 1.93 \\
$\alpha_0$ & 4 & $N(0, 1)$& 3.98 (3.66, 4.26) \\
$\alpha_1$ & -1 & $N(0, 1)$& -0.38 (-0.55, -0.21) \\
$\sigma_w$ & 1 &  & 1.38  \\
Model 3: \text{Bayesian standard deviations} &&& \\
$U^{obs}$ & $N(1, 1)$ & & \\
$U'$ & &$N(0, 3)$ & \\
$e_{ate}$ & 2 & $N(0, 1)$ & 2.14 (2.02, 2.27) \\ 
$\sigma_y$ & 1 &$N(1.93, 0.1)$  & 1.53 (1.34, 1.71) \\
$\alpha_0$ & 4 & $N(0, 1)$& 3.81 (3.48, 4.13) \\
$\alpha_1$ & -1 & $N(0, 1)$&-0.51 (-0.67, -0.35) \\
$\sigma_w$ & 1 & $N(1.38, 0.1)$ & 1.17 (1.00, 1.34) \\
\hline
\end{tabular}  
   \caption{Comparison of true values and estimates from three models with different estimation methods of two standard deviations.}
   \end{subtable}  
 \caption{Simulation results.}
 \label{tab:sim4-result-3sdmethods}
\end{table}

From table \ref{tab:sim4-result-3sdmethods}, Model 2 estimates the average treatment effect best while Model 1 overstimates the average treatment effect most. Sample standard deviations and Bayesian standard deviations with priors from sample standard deviations are both useful in this scenario. In all three models, we use a random intercept reparameterization in both the exposure and outcome distributions for compliers in the treatment arm and only in the outcome distribution for the other people. This approach seems to work well because the average treatment effect estimates from three models are close to true values or are not much biased.

\begin{figure}[htbp]
     \centering
     \includegraphics[width=0.5\textwidth]{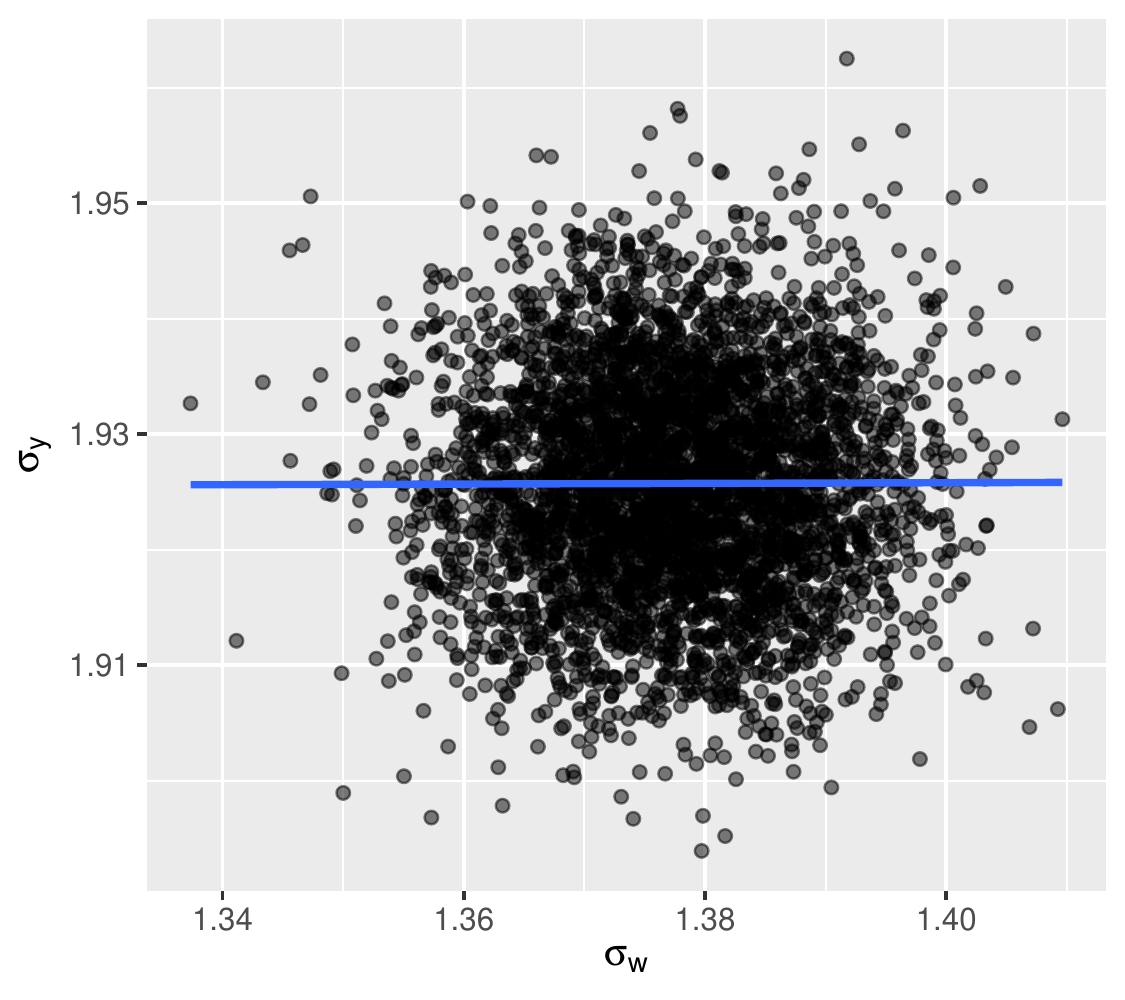}
        \caption{Scatter plot of posterior estimates of two standard deviations from Model 3. The blue line is a linear model smoothing line.}
        \label{fig:sim-u-sd-post-corr}
\end{figure}

Figure \ref{fig:sim-u-sd-post-corr} that comes from Model 3 with Bayesian standard deviations shows that there is no significant posterior correlation between exposure and outcome standard deviations $\sigma_w$ and $\sigma_y$, which indicates that model non-identifiability due to correlation between exposure and outcome coefficients of the same meaning does not occur. In the simulated data, $\sigma_w$ and $\sigma_y$ are independent a priori, and we do not find any reason to support they have natural posterior correlation from the model structure. Hence, we regard posterior correlation between standard deviations as a result from model non-identifiability. 

On the other hand, Model 2 with sample standard deviations cannot prove if there is posterior correlation between exposure and outcome parameters of the same meaning from coefficients. Unlike standard deviations, coefficients can have some natural posterior correlation because two means of the exposure distribution and the outcome distribution are linked through the reparameterized unmeasured confounder. For example, from a mathematical perspective, when the coefficient of $U^{obs}$ in the exposure distribution increases, estimates of $U^{obs}$ will change accordingly and in turn the coefficient of $U^{obs}$ and the intercept in the outcome distribution will also change accordingly. Hence, if there is posterior correlation between exposure and outcome coefficients of the same meaning, we cannot tell if it is natural posterior correlation, or a result of model non-identifiability, or a mixture of natural posterior correlation and a result of model non-identifiability. When we want to make sure our models are identified, we should use Bayesian standard deviations, though the effect estimates are plausible good but not as good as from sample standard deviations. However, when Bayesian standard deviations with very strongly informative priors cannot make the models fitted well, we may try using sample standard deviations. Fixing standard deviations is a very useful way to make the models fitted well. Essentially, sample standard deviations we use is a frequentist method. We combine Bayesian models with this frequentist method to help resolve modelling issues. Any more additional information from data whether from a frequentist method or not would be valuable, because Bayesian models involve too many parameters and parameter reduction can prevent appearance of model non-identifiability to some extent. 


We are curious about a question: how small should the standard deviations of the priors for $\sigma_w$ and $\sigma_y$ in Model 3 be, so that the priors are strongly informative? We try other prior standard deviations for comparison. Table \ref{tab:sim-1u-normal-sdpriorsd} shows posterior estimates of $\sigma_w$, $\sigma_y$ and $e_{ate}$ under different prior standard deviations of $\sigma_w$ and $\sigma_y$.
\begin{table}[htbp]
   \centering
   \begin{tabular}{rrrr}
  \hline
Prior SD & $\sigma_w$ & $\sigma_y$ & $e_{ate}$  \\ 
  \hline
0.01 &1.38 (1.36, 1.40)&1.93 (1.91, 1.94)&2.07 (1.93, 2.21) \\
0.10 &1.17 (1.00, 1.34) &1.53 (1.34, 1.71)&2.14 (2.02, 2.27) \\
0.15* &1.05 (0.84, 1.25)&1.20 (0.96, 1.44)& 2.19 (2.06, 2.31)\\
0.20 &-&-&-\\
0.50 &-&-&- \\
1.00 &-&-&- \\
   \hline
   \end{tabular}
   \caption{Posterior estimates of $\sigma_w$, $\sigma_y$ and $e_{ate}$ under different prior standard deviations of $\sigma_w$ and $\sigma_y$. The prior standard deviation is the same for $\sigma_w$ and $\sigma_y$. The prior means for $\sigma_w$ and $\sigma_y$ are their sample standard deviations. ``*'' indicates there appears to weak posterior correlation between $\sigma_w$ and $\sigma_y$. ``-'' indicates posterior estimates are unreliable because the model is divergent or sampled values are autocorrelated or sampling reaches the maximum tree depth.}
   \label{tab:sim-1u-normal-sdpriorsd}
\end{table}

\begin{figure}[htbp]
     \centering
     \includegraphics[width=0.5\textwidth]{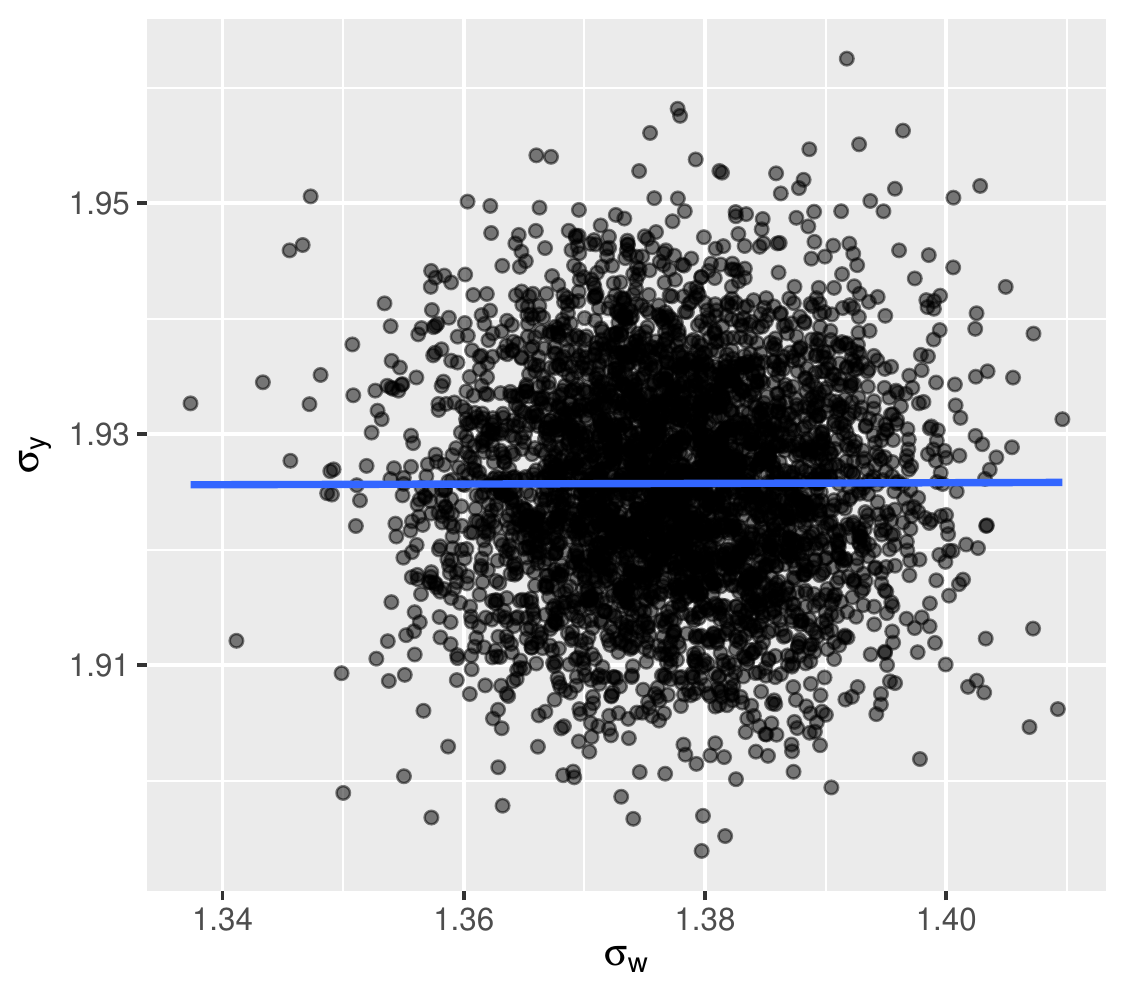}
        \caption{Scatter plot of posterior estimates of two standard deviations from the Bayesian model with a prior standard deviation of 0.01. The blue line is a linear model smoothing line.}
        \label{fig:sim-u-sd-post-corr-2}
\end{figure}

Posterior estimates of $\sigma_w$ and $\sigma_y$ are sensitive to the prior standard deviation of $\sigma_w$ and $\sigma_y$, while posterior estimates of $e_{ate}$ is slightly sensitive to the prior standard deviation of $\sigma_w$ and $\sigma_y$. When the prior standard deviation of $\sigma_w$ and $\sigma_y$ becomes larger even within a small threshold, the model can easily go into problems. Hence, to make the model fitted well, the priors of $\sigma_w$ and $\sigma_y$ should have a very small standard deviation. On the other hand, when the prior standard deviation is 0.01, the Bayesian model produces estimates that are very similar to Model 2 with sample standard deviations, and we check there is no posterior correlation between $\sigma_w$ and $\sigma_y$ in this Bayesian model from figure \ref{fig:sim-u-sd-post-corr-2}. Evidence from figure \ref{fig:sim-u-sd-post-corr-2} seems to even be stronger than evidence from figure \ref{fig:sim-u-sd-post-corr}. Essentially, $\sigma_w$ and $\sigma_y$ are estimated to be sample standard deviations. This indicates that in addition to using sample standard deviations as the prior means for $\sigma_w$ and $\sigma_y$ while letting data modify the posterior estimates, we can directly use a very strongly informative prior to make posterior estimates of $\sigma_w$ and $\sigma_y$ nearly equal to their sample standard deviations. Two advantages are that: (1) we still use sample information effectively but there is no need to nest the frequentist method into a Bayesian model, (2) we can check if model non-identifiability occurs due to correlation between exposure and outcome coefficients of the same meaning by checking posterior correlation between $\sigma_w$ and $\sigma_y$.

\begin{figure}[htbp]
     \centering
     \includegraphics[width=\textwidth]{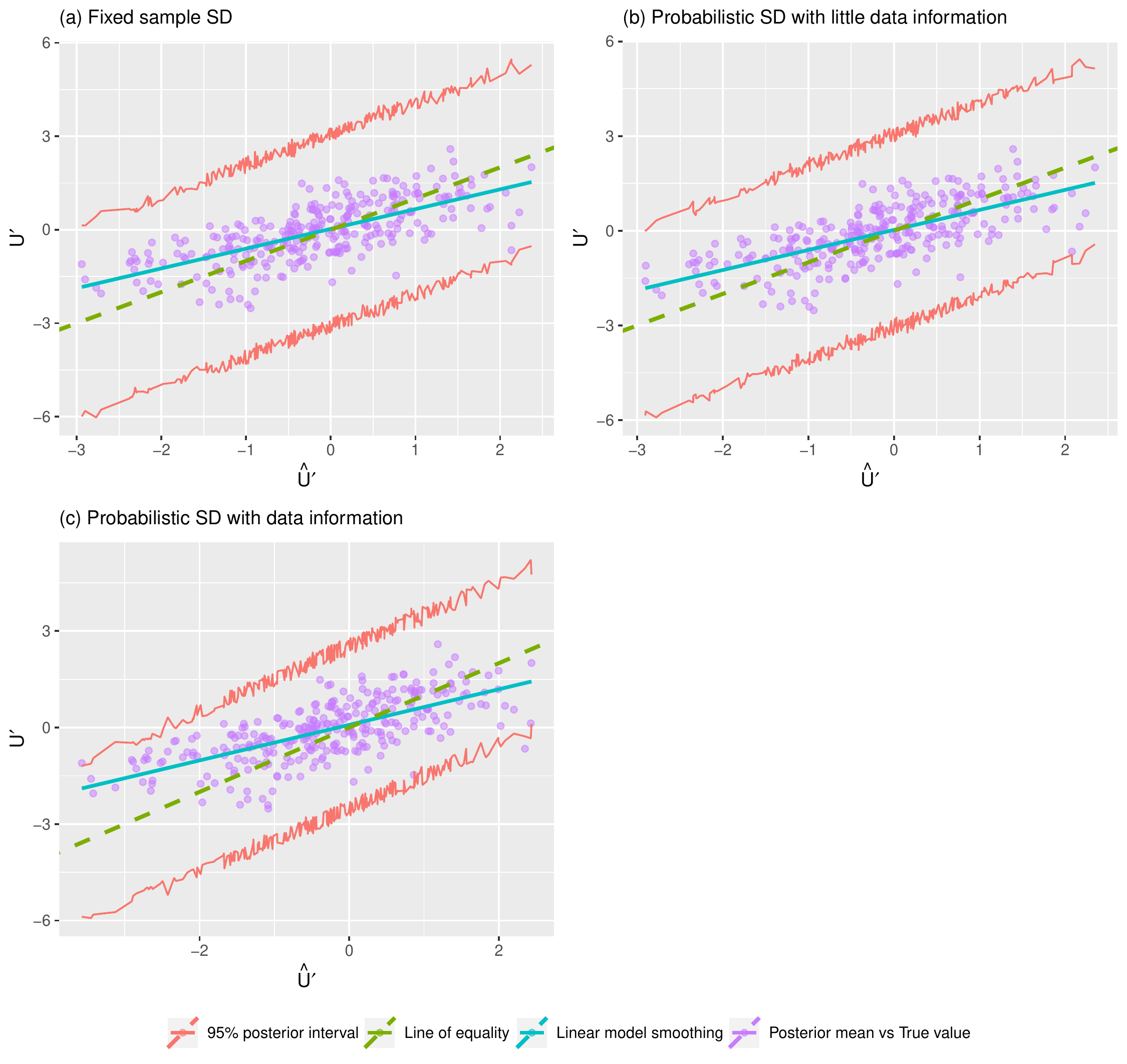}
        \caption{Comparison of posterior estimates and true values of the reparameterized unmeasured confounder $U'$ from three estimation methods of standard deviations (SD). True $U'$ is denoted by $U'$, while estimated $U'$ is denoted by $\hat U'$. True $U' = 1 - U^{obs}$, but due to reparameterization, estimates of $U^{obs}$ cannot be retrieved. Line of equality means that for points on this line, posterior means are equal to true values.}
        \label{fig:sim-u-scenario1-3sdmethods}
\end{figure}

Now we have three methods to estimate $\sigma_w$ and $\sigma_y$: (1) fixed sample standard deviations, (2) probabilistic standard deviations with little data information, (3) probabilistic standard deviations with data information. Method (2) means Bayesian standard deviations are estimated to be sample standard deviations as their prior means, while method (3) means Bayesian standard deviations use sample standard deviations as their prior means but a relatively larger prior standard deviation is used. Some modelling features about the three methods have been discussed. Then we want to compare how well the reparameterized unmeasured confounder is estimated from these three methods. Figure \ref{fig:sim-u-scenario1-3sdmethods} shows goodness of fit of the reparameterized unmeasured confounder $U'$ from three estimation methods of standard deviations. $U'$ is estimated plausibly well through three methods, meaning that the unmeasured confounding effect is well adjusted for. Method (1) and method (2) have better posterior means, while method (3) has narrower 95\% posterior intervals.

To sum up, three methods to estimate $\sigma_w$ and $\sigma_y$ are all useful and better than a weakly informative prior for $\sigma_w$ and $\sigma_y$ in the models. Sometimes, a weakly informative prior for $\sigma_w$ and $\sigma_y$ can work, but we now also have alternative methods to deal with bad models. We prefer probabilistic standard deviations to sample probabilistic standard deviations. Then we prefer probabilistic standard deviations using little data information in the Bayesian modelling process, because (1) Bayesian models are less likely to go into problems, (2) calculation of sample standard deviations is a relatively independent procedure that is less likely to be affected by other factors in Bayesian models and is almost always viable in practice. Hence, we use method (2) to do further explorations and present results. And for method (2), we think it is unnecessary to do prior sensitivity analysis of $\sigma_w$ and $\sigma_y$.

\begin{figure}[htbp]
     \centering
     \includegraphics[width=\textwidth]{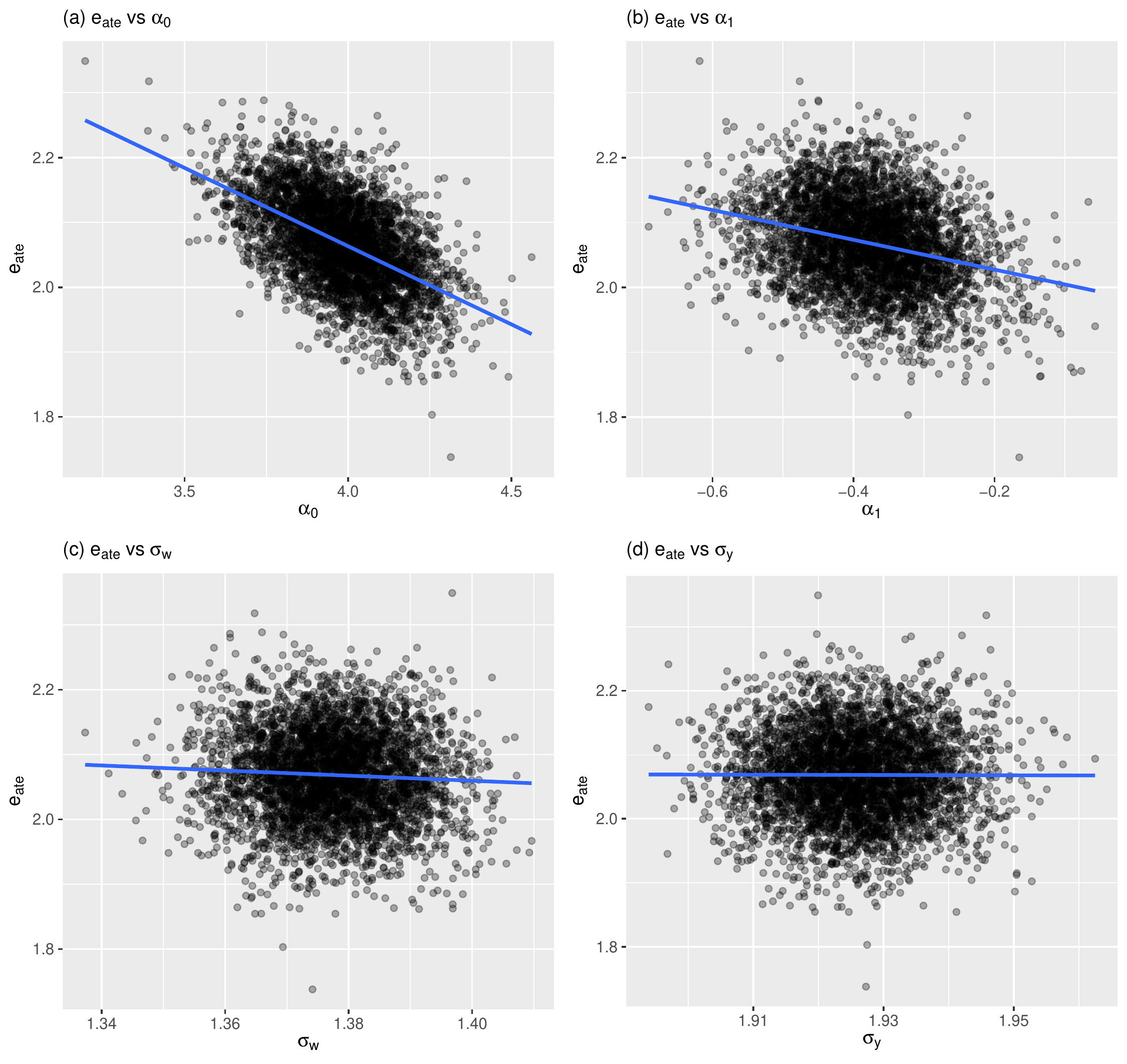}
        \caption{Scatter plots of posterior estimates of the average treatment effect $e_{ate}$, and two exposure coefficients $\alpha_0$ and $\alpha_1$ and two standard deviations $\sigma_w$ and $\sigma_y$, in the Bayesian model through method (2). Blue lines are linear model smoothing lines.}
        \label{fig:sim-u-scenario1-postcorr-coef}
\end{figure}

In the Bayesian model through method (2), we also find that  there is no significant posterior correlation between the average treatment effect $e_{ate}$ and $\sigma_w$, $\sigma_y$, but there is some posterior correlation between $e_{ate}$ and two exposure coefficients $\alpha_0$ and $\alpha_1$, as shown in figure \ref{fig:sim-u-scenario1-postcorr-coef}. Hence, we want to learn if the average treatment effect estimates are sensitive to the two exposure coefficients. 

Table \ref{tab:sim-1u-a1a2prior-method2} shows posterior estimates of $e_{ate}$ over different priors of $\alpha_0$ or $\alpha_1$, with the priors for the other parameters fixed as in the original Bayesian model. The reference prior of $\alpha_0$ and $\alpha_1$ is $N(0, 1)$. We think the prior mean of 0 and the prior standard deviation of 1 are both plausible. For $\alpha_0$, we try another two prior means, 4 and 8, and try another two prior standard deviations, 0.5 and 2. For $\alpha_1$, we try another two prior means, -2 and -1, and try another two prior standard deviations, 0.5 and 2.
\begin{table}[htbp]
\small
  \begin{subtable}{1\textwidth}
   \centering
   \begin{tabular}{rrrr}
  \hline
 \multirowcell{2}[0pt][r]{$\mu_{\alpha_0}$} & \multicolumn{3}{c}{$\sigma_{\alpha_0}$}  \\ 
  \cmidrule(l){2-4} 
& 0.5 & 1 & 2   \\ 
  \hline
0 &2.15 (2.00, 2.28)&2.07 (1.93, 2.21) &2.05 (1.91, 2.18) \\
4 &2.05 (1.91, 2.18) &2.05 (1.91, 2.18) & 2.05 (1.90, 2.19)\\
8 &1.98 (1.85, 2.11) &2.03 (1.89, 2.16) & 2.04 (1.91, 2.18)\\
   \hline
   \end{tabular}
   \caption{Posterior means (95\% intervals) of the average treatment effect $e_{ate}$ under different priors of $\alpha_0$. $\mu_{\alpha_0}$ is the prior mean of $\alpha_0$. $\sigma_{\alpha_0}$ is the prior standard deviation of $\alpha_0$. } 
   \end{subtable}
   
\vspace{1em}

  \begin{subtable}{1\textwidth}
   \centering
  \begin{tabular}{rrrr}
  \hline
 \multirowcell{2}[0pt][r]{$\mu_{\alpha_1}$} & \multicolumn{3}{c}{$\sigma_{\alpha_1}$}  \\ 
  \cmidrule(l){2-4} 
& 0.5 & 1 & 2   \\ 
  \hline
-2 &2.08 (1.94, 2.22)&2.08 (1.93, 2.21)& 2.07 (1.94, 2.21)\\
-1 &2.07 (1.94, 2.21)&2.07 (1.94, 2.21)&2.07 (1.94, 2.21) \\
0 &2.07 (1.93, 2.21)&2.07 (1.93, 2.21)& 2.07 (1.93, 2.19)\\
   \hline
   \end{tabular}
   \caption{Posterior means (95\% intervals) of the average treatment effect $e_{ate}$ under different priors of $\alpha_1$. $\mu_{\alpha_1}$ is the prior mean of $\alpha_1$. $\sigma_{\alpha_1}$ is the prior standard deviation of $\alpha_1$. } 
   \end{subtable}  
\caption{Posterior means (95\% intervals) of the average treatment effect $e_{ate}$ under different priors of $\alpha_0$ or $\alpha_1$.}
\label{tab:sim-1u-a1a2prior-method2}
\end{table}

All the models from table \ref{tab:sim-1u-a1a2prior-method2} are identified and show similar posterior correlation between $e_{ate}$ and $\alpha_0$ or $\alpha_1$. We have checked some models and found no posterior correlation between $e_{ate}$ and $\sigma_y$.

Posterior estimates of $e_{ate}$ are insensitive to the prior of $\alpha_0$ or $\alpha_1$, except when the prior of $\alpha_0$ becomes too extreme as $N(0, 0.5)$ and $N(8, 0.5)$. This indicates that even though there is posterior correlation between the average treatment effect and two exposure coefficients, the average treatment effect estimates are not affected by this posterior correlation. This also indicates that the average treatment effect and coefficients can have posterior correlation due to the model structure when the model is identified and posterior correlation between coefficients may not be used to diagnose whether the Bayesian model is identified or not. 

We also want to learn how the prior of the reparameterized unmeasured confounder $U'$ will affect posterior estimates of $e_{ate}$ and $U'$, when $U^{obs}$ comes from a Normal distribution. $U'$ has fixed true values and thus should not vary with its prior freely as non-identified parameters. Table \ref{tab:sim-1u-normal-uprior} shows how goodness of fit of $U'$ and posterior estimates of $e_{ate}$ over different priors of $U'$, with the priors for the other coefficients fixed. The reference prior of $U'$ is $N(0, 3)$. We think the prior mean of 0 and the prior standard deviation of 3 are both plausible. Then we try decreasing the prior mean to -1 and the prior standard deviation to 1, and try increasing the prior mean to 1 and the prior standard deviation to 5. 
\begin{table}[htbp]
   \centering
   \begin{tabular}{rrrrr}
  \hline
 \multirowcell{2}[0pt][r]{$\mu_{U'}$} & \multicolumn{3}{c}{$\sigma_{U'}$} & \\ 
  \cmidrule(l){2-4} 
& 1 & 3 & 5 &  \\ 
  \hline
\multirowcell{2}[0pt][r]{-1}&18.84 &19.87&27.86&  ~~RMSE$_{U'}$ \\
  &2.18 (2.10, 2.26)&2.29 (2.15, 2.42)&2.46 (2.27, 2.65)&  $e_{ate}$ \\
  \hline
\multirowcell{2}[0pt][r]{0}&12.90& 13.55&20.24&  \\
  &1.96 (1.87, 2.04) &2.07 (1.93, 2.21)&2.27 (2.06, 2.46)&\\
  \hline
\multirowcell{2}[0pt][r]{1}&22.59 & 17.64&-&\\
  &1.71 (1.62, 1.80)&1.79 (1.57, 1.96)&-& \\
   \hline
   \end{tabular}
   \caption{Root mean squared error estimates of $U'$ (RMSE$_{U'}$) and posterior means (95\% intervals) of the average treatment effect $e_{ate}$ under different priors of $U'$. $\mu_{U'}$ is the prior mean of $U'$. $\sigma_{U'}$ is the prior standard deviation of $U'$. RMSE$_{U'}$ is calculated from posterior means and true values of $U'$, as an indicator for goodness of fit of $U'$. RMSE$_{U'}$ and $e_{ate}$ are placed in two rows in order among all the combinations of $\mu_{U'}$ and $\sigma_{U'}$. They are only indicated in the first two rows. ``-'' indicates the model has issues such as divergence and sampling autocorrelation.} 
   \label{tab:sim-1u-normal-uprior}
\end{table}

In this case, the true distribution of $U'$ is $N(0, 1)$. For a fixed prior standard deviation of $U'$, as the prior mean of $U'$ is further away from zero that is the true mean of $U'$, RMSE$_{U'}$ increases, bias in the posterior mean of $e_{ate}$ increases, the 95\% posterior interval of $e_{ate}$ moves further away from the true effect. For a fixed prior mean of $U'$, as the prior standard deviation of $U'$ increases from 1 that is the true standard deviation of $U'$, RMSE$_{U'}$ also increases, bias in the posterior mean of $e_{ate}$ also increases, the 95\% posterior interval of $e_{ate}$ also moves further away from the true effect. We do not consider a prior standard deviation smaller than 1, because in practice we will choose a relatively large prior standard deviation to account for uncertainty in the unmeasured confounder. In addition, any prior deviation from the true distribution of $U'$ results in larger RMSE$_{U'}$ and more bias in posterior estimates of $e_{ate}$. This is plausible, but this does not mean posterior estimates are useless. Compared to the prior of $U'$ that is the true distribution of $U'$, the posterior estimates from the prior $N(0, 3)$ are also very good, and the posterior estimates from some nearby priors such as $N(-1, 1)$ are also relatively good even if there is some bias. The biggest range of $e_{ate}$ in table \ref{tab:sim-1u-normal-uprior} is $[1.57, 2.65]$, which has no extreme boundary points and seems plausible compared to the true effect of 2. We also note that there is a chance that the model can go wrong when the prior standard deviation of $U'$ takes a large value. Further, there seems to be negative correlation between goodness of fit of $U'$ and bias in posterior estimates of $e_{ate}$, which may indicate that better adjustment for the unmeasured confounding effect is associated with better average treatment effect estimates.

To sum up, posterior estimates of $e_{ate}$ and $U'$ are plausibly sensitive to the prior of $U'$. This sensitivity is plausible because it reflects different information from the prior but does not produce extreme, unreliable estimates. Another sensitivity is that the Bayesian model can fail when the prior of $U'$ becomes little informative, indicated by a relatively large prior standard deviation of $U'$. This may suggest that we should try using at least weakly informative priors for $U'$.

Finally, we want to build separate models with three unmeasured confounders that come from three different distributions and use the same Normal prior, in order to understand how posterior estimates will change when the prior distribution is different from the true distribution for the unmeasured confounder and how effective normal approximation of the unmeasured confounder will be. 

Before building Bayesian models, we need to calculated sample standard deviations. Pooling of sample standard deviations over different exposure levels needs attention. We are aware that it is crucial to obtain good pooled standard deviations from the sample to inform Bayesian models. However, we do not focus on this topic here, and thus pool standard deviations from two groups, one group with the treatment taken who is compliers in the treatment arm and the other group with the control intervention taken who is the other people. Better pooled standard deviations may be obtained through more careful methods. 

\begin{figure}[htbp]
     \centering
     \includegraphics[width=\textwidth]{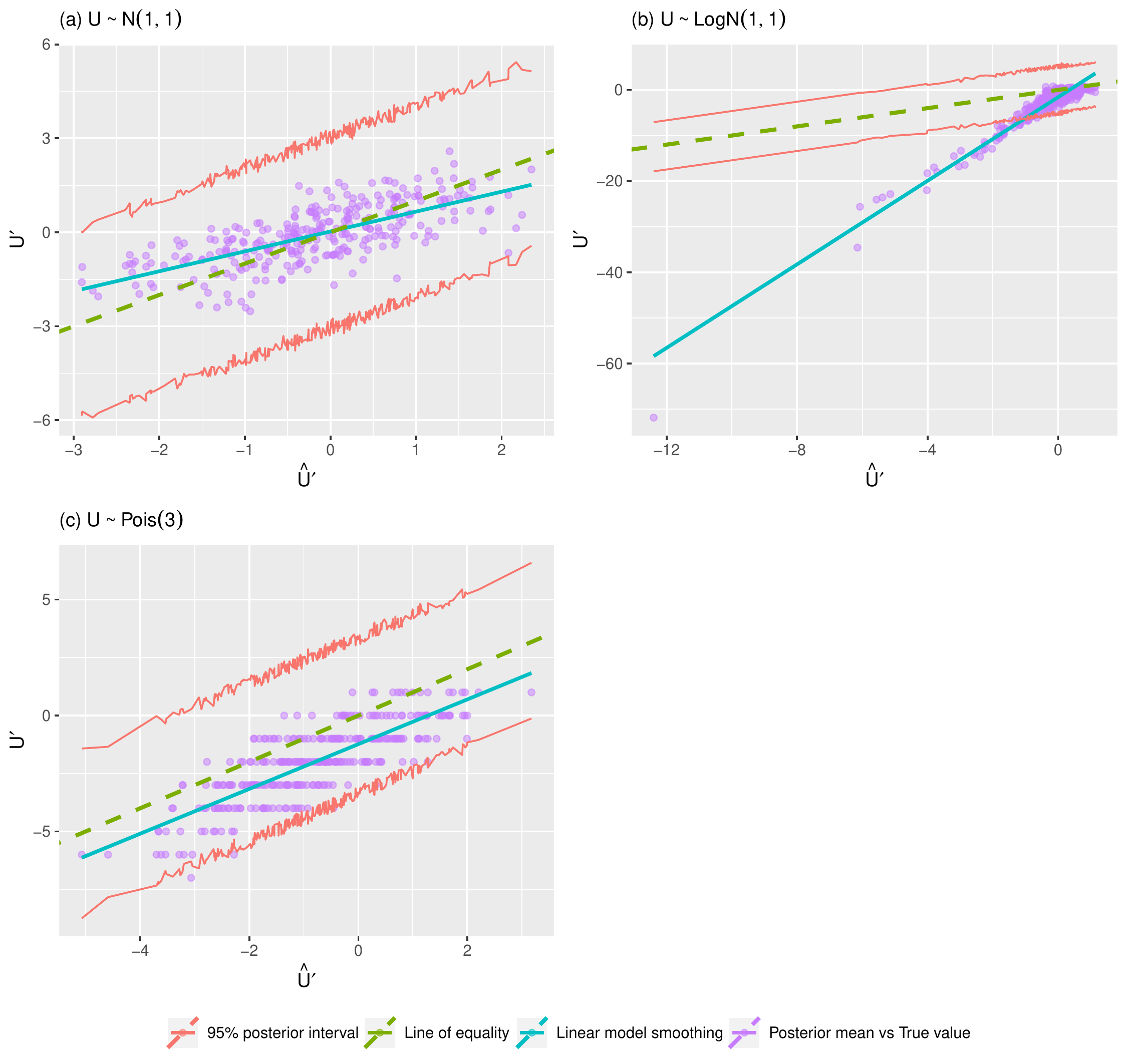}
        \caption{Comparison of posterior estimates and true values of the new unmeasured confounder $U'$. True $U'$ is denoted by $U'$, while estimated $U'$ is denoted by $\hat U'$. True $U' = 1 - U^{obs}$, but due to reparameterization, in reality estimates of $U^{obs}$ cannot be retrieved. Line of equality means that for points on this line, posterior means are equal to true values.}
        \label{fig:sim-u-scenario1-3true}
\end{figure}

Table \ref{tab:sim3-result-3distributions} shows posterior estimates from three identified models with three unmeasured confounders from three different distributions. Figure \ref{fig:sim-u-scenario1-3true} shows the goodness of fit on $U'$ by comparing true values and posterior estimates of $U'$.
\begin{table}[htbp]
\small
   \centering
\begin{tabular}{rrrr}
  \hline
Parameter & True value & Prior & Posterior mean (95\% interval)  \\ 
  \hline
Model 1 &&& \\
$U^{obs}$ & $N(1, 1)$ & & \\
$U'$ & &$N(0, 3)$ & \\
$e_{ate}$ & 2 & $N(0, 1)$ & 2.07 (1.93, 2.21)  \\ 
$\sigma_y$ & 1 &$N(1.93, 0.01)$ & 1.93 (1.91, 1.94)\\
$\alpha_0$ & 4 & $N(0, 1)$& 3.98 (3.67, 4.28) \\
$\alpha_1$ & -1 & $N(0, 1)$& -0.38 (-0.55, -0.21) \\
$\sigma_w$ & 1 & $N(1.38, 0.01)$ & 1.38 (1.36, 1.40) \\
Model 2 &&& \\
$U^{obs}$ & $LogN(1, 1)$ & & \\
$U'$ & &$N(0, 3)$ & \\
$e_{ate}$ & 2 & $N(0, 1)$ & 1.46 (1.33, 1.59)\\ 
$\sigma_y$ & 1 & $N(6.52, 0.01)$ & 6.52 (6.50, 6.54) \\
$\alpha_0$ & 4 & $N(0, 1)$& 6.17 (5.24, 7.03) \\
$\alpha_1$ & -1 & $N(0, 1)$& -0.90 (-1.37, -0.39) \\
$\sigma_w$ & 1 &  $N(5.24, 0.01)$ & 5.24 (5.22, 5.26) \\
Model 3 &&& \\
$U^{obs}$ & $Pois(3)$ & & \\
$U'$ & &$N(0, 3)$ & \\
$e_{ate}$ & 2 & $N(0, 1)$ & 1.72 (1.63, 1.82) \\ 
$\sigma_y$ & 1 & $N(2.27, 0.01)$ & 2.27 (2.25, 2.29) \\
$\alpha_0$ & 4 & $N(0, 1)$& 5.69 (5.28, 6.10) \\
$\alpha_1$ & -1 & $N(0, 1)$& -0.54 (-0.74, -0.36) \\
$\sigma_w$ & 1 &  $N(1.98, 0.01)$  & 1.98 (1.96, 2.00) \\
\hline
\end{tabular}  
   \caption{Comparison of true values and estimates from three models with different unmeasured confounder distributions.}
 \label{tab:sim3-result-3distributions}
\end{table}

When the unmeasured confounder $U^{obs}$ comes from a Normal distribution, the average treatment effect estimates are very good, and the unmeasured confounding effect is well adjusted for in this model because the reparameterized unmeasured confounder $U'$ is estimated well: posterior means of $U'$  are close to true values in a way that most points fall on or vary around the line of equality, and 95\% posterior intervals all cover true values. However, the coefficient of $U'$ in the exposure distribution is overestimated but has the right negative sign. 

When the Normal prior for $U'$ does not well approximate its true distribution reflected by the true distribution of $U^{obs}$, estimation of $U'$ can be insufficiently good so that there is residual confounding and the average treatment effect estimates are biased. For example, when $U^{obs}$ comes from a LogNormal distribution, posterior estimates of $U'$  deviate more from true values, which results in underestimated average treatment effects. However, the distribution of $U^{obs}$ that comes from a Poisson distribution does look like a Normal distribution, but there seems to be a shifting between $U'$ and $\hat{U}'$. From prior results, we learned that when the prior mean of $U'$ deviate from the true value, posterior estimates of $U'$ and $e_{ate}$ can become worse. The prior mean of $U'$ that corresponds to $U^{obs}$ from a Poisson distribution should be negative, but the prior mean of $U'$ we use in the model is 0. Hence, the shifting between $U'$ and $\hat{U}'$ is likely to be caused by the prior of $U'$.

These results indicate that both goodness of normal approximation on $U'$ and the prior of $U'$ are associated with goodness of fit on $U'$ and thus goodness of adjustment for the unmeasured confounding effect, which in turn affects accuracy of the average treatment effect estimates.

\begin{figure}[htbp]
     \centering
     \includegraphics[width=\textwidth, height = \textheight]{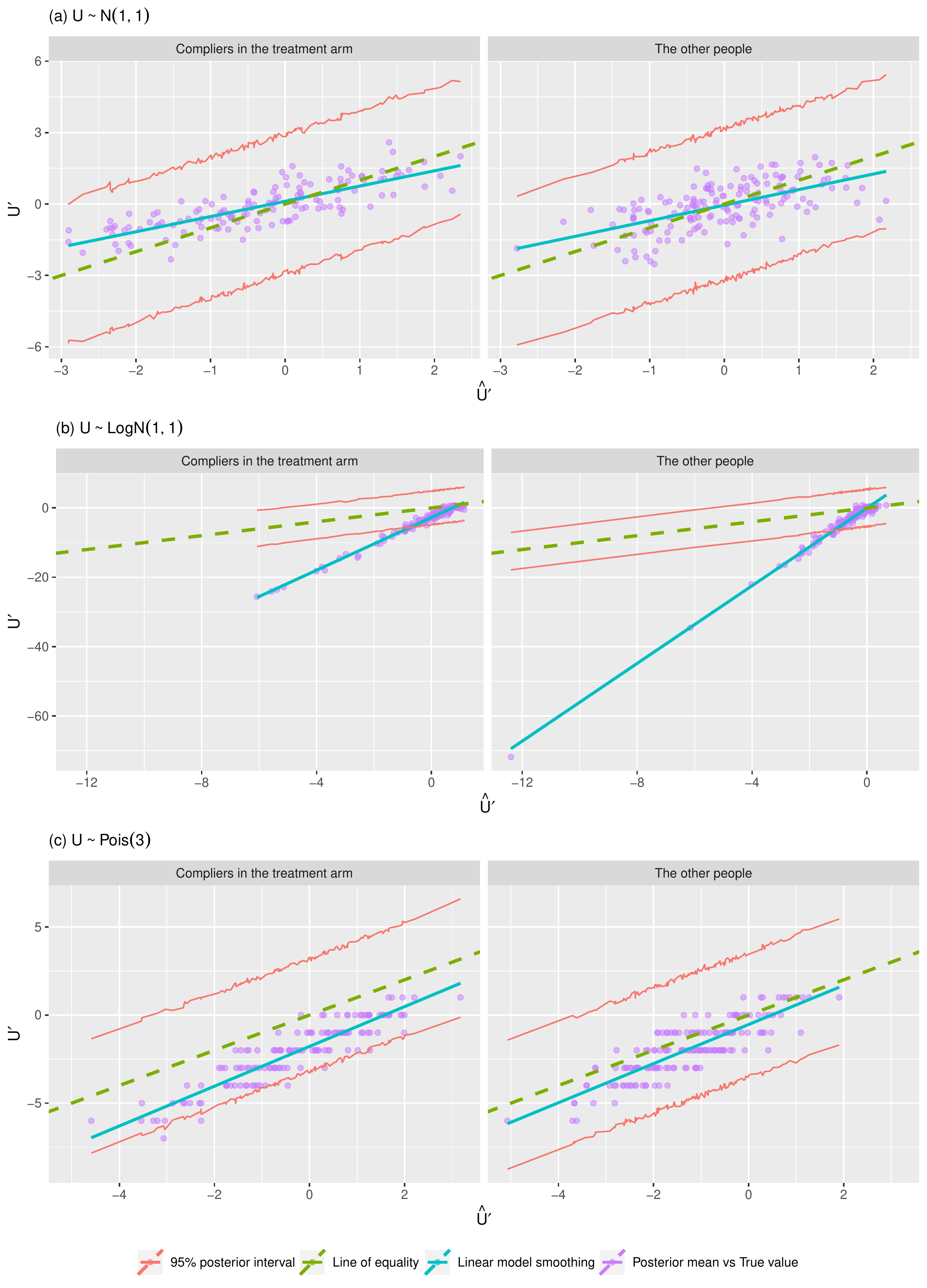}
        \caption{Comparison of posterior estimates and true values of the new unmeasured confounder $U'$ by group. True $U'$ is denoted by $U'$, while estimated $U'$ is denoted by $\hat U'$. Line of equality means that for points on this line, posterior means are equal to true values.}
        \label{fig:sim-u-scenario1-3true-grouping}
\end{figure}

In these three models, we also use a random intercept reparameterization only in the outcome distribution for the other people while in both the exposure and outcome distributions for compliers in the treatment arm. And we are curious about how well the random intercept can adjust for the unmeasured confounding effect under different true distributions of $U^{obs}$ in two groups. Figure \ref{fig:sim-u-scenario1-3true-grouping} shows posterior estimates and true values of $U'$ in two groups. 

Surprisingly, the random intercept reparameterization used in the other people nearly performs as well as the random intercept reparameterization used in compliers in the treatment arm. There is no much residual confounding from the other people when the true distribution of $U^{obs}$ approximate a Normal distribution well. When the true distribution of $U^{obs}$ comes from a Poisson distribution, estimation of $U'$ is even better in the other people and seems to be less affected by the prior of $U'$. These results indicate that the random intercept reparameterization used only in the outcome distribution can effectively adjust for the unmeasured confounding effect. However, we also recognize that the random intercept reparameterization used only in the outcome distribution of the other people is not defined definitely as a confounder and may be also affected by auxiliary variables, if any. In the next section, we will explore estimation of the reparameterized unmeasured confounder in the presence of unadjusted auxiliary variables.

In conclusion, the methods including estimation of the unmeasured confounder and use of sample standard deviations demonstrate usefulness and reliability in Bayesian causal models.

\subsection{When there are measured and unmeasured confounders with auxiliary variables}
\label{sec:ri-2}

This scenario has one-sided noncompliance, one measured confounder, one unmeasured confounder, one auxiliary variable on the exposure, one auxiliary variable on the outcome, no effect of the compliance behavior on the outcome. Auxiliary variables only affect either the exposure or the outcome. We still use the simplest framework on the simulated data and use the random intercept reparameterization in the missing variable model. We want to know how estimation of the unmeasured confounding effect and the average treatment effect be affected when auxiliary variables for the exposure and outcome are not adjusted for in the models.

The distributions used in data simulation are described in equation \ref{eqn:sim-6}. Some settings are identical to the previous simulation scenarios. The proportion of never-takers in the population is set to be 10\%. The measured confounder is denoted by $M^{obs}$. The auxiliary variable on the exposure is denoted by $X^{obs}_w$. The auxiliary variable on the outcome is denoted by $X^{obs}_y$. 
\begin{eqnarray}
Z^{obs} &\sim& Bern(0.5), \nonumber \\
G^{obs} &\sim& Bern(0.9), \nonumber \\
M^{obs} &\sim& Bern(0.7), \nonumber \\
U^{obs} &\sim& N(1, 1), \nonumber \\
X^{obs}_w &\sim& N(10, 2), \nonumber \\
X^{obs}_y &\sim& Bin(3, 0.5), \nonumber \\
W^{obs} &\sim& 
    \begin{cases}
      N(1+M^{obs}-0.5~U^{obs}+0.1~X^{obs}_w, 1)~{\rm T}[0.5,~], & \text{if $Z^{obs}=1$, $G^{obs}=co$} \\
      0, & \text{otherwise}  \nonumber 
    \end{cases} \\
Y^{obs} &\sim& N(1+2~W^{obs}+U^{obs}_1-0.5~U^{obs}_2, 1),
\label{eqn:sim-6}
\end{eqnarray}
where $Bin(n, p)$ is Binomial distribution with the number of trials to be $n$ and the probability of success to be $p$.

The Bayesian models are shown in equation \ref{eqn:sim6model}. Auxiliary variables are not included in the models.
\begin{eqnarray}
Y^{obs} ~|~ G^{obs}, Z^{obs} &\sim& N(\beta_1~W^{obs} + \beta_2~M^{obs} + U', \sigma_y), ~ U' = \beta_0+\beta_3~U^{obs}, \nonumber \\
W^{obs} ~|~ G^{obs} = co, Z^{obs} = 1 &\sim& N(\alpha_0 + \alpha_1~ M^{obs} + \alpha_2~U', \sigma_w).
\label{eqn:sim6model}
\end{eqnarray}

Sample standard deviations are calculated from two groups, compliers in the treatment arm and the other people, and are used to construct strongly informative priors for standard deviations of the exposure and the outcome in the Bayesian models. Table \ref{tab:sim6-result} shows posterior estimates from the identified models.

\begin{table}[htbp]
\small
  \begin{subtable}{0.9\textwidth}
   \centering
   \begin{tabular}{ccr}
  \hline
  $Z^{obs}$ & $G^{obs}$ & $N$ \\ 
  \hline
\begin{minipage}[t]{0.2\textwidth} \centering 0 \end{minipage} &  \begin{minipage}[t]{0.2\textwidth} \centering $nt$ \end{minipage} & \begin{minipage}[t]{0.15\textwidth} \raggedleft 16 \end{minipage}  \\ 
0 &   $co$ & 138 \\ 
1 &   $nt$ &  7 \\ 
1 &   $co$ & 139 \\ 
   \hline 
   \end{tabular}
   \caption{Simulated data. $N$ is the sample size in each category.}
   \end{subtable}
   
\vspace{1em}

 \begin{subtable}{1\textwidth}
   \centering
\begin{tabular}{rrrr}
  \hline
Parameter & True value & Prior & Posterior mean (95\% interval)  \\ 
  \hline
$U'$ & &$N(0, 3)$ & \\
$e_{ate}$ & 2 & $N(0, 1)$ &  1.88 (1.57, 2.23) \\ 
$\beta_2$ & 1 & $N(0, 1)$ &  1.82 (1.17, 2.46) \\ 
$\sigma_y$ & 1 & $N(2.29, 0.01)$ & 2.29 (2.27, 2.31) \\
$\alpha_0$ & 0.5 & $N(0, 1)$ &  1.79 (1.42, 2.14) \\
$\alpha_1$ & 1 & $N(0, 1)$ & 0.83 (0.43, 1.24) \\
$\alpha_2$ & 0.5  & $N(0, 1)$& 0.16 (0.01, 0.29)  \\
$\sigma_w$ & 1 & $N((1.01, 0.01))$ &  1.01 (0.99, 1.03) \\
\hline
\end{tabular}  
   \caption{Comparison of true values and posterior estimates.}
   \end{subtable}  
 \caption{Simulation results.}
 \label{tab:sim6-result}
\end{table}

Posterior estimates of $e_{ate}$ are a little biased but relatively accurate. This indicates that in the presence of the unadjusted auxiliary variables, the average treatment effect can also be estimated relatively well. $\beta_2$ is much overestimated. And we find that there is negative posterior correlation between $e_{ate}$ and $\beta_2$. We want to know if this posterior correlation would affect estimation of $e_{ate}$ and $\beta_2$ when they are in the same outcome distribution. Hence, we try different priors of $\beta_2$ to see how posterior estimates of $\beta_2$ change and how sensitive the average treatment effect estimates are to them. The reference prior is $N(0, 1)$. Other prior means are -2, 2. Other prior standard deviations are 0.5, 2. Table \ref{tab:sim6-b1-post-corr} shows posterior estimates of $e_{ate}$ and $\beta_2$ over different priors of $\beta_2$, with the priors for the other parameters fixed as in the original Bayesian models. 
\begin{table}[htbp]
\small
   \centering
   \begin{tabular}{rrrrr}
  \hline
 \multirowcell{2}[0pt][r]{$\mu_{\beta_2}$} & \multicolumn{3}{c}{$\sigma_{\beta_2}$}  \\ 
  \cmidrule(l){2-4} 
& 0.5 & 1 & 2   \\ 
  \hline
\multirowcell{2}[0pt][r]{-2}& 2.30 (1.91, 2.70) & 1.94 (1.62, 2.31) &1.85 (1.53, 2.20) & ~~$e_{ate}$ \\
&0.67 (0.06, 1.27) & 1.61 (0.98, 2.26) & 1.92 (1.22, 2.55) & $\beta_2$ \\
  \hline
\multirowcell{2}[0pt][r]{0}&  2.03 (1.71, 2.37) &1.88 (1.57, 2.23) & 1.81 (1.50, 2.15)&  \\
  & 1.38 (0.80, 1.92) &1.82 (1.17, 2.46) & 2.00 (1.35, 2.65) & \\
  \hline
\multirowcell{2}[0pt][r]{2}&1.81 (1.51, 2.11) & 1.80 (1.51, 2.13)& 1.80 (1.48, 2.13) & \\
  & 2.03 (1.48, 2.59)& 2.05 (1.44, 2.66)& 2.04 (1.41, 2.67)& \\
   \hline
   \end{tabular}
   \caption{Posterior means (95\% intervals) of the average treatment effect $e_{ate}$ and $\beta_2$ under different priors of $\beta_2$. $\mu_{\beta_2}$ is the prior mean of $\beta_2$. $\sigma_{\beta_2}$ is the prior standard deviation of $\beta_2$. } 
\label{tab:sim6-b1-post-corr}
\end{table}

All the models from table \ref{tab:sim6-b1-post-corr} are identified and show similar posterior correlation between the average treatment effect $e_{ate}$ and $\beta_2$. 

Posterior estimates of $e_{ate}$ and $\beta_2$ are both sensitive to the prior of $\beta_2$. For a fixed prior standard deviation, as the prior mean increases, the posterior mean of $e_{ate}$ decreases while the posterior mean of $\beta_2$ increases. For a fixed prior mean, as the prior standard deviation increases, the posterior mean of $e_{ate}$ decreases while the posterior mean of $\beta_2$ increases. Posterior estimates of $e_{ate}$ show some robustness: the greatest bias in posterior means is 0.3 that seems relatively small, and all 95\% posterior intervals cover the true value. However, posterior estimates of $e_{ate}$ are not robust enough: the posterior mean and interval of $\beta_2$ both shift as the prior of $\beta_2$ changes.

The average treatment effect estimates can be insensitive to some posterior correlations but also can be sensitive to other posterior correlations. We suggest doing prior sensitivity analysis of all coefficients that have posterior correlation with the average treatment effect.

\begin{figure}[htbp]
     \centering
     \includegraphics[width=\textwidth]{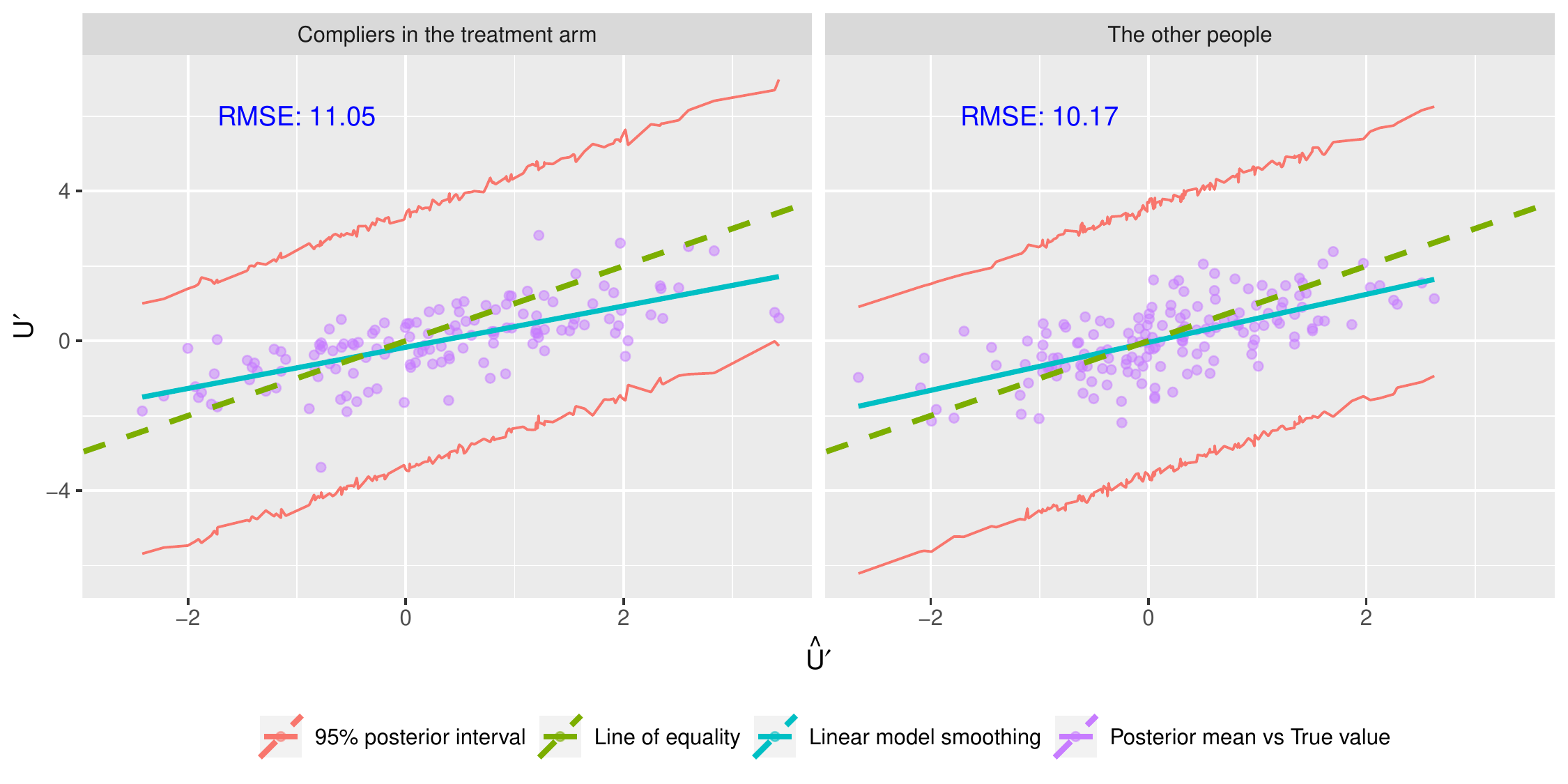}
        \caption{Comparison of posterior estimates and true values of the reparameterized unmeasured confounder $U'$ by group. True $U'$ is denoted by $U'$, while estimated $U'$ is denoted by $\hat U'$. Line of equality means that for points on this line, posterior means are equal to true values.}
        \label{fig:sim-u-scenario3-ufit}
\end{figure}

Then we want to explore how well the reparameterized unmeasured confounder $U'$ is estimated in the presence of unadjusted auxiliary variables. We compare posterior estimates of the reparameterized unmeasured confounder $U'$ with their true values calculated by $U' = 1 - U^{obs}$ in two groups, compliers in the treatment arm and the other people. Figure \ref{fig:sim-u-scenario3-ufit} shows goodness of fit of the reparameterized unmeasured confounder $U'$ compared to its true values by group. 

Goodness of fit on $U'$ is moderate: most points are around the line of equality but deviations are more apparent at two boundaries. In addition, $U'$ is estimated similarly in two groups, which indicates again adjusting for unmeasured confounding only in the outcome distribution can be effective comparably to adjusting for unmeasured confounding in both the exposure and outcome distributions.

We want to know if auxiliary variables affect estimation of $U'$ because $U'$ is not estimated well enough. We use a linear model to understand associations between the estimated reparameterized unmeasured confounder $\hat U'$ and two auxiliary variables $X^{obs}_w$ and $X^{obs}_y$ in two groups. The estimated reparameterized unmeasured confounder $\hat U'$ is posterior means of $U'$. In the linear model,  $\hat U'$ is the response variable, the true reparameterized unmeasured confounder $U'$ and $X^{obs}_w$, $X^{obs}_y$ and the group membership $L$ are the explanatory variables. The group membership is an indicator variable: $L=1$ means compliers in the treatment arm and $L=0$ in the other people. It is also a factor whose reference level is the other people. Interaction between $X^{obs}_w$, $X^{obs}_y$ and $L$ is included, in order to see group differences related to auxiliary variables. We treat the auxiliary variable on the outcome as a numeric variable. The linear model formula is $\hat U' \sim \upsilon_0+ \upsilon_1~U' + \upsilon_2~X^{obs}_w + \upsilon_3~X^{obs}_y + \upsilon_4~L + \upsilon_5~X^{obs}_w~L + \upsilon_6~X^{obs}_y~L$. Table \ref{tab:sim6-ulm} shows the linear model estimates. 95\% Wald intervals are calculated with sandwich variance estimates.
\begin{table}[htbp]
\small
   \centering
\begin{tabular}{rrr}
  \hline
Coefficient & True value & Point estimate (95\% interval) \\
  \hline
$\upsilon_0$ && -0.75(-1.22, -0.27) \\
 $\upsilon_1$&1 & 0.75(0.66, 0.83) \\
 $\upsilon_2$& & 0.05(0.00, 0.10) \\
 $\upsilon_3$& & 0.27(0.16, 0.39) \\
 $\upsilon_4$& & 0.08(-0.75, 0.91) \\
 $\upsilon_5$& & -0.02(-0.10, 0.06) \\
 $\upsilon_6$& & 0.16(-0.04, 0.36) \\ 
 $\upsilon_2+\upsilon_5$&& 0.03 (-0.03, 0.09)\\
 $\upsilon_3+\upsilon_6$&& 0.43 (0.27, 0.59) \\
  \hline
\end{tabular}  
   \caption{Point estimates (95\% interval) from the linear model between the reparameterized unmeasured confounder $U'$, two auxiliary variables and the group membership.}
 \label{tab:sim6-ulm}
\end{table}

$\upsilon_1$ is underestimated. This is why $\hat U'$ does not approximate $U'$ closely and the posterior mean of the average treatment effect is slightly underestimated due to residual confounding. $X^{obs}_w$ has no significant association with $\hat U'$ in either group, which indicates $X^{obs}_w$ may not affect estimation of $U'$. However, in the presence of $U'$ as the true target, $X^{obs}_y$ tends to be positively associated with $\hat U'$, more significantly in compliers in the treatment arm. This means that although most unmeasured confounding effect seems to be adjusted for by $\hat U'$, $\hat U'$ also contains part of the effect from the unadjusted auxiliary variable on the outcome. Hence, in the presence of unadjusted auxiliary variables including those on the outcome, adjustment for the unmeasured confounding effect is viable but becomes less effective. 

On the other hand, since the reparameterized unmeasured confounder is associated with the auxiliary variable on the outcome rather than the auxiliary variable on the exposure, it seems natural to hypothesize that the reparameterized unmeasured confounder may combine all the unadjusted effects on the outcome. Then can we only use a random intercept reparameterization on one bias variable $U$ in the outcome distribution, exclude the exposure distribution from the modelling process, and let the reparameterized bias variable adjust for all unadjusted effects including unmeasured confounding effects in the outcome distribution? The bias variable is a completely missing variable and is used to represent the combination of all the unadjusted variables. A random intercept reparameterization is to define a new bias variable $U'$ that is equal to the linear combination of the intercept and the bias variable in the outcome distribution, such that the reparameterized bias variable $U'$ looks like a random intercept in the outcome distribution. We call the outcome distribution with a reparameterized bias variable ``the random intercept outcome model''. We build this model, with the exposure distribution excluded.  Table \ref{tab:sim6-randomintercept} shows posterior estimates from the random intercept outcome model. 
\begin{table}[htbp]
\small
   \centering
\begin{tabular}{rrrr}
  \hline
Parameter & True value & Prior & Posterior mean (95\% interval)  \\ 
  \hline
$U'$ & &$N(0, 3)$ & \\
$e_{ate}$ & 2 & $N(0, 1)$ & 2.11 (1.82, 2.41) \\ 
$\beta_2$ & 1 & $N(0, 1)$ & 1.54 (0.95, 2.13) \\ 
$\sigma_y$ & 1 & $N(2.29, 0.01)$ & 2.29 (2.27, 2.31) \\
\hline
\end{tabular}  
   \caption{Comparison of true values and posterior estimates from the random intercept outcome model.}
\label{tab:sim6-randomintercept}
\end{table}

Posterior estimates of $e_{ate}$ are also relatively accurate, though the posterior mean is overestimated. $\beta_2$ is still overestimated but becomes less biased than in the original model. 

To understand the relationship between the estimated reparameterized bias variable $\hat U'$ and the auxiliary variable on the outcome, the true reparameterized unmeasured confounder $U'$, we build a linear model with $\hat U'$ as the response variable and the other two variables mentioned as the explanatory variables. $\hat U'$ is posterior means of $U'$. The linear model formula is $\hat U' \sim \upsilon_0+ \upsilon_1~U' + \upsilon_2~X^{obs}_y$. For comparison, we also build another linear model whose response variable $\hat U'$ comes from the original Bayesian models including the exposure distribution where the reparameterized bias variable is the reparameterized unmeasured confounder. Table \ref{tab:sim6-ulm-randomintercept} shows estimates from two linear models. 95\% Wald intervals are calculated with sandwich variance estimates.
\begin{table}[htbp]
\small
   \centering
\begin{tabular}{rrrr}
  \hline
 & $\upsilon_0$ & $\upsilon_1$ & $\upsilon_2$\\
  \hline
True & 0 & 1 & 0.5 \\
Use the exposure distribution &-0.30 (-0.46, -0.13)&0.74 (0.66, 0.83)&0.35 (0.26, 0.45) \\
Not use the exposure distribution &-0.33 (-0.49, -0.17)&0.67 (0.59, 0.75)& 0.34 (0.25, 0.44) \\
  \hline
\end{tabular}  
   \caption{Point estimates (95\% intervals) from two linear models between the estimated random intercept $U'$, the true reparameterized unmeasured confounder $U'$ and the auxiliary variables on the outcome.}
 \label{tab:sim6-ulm-randomintercept}
\end{table}

After the exposure distribution is excluded from the modelling process, the point estimate of $\upsilon_1$ slightly declines from 0.74 to 0.67, with the 95\% Wald interval is slightly shifted towards 0. This indicates that the reparameterized bias variable may adjust for less unmeasured confounding effect in the random intercept outcome model. Hence, only using a the random intercept outcome model without the exposure distribution may suffer more residual confounding. On the other hand, because of relatively accurate estimates of the average treatment effect, the random intercept outcome model shows some usefulness as a simpler surrogate causal model to full adjustment for unmeasured confounding. We do not conduct further research on the random intercept outcome model but suggest the it be studied more thoroughly as a technique to adjust for unmeasured confounding in the future.

\subsection{When there are two unmeasured confounders}

This scenario has one-sided noncompliance, two Normal unmeasured confounders, no effect of the compliance behavior on the outcome. We still use the simplest framework on the simulated data and use the random intercept reparameterization in the missing variable model. We consider three approaches: (1) use a single reparameterized unmeasured confounder and ignore the residual unmeasured confounder term, (2) use a reparameterized unmeasured confounder and another reparameterized random intercept in the exposure distribution to adjust for the residual unmeasured confounder term, (3) use two unmeasured confounders separately. We focus on approach (1). We use approaches (2) and (3) for simple comparison but do not intend to obtain estimates through these two approaches, where the models through approach (3) are non-identified.  We want to know how well the average treatment effect can be estimated in the presence of the mixed unmeasured confounding effects.

The distributions used in data simulation are described in equation \ref{eqn:sim-5}. Some settings are identical to the previous simulation scenarios. The proportion of never-takers in the population is set to be 10\%. Two unmeasured confounders are denoted by $U^{obs}_1$ and $U^{obs}_2$.
\begin{eqnarray}
Z^{obs} &\sim& Bern(0.5), \nonumber \\
G^{obs} &\sim& Bern(0.9), \nonumber \\
U^{obs}_1 &\sim& N(-1, 1), \nonumber \\
U^{obs}_2 &\sim& N(1, 2), \nonumber \\
W^{obs} &\sim& 
    \begin{cases}
      N(1+0.5~U^{obs}_1+U^{obs}_2, 1)~{\rm T}[0.5,~], & \text{if $Z^{obs}=1$, $G^{obs}=co$} \\
      0, & \text{otherwise}  \nonumber 
    \end{cases} \\
Y^{obs} &\sim& N(1+2~W^{obs}+U^{obs}_1-0.5~U^{obs}_2, 1).
\label{eqn:sim-5}
\end{eqnarray}

The Bayesian models through approach (1) are
\begin{eqnarray}
Y^{obs} ~|~ G^{obs}, Z^{obs} &\sim& N(\beta_1~W^{obs} + U', \sigma_y), ~ U' = \beta_0+\beta_2~U^{obs}_1+\beta_3~U^{obs}_2, \nonumber \\
W^{obs} ~|~ G^{obs} = co, Z^{obs} = 1 &\sim& N(\alpha_0 + \alpha_1~U', \sigma_w).
\label{eqn:2u-appro1}
\end{eqnarray}

The Bayesian models through approach (2) are
\begin{eqnarray}
Y^{obs} ~|~ G^{obs}, Z^{obs} &\sim& N(\beta_1~W^{obs} + U', \sigma_y), ~ U' = \beta_0+\beta_2~U^{obs}_1+\beta_3~U^{obs}_2, \nonumber \\
W^{obs} ~|~ G^{obs} = co, Z^{obs} = 1 &\sim& N(A + \alpha_1~U', \sigma_w),
\label{eqn:2u-appro2}
\end{eqnarray}
where $A$ is a parameter vector of length equal to the number of compliers in the treatment arm.

The Bayesian models through approach (3) are
\begin{eqnarray}
Y^{obs} ~|~ G^{obs}, Z^{obs} &\sim& N(\beta_1~W^{obs} + U_1' + U_2', \sigma_y),  \nonumber \\
&& ~ U_1' = \beta_0+\beta_2~U^{obs}_1, U_2' =\beta_3~U^{obs}_2, \nonumber \\
W^{obs} ~|~ G^{obs} = co, Z^{obs} = 1 &\sim& N(\alpha_0 + \alpha_1~U_1' + \alpha_2~U_2', \sigma_w).
\label{eqn:2u-appro3}
\end{eqnarray}

\begin{figure}[htbp]
     \centering
     \includegraphics[width=0.5\textwidth]{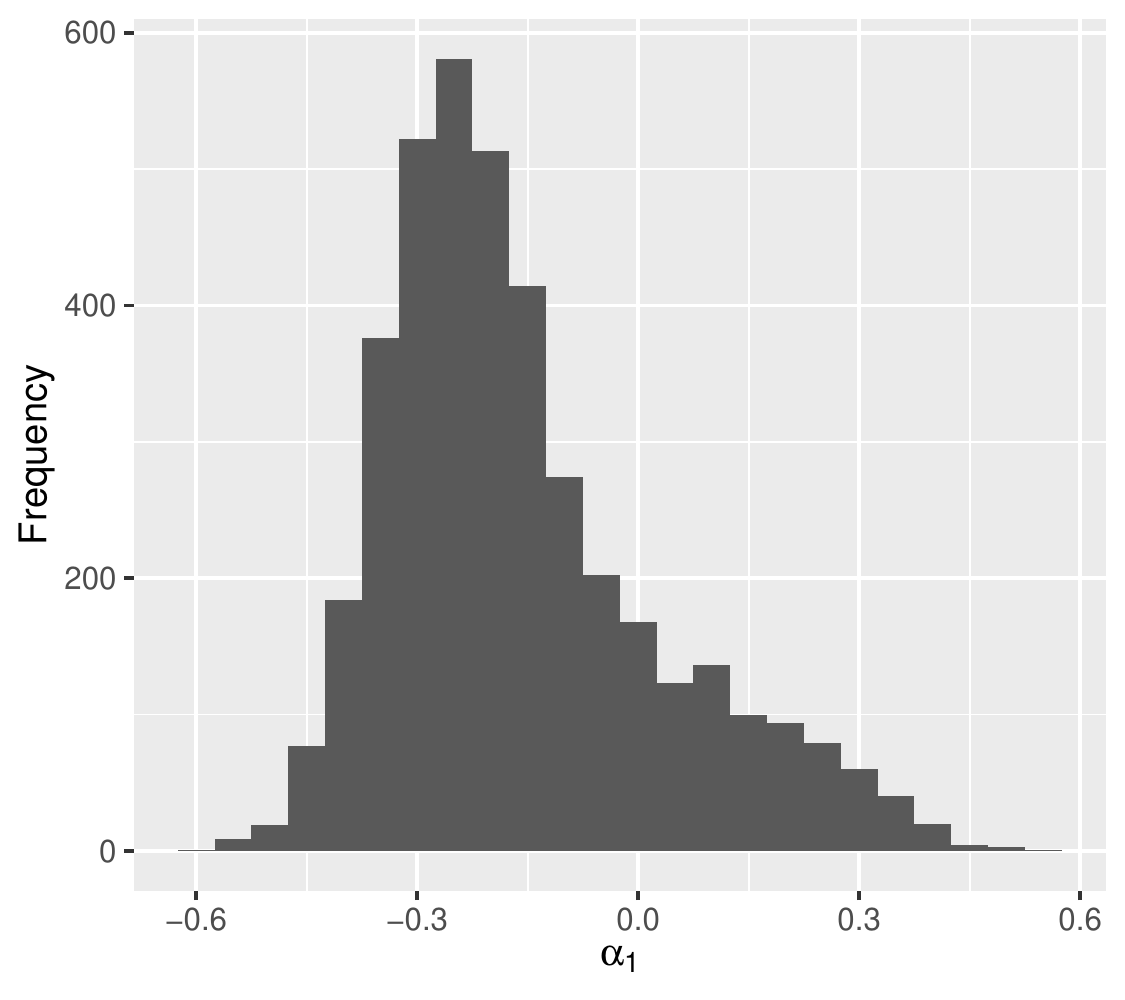}
        \caption{Histogram of posterior estimates of $\alpha_1$.}
        \label{fig:sim-u-scenario2-a2bimodal-original}
\end{figure}

We use sample standard deviations as the prior means for standard deviations of the exposure and the outcome and construct strongly informative priors for these two standard deviations in the models, as mentioned before. When we build Models \ref{eqn:2u-appro1} for the first time, we find that the posterior distribution of $\alpha_1$ tends to be very right-skewed, with a prior of $N(0, 1)$, shown in figure \ref{fig:sim-u-scenario2-a2bimodal-original}. Its bulk is negative but its tail greatly extends to a positive area. The model is problematic because of autocorrelated sampled values and the sign of $\alpha_1$ should only have one choice. Hence, we suspect the posterior distribution of $\alpha_1$ may be bimodal.

\begin{figure}[htbp]
     \centering
     \includegraphics[width=\textwidth, height = \textheight]{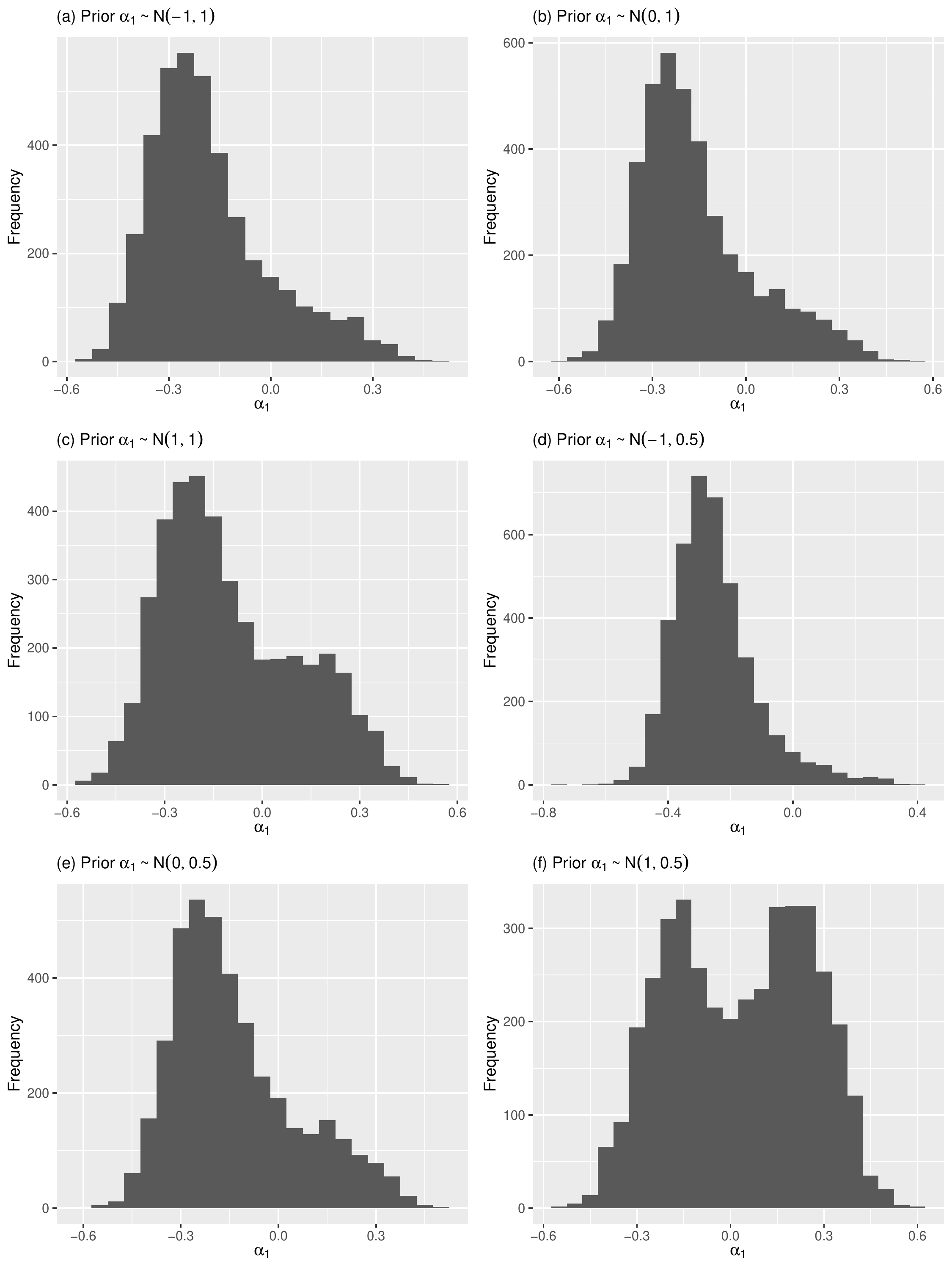}
        \caption{Histograms of posterior estimates of $\alpha_1$ under different priors.}
        \label{fig:sim-u-scenario2-a2bimodal-morepriors}
\end{figure}

Then we try different priors of $\alpha_1$ to see how its posterior distribution will change. At this stage, we do not require the model is fitted well because problems of the model are likely to be associated with bimodal posterior distributions. Figure \ref{fig:sim-u-scenario2-a2bimodal-morepriors} shows the posterior distribution of $\alpha_1$ under different priors. When the prior of $\alpha_1$ concentrates more on the positive area, the posterior distribution $\alpha_1$ shows a more clear positive mode alongside with a negative mode. It seems true that $\alpha_1$ does have a bimodal posterior distribution.

A bimodal posterior distribution of $\alpha_1$ indicates that there are two posterior solutions from Models \ref{eqn:2u-appro1}. We also try approach (2) and approach (3), and find the model similarly has two posterior solutions. Hence, we further build two models through approach (1) for two posterior solutions separately. $\alpha_1$ is set to be non-negative or non-positive in one model, so that we separate two posterior modes into two models. We also build an association model where the unmeasured confounding effects are not adjusted for, for comparison with two posterior solutions. Estimates from two posterior solutions and the association model are shown in table \ref{tab:sim4-a2-2postsol}.
\begin{table}[htbp]
\small
  \begin{subtable}{0.9\textwidth}
   \centering
   \begin{tabular}{ccr}
  \hline
  $Z^{obs}$ & $G^{obs}$ & $N$ \\ 
  \hline
\begin{minipage}[t]{0.2\textwidth} \centering 0 \end{minipage} &  \begin{minipage}[t]{0.2\textwidth} \centering $nt$ \end{minipage} & \begin{minipage}[t]{0.15\textwidth} \raggedleft 21 \end{minipage}  \\ 
0 &   $co$ & 147 \\ 
1 &   $nt$ &  18 \\ 
1 &   $co$ & 114 \\ 
   \hline 
   \end{tabular}
   \caption{Simulated data. $N$ is the sample size in each category.}
   \end{subtable}

\vspace{1em}

 \begin{subtable}{1\textwidth}
   \centering
\begin{tabular}{rrrr}
  \hline
Parameter & True value & Prior & Posterior mean (95\% interval)  \\ 
  \hline
Solution 1: $\alpha_1 \leq 0$ &&& \\
$U'$ & &$N(0, 3)$ & \\
$e_{ate}$ & 2 & $N(0, 1)$ & 1.92 (1.59, 2.24)  \\ 
$\sigma_y$ & 1 & $N(2.31, 0.01)$ &  2.31 (2.29, 2.33) \\
$\alpha_0$ & 2.5 & $N(0, 1)$& 2.07 (1.74, 2.38) \\
$\alpha_1$ & -1.5 & $N(-0.5, 1)$&  -0.24 (-0.45, -0.03) \\
$\sigma_w$ & 1 & $N((1.43, 0.01))$ & 1.43 (1.41, 1.45)  \\
Solution 2: $\alpha_1 \geq 0$ &&& \\
$U'$ & &$N(0, 3)$ & \\
$e_{ate}$ & 2 & $N(0, 1)$ & 1.44 (1.10, 1.79)  \\ 
$\sigma_y$ & 1 & $N(2.31, 0.01)$ & 2.31 (2.29, 2.33) \\
$\alpha_0$ & 2.5 & $N(0, 1)$& 2.05 (1.71, 2.36) \\
$\alpha_1$ & -1.5 & $N(1, 0.5)$& 0.20 (0.01, 0.43)  \\
$\sigma_w$ & 1 & $N((1.43, 0.01))$ & 1.43 (1.41, 1.45)  \\
Association model: no $U^{obs}_1$, $U^{obs}_2$ &&& \\
$e_{ate}$ & 2 & $N(0, 1)$ & 1.83 (1.68, 1.98) \\ 
$\sigma_y$ & 1 & $N(0, 1)$ & 1.85 (1.71, 2.00) \\
\hline
\end{tabular}  
   \caption{Comparison of true values and estimates from two posterior solutions.}
   \end{subtable}  
 \caption{Simulation results.}
 \label{tab:sim4-a2-2postsol}
\end{table}

In Solution 2, the model is not fitted well unless the prior of $\alpha_1$ becomes more informative with a smaller prior standard deviation. Compared to the association model, adjustment for the unmeasured confounding effects in both posterior solutions is informative. Solution 1 where $\alpha_1$ is estimated to be negative produces the most unbiased estimates of the average treatment effect. Posterior estimates of the average treatment effect from Solution 2 where $\alpha_1$ is estimated to be positive are more biased than the association estimates. We seem to obtain an unbiased solution as Solution 1 but also obtain a biased solution as Solution 2. To understand how the two solutions produce different estimates, we compare their posterior estimates of the reparameterized unmeasured confounder $U'$ with the true values calculated by $U' = 1 + U^{obs}_1 - 0.5~ U^{obs}_2$. Figure shows goodness of fit of the reparameterized unmeasured confounder $U'$ compared to its true values. 
\begin{figure}[htbp]
     \centering
     \includegraphics[width=\textwidth]{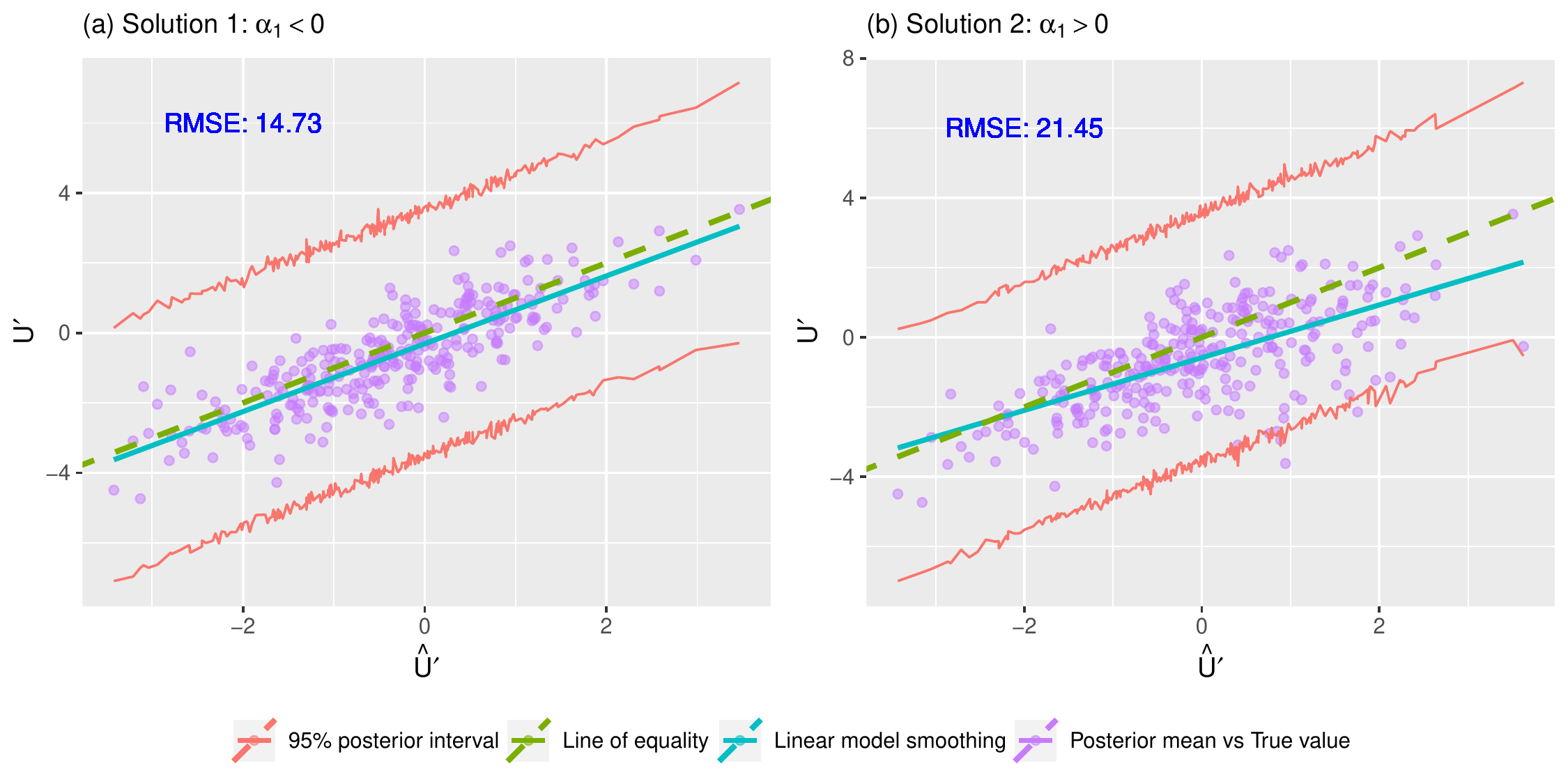}
        \caption{Comparison of posterior estimates and true values of the reparameterized unmeasured confounder $U'$. True $U'$ is denoted by $U'$, while estimated $U'$ is denoted by $\hat U'$. Line of equality means that for points on this line, posterior means are equal to true values.}
        \label{fig:sim-u-scenario2-2postsol-ufit}
\end{figure}

Solution 1 estimates the reparameterized unmeasured confounder $U'$ better than Solution 2. Posterior estimates of Solution 2 deviate from the true values for larger values. We want to further understand why goodness of fit of the reparameterized unmeasured confounder differs in two posterior solutions. Hence, we use linear models to understand the association between the estimated reparameterized unmeasured confounder $\hat U'$ and two true unmeasured confounders $U^{obs}_1$ and $U^{obs}_2$. The estimated reparameterized unmeasured confounder $\hat U'$ is posterior means of $U'$. In a linear model, $\hat U'$ is the response variable and $U^{obs}_1$ and $U^{obs}_2$ are the explanatory variables. The linear model formula is $\hat U' \sim \upsilon_0 + \upsilon_1~U^{obs}_1 + \upsilon_2~U^{obs}_2$. Table \ref{tab:sim5-2postsol-ulm} shows linear model estimates for two posterior solutions. 95\% Wald intervals are calculated with sandwich variance estimates.

\begin{table}[htbp]
\small
   \centering
\begin{tabular}{rrrr}
  \hline
 & $\upsilon_0$ & $\upsilon_1$ & $\upsilon_2$  \\ 
  \hline
True value & 1& 1&-0.5 \\
Solution 1: $\alpha_1 < 0$ &0.75 (0.63, 0.87) &0.60 (0.52, 0.67) &-0.40 (-0.44, -0.37)\\
Solution 2: $\alpha_1 > 0$ &0.93 (0.77, 1.10) & 0.69 (0.58, 0.80)& -0.23 (-0.29, -0.17)\\
\hline
\end{tabular}  
   \caption{Point estimates (95\% interval) from linear models between the reparameterized unmeasured confounder $U'$ and two true unmeasured confounders for two posterior solutions.}
 \label{tab:sim5-2postsol-ulm}
\end{table}

Both posterior solutions have residual confounding due to partially adjusting for two true unmeasured confounders, while Solution 2 adjusts for $U^{obs}_2$ less and thus has more residual confounding than Solution 1. We conclude several results. First, differences in posterior estimates of the average treatment effect from two posterior solutions are likely to be caused by residual confounding. Second, using the models with one reparameterized unmeasured confounder in the presence of multiple unmeasured confounders does lead to residual confounding. Third, even though there is residual confounding, the average treatment effect estimates can still be unbiased, as Solution 1 indicates. Fourth, partial adjustment for the mixed unmeasured confounding effects can produce multiple posterior solutions where unbiased solutions are mixed with biased solutions and it is hard to confirm which are unbiased solutions, as Solution 2 indicates. Finally, the reparameterized unmeasured confounder is estimated to be a combination of all true unmeasured confounders, rather than target on a single true unmeasured confounder; hence, it seems impossible to model either $U^{obs}_1$ or $U^{obs}_2$ through one reparameterized unmeasured confounder.

We give a formal definition of multiple posterior solutions to avoid ambiguity. Multiple posterior solutions are different posterior solutions that all make the same model identified and fitted well and that produce different average treatment effect estimates. We think multiple posterior solutions are another form of model non-identifiability due to residual confounding in the presence of multiple unknown confounders. The model can happen to find these solutions when combining different unmeasured confounding effects. We cannot control estimation flexibility of multiple true unmeasured confounders in one reparameterized unmeasured confounder because the reparameterized unmeasured confounder acts as a whole. There is a chance to obtain unbiased estimates of the average treatment effect, but there is also a risk to obtain overestimated or underestimated average treatment effect estimates. 

Although multiple posterior solutions are undesired, we should not avoid them by deliberately fixing the sign of the coefficient of the reparameterized unmeasured confounder. 
We suggest at first resolving the other types of model non-identifiability (sign reversion, shifting and scaling, model degeneration and zero variance, parameter correlation) and then reporting all multiple posterior solutions due to residual confounding if any. Due to restriction by the model structure, it seems likely, yet unproved, that we may at most obtain a finite set of posterior solutions. In addition, careful interpretation on the average treatment estimates should be necessary in the presence of multiple posterior solutions.

\section{Establishment of a Bayesian modelling procedure}
\label{sec:model-procedure}

From the methods of the Bayesian causal framework and estimation of unmeasured confounders and practical experience through simulation study, we propose a general, applicable Bayesian modelling procedure that implements our methods through RStan \autocite{stan_development_team_rstan_2022}. Steps of the Bayesian modelling procedure are:

(1) Create necessary assumptions in the data context.

(2) Choose a suitable Bayesian causal framework. Construct structural equations of counterfactual variables including unmeasured confounders and the complete data model of observed variables with missing data. Reparameterize structural equations of counterfactual variables or the complete data model.

(3) Build Bayesian models through RStan using the complete data model of observed variables. If standard deviations of the exposure and the outcome are estimated to be zero, incorporate good sample standard deviations in Bayesian models. If multiple posterior solutions occur, build separate Bayesian models for each posterior solution.

(4) Carry out necessary model diagnostics. Modify Bayesian models if any error occurs.

(5) Make inference on posterior distributions of the estimands of interest.

From these steps, we further explain how to choose the priors and how to carry out model diagnostics. Our prior choice and model diagnostics approach can be a good reference or starting point.

The role of priors through our methods is mainly to regularize Bayesian models. Hence, all priors are at least weakly informative. For standard deviations of the exposure and the outcome, their priors are strongly informative, with good sample standard deviations as the prior means and an extremely small value as the prior standard deviations, such as 0.01, in order to force posterior estimates to be the prior means. We do not elaborate on estimation methods of sample standard deviation. For the other parameters except the reparameterized unmeasured confounder such as coefficients, no strong restriction on the priors is needed. Usually we use the standard Normal prior for coefficients, but the scale of the prior should also depend on the scale of the variables being modelled. For the reparameterized unmeasured confounder, its prior should have a larger prior standard deviation than the priors for coefficients, to account for uncertainty in unmeasured confounders. We have used a $N(0, 3)$ prior for the reparameterized unmeasured confounder when sample standard deviations of the variables being modelled are about 1 or 2. However, the prior standard deviation should not be too large, otherwise Bayesian models can be hard to build. The threshold to determine if a prior standard deviation is large is not clear. We think a prior standard deviation slightly larger than sample standard deviations of the variables being modelled would be a good starting point. In addition, sometimes prior predictive checks can help find appropriate priors.

Model diagnostics include Stan diagnostics, check of model identifiability, posterior predictive checks and sensitivity analysis.

There are four main diagnostic measures about Bayesian models: divergence information, $\hat R$ value, Monte Carlo standard error, effective sample size. At first, Bayesian models should converge post warmup. Post-warmup divergence indicates the results may be biased and should not be used. A low $\hat R$ value indicates all Markov chains mix well and converge to a common distribution. It is suggested that $\hat R$ values for all parameters in Bayesian models should be below 1.1. Monte Carlo standard error measures the uncertainty associated with the Monte Carlo approximation. It is suggested that Monte Carlo standard errors for all parameters in Bayesian models should be smaller than 10\% of the posterior standard deviations. A high effective sample size indicates there is little autocorrelation in the draws within a Markov chain. It is suggested that effective sample sizes for all parameters in Bayesian models should be more than 10\% of the total sample size. The \texttt{shinystan} \texttt{R} package \autocite{gabry_shinystan_2022} is the source for descriptions and suggestions of four main diagnostic measures and is used to carry out Stan diagnostics.

Non-existence of four types of model non-identifiability should be confirmed in Bayesian models for a single posterior solutions. They are reversion of the sign of the unmeasured confounder, shifting and scaling of the unmeasured confounder, degeneration of the causal model and zero variances of the exposure and the outcome, correlation between exposure and outcome coefficients of the same meaning. Reparameterization and sample standard deviations should be used. Then we need to confirm that the coefficients of the reparameterized unmeasured confounder and the reparameterized unmeasured confounder itself are estimated to be proper. This usually means, posterior means of the coefficients of the reparameterized unmeasured confounder are non-zero, posterior estimates of these coefficients are finite, and posterior means of the reparameterized unmeasured confounder are not constant, posterior estimates of the reparameterized unmeasured confounder are finite. We also need to confirm that there is no or very slight, negligible posterior correlation between standard deviations of the variables being modelled. These steps make sure Bayesian models are identified for a single posterior solution.

Usually in posterior predictive checks, we generate samples from a model of one variable with estimated parameters to see if generated samples fit into data by comparing the distributions of minimal and maximal numbers, mean, standard deviation from generated samples with sample quantities. If sample quantities fall into a likely region in the sampling distributions, then the model of this variable is acceptable. But we do not set a hard principle in posterior predictive checks. One example of a hard principle is that all sample quantities must fall into the 95\% intervals of the sampling distributions. Usually Bayesian models we meet satisfy this hard principle, but we should determine if we will accept a Bayesian model by considering posterior predictive checks, modelling difficulty degree and model usefulness. In addition, we do not do posterior predictive checks strictly, which means we do not do posterior predictive checks on every variable that has a model specification in Bayesian models. Posterior predictive checks are carried out on some main variables or when they are needed to help make a decision on the model performance. 

Sensitivity analysis is to analyze sensitivity of the average treatment effect estimates to some part in the primary Bayesian models. Our approach is to change this part in primary Bayesian models to see how the average treatment effect estimates would change. Sensitivity analysis we will conduct in data analysis includes sensitivity analysis of missing data, sensitivity analysis of unmeasured confounders, sensitivity analysis of the priors, sensitivity analysis of sample standard deviations. The primary Bayesian models are the most complete models, which include missing data and unmeasured confounders.

Sensitivity analysis of missing data is to compare posterior estimates of the average treatment effect from the primary Bayesian models and the complete-case model that has the same model specification as the primary Bayesian models but only uses complete-case data. 

Sensitivity analysis of unmeasured confounders is to compare posterior estimates of the average treatment effect from the primary Bayesian models and another causal model that does not adjust for unmeasured confounders but has the same other parts in the model specification as the primary Bayesian models. If the average treatment effect estimates are insensitive to unmeasured confounders, there can be little unmeasured confounding effect. If the average treatment effect estimates are sensitive to unmeasured confounders, the causal model in comparison can be an association model that has residual confounding. 

Sensitivity analysis of the priors is to understand how posterior estimates of the average treatment effect change as the priors of some parameters in the primary Bayesian models change. To analyze the impact of the priors of the other parameters on the average treatment effect except its own prior, we change the priors of some parameters, such as the reparameterized unmeasured confounder and coefficients that have posterior correlation with the average treatment effect, in case there are a large amount of parameters. The prior mean and the prior standard deviation will be considered to be smaller and larger. This approach is fixed sensitivity analysis and similar to the one used by Rosenbaum and Rubin \autocite*{rosenbaum_assessing_1983}. If we want to understand how posterior estimates of the average treatment effect change with its own prior or how posterior estimates of a parameter change with its own prior, we use the prior sensitivity measure proposed by M\"{u}ller \autocite{muller_measuring_2012} in replacement of fixed sensitivity analysis for simplicity. The prior sensitivity measure is denoted by $S_p$. It is ``the derivative of the posterior mean relative to the prior mean'' (M\"{u}ller, 2012, p. 584) and is equal to ``the ratio of posterior and prior variance'' (M\"{u}ller, 2012, p. 584). We interpret this measure as the average difference in the posterior mean per one unit difference in the prior mean for one parameter with the prior standard deviation fixed.

Sensitivity analysis of sample standard deviations is to understand how posterior estimates of the average treatment effect change as sample standard deviations vary in a plausible interval. We construct strongly informative priors for standard deviations in Bayesian models so that posterior estimates of standard deviations are nearly sample standard deviations. Sample standard deviations we use to construct the priors are still point estimates and should have variability. We pick some other values from interval estimates of sample standard deviations and then build Bayesian models with each set of point estimates of standard deviations, to see how posterior estimates of the average treatment effect change. The approach to take the variability of sample standard deviations into account is also fixed sensitivity analysis. We do not do prior sensitivity analysis of standard deviations in Bayesian models and this sensitivity analysis is not necessary either, because we use sample standard deviations to avoid estimation of standard deviations in Bayesian models and model non-identifiability due to zero variances.



\chapter{Analysis of the PreventS trial data}
\label{sec:analysis-1}

\section{Preparation}

Health Wellness Coaching (HWC) is used as an intervention for primary prevention of stroke and cardiovascular disease in the PreventS trial. 15 HWC sessions were given to each participant in the treatment arm over 9 months post randomization. The PreventS trial data has one-sided noncompliance. Two sets of treatment are defined in compliers in the treatment arm, and each set of treatment will be analyzed independently. We first define non-zero treatments and then discuss how their natural zero level is treated. In the first set of treatment, we define a single treatment as the total number of HWC sessions attended at 9 months post randomization. This treatment is denoted by $W^{09, obs}$. In the second set of treatment, we define three sequential treatments as (1) the total number of HWC sessions attended at 3 months post randomization, (2) the total number of HWC sessions attended between 3 and 6 months post randomization, and (3) the total number of HWC sessions attended between 6 and 9 months post randomization, where the past total attendance numbers affect the future total attendance numbers. Three treatments are denoted by $W^{03, obs}$, $W^{36, obs}$, $W^{69, obs}$. For the first set of treatment, we want to estimate the average treatment effect on one outcome, that is, the average effect of one HWC session on one outcome. For the second set of treatment, we want to estimate three average treatment effects on one outcome, that is, the average effect of one HWC session on one outcome during three different periods post randomization. We will analyze three sequential treatments in the same models. Implicitly here, we assume a constant treatment effect on all the participants if one treatment is taken.

Four assessments were conducted at baseline and 3, 6, 9 months post randomization. The baseline assessment was conducted soon after randomization. Coaching evaluation forms completely record whether each of 15 HWC sessions was attended at 9 months post randomization. They also partially record when HWC sessions were attended. There is one implausible HWC session attendance date that was earlier than the randomization date. Since this HWC session was attended, this HWC session attendance date is set to be the randomization date rather than treated as missing. There are two implausible assessment dates at 3 and 6 months post randomization for the same participant because they are equal. The assessment date at 3 months post randomization is set to be missing because the duration between the recorded assessment date at 3 months post randomization and the assessment date at 9 months post randomization is about 2 months, which seems unlikely for another assessment to happen during this duration and thus supports that the recorded assessment date at 6 months post randomization is correct.

From records of coaching evaluation forms, it is straightforward to calculate the first set of treatment. To calculate the second set of treatment, we compare the HWC session attendance dates and the assessment dates. If an HWC session is recorded as attended by coaching evaluation forms and its attendance date is earlier than an assessment date, then this HWC session is regarded as attended before this assessment. If an HWC session is recorded as not-attended by coaching evaluation forms and its attendance date is the same as or later than an assessment date, then this HWC session is regarded as not-attended before this assessment. 

All the participants take their usual care that has been started before the PreventS trial. The HWC intervention is only available to participants assigned to the treatment arm. It is an addition to usual care if taken. Participants cannot attend any HWC session in the control arm. And no other form of control interventions is provided, such as placebo or other standard care. Hence, there is no always-taker, and the control intervention in the PreventS trial data can be regarded as zero HWC session attended that means the natural zero level of the treatments. Hence, never-takers in both arms can represent the natural zero level of the treatments. However, compliers in the control arm cannot represent the natural zero level of the treatments, because their exposure levels are suppressed by the assignment. Then we define two sets of exposure that can fit into our methods mentioned above. An exposure includes both a control and a treatment. For participants in the treatment arm and never-takers in the control arm, exposures are equal to treatments that now can take their natural zero value. For compliers in the control arm, exposures include treatments and the control intervention assigned by the assignment that cannot represent the natural zero level of the treatments. Different meanings of exposures lead to different ways of using the exposure distributions. For simplicity, we call the first set of exposure, the exposure at 9 months post randomization, and we call the second set of exposure, the exposure at 3 months post randomization, the exposure between 3 and 6 months post randomization, the exposure between 6 and 9 months post randomization.

We consider four outcomes. They are the change in the 5-year cardiovascular disease (CVD) risk score recalculated with PREDICT at 9 months post randomization, the change in the Life’s Simple Seven (LS7) total score at 9 months post randomization, the change in systolic blood pressure (SBP) at 9 months post randomization, the change in the Patient Health Questionnaire-9 (PHQ-9) score at 9 months post randomization. For simplicity, we call the four outcomes, the CVD risk outcome, the LS7 outcome, the SBP outcome and the PHQ-9 outcome. Four outcomes are denoted by $Y^{cvd, obs}$, $Y^{ls7, obs}$, $Y^{sbp, obs}$, $Y^{phq, obs}$. Each outcome is the difference between the risk score or level measured at 9 months post randomization and the baseline score or level. The first three outcomes measure cardiovascular risk. The fourth outcome measures mood. We think these four outcomes are useful to help us understand how the HWC intervention can benefit cardiovascular and mental health of the participants. The baseline scores or levels of the outcomes are restricted by inclusion and exclusion eligibility of the PreventS trial. Hence, we use the score or level at 9 months post randomization of the outcomes to explain the scale of each score or level. The 5-year CVD risk score at 9 months post randomization is left-truncated at 0 and right censored at 0.3. One baseline 5-year CVD risk score and three 5-year CVD risk score at 9 months post randomization are greater than 0.3, and they are then censored at 0.3 because 5-year CVD risk scores that are greater than 0.3 are unreliable. The LS7 total score at 9 months post randomization is bounded between 0 and 14. The SBP at 9 months post randomization is not bounded or censored. The PHQ-9 score at 9 months post randomization is bounded between 0 and 27.

Possible confounders are selected from pre-exposure, time-independent measured baseline covariates in the PreventS trial data based on our limited knowledge. Selected measured confounders are age, gender, ethnicity, education level, smoking status, alcohol frequency, baseline score or level of each outcome including the baseline 5-year CVD risk score, the baseline LS7 total score, the baseline SBP and the baseline PHQ-9 score. We also use generalized additive models on complete cases to explore functional forms of continuous confounders associated with each outcome. Exposures are included in generalized additive models and are calculated by HWC session attendance dates and assessment dates, from which missing dates are imputed by the \texttt{mice} \texttt{R} package \autocite{buuren_mice_2011}. Imputaions will not be used in Bayesian models. The CVD risk, SBP and PHQ-9 outcomes use Gaussian family and identity link. The LS7 outcome uses Poisson family and identity link. We find an insignificant quadratic term of age for the CVD risk and PHQ-9 outcomes, and a significant spline of the baseline 5-year CVD risk score for the CVD risk outcome. Hence, a B-spline of degree 3 of the baseline 5-year CVD risk score is used for the CVD risk outcome. The B-spline will be directly computed in Bayesian models \autocite{kharratzadeh_splines_2017}.

Table \ref{tab:descrip-stats-new} shows the meanings of measured confounders, descriptive statistics and comparisons of both measured confounders, exposures and outcomes in two arms. The empirical distributions of four outcomes approximate Normal distributions in two arms to some extent. Hence, two sample \textit{t}-tests are used for four outcomes. Kruskal-Wallis rank sum tests are used for the other numeric variables including exposures. Testing does not matter for exposures because difference in exposures between two arms is definite and we only want to show the range of each exposure. Exact tests are used for categorical variables. Measured confounders are distributed similarly in two arms. It seems natural to think the difference in the means of the outcomes is mainly attributed to the exposures. Hence, when we calculate sample standard deviations of the outcomes, we pool standard deviations of the outcomes over different exposure levels, without stratification on measured confounders. If deeper stratification on the exposures leads to different distributions of measured confounders over strata, we will then consider different pooling methods. We also notice that the scale of the CVD risk outcome is very small, which possibly makes it hard to identify significant average treatment effects. And the empirical distribution of the exposure at 9 months post randomization is left-skewed.
\begin{table}[htbp]
\small
  \begin{subtable}{1\textwidth}
   \centering
\begin{tabular}{lrrr}
  \hline
Variable& Treatment arm & Control arm & \textit{p}-value \\ 
  \hline
N &     159 &    161 &  \\ 
Age$^1$, years  &  60.71 [54.43, 66.87] &  60.57 [54.93, 65.51] &  0.75 \\ 
Baseline 5-year CVD risk score$^1$  &  0.14 [0.11, 0.18] &   0.14 [0.12, 0.17] &  0.49 \\ 
Baseline LS7 total score$^1$ &     7.00 [6.00, 8.00] &   7.00 [6.00, 9.00] &  0.49 \\ 
Baseline SBP$^1$ & 130.00 [120.00, 140.00] & 133.00 [122.00, 140.00] &  0.52 \\ 
Baseline PHQ-9 score$^1$ &  4.00 [1.00, 8.00] &   2.00 [1.00, 7.00] &  0.15 \\
Exposure at 9 months$^1$ & 13.00 [5.00, 15.00] &   0.00 [0.00, 0.00] & $<$0.01 \\ 
Exposure at 3 months$^1$  & 5.00 [3.00, 7.00] &   0.00 [0.00, 0.00] & $<$0.01 \\ 
Exposure between 3 and 6 months$^1$ &  3.00 [2.00, 4.00] &   0.00 [0.00, 0.00] & $<$0.01   \\ 
Exposure between 6 and 9 months$^1$ & 3.00 [1.00, 4.00] &   0.00 [0.00, 0.00] & $<$0.01 \\ 
CVD risk outcome$^2$ &  -0.02 (0.08) &  -0.01 (0.07) &  0.53 \\ 
LS7 outcome$^2$ &  0.35 (2.06) &  -0.41 (2.05) & $<$0.01 \\ 
SBP outcome$^2$ &   4.51 (22.04) &  10.58 (18.14) &  0.02 \\ 
PHQ-9 outcome$^2$ &  -1.88 (4.46) &  -1.50 (3.98) &  0.47 \\ 
   \hline
\multicolumn{4}{l}{$^1$ Statistic is median [IQR] for exact tests.} \\ 
\multicolumn{4}{l}{$^2$ Statistic is mean (SD) for \textit{t}-tests.} \\
\multicolumn{4}{l}{* Months in exposures mean months post randomization.} \\ 
\end{tabular}
   \caption{Median [interquartile range] or mean (standard deviation) from numeric variables.}
   \end{subtable}
\end{table}

\begin{table}[htbp]
\ContinuedFloat
\small
  \begin{subtable}{1\textwidth}
   \centering
\begin{tabular}{llrrr}
  \hline
Variable & Level & Treatment arm & Control arm & \textit{p}-value \\ 
  \hline
Gender & Female* &     61 (38.4)  &     62 (38.5)  &  1.00 \\ 
& Male &     98 (61.6)  &     99 (61.5)  &  \\ 
Ethnicity & European and others*&     60 (37.7)  &     62 (38.5)  &  1.00 \\ 
& Maori &     34 (21.4)  &     34 (21.1)  &  \\ 
 & Pasifika &     36 (22.6)  &     37 (23.0)  &  \\ 
 & Asian &     29 (18.2)  &     28 (17.4)  &  \\ 
Education level  & High school or below* &     77 (48.4)  &     80 (49.7)  &  0.97 \\ 
 & Polytechnic or college &     33 (20.8)  &     32 (19.9)  &  \\ 
 & University &     49 (30.8)  &     49 (30.4)  &  \\ 
Smoking status & Current smoker*  &    130 (81.8)  &    125 (77.6)  &  0.41 \\ 
 & Never or past smoker &     29 (18.2)  &     36 (22.4)  &  \\ 
Alcohol frequency & More than once a week* &     89 (56.0)  &     88 (54.7)  &  0.82 \\ 
 & Once a week or less &     70 (44.0)  &     73 (45.3)  &  \\ 
   \hline
\multicolumn{5}{l}{* is the reference level.} \\ 
\end{tabular}
\caption{Frequency (percentage\%) of levels from categorical variables.}
   \end{subtable}  
  \caption{Descriptive statistics of measured confounders, exposures and outcomes by arm.}
  \label{tab:descrip-stats-new}
\end{table}

Table \ref{tab:missing-stats-new} shows how much data is missing in measured confounders, exposures and outcomes in either arm or over two arms. All four outcomes have a moderate amount of missing data that takes up approximately about 20\% of all the data. There appears to be more missing data in outcomes in the treatment arm than in the control arm. However, we do not have evidence to support data is missing by arm. Hence, we think MCAR and MAR assumptions about missing data are still useful. The first set of exposure has no missing data, where there are 13 never-takers and 146 compliers. The second set of exposure has a moderate amount of missing data in the treatment arm. Either missing HWC session attendance dates or missing assessment dates or both will lead to missing attendances in the second set of exposure. There are four missing HWC session attendance dates in four participants, each from whom has one missing HWC session attendance date, with several or many observed HWC session attendances. But three exposures from the second set of exposure of these four participants are treated as missing, which means many observed HWC session attendances are also treated as missing. Data of the baseline variables is nearly complete, where the baseline LS7 total score and the baseline SBP have a few missing data. Although this moderate amount of missing data seems not to be a problem, existence of most missing data in the outcomes and the exposures in the treatment arm may lead to bias in the average treatment effect estimates.
\begin{table}[htbp]
\small
\centering
\begin{tabular}{lrrr}
  \hline
Variable & Treatment arm & Control arm & Overall \\ 
  \hline
LS7 outcome & 27.0 & 14.9 & 20.9 \\ 
PHQ-9 outcome& 26.4 & 12.4 & 19.4 \\ 
CVD risk outcome & 25.2 & 10.6 & 17.8 \\ 
SBP outcome & 25.2 & 10.6 & 17.8 \\ 
Exposure between 3 and 6 months post randomization& 23.3 & 0.0 & 11.6 \\ 
Exposure between 6 and 9 months post randomization& 20.8 & 0.0 & 10.3 \\ 
Exposure at 3 months post randomization& 16.4 & 0.0 & 8.1 \\
Baseline LS7 total score & 1.9 & 1.2 & 1.6 \\ 
Baseline SBP  & 0.6 & 0.0 & 0.3 \\ 
Exposure at 9 months post randomization& 0.0 & 0.0 & 0.0 \\ 
Age & 0.0 & 0.0 & 0.0 \\
Gender& 0.0 & 0.0 & 0.0 \\ 
Ethnicity & 0.0 & 0.0 & 0.0 \\ 
Education level & 0.0 & 0.0 & 0.0 \\ 
Smoking status & 0.0 & 0.0 & 0.0 \\ 
 Alcohol frequency  & 0.0 & 0.0 & 0.0 \\ 
Baseline 5-year CVD risk score  & 0.0 & 0.0 & 0.0 \\ 
Baseline PHQ-9 score & 0.0 & 0.0 & 0.0 \\ 
   \hline
\end{tabular}
  \caption{Percentages (\%) of missing data in measured confounders and outcomes by arm or over two arms.}
  \label{tab:missing-stats-new}
\end{table}

It is very likely that unmeasured confounders exist beyond our knowledge. We assume there are multiple unmeasured confounders between each exposure and each outcome. For three sequential exposures, we assume they have multiple unmeasured confounders in common. The missing variable model through the random intercept reparameterization will be used for each outcome. One reparameterized unmeasured confounder will be used, which means in the presence of multiple unmeasured confounders, the residual unmeasured confounder term is not adjusted for and thus there may be residual confounding. 

We build Bayesian models for each outcome separately for two reasons. The first reason is that we want to see if there is any association between the reparameterized unmeasured confounders from multiple outcomes. We think unmeasured confounders may be common for outcomes that measure the same thing and different for outcomes that measure different things. If the reparameterized unmeasured confounders from outcomes that measured the same thing are correlated, then there can be some common unmeasured confounders. If the reparameterized unmeasured confounders from outcomes that measured different things are uncorrelated, then there may be no common unmeasured confounders. The second reason is that building multiple outcomes together is more difficult, where construction of causal relationships among outcomes are needed. Since our methods are new, we want to first test them in a relatively simpler approach.

Other necessary assumptions not mentioned in this section include stable unit treatment value, identical condition and randomization for Bayesian causal frameworks, modelling assumptions from assumptions \ref{assump:compliance} to \ref{assump:mar}. We also assume the compliance behavior has a direct effect on exposures and only affects outcomes through exposures. Causal graphs are not drawn because causal graphs from former sections in the presence of one or three exposures can directly apply, where we only need to replace letter symbols by specific names of exposures, outcomes, confounders.

Bayesian models are built by the \texttt{rstan} \texttt{R} package version 2.21.5 \autocite{stan_development_team_rstan_2022} in \texttt{RStudio} version 2022.07.1+554. The \texttt{rstan} \texttt{R} package version 2.21.5 \autocite{stan_development_team_rstan_2022} uses \texttt{Stan} version 2.21.0. \texttt{RStudio} version 2022.07.1+554 is supported by \texttt{R} version 4.2.1 \autocite{r_core_team_r_2022}. Mdelling procedures are identical to those mentioned in section \ref{sec:model-procedure}, ``Establishment of a Bayesian modelling procedure''.

\section{Derivation of complete data models}

Since never-takers represent the natural zero level of the treatments, their exposure distributions can be used to estimate unmeasured confounders. For participants in the treatment arm and never-takers in the control arm, we adjust for unmeasured confounding in both the exposure and outcome distributions. For compliers in the control arm, we adjust for unmeasured confounding only in their outcome distributions, because their exposure distributions are fixed at zero. 

However, the compliance behavior in the control arm is unknown and thus needs estimation in Bayesian models to help us distinguish never-takers from compliers. Estimation of the compliance behavior as a binary variable is not straightforward in RStan because RStan does not support estimation of integers. Hence, marginalization is used, where we first need to find the complete data model with the missing compliance behavior given a Bernoulli distribution through a logistic model, then we marginalize or integrate the missing compliance behavior out of the complete data model, and finally build Bayesian models with the marginalized complete data model. This approach can avoid direct estimation of the missing compliance behavior. Next, we show how the marginalized complete data model is obtained.

In the presence of one exposure, we start with equation \ref{eqn:likelihood-with-missing}, the complete data model with the missing compliance behavior and missing data from the causal frameworks for observed variables, and further write it as
\begin{eqnarray}
&& f(y^{obs}, y^{obs}_{mis}, w^{obs}, w^{obs}_{mis}~|~Z^{obs}, G^{obs}, G^{obs}_{mis}, M^{obs}, M^{obs}_{mis}, U^{obs}, \theta) \nonumber \\
&=& \prod_{i \in \text{UC}} \left\{ ~f_i(y(0,w(0))^{obs}, y(0,w(0))^{mis}~|~W(0)^{obs}, W(0)^{mis}, G^{obs}_{mis}, M^{obs}, M^{obs}_{mis}, U^{obs}, \theta) \right. \nonumber \\
&& \left. \cdot~f_i(w(0)^{obs}, w(0)^{mis}~|~G^{obs}_{mis}, M^{obs}, M^{obs}_{mis}, U^{obs}, \theta) \right \} \nonumber \\
&& \cdot~ \prod_{j \in \text{HWC}}~\left \{ f_j(y(1,w(1))^{obs}, y(1,w(1))^{mis}~|~W(1)^{obs}, W(1)^{mis}, G^{obs}, M^{obs}, M^{obs}_{mis}, U^{obs}, \theta) \right. \nonumber \\
&&\left. \cdot~f_j(w(1)^{obs}, w(1)^{mis}~|~G^{obs}, M^{obs}, M^{obs}_{mis}, U^{obs}, \theta) \right \} \nonumber \\
&=& \prod_{i \in \text{UC}} \left\{ ~f_i(y(0,w(0))^{obs}, y(0,w(0))^{mis}~|~W(0)^{obs}, W(0)^{mis}, M^{obs}, M^{obs}_{mis}, U^{obs}, \theta) \right. \nonumber \\
&& \left. \cdot~f_i(w(0)^{obs}, w(0)^{mis}~|~G^{obs}_{mis}, M^{obs}, M^{obs}_{mis}, U^{obs}, \theta) \right \} \nonumber \\
&& \cdot~ \prod_{j \in \text{HWC}}~\left \{ f_j(y(1,w(1))^{obs}, y(1,w(1))^{mis}~|~W(1)^{obs}, W(1)^{mis}, M^{obs}, M^{obs}_{mis}, U^{obs}, \theta) \right. \nonumber \\
&&\left. \cdot~f_j(w(1)^{obs}, w(1)^{mis}~|~G^{obs}, M^{obs}, M^{obs}_{mis}, U^{obs}, \theta) \right \}, \\
&& \left (\text{$Y^{obs}, Y^{obs}_{mis} \upvDash G^{obs}, G^{obs}_{mis}~|~W^{obs}, W^{obs}_{mis}$} \right) \nonumber 
\label{eqn:likelihood-with-missing-1sided}
\end{eqnarray}
where ``UC'' represents the control arm and ``HWC'' represents the treatment arm, for simplicity. For generalizability, we include missing data for the exposure $W^{obs}_{mis}$ in the complete data model.

The observed compliance behavior is affected by the observed measured and unmeasured confounders. Hence, we denote the distribution of the observed compliance behavior by $f(g~|~M^{obs}, M^{obs}_{mis}, U^{obs}, \theta)$ that applies to both $G^{obs}$ and $G^{obs}_{mis}$. A Bernoulli distribution through a logistic model is used for the observed compliance behavior. The distribution of $G^{obs}_{mis}$ conditional on the confounders is thus written as
\begin{eqnarray}
P(G^{obs}_{mis}=0~|~M^{obs}, M^{obs}_{mis}, U^{obs}, \theta) &=& \frac{1}{1+e^l} \nonumber \\
P(G^{obs}_{mis}=1~|~M^{obs}, M^{obs}_{mis}, U^{obs}, \theta) &=& \frac{e^l}{1+e^l} \nonumber \\
l &=& \gamma_1~M^{obs} + \gamma_2~M^{obs}_{mis} + \gamma_3~U^{obs} + \zeta,
\label{eqn:logistic-g}
\end{eqnarray}
where $l$ is the linear predictor, $\gamma = (\gamma_1, \gamma_2, \gamma_3, \zeta)$ are unknown coefficients, and $E(\zeta)=0$, $Var(\zeta)=\sigma^2_\zeta$. The distribution of $G^{obs}$ is identical to equation \ref{eqn:logistic-g}.

Then we can obtain the joint distribution of the observed exposure, the observed outcome and the observed compliance behavior, as
\begin{eqnarray}
&& f(y^{obs}, y^{obs}_{mis}, w^{obs}, w^{obs}_{mis}, g^{obs}, g^{obs}_{mis}~|~Z^{obs}, M^{obs}, M^{obs}_{mis}, U^{obs}, \theta) \nonumber \\
&=& f(y^{obs}, y^{obs}_{mis}, w^{obs}, w^{obs}_{mis}~|~Z^{obs}, G^{obs}, G^{obs}_{mis}, M^{obs}, M^{obs}_{mis}, U^{obs}, \theta) \nonumber \\
&& \cdot~f(g^{obs}~|~M^{obs}, M^{obs}_{mis}, U^{obs}, \theta)~f(g^{obs}_{mis}~|~M^{obs}, M^{obs}_{mis}, U^{obs}, \theta) \nonumber \\
&& \left (\text{$G^{obs} \upvDash G^{obs}_{mis}$} \right) \nonumber \\
&=& \prod_{i \in \text{UC}}\left \{ ~f_i(y(0,w(0))^{obs}, y(0,w(0))^{mis}~|~W(0)^{obs}, W(0)^{mis}, M^{obs}, M^{obs}_{mis}, U^{obs}, \theta) \right. \nonumber \\
&& \left. \cdot~f_i(w(0)^{obs}, w(0)^{mis}~|~G^{obs}_{mis}, M^{obs}, M^{obs}_{mis}, U^{obs}, \theta)~f_i(g^{obs}_{mis}~|~M^{obs}, M^{obs}_{mis}, U^{obs}, \theta) \right \} \nonumber \\
&& \cdot~ \prod_{j \in \text{HWC}} ~\left \{ f_j(y(1,w(1))^{obs}, y(1,w(1))^{mis}~|~W(1)^{obs}, W(1)^{mis}, M^{obs}, M^{obs}_{mis}, U^{obs}, \theta) \right. \nonumber \\
&& \cdot~f_j(w(1)^{obs}, w(1)^{mis}~|~G^{obs}, M^{obs}, M^{obs}_{mis}, U^{obs}, \theta) \nonumber \\
&&\left. \cdot ~f_j(g^{obs}~|~M^{obs}, M^{obs}_{mis}, U^{obs}, \theta) \right \}
\label{eqn:likelihood-joint-g}
\end{eqnarray}

The joint distribution of all the variables is the product of equation \ref{eqn:likelihood-joint-g}, the distribution of $M^{obs}$ and $M^{obs}_{mis}$, and the priors of $U^{obs}$ and $\theta$. Since these additional distributions are independent of the observed compliance behavior, marginalization on the missing compliance behavior can be directly carried out in equation \ref{eqn:likelihood-joint-g}. We further assume the observed compliance behavior can only affect any outcome through the observed exposure(s). This means that conditional on the observed exposure, the outcome and the observed compliance behavior are independent. Equation \ref{eqn:marginal-g} shows the marginalized complete data model.
\begin{eqnarray}
&& f(y^{obs}, y^{obs}_{mis}, w^{obs}, w^{obs}_{mis}, g^{obs}~|~Z^{obs}, M^{obs}, M^{obs}_{mis}, U^{obs}, \theta) \nonumber \\
&=& \sum_{g^{obs}_{mis}=0}^1{f(y^{obs}, y^{obs}_{mis}, w^{obs}, w^{obs}_{mis}, g^{obs}, g^{obs}_{mis}~|~Z^{obs}, M^{obs}, M^{obs}_{mis}, U^{obs}, \theta)} \nonumber \\
&=& \prod_{i \in \text{UC}} ~\left\{f_i(y(0,w(0))^{obs}, y(0,w(0))^{mis}~|~W(0)^{obs}, W(0)^{mis}, M^{obs}, M^{obs}_{mis}, U^{obs}, \theta)\right\}  \nonumber \\
&& \cdot~ \left[ \prod_{i \in \text{UC}} \left\{ f_i(w(0)^{obs}, w(0)^{mis}~|~G^{obs}_{mis}=0, M^{obs}, M^{obs}_{mis}, U^{obs}, \theta) \right. \right.\nonumber \\
&& \left. \cdot~f_i(g^{obs}_{mis}=0~|~M^{obs}, M^{obs}_{mis}, U^{obs}, \theta) \right\} \nonumber \\
&& +~\prod_{i \in \text{UC}} \left\{f_i(w(0)^{obs}, w(0)^{mis}~|~G^{obs}_{mis}=1, M^{obs}, M^{obs}_{mis}, U^{obs}, \theta) \right. \nonumber \\
&& \cdot~\left. \left. f_i(g^{obs}_{mis}=1~|~M^{obs}, M^{obs}_{mis}, U^{obs}, \theta) \right\} \right ]  \nonumber \\
&& \cdot~ \prod_{j \in \text{HWC}}~ \left\{ f_j(y(1,w(1))^{obs}, y(1,w(1))^{mis}~|~W(1)^{obs}, W(1)^{mis}, M^{obs}, M^{obs}_{mis}, U^{obs}, \theta) \right. \nonumber \\
&& \cdot~f_j(w(1)^{obs}, w(1)^{mis}~|~G^{obs}, M^{obs}, M^{obs}_{mis}, U^{obs}, \theta) \nonumber \\
&& \left. \cdot~f_j(g^{obs}~|~M^{obs}, M^{obs}_{mis}, U^{obs}, \theta) \right\} \nonumber \\
&=& \prod_{i \in \text{UC}} ~\left\{f_i(y(0,w(0))^{obs}, y(0,w(0))^{mis}~|~W(0)^{obs}, W(0)^{mis}, M^{obs}, M^{obs}_{mis}, U^{obs}, \theta)\right\}  \nonumber \\
&& \cdot~ \left[ \prod_{i \in \text{UC}}\left \{ f_i(w(0)^{obs}, w(0)^{mis}~|~G^{obs}_{mis}=0, M^{obs}, M^{obs}_{mis}, U^{obs}, \theta) \right. \right. \nonumber \\
&& \left. \cdot~f_i(g^{obs}_{mis}=0~|~M^{obs}, M^{obs}_{mis}, U^{obs}, \theta) \right\} + \prod_{i \in \text{UC}} \left \{ \left. f_i(g^{obs}_{mis}=1~|~M^{obs}, M^{obs}_{mis}, U^{obs}, \theta) \right\} \right ]  \nonumber \\
&& \cdot~ \prod_{j \in \text{HWC}}~ \left\{ f_j(y(1,w(1))^{obs}, y(1,w(1))^{mis}~|~W(1)^{obs}, W(1)^{mis}, M^{obs}, M^{obs}_{mis}, U^{obs}, \theta) \right. \nonumber \\
&& \cdot~f_j(w(1)^{obs}, w(1)^{mis}~|~G^{obs}, M^{obs}, M^{obs}_{mis}, U^{obs}, \theta) \nonumber \\
&& \left. \cdot~f_j(g^{obs}~|~M^{obs}, M^{obs}_{mis}, U^{obs}, \theta) \right \} \nonumber \\
&& \left (P_i(W(0)=0~|~G^{obs}_{mis}=1)=1, W_i(0)=0, ~\text{for $i \in$ UC}\right )
\label{eqn:marginal-g}
\end{eqnarray}

Then a logarithm on the product of equation \ref{eqn:marginal-g}, the distribution of $M^{obs}$ and $M^{obs}_{mis}$ and the priors of $U^{obs}$ and $\theta$ produces the total log probability as the input in RStan. In this process, we do not estimate the missing observed compliance behavior in the control arm directly, but we obtain parameters related to its distribution.


In the presence of three exposures, the marginalized complete data model without the intervals involved can be written as
\begin{eqnarray}
&& f(y^{obs}, y^{obs}_{mis}, w^{1, obs}, w^{1, obs}_{mis}, w^{2, obs}, w^{2, obs}_{mis}, w^{3, obs}, w^{3, obs}_{mis}, g^{obs}~|~Z^{obs}, M^{obs}, M^{obs}_{mis}, U^{obs},  \nonumber \\
&& ~~~ \theta) \nonumber \\
&=& \prod_{i \in \text{UC}} ~\left\{f_i(y(0,w^1(0), w^2(0), w^3(0))^{obs}, y(0,w^1(0), w^2(0), w^3(0))^{mis}~|~W^1(0)^{obs}, \right. \nonumber \\
&&\left.  W^1(0)^{mis}, W^2(0)^{obs}, W^2(0)^{mis},W^3(0)^{obs}, W^3(0)^{mis},
M^{obs}, M^{obs}_{mis}, U^{obs}, \theta)\right\}  \nonumber \\
&& \cdot~ \left[ \prod_{i \in \text{UC}}\left \{ f_i(w^3(0)^{obs}, w^3(0)^{mis}~|~W^1(0)^{obs}, W^1(0)^{mis}, W^2(0)^{obs}, W^2(0)^{mis}, G^{obs}_{mis}=0,  \right. \right. \nonumber \\
&& M^{obs}, M^{obs}_{mis}, U^{obs}, \theta) ~\cdot~f_i(w^2(0)^{obs}, w^2(0)^{mis}~|~W^1(0)^{obs}, W^1(0)^{mis}, G^{obs}_{mis}=0, M^{obs},   \nonumber \\
&&  M^{obs}_{mis}, U^{obs}, \theta)~\cdot~f_i(w^1(0)^{obs}, w^1(0)^{mis}~|~ G^{obs}_{mis}=0, M^{obs}, M^{obs}_{mis}, U^{obs}, \theta)  \nonumber \\
&& \left. \cdot~f_i(g^{obs}_{mis}=0~|~M^{obs}, M^{obs}_{mis}, U^{obs}, \theta) \right\} + \prod_{i \in \text{UC}} \left \{ \left. f_i(g^{obs}_{mis}=1~|~M^{obs}, M^{obs}_{mis}, U^{obs}, \theta) \right\} \right ]  \nonumber \\
&& \cdot~ \prod_{j \in \text{HWC}}~ \left\{ f_j(y(1,w^1(1), w^2(1), w^3(1))^{obs}, y(1,w^1(1), w^2(1), w^3(1))^{mis}~|~W^1(1)^{obs},  \right. \nonumber \\
&& W^1(1)^{mis}, W^2(1)^{obs}, W^2(1)^{mis},W^3(1)^{obs}, W^3(1)^{mis},M^{obs}, M^{obs}_{mis}, U^{obs}, \theta) \nonumber \\
&& \cdot~f_j(w^3(1)^{obs}, w^3(1)^{mis}~|~W^1(1)^{obs}, W^1(1)^{mis}, W^2(1)^{obs}, W^2(1)^{mis}, G^{obs}, M^{obs}, M^{obs}_{mis}, \nonumber \\
&&  U^{obs}, \theta)~ \cdot~f_j(w^2(1)^{obs}, w^2(1)^{mis}~|~W^1(1)^{obs}, W^1(1)^{mis}, G^{obs}, M^{obs}, M^{obs}_{mis}, U^{obs}, \theta) \nonumber \\
&& \cdot~f_j(w^1(1)^{obs}, w^1(1)^{mis}~|~G^{obs}, M^{obs}, M^{obs}_{mis}, U^{obs}, \theta) \left. ~\cdot~f_j(g^{obs}~|~M^{obs}, M^{obs}_{mis}, U^{obs}, \theta) \right \} \nonumber \\
&& \left (P_i(W^h(0)=0~|~G^{obs}_{mis}=1)=1, ~\text{for $i \in$ UC, for $h = 1, 2, 3$}\right )
\label{eqn:marginal-g-3expo}
\end{eqnarray}

In the PreventS trial data, we assume the distributions of the outcomes are Normal distributions and the distributions of the exposures are truncated Normal distributions. The empirical distributions of the outcomes approximate Normal distributions to some extent. Hence, the normality assumption for the outcomes is plausible. On the other hand, each exposure is bounded, with a lower bound of 0 and an upper bound of 15. Hence, their true distributions should be truncated distributions. We assume the large sample distributions of the exposures are truncated Normal distributions and use Normal distributions to build the exposure distributions, because (1) the empirical distributions of some exposures in compliers in the treatment arm look like Normal distributions, or possibly \textit{t} distributions, and thus the normality assumption has some plausibility, (2) truncated Normal distributions and Normal distributions are not different in the Bayesian modelling process because no exposure falls out of the truncation bounds, (3) most importantly, the means and the variances from Normal distributions are relatively independent, so we can use sample standard deviations to estimate standard deviations in Bayesian models, without considering the relationship between the means and the variances.


\section{Calculation of sample standard deviations}

If standard deviations of exposures and outcomes are estimated well directly in Bayesian models, sample standard deviations may not be necessary. Sample standard deviations should be used in case Bayesian models produce zero standard deviations of exposures or outcomes, or run into other model non-identifiability issues related to standard deviations, or are hard to fit. Hence, we first calculate sample standard deviations.

The known observed compliance behavior in the treatment arm is defined based on $W^{09, obs}$, which applies to both calculation of sample standard deviations and Bayesian models with both sets of exposure. If $W^{09, obs} > 0$ in the treatment arm, the known observed compliance behavior is complier. If $W^{09, obs} = 0$ in the treatment arm, the known observed compliance behavior is never-taker. 

We use the treatment arm to calculate sample standard deviation of each exposure, because assigning compliers in the control arm with the control intervention does not reflect true variation in exposures. Since we assume the observed compliance behavior has a direct effect on all observed exposures, the means of exposures of never-takers in the treatment arm  and the means of exposures of compliers in the treatment arm are assumed to be different, with the other confounders fixed. Hence, the sample standard deviation of each exposure, including two sets of exposure, is pooled from sample standard deviations in two groups, compliers in the treatment arm and never-takers in the treatment arm, through weighted averaging.

We use both arms to calculate sample standard deviation of each outcome. Since we assume the observed compliance behavior has no direct effect on all observed outcomes and each exposure has a direct effect on all observed outcomes, differences between the means of outcomes of compliers in the treatment arm  and the means of outcomes of the other participants should mainly be attributed to different exposure levels, with the other confounders fixed. Hence, the sample standard deviation of each outcome is pooled from sample standard deviations in two groups, compliers in the treatment arm and the other participants, through weighted averaging. We also try stratifying compliers in the treatment arm into more groups with different exposure levels based on complete cases. Stratification we have considered includes (1) $0 < W^{09, obs} \leq 5$ and $ W^{09, obs} > 5$, (2) $0 < W^{09, obs} \leq 10$ and $ W^{09, obs} > 10$, (3) $0 < W^{09, obs} \leq 5$, $5 < W^{09, obs} \leq 10$, $ W^{09, obs} > 10$. None of these stratification approaches produces significantly different sample standard deviations of the outcomes. Hence, we do not further consider deeper stratification on the exposure levels.

Missing data is a major problem in calculation of sample standard deviations. Table \ref{tab:missing-stats-new} shows that most missing data exists in the exposures in the treatment arm and the outcomes. Due to uncertainty of missing data, calculation of sample standard deviations of the exposures and the outcome based on complete cases may involve bias. Particularly, we have noticed that some participants who attended some HWC sessions but then had one missing HWC attendance due to a missing HWC attendance or assessment date are treated as missing observations, which makes many useful HWC session attendance missing and thus may greatly bias sample standard deviations of the exposures from the second set of exposure. Hence, we calculate sample standard deviations based on the original data set including missing data, and compare the resultant estimates with sample standard deviations calculated based on complete cases. 

For calculation of sample standard deviations with missing data, we use multiple imputation on missing data through the \texttt{mice} \texttt{R} package \autocite{buuren_mice_2011} that uses chained equations. Missing HWC session attendance dates and assessment dates are imputed within their own blocks, alongside with measured confounders and outcomes. This means the predictors of missing HWC session attendance dates are HWC session attendance dates and the predictors of missing assessment dates are assessment dates. Three exposures from the second set of exposure are calculated by deterministic functions of HWC session attendance dates and assessment dates through the passive imputation method provided by the \texttt{mice} \texttt{R} package \autocite{buuren_mice_2011}. The first set of exposure and the second set of exposures have some linearly dependent relationships, because the sum of the second set of exposures is expected to be the first set of exposure. When four exposures are used as predictors for missing measured confounders and outcomes, the first set of exposure will be forced out of the imputation process by the \texttt{mice} \texttt{R} package \autocite{buuren_mice_2011}. Hence, only the first set of exposure is used as a predictor to impute missing measured confounders and outcomes. 10 imputed data sets are created, and each imputed data set runs 10 iterations. We have checked that overall the \texttt{mice} algorithm converges well and there is no unusual distributional discrepancy between the observed and imputed data. We also have checked that the assessment date at 3 months post randomization that is set to be missing is imputed well. In each imputed data set, we calculate pooled standard deviations of exposures and outcomes. Then, we pool pooled standard deviations from multiple imputed data sets through Rubin's rule \autocite{rubin_multiple_1987}. We assume standard deviations of exposures and outcomes are Normally distributed. Hence, sample standard deviations of exposures and outcomes pooled from multiple imputed data sets are averages of pooled standard deviations of exposures and outcomes from multiple imputed data sets. For calculation of sample standard deviations based on complete cases, we do not omit missing data from all the variables at the beginning. Then for each exposure or outcome, we omit its own missing data and pool standard deviations from two groups.

To obtain more realistic interval estimates of sample standard deviations without the normality assumption, we use the bootstrap method to obtain the 95\% interval estimates of sample standard deviations. Steps in the bootstrap approach are: (1) in either arm, sample participants with replacement, where the sampling size is the original sample size of either arm and missing data is included, (2) repeat step (1) 100 times and obtain 100 bootstrap samples, (3) for each bootstrap sample, calculate sample standard deviations of exposures and outcomes in two groups and pool them through weighted averaging, (4) find the 95\% interval estimates of sample standard deviations of exposures and outcomes across 100 bootstrap samples. For calculation based on complete cases, step (3) omits missing data of each variable before calculating the pooled standard deviation of this variable. For calculation with missing data, we need to consider how to combine bootstrap with multiple imputation. We choose the ``first bootstrap, then multiple imputation'' approach, which is proven to be correct \autocite{schomaker_bootstrap_2018}. This means, for each bootstrap sample, step (3) first produces 10 imputed data sets, then calculates pooled standard deviations of exposures and outcomes in each imputed data set, and finally pool pooled standard deviations of exposures and outcomes from 10 imputed data sets through Rubin's rule. We do not do imputation diagnostics for each imputation model when the bootstrap procedure and the multiple imputation procedure are programmed together. No imputation model runs into error.

Table \ref{tab:sd-expoout-boot} shows sample standard deviations from imputed data and complete cases. Sample standard deviations of the outcomes and the first set of exposure and their 95\% bootstrap intervals are comparable between complete cases and imputed data. Sample standard deviations of three exposures from the second set of exposure and their 95\% bootstrap intervals are slightly larger from imputed data than from complete cases. We have checked the first two imputed data sets and found that, for the four participants who have one missing HWC session attendance date and thus have missing exposures, three of them have high imputed exposure levels and one of them have low imputed exposure levels. For example, for the exposure at 9 months post randomization, the four imputed values are 3, 15, 15, 15. Hence, multiple imputation results in larger variation in the exposures. We think this is plausible because multiple imputation recovers many observed HWC session attendances by imputing only one missing HWC session attendance date and thus recovers some variation that already exist in the exposures. Multiple imputation should be helpful in this case. Hence, we use sample standard deviations and their 95\% bootstrap intervals from imputed data in Bayesian models. Sample standard deviations of the outcomes will be used for both sets of exposure.

\begin{table}[htbp]
\small
   \centering
\begin{tabular}{rrr}
  \hline
Variable & Complete cases & Imputed data \\
  \hline 
$W^{09, obs}$  &4.83 (4.249, 5.301) & 4.83 (4.249, 5.301) \\
$W^{03, obs}$  &2.30 (2.008, 2.606) &2.43 (2.156, 2.659) \\
$W^{36, obs}$  &1.57 (1.373, 1.788) &1.80 (1.601, 1.990) \\
$W^{69, obs}$  & 1.50 (1.310, 1.640) &1.64 (1.501, 1.777) \\
$Y^{cvd, obs}$  & 0.08 (0.068, 0.085)& 0.08 (0.068, 0.083)\\
$Y^{ls7, obs}$ & 2.06 (1.918, 2.242) & 2.06 (1.954, 2.255)\\
$Y^{sbp, obs}$  &19.94 (18.364, 21.723)& 19.99 (18.608, 22.081)\\
$Y^{phq, obs}$ &4.20 (3.673, 4.685) &4.20 (3.841, 4.760) \\
  \hline
\end{tabular}  
   \caption{Sample standard deviations (95\% bootstrap intervals) for exposures and outcomes from complete cases and imputed data.}
 \label{tab:sd-expoout-boot}
\end{table}

\section{Analysis of one exposure}
\label{sec:analysis-2}

The single exposure comes from the first set of exposure, and we build Bayesian models for this exposure and each of four outcomes, in order to understand how one HWC session attendance would affect each outcome on average during 9 months post randomization.

\subsection{Results of Life's Simple Seven total score}

Distributions for the observed compliance behavior, the exposure from the first set of exposure, the LS7 outcome in a Bayesian model are shown in equation \ref{eqn:results-model-formula-ls7}. Missing data is not shown but uses the models for the corresponding observed variables. Bayesian models for the exposure from the first set of exposure and the LS7 outcome have already been reparameterized through the random intercept reparameterization. The reparameterized unmeasured confounder is denoted by $U'$. The symbol ``${\rm logit}^{-1}$'' is the logistic function. The symbol ``Age$^{std}$'' means the measured confounder, age, has been standardized, because ages are relatively big numbers and it is easier to find an appropriate scale for the prior of the standardized age. Reference levels of categorical measured confounders are shown in table \ref{tab:descrip-stats-new}. Each categorical measured confounder has one coefficient in one Bayesian model. This is only for display simplicity, because the levels except the reference level should have separate coefficients. For each categorical measured confounder, the coefficients of the levels except the reference level are presented in the format ``coefficient of the confounder, the level name''. For example, the coefficient of the ``Male'' level in the confounder ``Gender'' from the model of $Y^{ls7, obs}$ is presented as ``$\beta_2$, Male''. $\beta_8$ is the average effect of the exposure $W^{09, obs}$, and we denote it by $e_{ate}$ when presenting results. 
\begin{eqnarray}
G^{obs} &\sim& Bern({\rm logit}^{-1}(\gamma_0 + \gamma_1~\text{Age$^{std}$} + \gamma_2~\text{Gender} + \gamma_3~\text{Ethnicity}  \nonumber \\
&& + ~\gamma_4~\text{Education level} + \gamma_5~\text{Smoking status} + \gamma_6~\text{Alcohol frequency} \nonumber \\
&& +~ \gamma_7~\text{Baseline LS7 total score})),  \nonumber \\
W^{09, obs} &\sim& 
    \begin{cases}
      0, & \text{if $Z^{obs}=0$, $G^{obs}=co$} \\
     N(\alpha_0 + \alpha_1~\text{Age$^{std}$} + \alpha_2~\text{Gender} + \alpha_3~\text{Ethnicity} \\
     + ~\alpha_4~\text{Education level} + \alpha_5~\text{Smoking status} \\
     +~ \alpha_6~\text{Alcohol frequency} \\
     +~ \alpha_7~\text{Baseline LS7 total score}\\
     +~ \alpha_8~G^{obs} + \alpha_9~U', \sigma_{w, 09}), & \text{otherwise}  \nonumber 
    \end{cases} \\
Y^{ls7, obs} &\sim& N(\beta_1~\text{Age$^{std}$} + \beta_2~\text{Gender} + \beta_3~\text{Ethnicity}  + \beta_4~\text{Education level}  \nonumber \\
&& +~ \beta_5~\text{Smoking status} + \beta_6~\text{Alcohol frequency} + \beta_7~\text{Baseline LS7 total score}  \nonumber \\
&& +~ \beta_8~W^{09, obs} + U', \sigma_{y, ls7}), \nonumber \\
U' &\sim& N(0, 3)
\label{eqn:results-model-formula-ls7}
\end{eqnarray}

Unmeasured confounders should also have a direct effect on the observed compliance behavior $G^{obs}$, but we do not include $U'$ in the distribution for $G^{obs}$, because (1) Bayesian models may not be fitted well when we use $U'$ in the distribution for $G^{obs}$, (2) without $U'$ the distribution for $G^{obs}$ is still valid and useful, (3) there is no residual confounding from excluding $U'$ from the distribution for $G^{obs}$. The observed baseline LS7 total score have missing values. The observed baseline LS7 total score with no missing values is denoted by $M^{ls7bl, obs}$ and the observed baseline LS7 total score with missing values is denoted by $M^{ls7bl, obs}_{mis}$. Their distributions are shown in equation \ref{eqn:results-model-formula-ls7bl-missing}, where we assume age, gender, ethnicity, education level, smoking status and alcohol frequency all have a direct effect on the baseline LS7 total score. Unmeasured confounders can also have a direct effect on the baseline LS7 total score, but we do not include $U'$ in the distribution for the baseline LS7 total score, because (1) Bayesian models can be simpler, (2) the distributions for $M^{ls7bl, obs}$ and $M^{ls7bl, obs}_{mis}$ are still valid and useful, (3) there is no residual confounding after conditioning on the baseline LS7 total score in the exposure and outcome distributions. 
\begin{eqnarray}
M^{ls7bl, obs} &\sim& N(\upsilon_1~\text{Age$^{std}$} + \upsilon_2~\text{Gender} + \upsilon_3~\text{Ethnicity}  + \upsilon_4~\text{Education level}  \nonumber \\
&& +~ \upsilon_5~\text{Smoking status} + \upsilon_6~\text{Alcohol frequency}, \sigma_{m, ls7bl}), \nonumber \\
M^{ls7bl, obs}_{mis} &\sim& N(\upsilon_1~\text{Age$^{std}$} + \upsilon_2~\text{Gender} + \upsilon_3~\text{Ethnicity}  + \upsilon_4~\text{Education level}  \nonumber \\
&& +~ \upsilon_5~\text{Smoking status} + \upsilon_6~\text{Alcohol frequency}, \sigma_{m, ls7bl})
\label{eqn:results-model-formula-ls7bl-missing}
\end{eqnarray}

We use sample standard deviations from table \ref{tab:sd-expoout-boot} as the prior means for standard deviations of the exposure and the LS7 outcome, $\sigma_{w, 09}$ and $\sigma_{y, ls7}$, and construct strongly informative priors for $\sigma_{w, 09}$ and $\sigma_{y, ls7}$ with a prior standard deviation of 0.01, so that their posterior estimates will become sample standard deviations. 

\begin{figure}[htbp]
     \centering
     \includegraphics[width=1\textwidth]{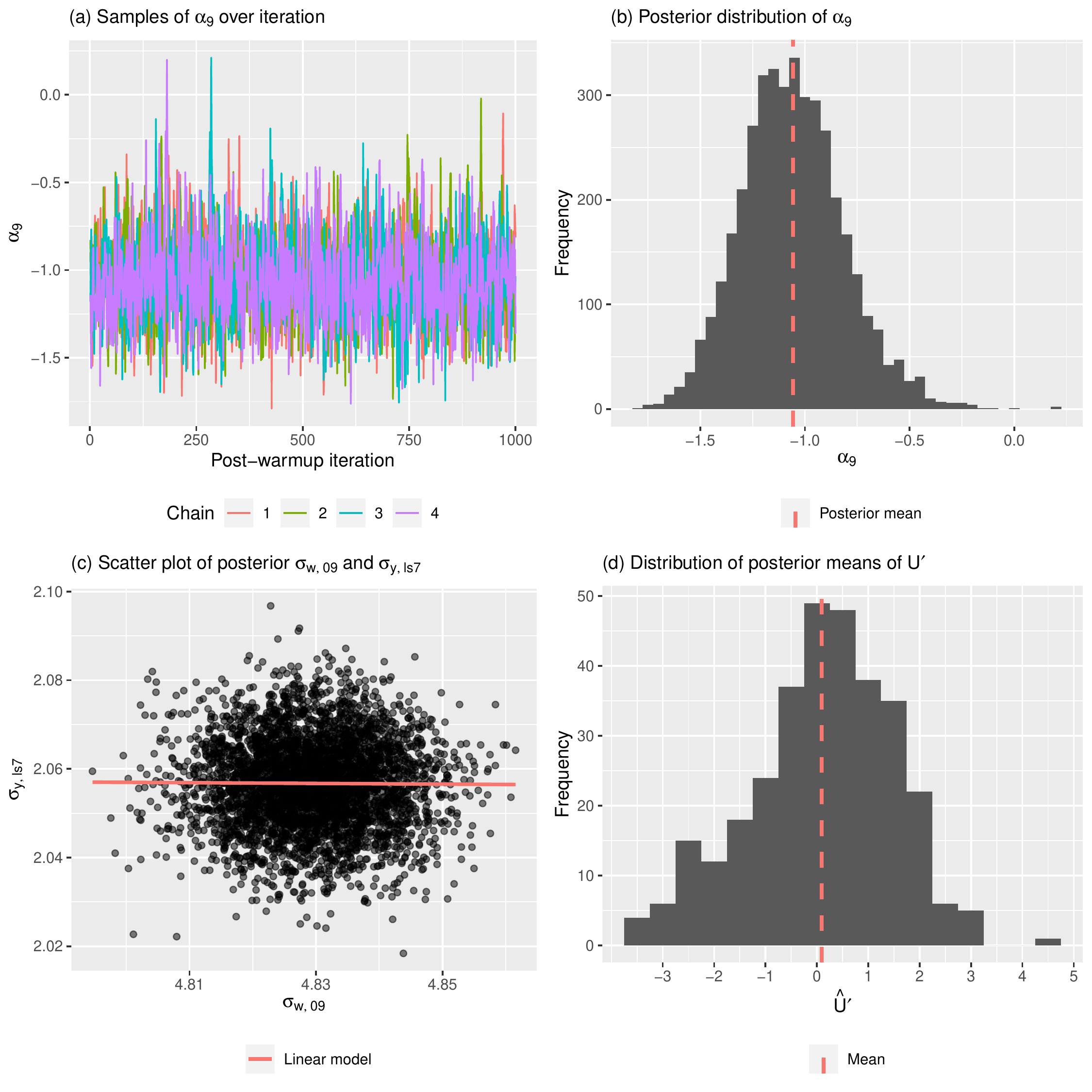}
        \caption{Identifiability diagnostics of the Bayesian model for the LS7 outcome. Figure (a) shows the Monte Carlo samples of $\alpha_9$ over iteration from 4 Markov chains. The samples do not show unusual behaviors and all chains are mixed well. Figure (b) shows the posterior distribution of $\alpha_9$ is proper. Figure (c) shows there is little posterior correlation between $\sigma_{w, 09}$ and $\sigma_{y, ls7}$ from a linear model between them. In Figure (d), the estimated $U'$ is denoted by $\hat U'$. We use posterior means of $U'$ as the point estimates of $U'$. Figure (d) shows that $U'$ is not estimated to be zero or constant.}
        \label{fig:res-ls7-diag}
\end{figure}

The Bayesian model for the LS7 outcome is fitted well. It is convergent, all $\hat R$ values are below 1.1, all Monte Carlo standard errors are small and all effective sample sizes are high. In addition, this model is identified, where the coefficient of $U'$ and $U'$ itself are estimated well and the standard deviations of the exposure and the outcome have little posterior correlation. Diagnostic figures and comments are shown in figure \ref{fig:res-ls7-diag}.

 Table \ref{tab:res-ls7-post-est} shows posterior estimates of parameters related to $W^{09, obs}$ and $Y^{ls7, obs}$ from the Bayesian model for the LS7 outcome and two sensitivity analysis models. Two sensitivity analysis models are a Bayesian model that adjusts for unmeasured confounding but is fitted on complete cases and a Bayesian association model that is fitted with missing data but does not adjust for unmeasured confounding. To avoid confusion with the Bayesian model for the LS7 outcome, we call the sensitivity analysis models, a complete-case model and an association model. The complete-case model uses all distributions in equation \ref{eqn:results-model-formula-ls7}. The association model uses the distribution for $G^{obs}$ in equation \ref{eqn:results-model-formula-ls7} and its distributions for the exposure and the outcome are shown in equation \ref{eqn:results-ls7-sa-association}.
 
 \begin{eqnarray}
W^{09, obs} &\sim& 
    \begin{cases}
      0, & \text{if $Z^{obs}=0$, $G^{obs}=co$} \\
     N(\alpha_0 + \alpha_1~\text{Age$^{std}$} + \alpha_2~\text{Gender} + \alpha_3~\text{Ethnicity} \\
     + ~\alpha_4~\text{Education level} + \alpha_5~\text{Smoking status} \\
     +~ \alpha_6~\text{Alcohol frequency} \\
     +~ \alpha_7~\text{Baseline LS7 total score}\\
     +~ \alpha_8~G^{obs}, \sigma_{w, 09}), & \text{otherwise}  \nonumber 
    \end{cases} \\
Y^{ls7, obs} &\sim& N(\beta_0 + \beta_1~\text{Age$^{std}$} + \beta_2~\text{Gender} + \beta_3~\text{Ethnicity}  + \beta_4~\text{Education level}  \nonumber \\
&& +~ \beta_5~\text{Smoking status} + \beta_6~\text{Alcohol frequency} + \beta_7~\text{Baseline LS7 total score}  \nonumber \\
&& +~ \beta_8~W^{09, obs}, \sigma_{y, ls7})
\label{eqn:results-ls7-sa-association}
\end{eqnarray}

\begin{table}[htbp]
\small
   \centering
\begin{tabular}{lrrrr}
  \hline
\multirowcell{2}[0pt][l]{Parameter} & \multirowcell{2}[0pt][r]{Prior} & \multicolumn{3}{c}{Posterior mean (95\% interval)}\\ 
  \cmidrule(l){3-5} 
& & Bayesian & Complete cases & Association \\ 
  \hline
$\beta_0$ &$N(0, 3)$  &  &  & 3.44 (2.30, 4.57) \\ 
$\beta_1$ &$N(0, 1)$ & 0.22 (-0.23, 0.67) & 0.23 (-0.25, 0.73) & 0.36 (0.10, 0.63) \\ 
$\beta_2$, Male &$N(0, 1)$ & 0.23 (-0.56, 1.05) & 0.19 (-0.68, 1.05) & 0.01 (-0.51, 0.52) \\ 
$\beta_3$, & &&& \\
\quad~ Maori &$N(0, 1)$  & 0.26 (-0.74, 1.23) & 0.28 (-0.82, 1.37) & 0.00 (-0.70, 0.69) \\ 
\quad~ Pasifika &$N(0, 1)$ & 0.25 (-0.75, 1.26) & 0.33 (-0.70, 1.33) & 0.07 (-0.57, 0.75) \\ 
\quad~ Asian &$N(0, 1)$ & 0.34 (-0.73, 1.39) & 0.35 (-0.79, 1.52) & 0.32 (-0.42, 1.07) \\ 
$\beta_4$,  & &&& \\
\quad~ College &$N(0, 1)$ & 0.57 (-0.34, 1.52) & 0.65 (-0.34, 1.66) & 0.62 (-0.03, 1.24) \\ 
\quad~ University &$N(0, 1)$ & 0.46 (-0.48, 1.37) & 0.48 (-0.53, 1.48) & 0.44 (-0.17, 1.03) \\ 
$\beta_5$, Non-current &$N(0, 1)$ & 0.06 (-0.97, 1.06) & 0.05 (-1.07, 1.16) & -0.73 (-1.43, 0.01) \\ 
$\beta_6$, More than &$N(0, 1)$ & 0.23 (-0.52, 1.01) & 0.22 (-0.61, 1.06) & -0.03 (-0.52, 0.47) \\ 
$\beta_7$ &$N(0, 1)$  & -0.22 (-0.34, -0.11) & -0.19 (-0.32, -0.06) & -0.53 (-0.66, -0.41) \\ 
$e_{ate}$ &$N(0, 1)$ & 0.16 (0.09, 0.22) & 0.08 (0.01, 0.15) & 0.03 (-0.01, 0.07) \\ 
$\sigma_{y, ls7}$ &$N(2.06, 0.01)$  & 2.06 (2.04, 2.08) & 2.06 (2.04, 2.08) & 2.06 (2.04, 2.08) \\ 
$\alpha_0$ &$N(10, 1)$ & 10.45 (8.75, 12.11) & 10.33 (8.51, 12.12) & 10.08 (8.31, 11.85) \\ 
$\alpha_1$ &$N(0, 1)$ & 0.26 (-0.53, 1.05) & 0.02 (-0.9, 0.88) & 0.25 (-0.45, 0.93) \\ 
$\alpha_2$, Male &$N(0, 1)$ & -1.14 (-2.44, 0.20) & -0.50 (-1.87, 0.88) & -1.44 (-2.69, -0.25) \\
$\alpha_3$, & &&& \\
\quad~ Maori &$N(0, 1)$ & -0.37 (-1.85, 1.08) & -0.27 (-1.87, 1.36) & -0.58 (-1.98, 0.81) \\ 
\quad~ Pasifika &$N(0, 1)$ & -0.36 (-1.79, 1.02) & -0.26 (-1.78, 1.30) & -0.59 (-1.97, 0.80) \\ 
\quad~ Asian &$N(0, 1)$ & -0.32 (-1.85, 1.18) & -0.31 (-1.95, 1.32) & -0.49 (-1.89, 0.92) \\ 
$\alpha_4$,  & &&& \\
\quad~ College &$N(0, 1)$ & 1.00 (-0.40, 2.40) & 0.28 (-1.26, 1.74) & 1.18 (-0.26, 2.61) \\ 
\quad~ University &$N(0, 1)$  & -0.22 (-1.69, 1.17) & 0.49 (-1.03, 1.97) & -0.16 (-1.47, 1.12) \\ 
$\alpha_5$, Non-current &$N(0, 1)$ & -1.08 (-2.55, 0.42) & -0.11 (-1.70, 1.52) & -1.31 (-2.67, 0.08) \\ 
$\alpha_6$, More than &$N(0, 1)$  & -0.02 (-1.38, 1.25) & 0.10 (-1.26, 1.48) & 0.15 (-1.06, 1.35) \\ 
$\alpha_7$ &$N(0, 1)$& -0.36 (-0.66, -0.06) & 0.10 (-0.24, 0.44) & -0.30 (-0.58, -0.02) \\ 
$\alpha_8$ &$N(0, 1)$& 3.01 (1.36, 4.60) & 1.64 (-0.11, 3.39) & 3.48 (2.04, 4.98) \\ 
$\alpha_9$ &$N(0, 1)$ & -1.06 (-1.51, -0.53) & -0.22 (-0.81, 0.37) &  \\ 
  $\sigma_{w, 09}$ &$N(4.83, 0.01)$  & 4.83 (4.81, 4.85) & 4.83 (4.81, 4.85) & 4.83 (4.81, 4.85) \\
\hline
\end{tabular}  
   \caption{Posterior estimates from the Bayesian model for the LS7 outcome, complete-case and association models.}
 \label{tab:res-ls7-post-est}
\end{table}

From table \ref{tab:res-ls7-post-est}, the posterior mean of $e_{ate}$ is 0.16, which means that attending one HWC session can increase the LS7 total score by 0.16 on average. The 95\% posterior interval of $e_{ate}$ is (0.09, 0.22) and is fully positive, which indicates that the average effect of one HWC session on the LS7 total score is statistically significantly positive. However, the posterior estimates of $e_{ate}$ are relatively small. Attending a few HWC sessions may not improve the LS7 total score obviously nor improve cardiovascular health obviously. The LS7 total score ranges from 0 to 14, containing 3 categories: 0-4 as inadequate cardiovascular health, 5-9 as average cardiovascular health, 10-14 as optimum cardiovascular health. Hence, there can be a chance that the cardiovascular health status is upgraded only after many HWC sessions are attended, such as 10 HWC sessions. 

There seems to be significant unmeasured confounding effects between the exposure at 9 months post randomization and the LS7 outcome. First, $U'$ is not estimated to be zero or constant, and the posterior mean of $\alpha_9$ is relatively large, though its 95\% posterior interval covers 0. Second, compared to the original Bayesian model for the LS7 outocme, the posterior mean of $e_{ate}$ from the association model is smaller and the 95\% posterior interval of $e_{ate}$ from the association model is shifted towards zero. The association model seems to indicate that the average effect of one HWC session is statistically insignificant. Hence, adjusting for unmeasured confounding seems to help reveal the beneficial effect of the HWC intervention. On the other hand, the posterior estimates of $e_{ate}$ from the complete-case model are also smaller than those from the Bayesian model but are still statistically significant, which indicates that missingness in the baseline LS7 total score and the LS7 outcome may slightly bias the average treatment effect estimates.

Sensitivity of the posterior estimates of $e_{ate}$ is tested against informativeness of the priors of $e_{ate}$, $U'$, $\alpha_8$, $\alpha_9$ and $\beta_7$. There is little posterior correlation between $e_{ate}$ and all coefficients except $\alpha_8$, $\alpha_9$ and $\beta_7$. Weakly informative priors are used in the original Bayesian model for the LS7 outcome. However, we do not know the threshold for a prior to be weakly informative, compared to informative and uninformative priors. Slightly more informative and less informative priors are considered in sensitivity analysis, with the prior means held fixed and the prior standard deviations varying, in order to see how the posterior estimates of $e_{ate}$ would change with different informativeness from the priors to be tested. There is no indication about whether the prior means of $e_{ate}$, $U'$, $\alpha_8$, $\alpha_9$ and $\beta_7$ should be positive or not. Hence, zero prior means would be suitable. Table \ref{tab:results-ls7-sa-prior} shows the results of prior sensitivity analysis.

\begin{table}[htbp]
\small
   \begin{subtable}{1\textwidth}
   \centering
   \begin{tabular}{rrrr}
  \hline
Prior & $N(0, 0.5)$ & $N(0, 1)$ & $N(0, 10)$   \\ 
  \hline
$S_p$ & 0.01 & 0.00 & 0.00 \\
$e_{ate}$ &0.16 (0.08, 0.22) &0.16 (0.09, 0.22) & 0.16 (0.08, 0.22) \\
   \hline
   \end{tabular}
   \caption{Prior sensitivity measures and posterior mean (95\% interval) of $e_{ate}$ under priors of $e_{ate}$ with different informativeness. $S_p$ is the prior sensitivity measure.} 
   \end{subtable}
   
\vspace{1em}

    \begin{subtable}{1\textwidth}
   \centering
   \begin{tabular}{rrrrr}
  \hline
Prior & $N(0, 1)$ & $N(0, 3)$ & $N(0, 5)$ & $N(0, 10)$   \\ 
  \hline
$e_{ate}$ &0.07 (0.03, 0.12)&0.16 (0.09, 0.22)&0.27 (0.18, 0.36)&- \\
$\alpha_9$ &-0.80 (-2.28, 1.03)&-1.06 (-1.51, -0.53)&-1.04 (-1.34, -0.73)&-\\ 
   \hline
   \end{tabular}
   \caption{Posterior mean (95\% interval) of $e_{ate}$ and $\alpha_9$ under priors of $U'$ with different informativeness. ``-'' indicates the Bayesian model built in sensitivity analysis is bad. $\alpha_9$ is used as a reference for the strength of unmeasured confounding.} 
   \end{subtable}
 
 \vspace{1em}
 
       \begin{subtable}{1\textwidth}
   \centering
   \begin{tabular}{rrrr}
  \hline
Prior & $N(0, 0.5)$ & $N(0, 1)$ & $N(0, 10)$   \\ 
  \hline
$e_{ate}$ &0.17 (0.11, 0.24) &0.16 (0.09, 0.22)& 0.13 (0.06, 0.19) \\
   \hline
   \end{tabular}
   \caption{Posterior mean (95\% interval) of $e_{ate}$ under priors of $\alpha_8$ with different informativeness.} 
   \end{subtable}
\end{table}

\begin{table}[htbp]
\ContinuedFloat
\small
   
         \begin{subtable}{1\textwidth}
   \centering
   \begin{tabular}{rrrr}
  \hline
Prior & $N(0, 0.5)$ & $N(0, 1)$ & $N(0, 10)$   \\ 
  \hline
   $e_{ate}$ &0.15 (0.07, 0.21) &0.16 (0.09, 0.22) & 0.16 (0.09, 0.22) \\
   \hline
   \end{tabular}
   \caption{Posterior mean (95\% interval) of $e_{ate}$ under priors of $\alpha_9$ with different informativeness.} 
   \end{subtable}

   \vspace{1em}
   
       \begin{subtable}{1\textwidth}
   \centering
   \begin{tabular}{rrrr}
  \hline
Prior & $N(0, 0.5)$ & $N(0, 1)$ & $N(0, 10)$   \\ 
  \hline
$e_{ate}$ &0.16 (0.08, 0.22) &0.16 (0.09, 0.22)& 0.16 (0.09, 0.22) \\
   \hline
   \end{tabular}
   \caption{Posterior mean (95\% interval) of $e_{ate}$ under  priors of $\beta_7$ with different informativeness.} 
   \end{subtable}
 \caption{Results of prior sensitivity analysis.}
 \label{tab:results-ls7-sa-prior}
\end{table}

From table \ref{tab:results-ls7-sa-prior}, the posterior estimates of $e_{ate}$ are insensitive to informativeness of the priors of $e_{ate}$, $\alpha_8$, $\alpha_9$ and $\beta_7$, but they are sensitive to informativeness of the prior of $U'$. When the prior of $U'$ becomes less informative to an extent that the Bayesian model built in sensitivity analysis is good, the posterior estimates of $e_{ate}$ tend to be more positive while the posterior estimates of $\alpha_9$ tend to be more negative. It seems that more unmeasured confounding is adjusted for, but the changes in the posterior estimates of $e_{ate}$ are relatively small, which does not invalidate any conclusion made on the average effect of one HWC session.

\begin{figure}[htbp]
     \centering
     \includegraphics[width=1\textwidth]{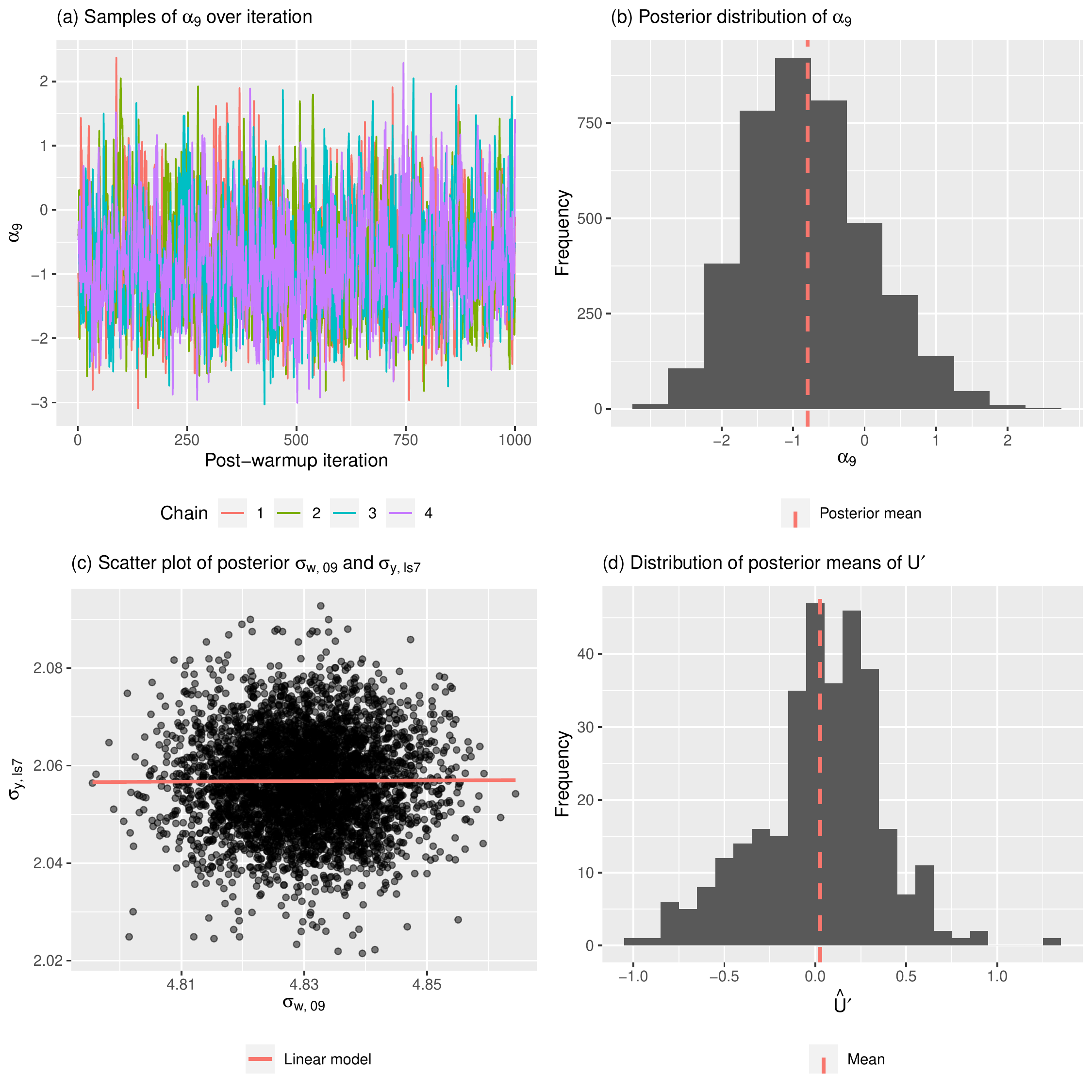}
        \caption{Identifiability diagnostics of the Bayesian model for the LS7 outcome when $N(0, 1)$ is the prior of $U'$. Figure (a) shows the Monte Carlo samples of $\alpha_9$ over iteration from 4 Markov chains. The samples do not show unusual behaviors. Figure (b) shows the posterior distribution of $\alpha_9$ is proper. Figure (c) shows there is little posterior correlation between $\sigma_{w, 09}$ and $\sigma_{y, ls7}$ from a linear model between them. In Figure (d), the estimated $U'$ is denoted by $\hat U'$. We use posterior means of $U'$ as the point estimates of $U'$. Figure (d) shows that $U'$ is not estimated to be zero or constant.}
        \label{fig:res-ls7-sa-u01-diag}
\end{figure}

When the prior of $U'$ becomes more informative, the posterior estimates of $e_{ate}$ tend to decrease. We build a good identified Bayesian model with $N(0, 1)$ as the prior of $U'$. Diagnostic figures and comments are shown in figure \ref{fig:res-ls7-sa-u01-diag}. Compared to figure \ref{fig:res-ls7-diag} for the original Bayesian model for the LS7 outcome, the posterior means of $U'$ become smaller and concentrate more on around zero because the informative prior of $U'$ makes the values around zero more possible, and the posterior distribution of $\alpha_9$ is shifted towards zero, which adjusts for less unmeasured confounding. Then this new Bayesian model with $N(0, 1)$ as the prior of $U'$ is pulled towards the association model. It seems appropriate to suggest that the prior of $U'$ should not be too informative.

\begin{figure}[htbp]
     \centering
     \includegraphics[width=1\textwidth]{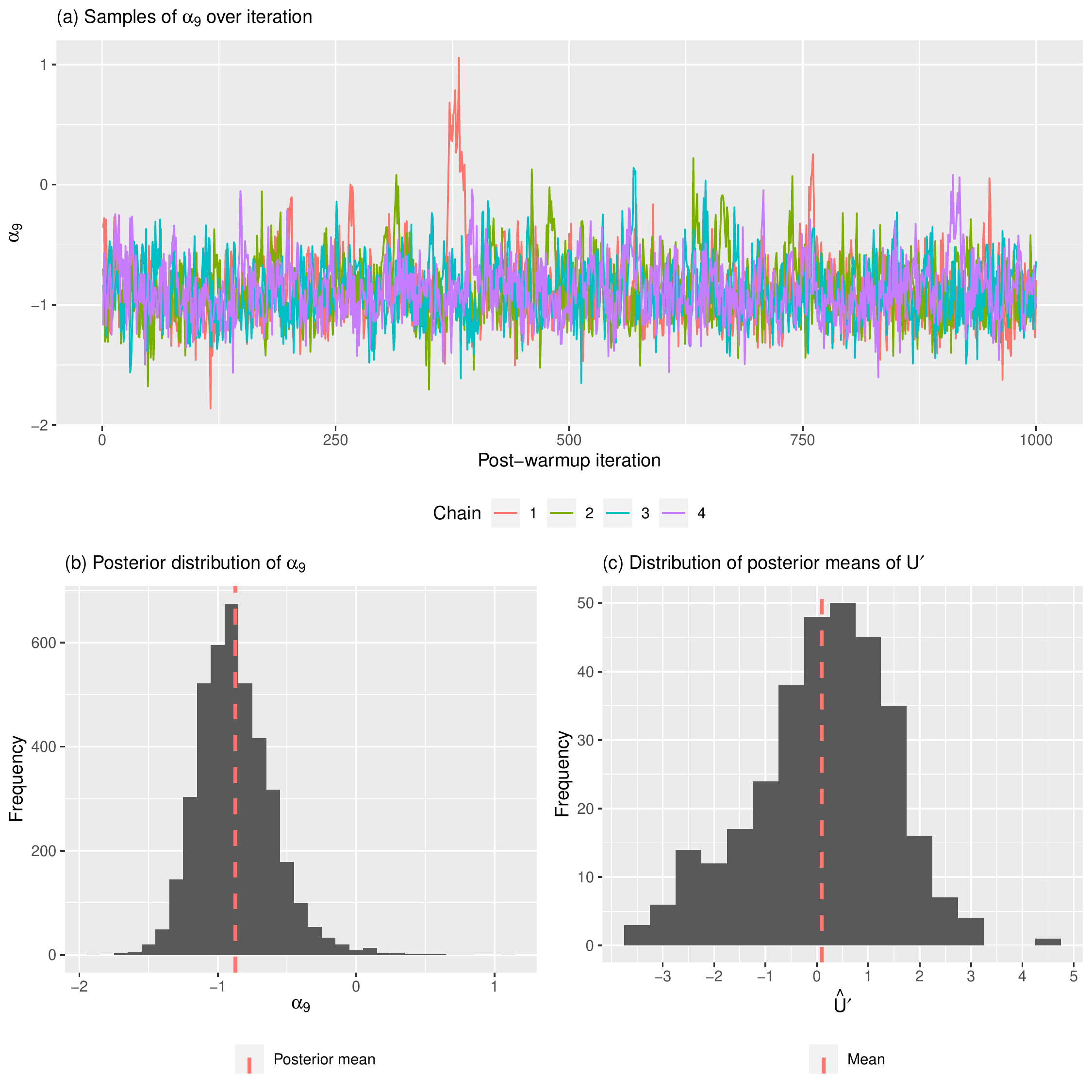}
        \caption{Identifiability diagnostics of the Bayesian model for the LS7 outcome when $N(0, 0.5)$ is the prior of $\alpha_9$. Figure (a) shows the Monte Carlo samples of $\alpha_9$ over iteration from 4 Markov chains. The samples show one unusual transition above zero. Figure (b) shows the posterior distribution of $\alpha_9$ is proper. In Figure (c), the estimated $U'$ is denoted by $\hat U'$. We use posterior means of $U'$ as the point estimates of $U'$. Figure (c) shows that $U'$ is not estimated to be zero or constant.}
        \label{fig:res-ls7-sa-au05-diag}
\end{figure}

We also notice that when the prior of $\alpha_9$ becomes more informative as $N(0, 0.5)$, there is an unusual transition above zero from one Markov chain during Monte Carlo sampling of $\alpha_9$, where this chain stays in the positive area for several iterations. There is no strong evidence that the posterior distribution of $\alpha_9$ is bimodal, and the posterior estimates of $\alpha_9$, $U'$ and $e_{ate}$ are quite similar to those from the original Bayesian model. Figure \ref{fig:res-ls7-sa-au05-diag} shows these results.

\begin{figure}[htbp]
     \centering
     \includegraphics[width=1\textwidth, height=1\textheight]{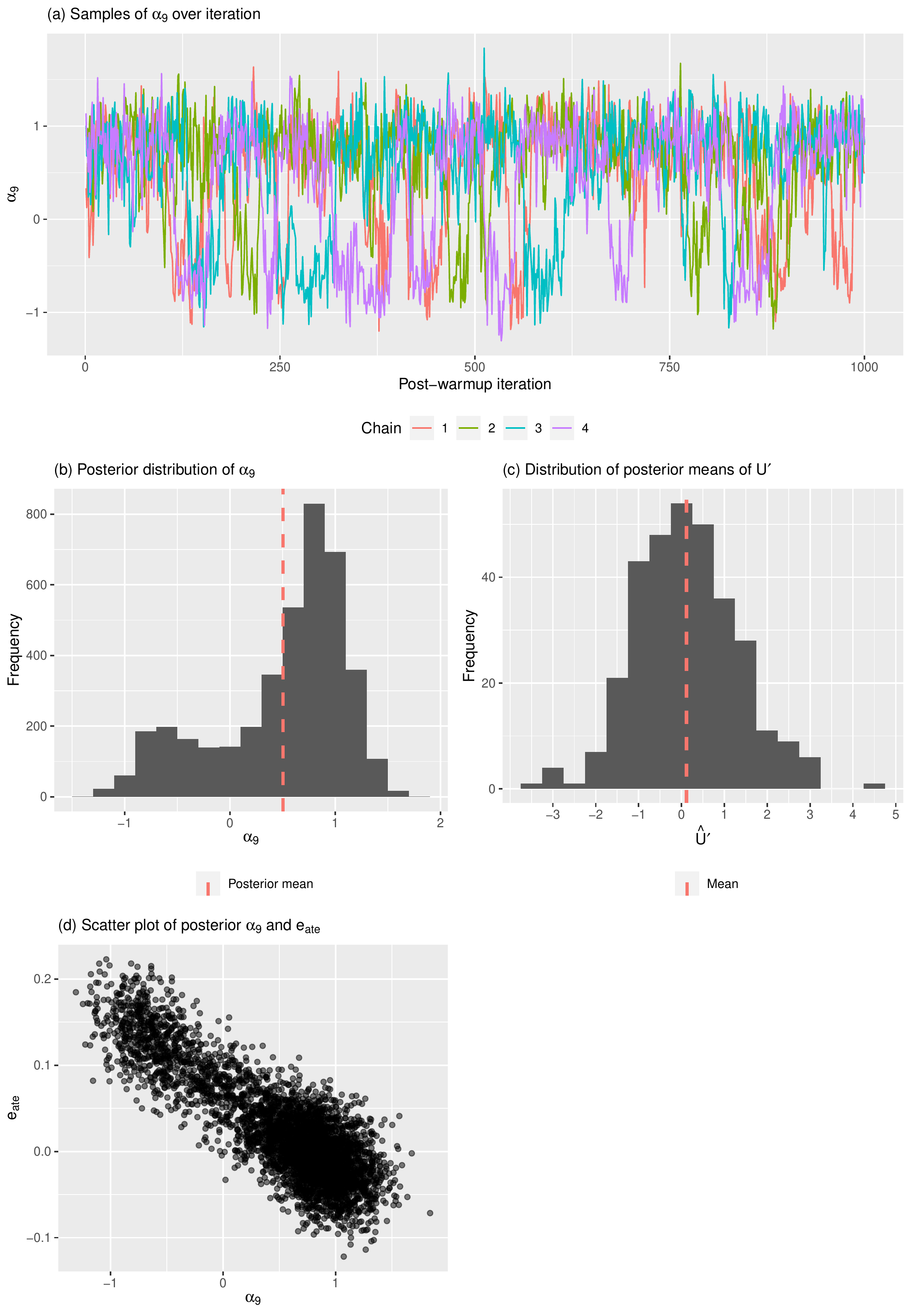}
        \caption{Identifiability diagnostics of the Bayesian model for the LS7 outcome when $N(0, 0.5)$ is the prior of $\alpha_9$. Figure (a) shows the Monte Carlo samples of $\alpha_9$ over iteration from 4 Markov chains. The chains visit both positive and negative areas frequently. Figure (b) shows the posterior distribution of $\alpha_9$ is bimodal. In Figure (c), the estimated $U'$ is denoted by $\hat U'$. We use posterior means of $U'$ as the point estimates of $U'$. Figure (c) shows that $U'$ is not estimated to be zero or constant. Figure (d) shows the association between $\alpha_9$ and $e_{ate}$, indicating two different sets of average treatment effect estimates.}
        \label{fig:res-ls7-sa-au105-diag}
\end{figure}

However, we also build a Bayesian model with $N(1, 0.5)$ as the prior of $\alpha_9$. Figure \ref{fig:res-ls7-sa-au105-diag} shows these results. This new Bayesian model with $N(1, 0.5)$ as the prior of $\alpha_9$ produces a bimodal posterior distribution of $\alpha_9$ with a negative mode as in the original Bayesian model and a positive mode identified by the positively informative prior. This indicates this new Bayesian model has two posterior solutions. The posterior estimates of $e_{ate}$ now decrease to 0.03 (-0.07, 0.17). One posterior solution produces statistically significant average treatment effects similar to the original Bayesian model, while the other posterior solution produces statistically insignificant average treatment effects similar to the association model. We also build a Bayesian model for the posterior solution that produces statistically insignificant average treatment effects by restricting $\alpha_9$ to be positive, and find that the posterior means of $U'$ range from -4 to 4, which indicates $U'$ is not estimated to be zero and thus the posterior solution that produces statistically insignificant average treatment effects is not exactly the association model.

Hence, there is a possibility that the posterior solution from the original Bayesian model is biased, because we cannot distinguish which posterior solution is the right one from the Bayesian model with $N(1, 0.5)$ as the prior of $\alpha_9$. To understand whether the original Bayesian model is useful for inference, we build a Bayesian instrumental variable model through two-stage least squares \autocite{angrist_two-stage_1995}, using equation \ref{eqn:results-ls7-sa-iv}. The assignment $Z^{obs}$ is a valid instrument after conditioning on measured confounders. We still use sample standard deviations in the Bayesian instrumental variable model, otherwise standard deviations are estimates to be zero by the Bayesian instrumental variable model itself. The prior for $\alpha_0$ is $N(10, 1)$ and the prior for $\beta_0$ is $N(0, 3)$. Priors for the other parameters are $N(0, 1)$. 
 \begin{eqnarray}
W^{09, obs} &\sim& N(\alpha_0 + \alpha_1~\text{Age$^{std}$} + \alpha_2~\text{Gender} + \alpha_3~\text{Ethnicity}  + \alpha_4~\text{Education level}  \nonumber \\
&& +~ \alpha_5~\text{Smoking status} + \alpha_6~\text{Alcohol frequency} + \alpha_7~\text{Baseline LS7 total score}  \nonumber \\
&& +~ \alpha_8~Z^{obs}, \sigma_{w, 09}) \\
Y^{ls7, obs} &\sim& N(\beta_0 + \beta_1~\text{Age$^{std}$} + \beta_2~\text{Gender} + \beta_3~\text{Ethnicity}  + \beta_4~\text{Education level}  \nonumber \\
&& +~ \beta_5~\text{Smoking status} + \beta_6~\text{Alcohol frequency} + \beta_7~\text{Baseline LS7 total score}  \nonumber \\
&& +~ \beta_8~Z^{obs}, \sigma_{y, ls7}) \\
e_{ate} &=& \frac{\beta_8}{\alpha_8}
\label{eqn:results-ls7-sa-iv}
\end{eqnarray}

In addition, we build a Bayesian model where the exposure distribution is excluded and only the outcome distribution adjusts for unmeasured confounding, which we call the random intercept outcome model. The priors for unknown parameters including $U'$ are identical to those from the original Bayesian model correspondingly. Adjusting for unmeasured confounding only in the outcome distribution does not produce multiple posterior solutions. Table \ref{tab:results-ls7-sa-iv-rt} shows the posterior estimates of $e_{ate}$ from the instrumental variable model and the random intercept outcome model.
\begin{table}[htbp]
\small
   \centering
  \begin{tabular}{rrr}
  \hline
  & Instrumental variable model & Random intercept outcome model \\ 
  \hline
 $e_{ate}$ &0.07 (0.02, 0.13)&0.07 (0.01, 0.13) \\
   \hline
   \end{tabular}
 \caption{Posterior mean (95\% interval) of $e_{ate}$ from the instrumental variable model and the random intercept outcome model.}
 \label{tab:results-ls7-sa-iv-rt}
\end{table}

The average treatment effect estimates from the instrumental variable model and the random intercept outcome model are nearly identical. They are smaller than those from the original Bayesian model, but they are slightly statistically significant, which makes the original Bayesian model more convincing. Since the average treatment effect estimates from the instrumental variable model and the random intercept outcome model can also be biased, it may be because the original Bayesian model actually adjusts for more unmeasured confounding that the average treatment effect estimates from the original Bayesian model are larger and less biased. On the other hand, it is also a little problematic that we use a likely informative prior, $N(1, 0.5)$, for $\alpha_9$, when we do not have this information in reality. Hence, it may be safer to conclude that the HWC intervention is statistically effective, but it is not easy to conclude how effective the HWC intervention is. The original Bayesian model is useful, but its results are sensitive to adjustment for unmeasured confounding.

Then sensitivity of the posterior estimates of $e_{ate}$ is tested against variations in sample standard deviations of the exposure and the LS7 outcome. Variations in sample standard deviations are represented by different locations of the priors of $\sigma_{w, 09}$ and $\sigma_{y, ls7}$, where sample standard deviations are used as the prior means. Prior locations that are considered include point estimates of sample standard deviations and two boundary points from the 95\% bootstrap intervals of sample standard deviations from table \ref{tab:sd-expoout-boot}, possibly other values if necessary. Table \ref{tab:results-ls7-sa-sd} shows the results of sensitivity analysis of sample standard deviations. 

\begin{table}[htbp]
\small
    \begin{subtable}{1\textwidth}
   \centering
   \begin{tabular}{rrrrr}
  \hline
Prior & $N(4.25, 0.01)$ & $N(4.83, 0.01)$ & $N(5.3, 0.01)$ & $N(10, 0.01)$\\ 
  \hline
$e_{ate}$&-&0.16 (0.09, 0.22) &0.13 (0.03, 0.20)&0.07 (0.01, 0.14) \\
   \hline
   \end{tabular}
   \caption{Posterior mean (95\% interval) of $e_{ate}$ under priors of $\sigma_{w, 09}$ with different locations. ``-'' indicates the Bayesian model built in sensitivity analysis is bad.} 
   \end{subtable}
   
\vspace{1em}

    \begin{subtable}{1\textwidth}
   \centering
   \begin{tabular}{rrrrrr}
  \hline
Prior & $N(1.95, 0.01)$ & $N(2, 0.01)$ & $N(2.06, 0.01)$ & $N(2.15, 0.01)$ & $N(2.26, 0.01)$\\ 
  \hline
$e_{ate}$&- &0.16 (0.09, 0.22) &0.16 (0.09, 0.22) & 0.16 (0.08, 0.22) &-  \\
   \hline
   \end{tabular}
   \caption{Posterior mean (95\% interval) of $e_{ate}$ under priors of $\sigma_{y, ls7}$ with different locations. ``-'' indicates the Bayesian model built in sensitivity analysis is bad.} 
   \end{subtable}
 \caption{Results of sensitivity analysis of sample standard deviations.}
 \label{tab:results-ls7-sa-sd}
\end{table}

Overall, the posterior estimates of $e_{ate}$ are sensitive to variations in sample standard deviations of the exposure and the outcome. When sample standard deviations become smaller or larger, there is a possibility that the Bayesian model is not fitted well in sensitivity analysis or the average treatment effect estimates are changed. The posterior estimates of $e_{ate}$ are insensitive to small variations in sample standard deviations around point estimates of sample standard deviations. These results show that the average treatment effect estimates have slight robustness against estimation accuracy of standard deviations, and less biased estimation of standard deviations seem necessary to obtain less biased average treatment effect estimates.

In conclusion, there is evidence that the HWC intervention is slightly beneficial to cardiovascular health by increasing the LS7 total score. On completion of the HWC intervention, on average the LS7 total score will be improved by about 2.4 (1.4, 3.3), but due to model sensitivity, the actual improvement may be smaller. Since the beneficial effect of one HWC session is small or possibly negligible, strong compliance to the HWC intervention may be necessary to see an obvious improvement in cardiovascular health.

\subsection{Results of systolic blood pressure}

Distributions for the observed compliance behavior, the exposure from the first set of exposure, the SBP outcome in a Bayesian model are shown in equation \ref{eqn:results-model-formula-sbp}. Missing data is not shown but uses the models for the corresponding observed variables. Bayesian models for the exposure from the first set of exposure and the SBP outcome have already been reparameterized through the random intercept reparameterization. The reparameterized unmeasured confounder is denoted by $U'$. The symbol ``${\rm logit}^{-1}$'' is the logistic function.  The symbol ``Age$^{std}$'' means the standardized age. The symbol ``Baseline SBP$^{std}$'' means the standardized baseline SBP. For each categorical measured confounder, the coefficients of the levels except the reference level are still presented in the format ``coefficient of the confounder, the level name''. $\beta_8$ is the average effect of the exposure $W^{09, obs}$, and we denote it by $e_{ate}$ when presenting results. 
\begin{eqnarray}
G^{obs} &\sim& Bern({\rm logit}^{-1}(\gamma_0 + \gamma_1~\text{Age$^{std}$} + \gamma_2~\text{Gender} + \gamma_3~\text{Ethnicity}  \nonumber \\
&& + ~\gamma_4~\text{Education level} + \gamma_5~\text{Smoking status} + \gamma_6~\text{Alcohol frequency} \nonumber \\
&& +~ \gamma_7~\text{Baseline SBP$^{std}$})),  \nonumber \\
W^{09, obs} &\sim& 
    \begin{cases}
      0, & \text{if $Z^{obs}=0$, $G^{obs}=co$} \\
     N(\alpha_0 + \alpha_1~\text{Age$^{std}$} + \alpha_2~\text{Gender} + \alpha_3~\text{Ethnicity} \\
     + ~\alpha_4~\text{Education level} + \alpha_5~\text{Smoking status} \\
     +~ \alpha_6~\text{Alcohol frequency} \\
     +~ \alpha_7~\text{Baseline SBP$^{std}$}\\
     +~ \alpha_8~G^{obs} + \alpha_9~U', \sigma_{w, 09}), & \text{otherwise}  \nonumber 
    \end{cases} \\
Y^{sbp, obs} &\sim& N(\beta_1~\text{Age$^{std}$} + \beta_2~\text{Gender} + \beta_3~\text{Ethnicity}  + \beta_4~\text{Education level}  \nonumber \\
&& +~ \beta_5~\text{Smoking status} + \beta_6~\text{Alcohol frequency} + \beta_7~\text{Baseline SBP}  \nonumber \\
&& +~ \beta_8~W^{09, obs} + U', \sigma_{y, sbp}), \nonumber \\
U' &\sim& N(0, 3)
\label{eqn:results-model-formula-sbp}
\end{eqnarray}

The observed baseline SBP with no missing values is denoted by $M^{sbpbl, obs}$ and the observed baseline SBP with missing values is denoted by $M^{sbpbl, obs}_{mis}$. Their distributions are shown in equation \ref{eqn:results-model-formula-sbpbl-missing}, where we assume age, gender, ethnicity, education level, smoking status and alcohol frequency all have a direct effect on the baseline SBP. 
\begin{eqnarray}
M^{sbpbl, obs} &\sim& N(\upsilon_1~\text{Age$^{std}$} + \upsilon_2~\text{Gender} + \upsilon_3~\text{Ethnicity}  + \upsilon_4~\text{Education level}  \nonumber \\
&& +~ \upsilon_5~\text{Smoking status} + \upsilon_6~\text{Alcohol frequency}, \sigma_{m, sbpbl}), \nonumber \\
M^{sbpbl, obs}_{mis} &\sim& N(\upsilon_1~\text{Age$^{std}$} + \upsilon_2~\text{Gender} + \upsilon_3~\text{Ethnicity}  + \upsilon_4~\text{Education level}  \nonumber \\
&& +~ \upsilon_5~\text{Smoking status} + \upsilon_6~\text{Alcohol frequency}, \sigma_{m, sbpbl})
\label{eqn:results-model-formula-sbpbl-missing}
\end{eqnarray}

We use sample standard deviations from table \ref{tab:sd-expoout-boot} as the prior means for standard deviations of the exposure and the SBP outcome, $\sigma_{w, 09}$ and $\sigma_{y, sbp}$, and construct strongly informative priors for $\sigma_{w, 09}$ and $\sigma_{y, sbp}$ with a prior standard deviation of 0.01, so that their posterior estimates will become sample standard deviations. 

\begin{figure}[htbp]
     \centering
     \includegraphics[width=1\textwidth]{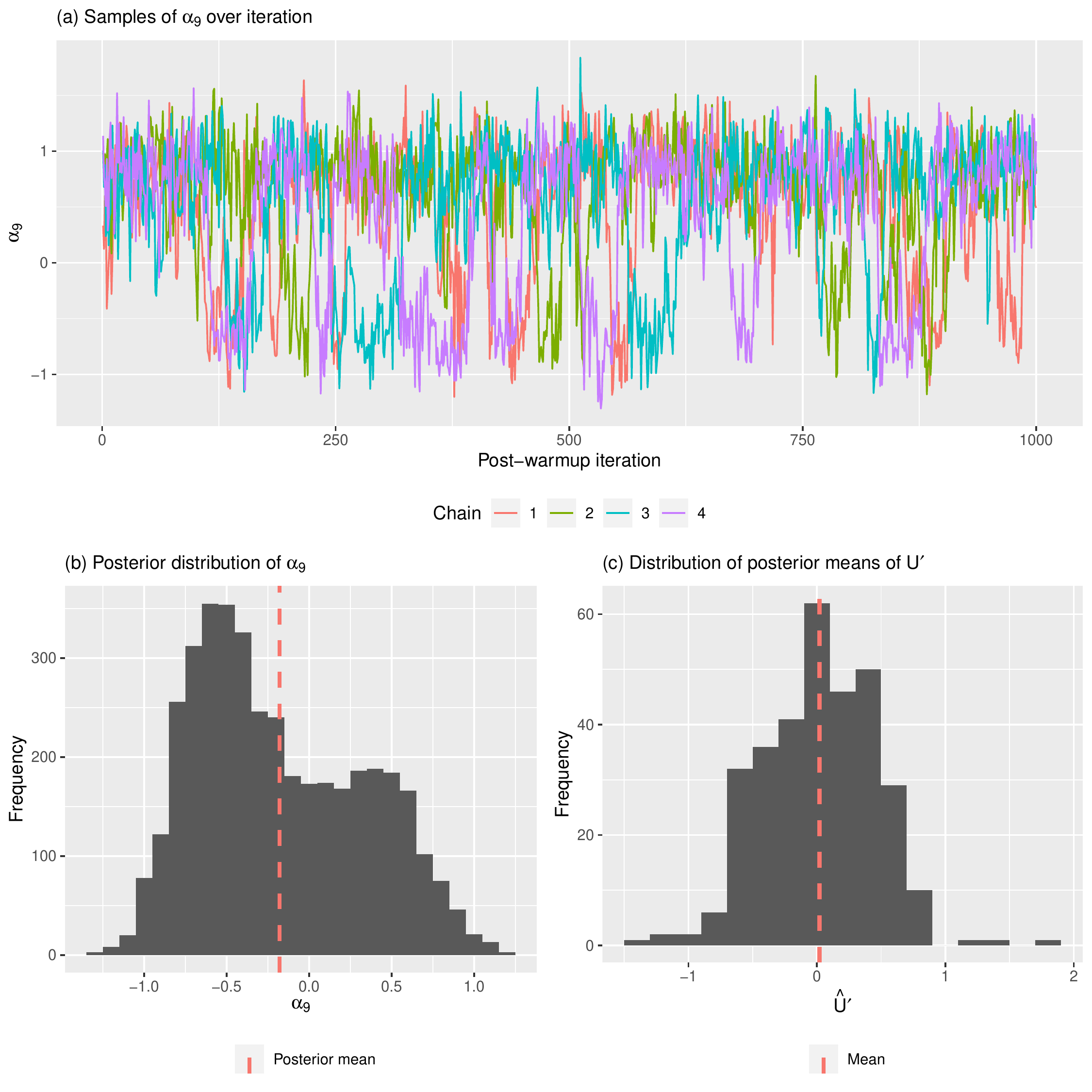}
        \caption{Diagnostics of the Bayesian model for the SBP outcome. Figure (a) shows the Monte Carlo samples of $\alpha_9$ over iteration from 4 Markov chains. The chains are mixed well. Figure (b) shows the posterior distribution of $\alpha_9$ is bimodal. In Figure (c), the estimated $U'$ is denoted by $\hat U'$. We use posterior means of $U'$ as the point estimates of $U'$. Figure (c) shows that $U'$ is not estimated to be zero or constant, but it is small.}
        \label{fig:res-sbp-diag}
\end{figure}

The initial Bayesian model for the SBP outcome shows two posterior solutions because the posterior distribution of $\alpha_9$ is bimodal around 0.5 and -0.5, when the prior of $U'$ is $N(0, 3)$ and the priors of the coefficients are $N(0, 1)$. Some diagnostic results are shown in figure \ref{fig:res-sbp-diag}.

Then we build a Bayesian model for each posterior solution separately. $\alpha_9$ is restricted to be non-negative in one Bayesian model and non-positive in the other Bayesian model. To avoid ambiguity, we call the two new Bayesian models, the restricted Bayesian models. Both restricted Bayesian models are fitted well  They are convergent, all $\hat R$ values are below 1.1, all Monte Carlo standard errors are small and all effective sample sizes are high. In addition, they are identified, where the coefficient of $U'$ and $U'$ itself are estimated well and the standard deviations of the exposure and the outcome have little posterior correlation. Diagnostic figures and comments for two restricted Bayesian models are shown in figure \ref{fig:res-sbp-diag-2}.

\begin{figure}[htbp]
     \centering
     \begin{subfigure}[htbp]{1\textwidth}
         \centering
         \includegraphics[width=1\textwidth]{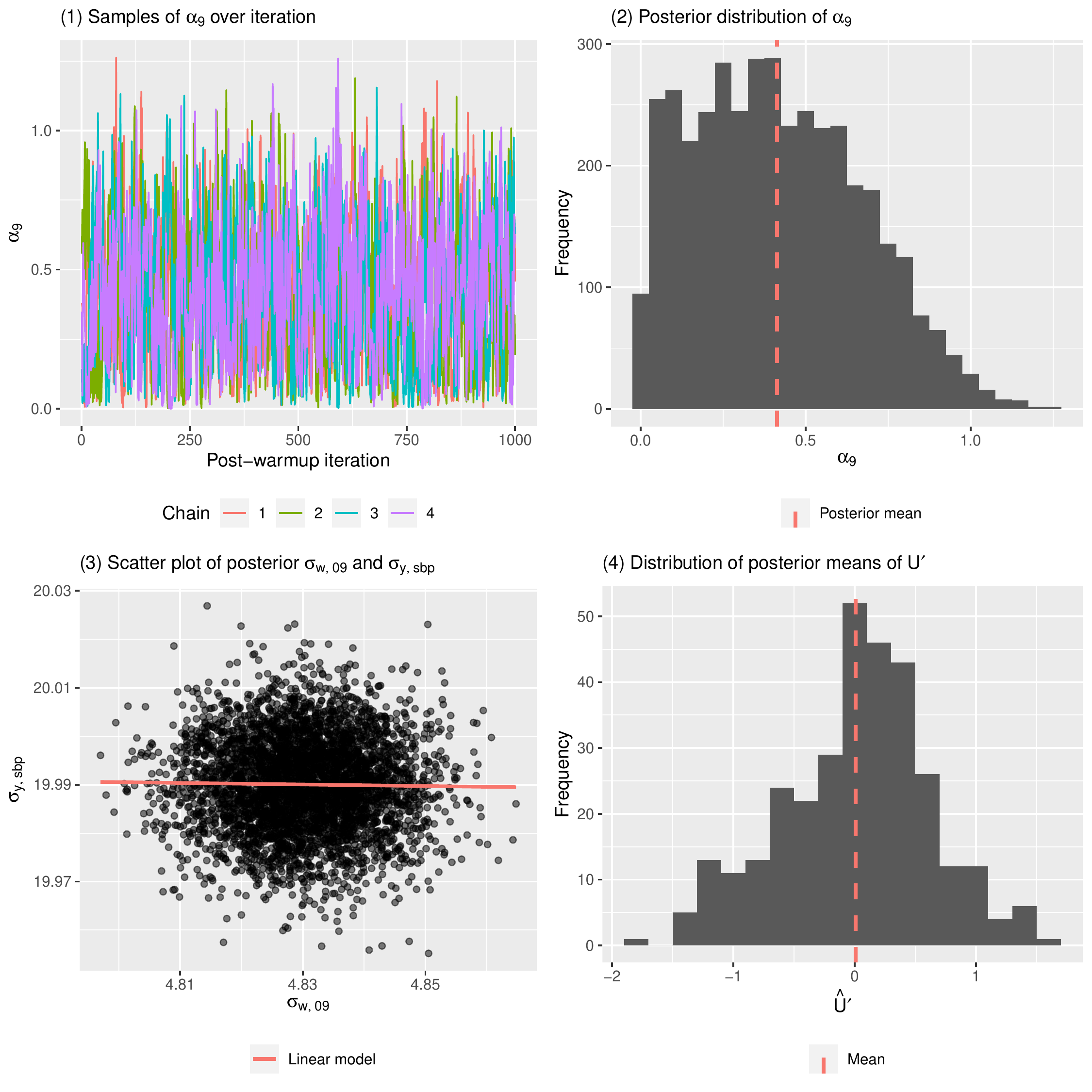}
         \caption{$\alpha_9 \geq 0$.}
     \end{subfigure}
\end{figure}

\begin{figure}[htbp]
\ContinuedFloat
     \centering
     \begin{subfigure}[htbp]{1\textwidth}
         \centering
         \includegraphics[width=1\textwidth]{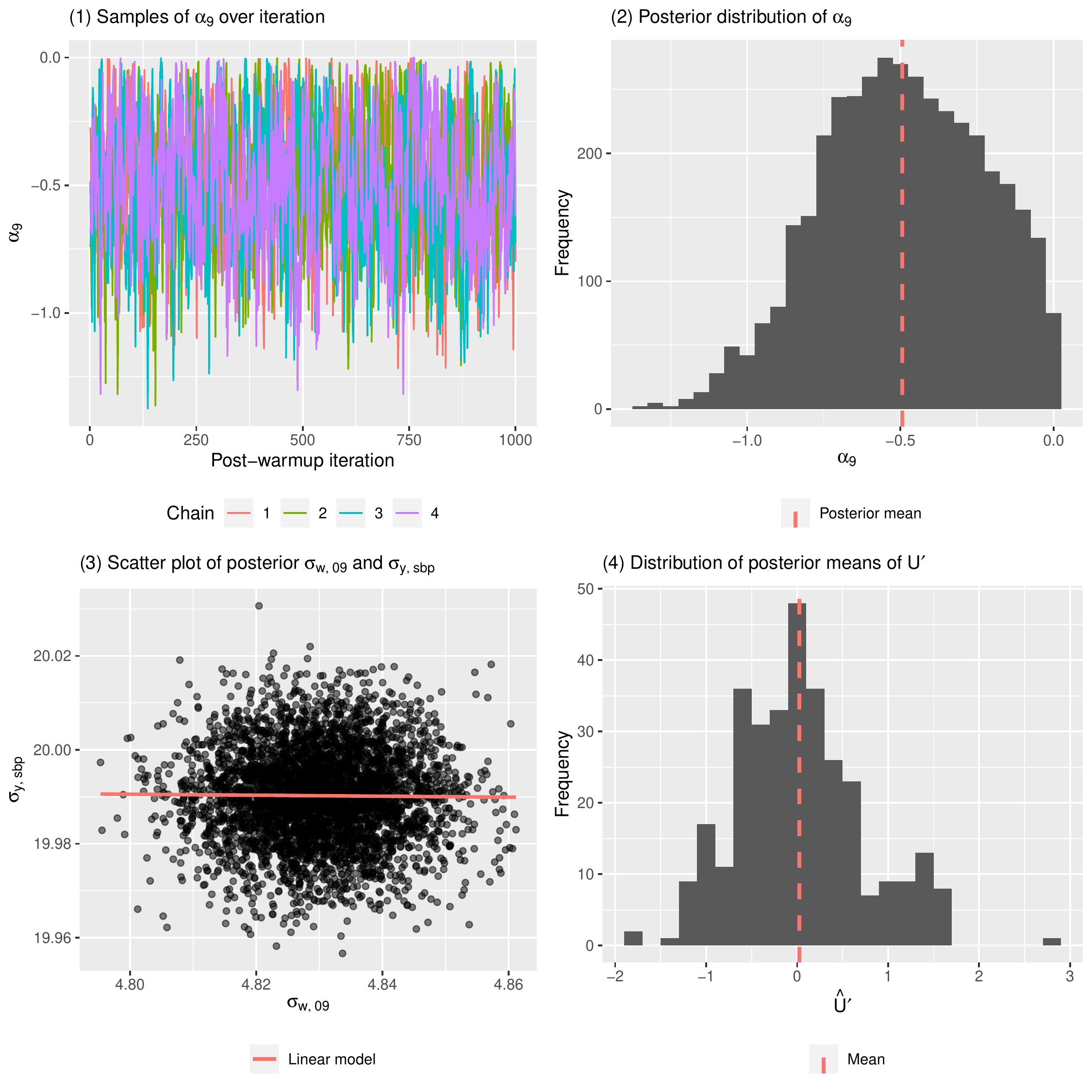}
         \caption{$\alpha_9 \leq 0$.}
     \end{subfigure}
        \caption{Identifiability diagnostics of two restricted Bayesian model for the SBP outcome. Figure (a) shows the Monte Carlo samples of $\alpha_9$ over iteration from 4 Markov chains. The samples do not show unusual behaviors and all chains are mixed well. Figure (b) shows the posterior distribution of $\alpha_9$ is proper. Figure (c) shows there is little posterior correlation between $\sigma_{w, 09}$ and $\sigma_{y, ls7}$ from a linear model between them. In Figure (d), the estimated $U'$ is denoted by $\hat U'$. We use posterior means of $U'$ as the point estimates of $U'$. Figure (d) shows that $U'$ is not estimated to be zero or constant.}
        \label{fig:res-sbp-diag-2}
\end{figure}

Table \ref{tab:res-sbp-post-est} shows posterior estimates of parameters related to $W^{09, obs}$ and $Y^{sbp, obs}$ from two restricted Bayesian models for the SBP outcome and two sensitivity analysis models. Two sensitivity analysis models include a complete-case model and an association model. Two restricted Bayesian models have the same association model, but their complete-case models are built separately. The complete-case model uses all distributions in equation \ref{eqn:results-model-formula-sbp}. The association model uses the distribution for $G^{obs}$ in equation \ref{eqn:results-model-formula-sbp} and its distributions for the exposure and the outcome are shown in equation \ref{eqn:results-sbp-sa-association}.
 
 \begin{eqnarray}
W^{09, obs} &\sim& 
    \begin{cases}
      0, & \text{if $Z^{obs}=0$, $G^{obs}=co$} \\
     N(\alpha_0 + \alpha_1~\text{Age$^{std}$} + \alpha_2~\text{Gender} + \alpha_3~\text{Ethnicity} \\
     + ~\alpha_4~\text{Education level} + \alpha_5~\text{Smoking status} \\
     +~ \alpha_6~\text{Alcohol frequency} \\
     +~ \alpha_7~\text{Baseline SBP$^{std}$}\\
     +~ \alpha_8~G^{obs}, \sigma_{w, 09}), & \text{otherwise}  \nonumber 
    \end{cases} \\
Y^{sbp, obs} &\sim& N(\beta_0 + \beta_1~\text{Age$^{std}$} + \beta_2~\text{Gender} + \beta_3~\text{Ethnicity}  + \beta_4~\text{Education level}  \nonumber \\
&& +~ \beta_5~\text{Smoking status} + \beta_6~\text{Alcohol frequency} + \beta_7~\text{Baseline SBP}  \nonumber \\
&& +~ \beta_8~W^{09, obs}, \sigma_{y, sbp})
\label{eqn:results-sbp-sa-association}
\end{eqnarray}

\begin{table}[htbp]
\small
   \begin{subtable}{1\textwidth}
   \centering
\begin{tabular}{lrrrr}
  \hline
\multirowcell{2}[0pt][l]{Parameter} & \multirowcell{2}[0pt][r]{Prior} & \multicolumn{3}{c}{Posterior mean (95\% interval)}\\ 
  \cmidrule(l){3-5} 
& & Bayesian & Complete cases & Association \\ 
  \hline
$\beta_0$ &$N(0, 3)$&  &  & 5.10 (-0.74, 10.66) \\ 
$\beta_1$ &$N(0, 1)$ & 0.30 (-1.24, 1.83) & 0.08 (-1.47, 1.64) & 0.34 (-1.21, 1.90) \\ 
$\beta_2$, Male &$N(0, 1)$ & 1.16 (-0.72, 2.98) & 0.97 (-0.89, 2.84) & 1.14 (-0.75, 2.95) \\ 
$\beta_3$, & &&& \\
\quad~ Maori &$N(0, 1)$ & 0.09 (-1.77, 1.93) & -0.02 (-1.91, 1.83) & 0.10 (-1.74, 1.92) \\ 
\quad~ Pasifika &$N(0, 1)$ & 0.16 (-1.61, 1.98) & 0.32 (-1.56, 2.21) & 0.15 (-1.70, 2.05) \\ 
\quad~ Asian &$N(0, 1)$ & -0.07 (-2.00, 1.82) & -0.10 (-1.93, 1.75) & -0.05 (-1.98, 1.92) \\ 
$\beta_4$,  & &&& \\
\quad~ College &$N(0, 1)$ & -0.64 (-2.41, 1.21) & -0.68 (-2.44, 1.10) & -0.63 (-2.48, 1.18) \\ 
\quad~ University &$N(0, 1)$ & 0.28 (-1.65, 2.21) & 0.25 (-1.57, 2.05) & 0.31 (-1.46, 2.15) \\ 
$\beta_5$, Non-current &$N(0, 1)$ & 0.02 (-1.79, 1.92) & -0.01 (-1.92, 1.86) & 0.02 (-1.83, 1.83) \\ 
$\beta_6$, More than &$N(0, 1)$ & -0.23 (-2.06, 1.63) & -0.05 (-1.87, 1.76) & -0.25 (-2.03, 1.55) \\ 
$\beta_7$ &$N(0, 1)$ & 0.07 (0.04, 0.10) & 0.07 (0.05, 0.10) & 0.03 (-0.02, 0.08) \\ 
$e_{ate}$ &$N(0, 1)$ & -0.54 (-0.90, -0.19) & -0.62 (-1.00, -0.23) & -0.51 (-0.86, -0.16) \\ 
$\sigma_{y, sbp}$ &$N(19.99, 0.01)$ & 19.99 (19.97, 20.01) & 19.99 (19.97, 20.01) & 19.99 (19.97, 20.01) \\ 
$\alpha_0$ &$N(10, 1)$ & 9.13 (7.64, 10.59) & 10.60 (9.07, 12.11) & 9.06 (7.60, 10.52) \\ 
$\alpha_1$ &$N(0, 1)$ & 0.07 (-0.66, 0.80) & 0.08 (-0.73, 0.88) & 0.08 (-0.62, 0.78) \\ 
$\alpha_2$, Male &$N(0, 1)$  & -1.73 (-2.93, -0.50) & -0.44 (-1.79, 0.89) & -1.77 (-2.97, -0.61) \\ 
$\alpha_3$, & &&& \\
\quad~ Maori &$N(0, 1)$ & -0.56 (-2.04, 0.86) & -0.25 (-1.87, 1.39) & -0.61 (-1.91, 0.71) \\ 
\quad~ Pasifika &$N(0, 1)$ & -0.71 (-2.11, 0.69) & -0.25 (-1.72, 1.23) & -0.75 (-2.11, 0.59) \\ 
\quad~ Asian &$N(0, 1)$ & -0.59 (-2.08, 0.92) & -0.29 (-1.84, 1.30) & -0.60 (-2.01, 0.80) \\ 
$\alpha_4$,  & &&& \\
\quad~ College &$N(0, 1)$ & 0.97 (-0.35, 2.34) & 0.33 (-1.12, 1.85) & 1.00 (-0.38, 2.37) \\ 
\quad~ University &$N(0, 1)$ & -0.40 (-1.76, 0.96) & 0.61 (-0.82, 2.06) & -0.41 (-1.75, 0.89) \\ 
$\alpha_5$, Non-current &$N(0, 1)$ & -1.10 (-2.52, 0.39) & -0.19 (-1.84, 1.45) & -1.15 (-2.60, 0.24) \\ 
$\alpha_6$, More than &$N(0, 1)$ & -0.03 (-1.23, 1.14) & 0.24 (-1.09, 1.57) & -0.05 (-1.24, 1.16) \\ 
$\alpha_7$ &$N(0, 1)$ & 0.30 (-0.42, 1.05) & 0.21 (-0.62, 1.04) & 0.22 (-0.45, 0.90) \\ 
$\alpha_8$ &$N(0, 1)$ & 2.71 (1.30, 4.12) & 1.88 (0.42, 3.36) & 2.86 (1.50, 4.23) \\ 
$\alpha_9$ &$N(0, 1)$ & 0.41 (0.03, 0.93) & 0.18 (0.01, 0.51) &  \\ 
$\sigma_{w, 09}$ &$N(4.83, 0.01)$& 4.83 (4.81, 4.85) & 4.83 (4.81, 4.85) & 4.83 (4.81, 4.85) \\ 
\hline
\end{tabular}  
   \caption{$\alpha_9 \geq 0$.} 
   \end{subtable}
\end{table}

\begin{table}[htbp]
\ContinuedFloat
\small
   \begin{subtable}{1\textwidth}
   \centering
\begin{tabular}{lrrrr}
  \hline
\multirowcell{2}[0pt][l]{Parameter} & \multirowcell{2}[0pt][r]{Prior} & \multicolumn{3}{c}{Posterior mean (95\% interval)}\\ 
  \cmidrule(l){3-5} 
& & Bayesian & Complete cases & Association \\ 
  \hline
$\beta_0$ &$N(0, 3)$ &  &  & 5.10 (-0.74, 10.66) \\ 
$\beta_1$ &$N(0, 1)$ & 0.33 (-1.22, 1.86) & 0.11 (-1.47, 1.66) & 0.34 (-1.21, 1.90) \\ 
$\beta_2$, Male &$N(0, 1)$ & 1.19 (-0.62, 3.10) & 0.96 (-0.88, 2.77) & 1.14 (-0.75, 2.95) \\ 
$\beta_3$, & &&& \\
\quad~ Maori &$N(0, 1)$ & 0.10 (-1.71, 1.95) & 0.00 (-1.92, 1.92) & 0.10 (-1.74, 1.92) \\ 
\quad~ Pasifika &$N(0, 1)$ & 0.14 (-1.70, 2.04) & 0.32 (-1.66, 2.28) & 0.15 (-1.70, 2.05) \\ 
\quad~ Asian &$N(0, 1)$ & -0.03 (-1.86, 1.87) & -0.11 (-1.95, 1.77) & -0.05 (-1.98, 1.92) \\ 
$\beta_4$,  & &&& \\
\quad~ College &$N(0, 1)$ & -0.62 (-2.48, 1.24) & -0.67 (-2.49, 1.16) & -0.63 (-2.48, 1.18) \\ 
\quad~ University &$N(0, 1)$& 0.31 (-1.54, 2.18) & 0.24 (-1.65, 2.13) & 0.31 (-1.46, 2.15) \\ 
$\beta_5$, Non-current &$N(0, 1)$ & 0.04 (-1.78, 1.89) & 0.00 (-1.93, 1.97) & 0.02 (-1.83, 1.83) \\ 
$\beta_6$, More than &$N(0, 1)$ & -0.23 (-2.06, 1.61) & -0.05 (-1.89, 1.81) & -0.25 (-2.03, 1.55) \\ 
$\beta_7$ &$N(0, 1)$ & 0.07 (0.04, 0.09) & 0.07 (0.05, 0.10) & 0.03 (-0.02, 0.08) \\ 
$e_{ate}$ &$N(0, 1)$ & -0.45 (-0.81, -0.09) & -0.60 (-0.96, -0.23) & -0.51 (-0.86, -0.16) \\ 
$\sigma_{y, sbp}$ &$N(19.99, 0.01)$ & 19.99 (19.97, 20.01) & 19.99 (19.97, 20.01) & 19.99 (19.97, 20.01) \\ 
$\alpha_0$ &$N(10, 1)$ & 9.15 (7.73, 10.62) & 10.59 (9.04, 12.16) & 9.06 (7.60, 10.52) \\ 
$\alpha_1$ &$N(0, 1)$ & 0.10 (-0.62, 0.82) & 0.09 (-0.77, 0.93) & 0.08 (-0.62, 0.78) \\ 
$\alpha_2$, Male &$N(0, 1)$ & -1.68 (-2.89, -0.48) & -0.43 (-1.80, 0.91) & -1.77 (-2.97, -0.61) \\ 
$\alpha_3$, & &&& \\
\quad~ Maori &$N(0, 1)$ & -0.58 (-2.02, 0.91) & -0.28 (-1.92, 1.31) & -0.61 (-1.91, 0.71) \\ 
\quad~ Pasifika &$N(0, 1)$ & -0.70 (-2.12, 0.66) & -0.24 (-1.72, 1.27) & -0.75 (-2.11, 0.59) \\ 
\quad~ Asian &$N(0, 1)$ & -0.57 (-2.03, 0.89) & -0.28 (-1.85, 1.28) & -0.60 (-2.01, 0.80) \\ 
$\alpha_4$,  & &&& \\
\quad~ College &$N(0, 1)$  & 0.94 (-0.45, 2.33) & 0.32 (-1.21, 1.80) & 1.00 (-0.38, 2.37) \\ 
\quad~ University &$N(0, 1)$ & -0.40 (-1.74, 0.92) & 0.62 (-0.92, 2.16) & -0.41 (-1.75, 0.89) \\ 
$\alpha_5$, Non-current &$N(0, 1)$& -1.07 (-2.50, 0.40) & -0.17 (-1.84, 1.50) & -1.15 (-2.60, 0.24) \\ 
$\alpha_6$, More than &$N(0, 1)$ & -0.05 (-1.22, 1.15) & 0.24 (-1.12, 1.64) & -0.05 (-1.24, 1.16) \\ 
$\alpha_7$ &$N(0, 1)$ & 0.14 (-0.60, 0.87) & 0.12 (-0.66, 0.92) & 0.22 (-0.45, 0.90) \\ 
$\alpha_8$ &$N(0, 1)$  & 2.67 (1.26, 4.07) & 1.88 (0.31, 3.46) & 2.86 (1.50, 4.23) \\ 
$\alpha_9$ &$N(0, 1)$  & -0.49 (-1.03, -0.03) & -0.21 (-0.58, -0.01) &   \\ 
$\sigma_{w, 09}$ &$N(4.83, 0.01)$ & 4.83 (4.81, 4.85) & 4.83 (4.81, 4.85) & 4.83 (4.81, 4.85) \\ 
\hline
\end{tabular}  
   \caption{$\alpha_9 \leq 0$.} 
   \end{subtable}
      \caption{Posterior estimates from two restricted Bayesian models for the SBP outcome, complete-case and association models.}
 \label{tab:res-sbp-post-est}
\end{table}

From two restricted Bayesian models, the larger posterior estimates of $e_{ate}$ are -0.54 (-0.90, -0.19) and the smaller posterior estimates of $e_{ate}$ are -0.45 (-0.81, -0.09). Two sets of posterior estimates of $e_{ate}$ are similar, with a difference about 0.1. For each restricted Bayesian model, the posterior estimates of $e_{ate}$ are similar to those from the corresponding complete-case and association models, which indicates that there is possibly little unmeasured confounding and missingness does not affect estimation of the average treatment effect significantly. Little unmeasured confounding can also explain why two sets of posterior estimates of $e_{ate}$ are similar. In this case, multiple posterior solutions seem not to cause inferential problems. Since two sets of posterior estimates of $e_{ate}$ are similar, we can use both of them to draw inference about the average treatment effect. 

From the restricted Bayesian model with $\alpha_9 \geq 0$, the posterior mean of $e_{ate}$ is -0.54 mm Hg, which means that attending one HWC session can decrease systolic blood pressure by 0.54 mm Hg on average. The 95\% posterior interval of $e_{ate}$ is (-0.90, -0.19) and is fully negative, which indicates that the average effect of one HWC session on systolic blood pressure is statistically significantly negative. From the restricted Bayesian model with $\alpha_9 \leq 0$, the average effect of one HWC session can be slightly weaker and slightly less statistically significant.

The average effect of one HWC session on systolic blood pressure is relatively moderate. Normal systolic blood pressure is less than 120 mm Hg, elevated systolic blood pressure is 120-129 mm Hg, high blood pressure is 130-139 mm Hg and above. Attending a few HWC sessions may not improve systolic blood pressure obviously, but it is likely that systolic blood pressure is significantly decreased after many HWC sessions are attended, such as 10 HWC sessions. The HWC intervention may effectively improve systolic blood pressure and further cardiovascular health. 

In each restricted Bayesian model, sensitivity of the posterior estimates of $e_{ate}$ is tested against informativeness of the priors of $e_{ate}$, $U'$ and $\alpha_9$. Slightly more informative and less informative priors are considered in sensitivity analysis, with the prior means held fixed, in order to see how the posterior estimates of $e_{ate}$ would change with different informativeness from the priors to be tested. Zero prior means of $e_{ate}$, $U'$ and $\alpha_9$ would be suitable, as we do not have information about their means in reality. Table \ref{tab:results-sbp-sa-prior} shows the results of prior sensitivity analysis.

\begin{table}[htbp]
\small
   \begin{subtable}{1\textwidth}
   \centering
   \begin{tabular}{rrrr}
  \hline
\multirowcell{2}[0pt][r]{Model} & \multicolumn{3}{c}{Prior}\\ 
  \cmidrule(l){2-4} 
& $N(0, 0.5)$ & $N(0, 1)$ & $N(0, 10)$   \\   
  \hline
$\alpha_9 \geq 0$ &&& \\
$S_p$ &0.13&0.03&0.00 \\
$e_{ate}$ & -0.49 (-0.84, -0.14)&-0.54 (-0.90, -0.19)& -0.55 (-0.91, -0.19)\\
$\alpha_9 \leq 0$ &&& \\
 $S_p$ & 0.13&0.03&0.00 \\
$e_{ate}$ &-0.45 (-0.81, -0.09)&-0.45 (-0.81, -0.09)&-0.47 (-0.83, -0.11)\\
   \hline
   \end{tabular}
   \caption{Prior sensitivity measures and posterior mean (95\% interval) of $e_{ate}$ under priors of $e_{ate}$ with different informativeness from two restricted Bayesian models built in sensitivity analysis. $S_p$ is the prior sensitivity measure.} 
   \end{subtable}
   
\vspace{1em}

    \begin{subtable}{1\textwidth}
   \centering
   \begin{tabular}{rrrrr}
  \hline
\multirowcell{2}[0pt][r]{Model} & \multicolumn{4}{c}{Prior}\\ 
  \cmidrule(l){2-5} 
& $N(0, 1)$ & $N(0, 3)$ & $N(0, 5)$ & $N(0, 10)$ \\ 
  \hline
$\alpha_9 \geq 0$ &&&& \\
$e_{ate}$ &-0.51 (-0.87, -0.15)&-0.54 (-0.90, -0.19)&-0.56 (-0.94, -0.18)&-0.60 (-1.02, -0.19) \\
$\alpha_9$ &0.82 (0.02, 2.09)&0.41 (0.03, 0.93)&0.25 (0.01, 0.57)&0.10 (0.00, 0.27)\\ 
$\alpha_9 \leq 0$ &&&& \\
  $e_{ate}$ &-0.49 (-0.85, -0.14)&-0.45 (-0.81, -0.09)&-0.42 (-0.79, -0.06)&-0.33 (-0.74, 0.06) \\
$\alpha_9$ &-0.86 (-2.18, -0.04)&-0.49 (-1.03, -0.03)&-0.32 (-0.61, -0.03)&-0.18 (-0.32, -0.02)\\ 
   \hline
   \end{tabular}
   \caption{Posterior mean (95\% interval) of $e_{ate}$ and $\alpha_9$ under priors of $U'$ with different informativeness from two restricted Bayesian models built in sensitivity analysis. $\alpha_9$ is used as a reference for the strength of unmeasured confounding.} 
   \end{subtable}
 
 \vspace{1em}
 
       \begin{subtable}{1\textwidth}
   \centering
   \begin{tabular}{rrrr}
  \hline
\multirowcell{2}[0pt][r]{Model} & \multicolumn{3}{c}{Prior}\\ 
  \cmidrule(l){2-4} 
& $N(0, 0.5)$ & $N(0, 1)$ & $N(0, 10)$   \\   
  \hline
  $\alpha_9 \geq 0$ & -0.53 (-0.90, -0.18)&-0.54 (-0.90, -0.19)& -0.54 (-0.90, -0.18)\\
$\alpha_9 \leq 0$ &-0.46 (-0.82, -0.11)&-0.45 (-0.81, -0.09)& -0.45 (-0.80, -0.10) \\
   \hline
   \end{tabular}
   \caption{Posterior mean (95\% interval) of $e_{ate}$ under priors of $\alpha_9$ with different informativeness from two restricted Bayesian models built in sensitivity analysis.} 
   \end{subtable}
    \caption{Results of prior sensitivity analysis.}
 \label{tab:results-sbp-sa-prior}
\end{table}

From table \ref{tab:results-sbp-sa-prior}, the posterior estimates of $e_{ate}$ are insensitive to informativeness of the priors of $e_{ate}$ and $\alpha_9$, but they are slightly sensitive to informativeness of the prior of $U'$. When the prior of $U'$ becomes less informative to an extent that the Bayesian model built in sensitivity analysis is good, the posterior estimates of $e_{ate}$ from the restricted Bayesian model with $\alpha_9 \geq 0$ in sensitivity analysis tend to decrease a little while the posterior estimates of $e_{ate}$ from the restricted Bayesian model with $\alpha_9 \leq 0$ in sensitivity analysis tend to increase a little and may become statistically insignificant.

We also build a Bayesian instrumental variable model through two-stage least squares \autocite{angrist_two-stage_1995}, using equation \ref{eqn:results-sbp-sa-iv}. The assignment $Z^{obs}$ is a valid instrument after conditioning on measured confounders. We still use sample standard deviations in the Bayesian instrumental variable model. The prior for $\alpha_0$ is $N(10, 1)$ and the prior for $\beta_0$ is $N(0, 3)$. Priors for the other parameters are $N(0, 1)$.
 \begin{eqnarray}
W^{09, obs} &\sim& N(\alpha_0 + \alpha_1~\text{Age$^{std}$} + \alpha_2~\text{Gender} + \alpha_3~\text{Ethnicity}  + \alpha_4~\text{Education level}  \nonumber \\
&& +~ \alpha_5~\text{Smoking status} + \alpha_6~\text{Alcohol frequency} + \alpha_7~\text{Baseline SBP}  \nonumber \\
&& +~ \alpha_8~Z^{obs}, \sigma_{w, 09}) \\
Y^{sbp, obs} &\sim& N(\beta_0 + \beta_1~\text{Age$^{std}$} + \beta_2~\text{Gender} + \beta_3~\text{Ethnicity}  + \beta_4~\text{Education level}  \nonumber \\
&& +~ \beta_5~\text{Smoking status} + \beta_6~\text{Alcohol frequency} + \beta_7~\text{Baseline SBP}  \nonumber \\
&& +~ \beta_8~Z^{obs}, \sigma_{y, sbp}) \\
e_{ate} &=& \frac{\beta_8}{\alpha_8}
\label{eqn:results-sbp-sa-iv}
\end{eqnarray}

In addition, we build a Bayesian random intercept outcome model. The priors for unknown parameters including $U'$ in the random intercept outcome model are identical to those from two original restricted Bayesian models correspondingly. Table \ref{tab:results-sbp-sa-iv-ri} shows the posterior estimates of $e_{ate}$ from the instrumental variable model and the random intercept outcome model.
\begin{table}[htbp]
\small
   \centering
  \begin{tabular}{rrr}
  \hline
  & Instrumental variable model & Random intercept outcome model \\ 
  \hline
 $e_{ate}$ &-0.01 (-0.08, 0.07) & -0.50 (-0.85, -0.14) \\
   \hline
   \end{tabular}
 \caption{Posterior mean (95\% interval) of $e_{ate}$ from the instrumental variable model and the random intercept outcome model.}
 \label{tab:results-sbp-sa-iv-ri}
\end{table}

The average treatment effect estimates from the random intercept outcome model are similar to those from two original restricted Bayesian models, but the average treatment effect estimates from the instrumental variable model are quite different from those from two original restricted Bayesian models and the random intercept outcome model. The average treatment effect estimates from the instrumental variable model are statistically insignificant, but they may be biased, possibly because the coefficient of the instrument in the linear model of the SBP outcome is small and thus the instrument is weakly correlated with the SBP outcome \autocite{bound_problems_1995}. It is hard to determine if the average treatment effect estimates from two original restricted Bayesian models are reliable, by comparison with the the instrumental variable model and the random intercept outcome model. It seems more appropriate to conclude that due to model sensitivity and model choices, the average treatment effect estimates from two original restricted Bayesian models may become statistically insignificant.

Then sensitivity of the posterior estimates of $e_{ate}$ is tested against variations in sample standard deviations of the exposure and the SBP outcome. Variations in sample standard deviations are represented by different locations of the priors of $\sigma_{w, 09}$ and $\sigma_{y, sbp}$, where sample standard deviations are used as the prior means. Prior locations that are considered include point estimates of sample standard deviations and two boundary points from the 95\% bootstrap intervals of sample standard deviations from table \ref{tab:sd-expoout-boot},possibly other values if necessary. Table \ref{tab:results-sbp-sa-sd} shows the results of sensitivity analysis of sample standard deviations. 
\begin{table}[htbp]
\small
    \begin{subtable}{1\textwidth}
   \centering
     \begin{tabular}{rrrrrr}
  \hline
\multirowcell{2}[0pt][r]{Model} & \multicolumn{5}{c}{Prior}\\ 
  \cmidrule(l){2-6} 
&$N(0.5, 0.01)$& $N(4.25, 0.01)$ & $N(4.83, 0.01)$ & $N(5.3, 0.01)$ & $N(10, 0.01)$  \\   
  \hline
  $\alpha_9 \geq 0$ &-&-0.58 (-0.93, -0.23)&-0.54 (-0.90, -0.19)&-0.52 (-0.88, -0.16)&-0.51 (-0.87, -0.15)\\
$\alpha_9 \leq 0$ &-&-&-0.45 (-0.81, -0.09)&-0.47 (-0.82, -0.12)&-0.49 (-0.85, -0.14) \\
   \hline
   \end{tabular}
   
   \caption{Posterior mean (95\% interval) of $e_{ate}$ under priors of $\sigma_{w, 09}$ with different locations. ``-'' indicates the Bayesian model built in sensitivity analysis is bad.} 
   \end{subtable}
   
\vspace{1em}

    \begin{subtable}{1\textwidth}
   \centering
    \begin{tabular}{rrrrrr}
  \hline
\multirowcell{2}[0pt][r]{Model} & \multicolumn{5}{c}{Prior}\\ 
  \cmidrule(l){2-6} 
& $N(1, 0.01)$ & $N(18.61, 0.01)$ & $N(19.99, 0.01)$ & $N(22.08, 0.01)$ & $N(50, 0.01)$   \\   
  \hline
  $\alpha_9 \geq 0$ & -0.48 (-0.53, -0.42)&-0.54 (-0.87, -0.20)&-0.54 (-0.90, -0.19)&-0.53 (-0.93, -0.13)&-\\
$\alpha_9 \leq 0$ &-&-0.46 (-0.80, -0.13)&-0.45 (-0.81, -0.09)&-0.45 (-0.86, -0.05)&- \\
   \hline
   \end{tabular}
   \caption{Posterior mean (95\% interval) of $e_{ate}$ under priors of $\sigma_{y, sbp}$ with different locations.} 
   \end{subtable}
 \caption{Results of sensitivity analysis of sample standard deviations.}
 \label{tab:results-sbp-sa-sd}
\end{table}

Overall, the posterior estimates of $e_{ate}$ from the original restricted Bayesian model with $\alpha_9 \geq 0$ are insensitive to variations in sample standard deviations. The posterior estimates of $e_{ate}$ from the original restricted Bayesian model with $\alpha_9 \leq 0$ are more sensitive to variations in sample standard deviations, which leads to bad models rather than changes in the posterior estimates of $e_{ate}$. This indicates estimation of standard deviations affects modelling more than estimation accuracy of the average treatment effect.

In conclusion, there is evidence that the HWC intervention is beneficial to cardiovascular health by effectively reducing systolic blood pressure. On completion of the HWC intervention, on average systolic blood pressure will be reduced by about 8.1 (2.9, 13.5) mm Hg, but due to model sensitivity and model choices, the actual decrease in systolic blood pressure may become smaller or even statistically insignificant. Strong compliance to the HWC intervention is recommended to see a likely significant decrease in systolic blood pressure and a likely significant improvement in cardiovascular health.


\subsection{Results of 5-year cardiovascular disease risk score}

Distributions for the observed compliance behavior, the exposure from the first set of exposure, the CVD risk outcome in a Bayesian model are shown in equation \ref{eqn:results-model-formula-cvd}. Missing data is not shown but uses the models for the corresponding observed variables. Bayesian models for the exposure from the first set of exposure and the CVD risk outcome have already been reparameterized through the random intercept reparameterization. The reparameterized unmeasured confounder is denoted by $U'$. The symbol ``${\rm logit}^{-1}$'' is the logistic function.  The symbol ``Age$^{std}$'' means the standardized age.  The symbol ``Baseline CVD risk score$^{BS}$'' means the B-spline of the baseline CVD risk score that involves 5 coefficients in a Bayesian model. For each categorical measured confounder, the coefficients of the levels except the reference level are still presented in the format ``coefficient of the confounder, the level name''. $\beta_7$ is the average effect of the exposure $W^{09, obs}$, and we denote it by $e_{ate}$ when presenting results. 
\begin{eqnarray}
G^{obs} &\sim& Bern({\rm logit}^{-1}(\gamma_0 + \gamma_1~\text{Age$^{std}$} + \gamma_2~\text{Gender} + \gamma_3~\text{Ethnicity}  \nonumber \\
&& + ~\gamma_4~\text{Education level} + \gamma_5~\text{Smoking status} + \gamma_6~\text{Alcohol frequency} \nonumber \\
&& +~ \gamma_7~\text{Baseline CVD risk score})),  \nonumber \\
W^{09, obs} &\sim& 
    \begin{cases}
      0, & \text{if $Z^{obs}=0$, $G^{obs}=co$} \\
     N(\alpha_0 + \alpha_1~\text{Age$^{std}$} + \alpha_2~\text{Gender} + \alpha_3~\text{Ethnicity} \\
     + ~\alpha_4~\text{Education level} + \alpha_5~\text{Smoking status} \\
     +~ \alpha_6~\text{Alcohol frequency} \\
     +~ \alpha_7~\text{Baseline CVD risk score}\\
     +~ \alpha_8~G^{obs} + \alpha_9~U', \sigma_{w, 09}), & \text{otherwise}  \nonumber 
    \end{cases} \\
Y^{cvd, obs} &\sim& N(\beta_1~\text{Age$^{std}$} + \beta_2~\text{Gender} + \beta_3~\text{Ethnicity}  + \beta_4~\text{Education level}  \nonumber \\
&& +~ \beta_5~\text{Smoking status} + \beta_6~\text{Alcohol frequency} + \beta_7~W^{09, obs}  \nonumber \\
&& +~ \text{Baseline CVD risk score$^{BS}$} + U', \sigma_{y, cvd}), \nonumber \\
U' &\sim& N(0, 1)
\label{eqn:results-model-formula-cvd}
\end{eqnarray}

We use sample standard deviations from table \ref{tab:sd-expoout-boot} as the prior means for standard deviations of the exposure and the CVD risk outcome, $\sigma_{w, 09}$ and $\sigma_{y, cvd}$, and construct strongly informative priors for $\sigma_{w, 09}$ and $\sigma_{y, cvd}$ with a prior standard deviation of 0.01, so that their posterior estimates will become sample standard deviations. 

\begin{figure}[htbp]
     \centering
     \includegraphics[width=1\textwidth]{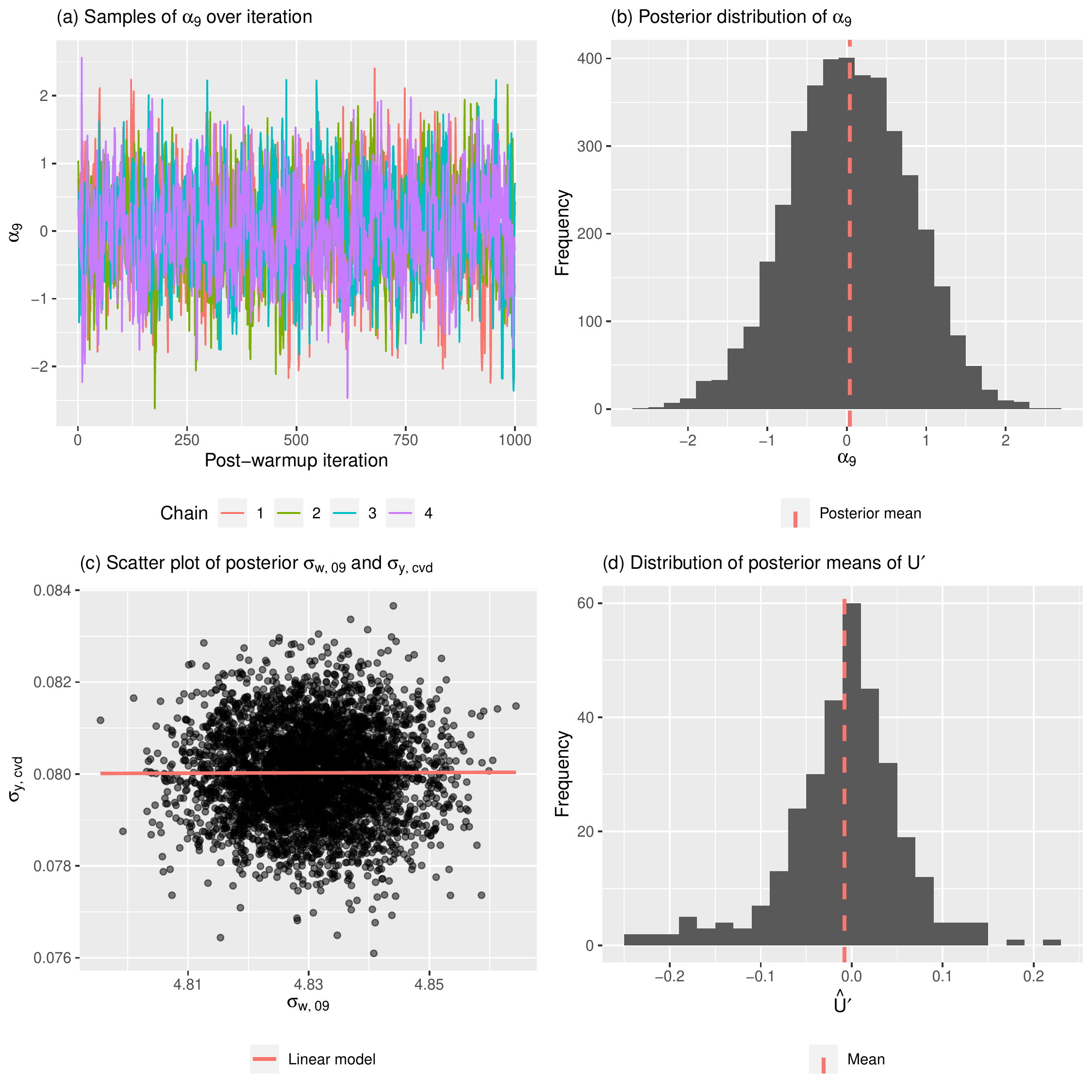}
        \caption{Identifiability diagnostics of the Bayesian model for the CVD risk outcome. Figure (a) shows the Monte Carlo samples of $\alpha_9$ over iteration from 4 Markov chains. The samples do not show unusual behaviors and all chains are mixed well. Figure (b) shows the posterior distribution of $\alpha_9$ is proper. Figure (c) shows there is little posterior correlation between $\sigma_{w, 09}$ and $\sigma_{y, cvd}$ from a linear model between them. In Figure (d), the estimated $U'$ is denoted by $\hat U'$. We use posterior means of $U'$ as the point estimates of $U'$. Figure (d) shows that $U'$ is not estimated to be zero or constant.}
        \label{fig:res-cvd-diag}
\end{figure}

The CVD risk outcome has a small magnitude. Hence, compared to the other outcomes, the priors of unknown parameters for the CVD risk outcome have smaller scales. The Bayesian model for the CVD risk outcome is fitted relatively well. It is convergent, all $\hat R$ values are below 1.1, all Monte Carlo standard errors are small and all effective sample sizes are high. In addition, this model is identified, where the coefficient of $U'$ and $U'$ itself are estimated well and the standard deviations of the exposure and the outcome have little posterior correlation. Diagnostic figures and comments are shown in figure \ref{fig:res-cvd-diag}. The posterior means of $U'$ are relatively small.

Table \ref{tab:res-cvd-post-est} shows posterior estimates of parameters related to $W^{09, obs}$ and $Y^{cvd, obs}$ from the Bayesian model for the CVD risk outcome and an association model. The association model uses the distribution for $G^{obs}$ in equation \ref{eqn:results-model-formula-cvd} and its distributions for the exposure and the outcome are shown in equation \ref{eqn:results-cvd-sa-association}. 
 \begin{eqnarray}
W^{09, obs} &\sim& 
    \begin{cases}
      0, & \text{if $Z^{obs}=0$, $G^{obs}=co$} \\
     N(\alpha_0 + \alpha_1~\text{Age$^{std}$} + \alpha_2~\text{Gender} + \alpha_3~\text{Ethnicity} \\
     + ~\alpha_4~\text{Education level} + \alpha_5~\text{Smoking status} \\
     +~ \alpha_6~\text{Alcohol frequency} \\
     +~ \alpha_7~\text{Baseline CVD risk score}\\
     +~ \alpha_8~G^{obs}, \sigma_{w, 09}), & \text{otherwise}  \nonumber 
    \end{cases} \\
Y^{cvd, obs} &\sim& N(\beta_0 + \beta_1~\text{Age$^{std}$} + \beta_2~\text{Gender} + \beta_3~\text{Ethnicity}  + \beta_4~\text{Education level}  \nonumber \\
&& +~ \beta_5~\text{Smoking status} + \beta_6~\text{Alcohol frequency} + \beta_7~W^{09, obs} \nonumber \\
&& +~ \text{Baseline CVD risk score$^{BS}$}, \sigma_{y, cvd})
\label{eqn:results-cvd-sa-association}
\end{eqnarray}

\begin{table}[htbp]
\small
   \centering
\begin{tabular}{lrrr}
  \hline
\multirowcell{2}[0pt][l]{Parameter} & \multirowcell{2}[0pt][r]{Prior} & \multicolumn{2}{c}{Posterior mean (95\% interval)}\\ 
  \cmidrule(l){3-4} 
& & Bayesian & Association \\ 
  \hline
$\beta_0$ &$N(0, 1)$ &  & -0.16 (-0.19, -0.13) \\ 
$\beta_1$ &$N(0, 0.1)$  & 0.01 (-0.09, 0.11) & 0.02 (0.01, 0.03) \\ 
$\beta_2$, Male &$N(0, 0.1)$ & -0.01 (-0.15, 0.14) & 0.01 (-0.01, 0.03) \\ 
$\beta_3$, & && \\
\quad~ Maori &$N(0, 0.1)$  & 0.00 (-0.16, 0.16) & 0.02 (-0.01, 0.04) \\ 
\quad~ Pasifika &$N(0, 0.1)$ & -0.01 (-0.16, 0.15) & 0.02 (-0.01, 0.05) \\ 
\quad~ Asian &$N(0, 0.1)$ & 0.00 (-0.16, 0.16) & 0.02 (-0.01, 0.04) \\ 
$\beta_4$,  & && \\
\quad~ College &$N(0, 0.1)$ & -0.01 (-0.17, 0.15) & 0.00 (-0.03, 0.02) \\ 
\quad~ University &$N(0, 0.1)$ & -0.01 (-0.17, 0.15) & 0.00 (-0.02, 0.03) \\ 
$\beta_5$, Non-current &$N(0, 0.1)$ & 0.01 (-0.15, 0.17) & 0.05 (0.02, 0.07) \\ 
$\beta_6$, More than &$N(0, 0.1)$ & -0.01 (-0.16, 0.13) & -0.01 (-0.02, 0.01) \\ 
$e_{ate}$ &$N(0, 0.1)$ & 0.00 (-0.03, 0.02) & 0.00 (0.00, 0.00) \\ 
$\sigma_{y, cvd}$ &$N(0.08, 0.001)$ & 0.08 (0.08, 0.08) & 0.08 (0.08, 0.08) \\ 
$\alpha_0$ &$N(10, 1)$ & 9.04 (7.63, 10.45) & 9.07 (7.66, 10.56) \\ 
$\alpha_1$ &$N(0, 1)$ & 0.08 (-0.63, 0.80) & 0.07 (-0.62, 0.80) \\ 
$\alpha_2$, Male &$N(0, 1)$ & -1.75 (-2.88, -0.58) & -1.76 (-2.89, -0.62) \\ 
$\alpha_3$, & && \\
\quad~ Maori &$N(0, 1)$ & -0.59 (-1.97, 0.82) & -0.60 (-1.93, 0.77) \\ 
\quad~ Pasifika &$N(0, 1)$  & -0.78 (-2.11, 0.59) & -0.77 (-2.14, 0.60) \\ 
\quad~ Asian &$N(0, 1)$ & -0.58 (-1.99, 0.84) & -0.61 (-2.05, 0.82) \\
$\alpha_4$,  & && \\
\quad~ College &$N(0, 1)$ & 1.02 (-0.38, 2.39) & 1.03 (-0.32, 2.34) \\ 
\quad~ University &$N(0, 1)$  & -0.46 (-1.82, 0.84) & -0.46 (-1.79, 0.85) \\ 
$\alpha_5$, Non-current &$N(0, 1)$ & -1.12 (-2.51, 0.32) & -1.16 (-2.55, 0.19) \\ 
$\alpha_6$, More than &$N(0, 1)$& -0.07 (-1.24, 1.10) & -0.06 (-1.23, 1.11) \\ 
$\alpha_7$ &$N(0, 1)$ & -0.02 (-1.90, 1.86) & 0.00 (-1.84, 1.92) \\ 
$\alpha_8$ &$N(0, 1)$ & 2.83 (1.46, 4.15) & 2.86 (1.53, 4.17) \\ 
$\alpha_9$ &$N(0, 0.5)$ & 0.04 (-1.44, 1.46) &  \\ 
$\sigma_{w, 09}$ &$N(4.83, 0.01)$  & 4.83 (4.81, 4.85) & 4.83 (4.81, 4.85) \\ 
\hline
\end{tabular}  
      \caption{Posterior estimates from the Bayesian model for the CVD risk outcome, complete-case and association models.}
 \label{tab:res-cvd-post-est}
\end{table}

From table \ref{tab:res-cvd-post-est}, the posterior mean of $e_{ate}$ is 0.00, which means that attending one HWC session does not increase or decrease the 5-year CVD risk score on average. The 95\% posterior interval of $e_{ate}$ is (-0.03, 0.02). This interval covers zero and narrow, which indicates that the average effect of one HWC session on the 5-year CVD risk score is not statistically significant. The 5-year CVD risk score seems not to be changed after many HWC sessions are attended, such as 10 HWC sessions.

There may be little unmeasured confounding in the CVD risk outcome. First, the average effect estimates of one HWC session from the original Bayesian model for the CVD risk outcome are similar to those from the association model. Second, $U'$ is estimated to be very small, possibly associated with the small range of the CVD risk outcome, and the coefficient of $U'$ in the exposure distribution, $\alpha_9$, is estimated to be insignificant. However, there is also a possibility that unmeasured confounding cannot be well estimated in a very small outcome after many measured confounders have been adjusted for.

A complete-case model that uses all distributions in equation \ref{eqn:results-model-formula-cvd} is not fitted well with the same priors from the original Bayesian model for the CVD risk outcome. Failure in building the complete-case model with the priors unchanged may indicate that the original Bayesian model for the CVD risk outcome relies on imputation on missing 5-year CVD risk scores at 9 months post randomization, and its model structure is sensitive to data changes. From the original Bayesian model for the CVD risk outcome, imputation on missing 5-year CVD risk scores at 9 months post randomization is bounded between 0 and 0.3, and the imputed values do not look like Normally distributed. Hence, there is a possibility that the original Bayesian model for the CVD risk outcome does not well impute missing 5-year CVD risk scores at 9 months post randomization. Since missing 5-year CVD risk scores at 9 months post randomization are imputed by the outcome distribution directly, the priors of parameters associated with the outcome distribution may also affect imputation. Then a complete-case model with modified priors including $N(0, 0.5)$ as the prior of $U'$ is fitted, and its posterior estimates of $e_{ate}$ are 0.00 (-0.01, 0.01). This may indicate missingness does not affect estimation of the average treatment effect much.

Sensitivity of the posterior estimates of $e_{ate}$ is tested against informativeness of the priors of $e_{ate}$, $U'$ and $\alpha_9$. Slightly more informative and less informative priors are considered in sensitivity analysis, with the prior means held fixed, in order to see how the posterior estimates of $e_{ate}$ would change with different informativeness from the priors to be tested. Zero prior means of $e_{ate}$, $U'$ and $\alpha_9$ would be suitable, as we do not have information about their means in reality. Table \ref{tab:results-cvd-sa-prior} shows the results of prior sensitivity analysis. No suspicious multiple posterior solutions occur.

\begin{table}[htbp]
\small
   \begin{subtable}{1\textwidth}
   \centering
   \begin{tabular}{rrrr}
  \hline
Prior & $N(0, 0.01)$ & $N(0, 0.1)$ & $N(0, 1)$   \\ 
  \hline
$S_p$ &0.66&0.00& -\\
$e_{ate}$ & 0.00 (-0.02, 0.01)&0.00 (-0.03, 0.02)&- \\
   \hline
   \end{tabular}
   \caption{Prior sensitivity measures and posterior mean (95\% interval) of $e_{ate}$ under priors of $e_{ate}$ with different informativeness. $S_p$ is the prior sensitivity measure.} 
   \end{subtable}
   
\vspace{1em}

    \begin{subtable}{1\textwidth}
   \centering
   \begin{tabular}{rrrrr}
  \hline
Prior & $N(0, 0.5)$ & $N(0, 1)$ &$N(0, 2)$& $N(0, 3)$  \\ 
  \hline
$e_{ate}$ &- &0.00 (-0.03, 0.02)&-&- \\
$\alpha_9$ &- &0.04 (-1.44, 1.46)&-&-\\ 
   \hline
   \end{tabular}
   \caption{Posterior mean (95\% interval) of $e_{ate}$ and $\alpha_9$ under priors of $U'$ with different informativeness. ``-'' indicates the Bayesian model built in sensitivity analysis is bad. $\alpha_9$ is used as a reference for the strength of unmeasured confounding.} 
   \end{subtable}
 
 \vspace{1em}
 
       \begin{subtable}{1\textwidth}
   \centering
   \begin{tabular}{rrrr}
  \hline
Prior & $N(0, 0.1)$ & $N(0, 0.5)$ & $N(0, 1)$   \\ 
  \hline
$e_{ate}$ & 0.00 (-0.02, 0.02) &0.00 (-0.03, 0.02)& - \\
   \hline
   \end{tabular}
   \caption{Posterior mean (95\% interval) of $e_{ate}$ under priors of $\alpha_9$ with different informativeness.} 
   \end{subtable}
    \caption{Results of prior sensitivity analysis.}
 \label{tab:results-cvd-sa-prior}
\end{table}

From table \ref{tab:results-cvd-sa-prior}, the posterior estimates of $e_{ate}$ are insensitive to informativeness of the priors of $e_{ate}$ and $\alpha_9$ when they are in a well fitted Bayesian model in sensitivity analysis, but the posterior mean of $e_{ate}$ is sensitive to the shifting in the prior of $e_{ate}$ with a prior standard deviation of 0.01, possibly because this small prior standard deviation is too informative. Modelling is sensitive to informativeness of the prior of $e_{ate}$, $U'$ and $\alpha_9$. When the prior of $e_{ate}$ or $\alpha_9$ becomes less informative, or when the prior of $U'$ is strongly informative as $N(0, 0.5)$ or less informative than $N(0, 1)$, a Bayesian model in sensitivity analysis may not be fitted well, where the average treatment effect estimates may be unreliable.

We also build a Bayesian instrumental variable model through two-stage least squares \autocite{angrist_two-stage_1995}, using equation \ref{eqn:results-cvd-sa-iv}. The assignment $Z^{obs}$ is a valid instrument after conditioning on measured confounders. We still use sample standard deviations in the Bayesian instrumental variable model. The prior of $\alpha_8$ is $N(0, 1)$. The priors of $\beta_0$ and $\beta_7$ are $N(0, 0.1)$. Priors for the other parameters are identical to those from the original Bayesian model for the CVD risk outcome correspondingly.
 \begin{eqnarray}
W^{09, obs} &\sim& N(\alpha_0 + \alpha_1~\text{Age$^{std}$} + \alpha_2~\text{Gender} + \alpha_3~\text{Ethnicity}  + \alpha_4~\text{Education level}  \nonumber \\
&& +~ \alpha_5~\text{Smoking status} + \alpha_6~\text{Alcohol frequency} + \alpha_7~\text{Baseline CVD risk score}  \nonumber \\
&& +~ \alpha_8~Z^{obs}, \sigma_{w, 09}) \\
Y^{cvd, obs} &\sim& N(\beta_0 + \beta_1~\text{Age$^{std}$} + \beta_2~\text{Gender} + \beta_3~\text{Ethnicity}  + \beta_4~\text{Education level}  \nonumber \\
&& +~ \beta_5~\text{Smoking status} + \beta_6~\text{Alcohol frequency} + \beta_7~Z^{obs}  \nonumber \\
&& +~ \text{Baseline CVD risk score$^{BS}$}, \sigma_{y, cvd}) \\
e_{ate} &=& \frac{\beta_7}{\alpha_8}
\label{eqn:results-cvd-sa-iv}
\end{eqnarray}

In addition, we build a Bayesian random intercept outcome model. The priors of the coefficients of measured confounders are $N(0, 0.05)$ and the prior of $U'$ is $N(0, 0.5)$. The prior of the standard deviation of the CVD risk outcome is identical to that in the original Bayesian model. Table \ref{tab:results-cvd-sa-iv-ri} shows the posterior estimates of $e_{ate}$ from the instrumental variable model and the random intercept outcome model. The average treatment effect estimates from the instrumental variable model, the random intercept outcome model the original Bayesian model and are quite similar. They all indicate there is no statistically significant effect of the HWC intervention on the 5-year CVD risk score.

\begin{table}[htbp]
\small
   \centering
  \begin{tabular}{rrr}
  \hline
  & Instrumental variable model & Random intercept outcome model \\ 
  \hline
 $e_{ate}$  & 0.00 (0.00, 0.00) & 0.00 (-0.01, 0.01) \\
   \hline
   \end{tabular}
 \caption{Posterior mean (95\% interval) of $e_{ate}$ from the instrumental variable model and the random intercept outcome model.}
 \label{tab:results-cvd-sa-iv-ri}
\end{table}

Then sensitivity of the posterior estimates of $e_{ate}$ is tested against variations in sample standard deviations of the exposure and the CVD risk outcome. Variations in sample standard deviations are represented by different locations of the priors of $\sigma_{w, 09}$ and $\sigma_{y, cvd}$, where sample standard deviations are used as the prior means. Prior locations that are considered include point estimates of sample standard deviations and two boundary points from the 95\% bootstrap intervals of sample standard deviations from table \ref{tab:sd-expoout-boot}, possibly other values if necessary. Table \ref{tab:results-cvd-sa-sd} shows the results of sensitivity analysis of sample standard deviations.  The posterior estimates of $e_{ate}$ are insensitive to variations in sample standard deviations, but modelling is sensitive to variations in sample standard deviations.

\begin{table}[htbp]
\small
    \begin{subtable}{1\textwidth}
   \centering
   \begin{tabular}{rrrrrr}
  \hline
Prior $N(4.25, 0.01)$ & $N(4.83, 0.01)$ & $N(5.3, 0.01)$ & $N(10, 0.01)$\\ 
  \hline
$e_{ate}$& -&0.00 (-0.03, 0.02)&0.00 (-0.02, 0.02)&- \\
   \hline
   \end{tabular}
   \caption{Posterior mean (95\% interval) of $e_{ate}$ under priors of $\sigma_{w, 09}$ with different locations. ``-'' indicates the Bayesian model fitted in sensitivity analysis is bad.} 
   \end{subtable}
   
\vspace{1em}

    \begin{subtable}{1\textwidth}
   \centering
   \begin{tabular}{rrrrr}
  \hline
Prior & $N(0.068, 0.01)$ & $N(0.08, 0.01)$ & $N(0.083, 0.01)$ & $N(1, 0.01)$\\ 
  \hline
$e_{ate}$ &-&0.00 (-0.03, 0.02)&0.00 (-0.03, 0.02)&0.00 (-0.03, 0.03) \\
   \hline
   \end{tabular}
   \caption{Posterior mean (95\% interval) of $e_{ate}$ under priors of $\sigma_{y, cvd}$ with different locations. ``-'' indicates the Bayesian model built in sensitivity analysis is bad.} 
   \end{subtable}
 \caption{Results of sensitivity analysis of sample standard deviations.}
 \label{tab:results-cvd-sa-sd}
\end{table}

In conclusion, there is evidence that the HWC intervention has no effect on cardiovascular health by changing the 5-year CVD risk score. On completion of the HWC intervention, on average the 5-year CVD risk score will not be changed. There is a possibility that there is no statistically significant effect of the HWC intervention on the 5-year CVD risk score because changes in the 5-year CVD risk score are very small and thus the average treatment effect is hard to detect.

\subsection{Results of Patient Health Questionnaire-9 score}

Distributions for the observed compliance behavior, the exposure from the first set of exposure, the PHQ-9 outcome in a Bayesian model are shown in equation \ref{eqn:results-model-formula-phq}. Missing data is not shown but uses the models for the corresponding observed variables. Bayesian models for the exposure from the first set of exposure and the PHQ-9 outcome have already been reparameterized through the random intercept reparameterization. The reparameterized unmeasured confounder is denoted by $U'$. The symbol ``${\rm logit}^{-1}$'' is the logistic function.  The symbol ``Age$^{std}$'' means the standardized age. For each categorical measured confounder, the coefficients of the levels except the reference level are still presented in the format ``coefficient of the confounder, the level name''. $\beta_8$ is the average effect of the exposure $W^{09, obs}$, and we denote it by $e_{ate}$ when presenting results. 
\begin{eqnarray}
G^{obs} &\sim& Bern({\rm logit}^{-1}(\gamma_0 + \gamma_1~\text{Age$^{std}$} + \gamma_2~\text{Gender} + \gamma_3~\text{Ethnicity}  \nonumber \\
&& + ~\gamma_4~\text{Education level} + \gamma_5~\text{Smoking status} + \gamma_6~\text{Alcohol frequency} \nonumber \\
&& +~ \gamma_7~\text{Baseline PHQ-9 score})),  \nonumber \\
W^{09, obs} &\sim& 
    \begin{cases}
      0, & \text{if $Z^{obs}=0$, $G^{obs}=co$} \\
     N(\alpha_0 + \alpha_1~\text{Age$^{std}$} + \alpha_2~\text{Gender} + \alpha_3~\text{Ethnicity} \\
     + ~\alpha_4~\text{Education level} + \alpha_5~\text{Smoking status} \\
     +~ \alpha_6~\text{Alcohol frequency} \\
     +~ \alpha_7~\text{Baseline PHQ-9 score}\\
     +~ \alpha_8~G^{obs} + \alpha_9~U', \sigma_{w, 09}), & \text{otherwise}  \nonumber 
    \end{cases} \\
Y^{phq, obs} &\sim& N(\beta_1~\text{Age$^{std}$} + \beta_2~\text{Gender} + \beta_3~\text{Ethnicity}  + \beta_4~\text{Education level}  \nonumber \\
&& +~ \beta_5~\text{Smoking status} + \beta_6~\text{Alcohol frequency} + \beta_7~\text{Baseline PHQ-9 score}  \nonumber \\
&& +~ \beta_8~W^{09, obs} + U', \sigma_{y, phq}), \nonumber \\
U' &\sim& N(0, 3)
\label{eqn:results-model-formula-phq}
\end{eqnarray}

We use sample standard deviations from table \ref{tab:sd-expoout-boot} as the prior means for standard deviations of the exposure and the PHQ-9 outcome, $\sigma_{w, 09}$ and $\sigma_{y, phq}$, and construct strongly informative priors for $\sigma_{w, 09}$ and $\sigma_{y, phq}$ with a prior standard deviation of 0.01, so that their posterior estimates will become sample standard deviations. 

\begin{figure}[htbp]
     \centering
     \includegraphics[width=1\textwidth]{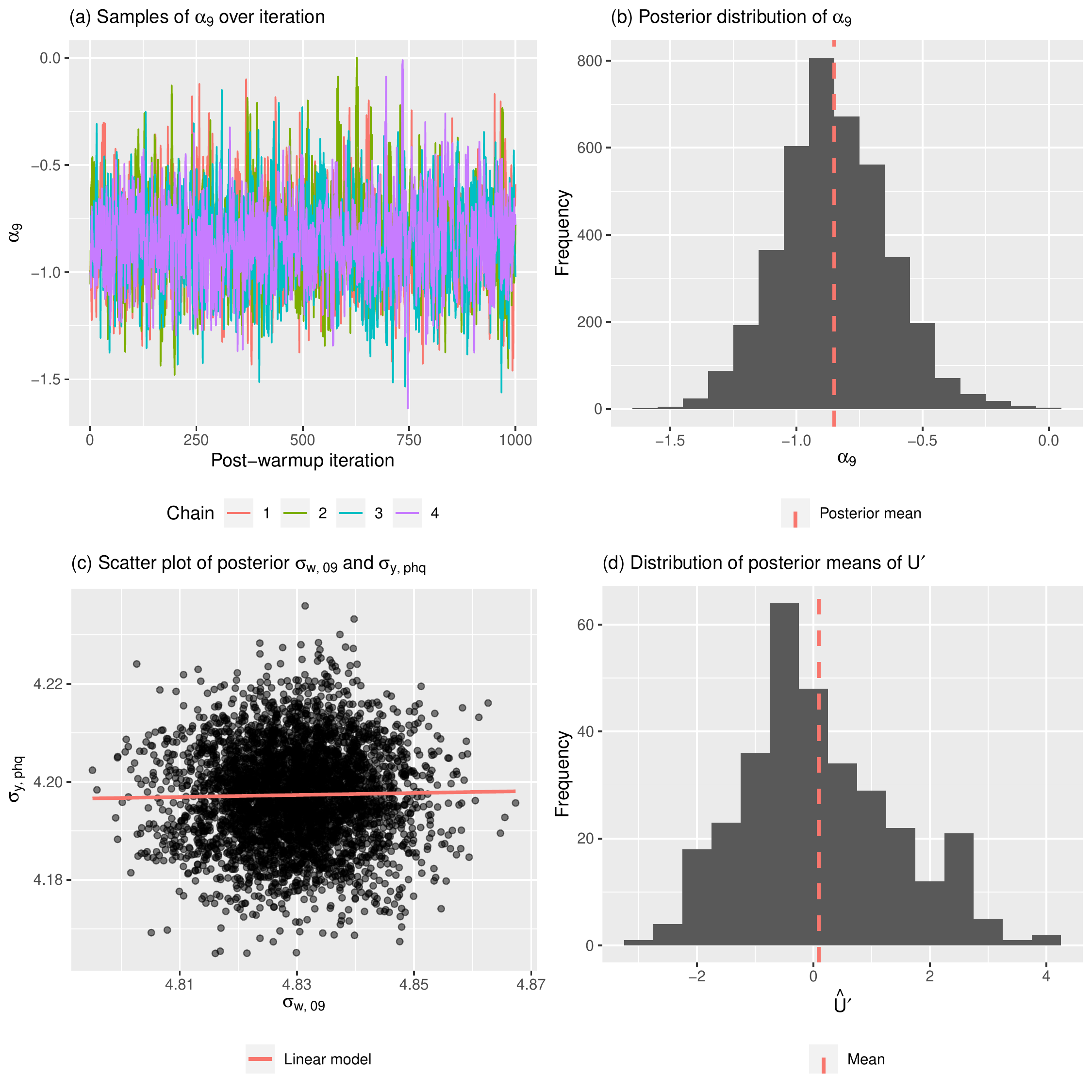}
        \caption{Identifiability diagnostics of the Bayesian model for the PHQ-9 outcome. Figure (a) shows the Monte Carlo samples of $\alpha_9$ over iteration from 4 Markov chains. The samples do not show unusual behaviors and all chains are mixed well. Figure (b) shows the posterior distribution of $\alpha_9$ is proper. Figure (c) shows there is little posterior correlation between $\sigma_{w, 09}$ and $\sigma_{y, phq}$ from a linear model between them. In Figure (d), the estimated $U'$ is denoted by $\hat U'$. We use posterior means of $U'$ as the point estimates of $U'$. Figure (d) shows that $U'$ is not estimated to be zero or constant.}
        \label{fig:res-phq-diag}
\end{figure}

The Bayesian model for the PHQ-9 outcome is fitted well. It is convergent, all $\hat R$ values are below 1.1, all Monte Carlo standard errors are small and all effective sample sizes are high. In addition, this model is identified, where the coefficient of $U'$ and $U'$ itself are estimated well and the standard deviations of the exposure and the outcome have little posterior correlation. Diagnostic figures and comments are shown in figure \ref{fig:res-phq-diag}.

Table \ref{tab:res-phq-post-est} shows posterior estimates of parameters related to $W^{09, obs}$ and $Y^{phq, obs}$ from the Bayesian model for the PHQ-9 outcome and two sensitivity analysis models. Two sensitivity analysis models include a complete-case model and an association model. The complete-case model uses all distributions in equation \ref{eqn:results-model-formula-phq}. The association model uses the distribution for $G^{obs}$ in equation \ref{eqn:results-model-formula-phq} and its distributions for the exposure and the outcome are shown in equation \ref{eqn:results-phq-sa-association}.
 
 \begin{eqnarray}
W^{09, obs} &\sim& 
    \begin{cases}
      0, & \text{if $Z^{obs}=0$, $G^{obs}=co$} \\
     N(\alpha_0 + \alpha_1~\text{Age$^{std}$} + \alpha_2~\text{Gender} + \alpha_3~\text{Ethnicity} \\
     + ~\alpha_4~\text{Education level} + \alpha_5~\text{Smoking status} \\
     +~ \alpha_6~\text{Alcohol frequency} \\
     +~ \alpha_7~\text{Baseline PHQ-9 score}\\
     +~ \alpha_8~G^{obs}, \sigma_{w, 09}), & \text{otherwise}  \nonumber 
    \end{cases} \\
Y^{phq, obs} &\sim& N(\beta_0 + \beta_1~\text{Age$^{std}$} + \beta_2~\text{Gender} + \beta_3~\text{Ethnicity}  + \beta_4~\text{Education level}  \nonumber \\
&& +~ \beta_5~\text{Smoking status} + \beta_6~\text{Alcohol frequency} + \beta_7~\text{Baseline PHQ-9 score}  \nonumber \\
&& +~ \beta_8~W^{09, obs}, \sigma_{y, phq})
\label{eqn:results-phq-sa-association}
\end{eqnarray}

\begin{table}[htbp]
\small
   \centering
\begin{tabular}{lrrrr}
  \hline
\multirowcell{2}[0pt][l]{Parameter} & \multirowcell{2}[0pt][r]{Prior} & \multicolumn{3}{c}{Posterior mean (95\% interval)}\\ 
  \cmidrule(l){3-5} 
& & Bayesian & Complete cases & Association \\ 
  \hline
$\beta_0$ &$N(0, 3)$ &  &  & 1.45 (0.20, 2.66) \\ 
$\beta_1$ &$N(0, 1)$ & -0.07 (-0.68, 0.56) & 0.08 (-0.59, 0.76) & -0.17 (-0.68, 0.34) \\ 
$\beta_2$, Male &$N(0, 1)$ & 1.00 (0.09, 1.92) & 0.60 (-0.42, 1.56) & 0.53 (-0.37, 1.48) \\ 
$\beta_3$, & &&& \\
\quad~ Maori &$N(0, 1)$ & -0.04 (-1.26, 1.12) & 0.05 (-1.29, 1.44) & -0.39 (-1.54, 0.78) \\ 
\quad~ Pasifika &$N(0, 1)$ & 0.00 (-1.20, 1.17) & -0.36 (-1.66, 0.91) & -0.39 (-1.53, 0.73) \\ 
\quad~ Asian &$N(0, 1)$  & -0.34 (-1.62, 0.91) & -0.40 (-1.77, 1.00) & -0.61 (-1.82, 0.61) \\ 
$\beta_4$,  & &&& \\
\quad~ College &$N(0, 1)$  & 0.04 (-1.14, 1.22) & 0.26 (-0.92, 1.50) & -0.24 (-1.34, 0.82) \\ 
\quad~ University &$N(0, 1)$& 0.19 (-0.91, 1.28) & -0.04 (-1.25, 1.17) & -0.20 (-1.21, 0.83) \\ 
$\beta_5$, Non-current &$N(0, 1)$ & 0.55 (-0.74, 1.78) & 0.34 (-1.01, 1.73) & 0.41 (-0.74, 1.54) \\ 
$\beta_6$, More than &$N(0, 1)$ & 0.64 (-0.35, 1.61) & 0.48 (-0.55, 1.53) & 0.40 (-0.50, 1.30) \\ 
$\beta_7$ &$N(0, 1)$ & -0.56 (-0.68, -0.44) & -0.54 (-0.67, -0.40) & -0.60 (-0.71, -0.49) \\ 
$e_{ate}$ &$N(0, 1)$ & 0.08 (-0.01, 0.16) & 0.04 (-0.06, 0.13) & -0.01 (-0.09, 0.06) \\ 
$\sigma_{y, phq}$ &$N(4.2, 0.01)$ & 4.20 (4.18, 4.22) & 4.20 (4.18, 4.22) & 4.20 (4.18, 4.22) \\ 
$\alpha_0$ &$N(10, 1)$ & 9.14 (7.55, 10.67) & 10.44 (8.82, 12.10) & 8.77 (7.32, 10.23) \\ 
$\alpha_1$ &$N(0, 1)$  & 0.10 (-0.67, 0.87) & 0.12 (-0.73, 0.95) & 0.13 (-0.55, 0.81) \\ 
$\alpha_2$, Male &$N(0, 1)$ & -1.51 (-2.71, -0.27) & -0.47 (-1.81, 0.87) & -1.75 (-2.94, -0.55) \\ 
$\alpha_3$, & &&& \\
\quad~ Maori &$N(0, 1)$ & -0.67 (-2.10, 0.79) & -0.31 (-1.94, 1.28) & -0.76 (-2.17, 0.67) \\ 
\quad~ Pasifika &$N(0, 1)$ & -0.86 (-2.28, 0.57) & -0.35 (-1.92, 1.17) & -0.83 (-2.20, 0.60) \\ 
\quad~ Asian &$N(0, 1)$ & -0.59 (-2.04, 0.94) & -0.29 (-1.81, 1.30) & -0.62 (-2.10, 0.76) \\ 
$\alpha_4$,  & &&& \\
\quad~ College &$N(0, 1)$  & 0.89 (-0.54, 2.32) & 0.39 (-1.04, 1.89) & 1.06 (-0.32, 2.53) \\ 
\quad~ University &$N(0, 1)$ & -0.35 (-1.70, 0.99) & 0.63 (-0.86, 2.11) & -0.37 (-1.68, 0.96) \\ 
$\alpha_5$, Non-current &$N(0, 1)$  & -1.23 (-2.73, 0.33) & -0.24 (-1.82, 1.35) & -1.29 (-2.77, 0.18) \\ 
$\alpha_6$, More than &$N(0, 1)$ & -0.16 (-1.46, 1.12) & 0.15 (-1.21, 1.51) & -0.12 (-1.35, 1.10) \\ 
$\alpha_7$ &$N(0, 1)$ & 0.11 (-0.06, 0.28) & 0.08 (-0.14, 0.30) & 0.13 (-0.03, 0.29) \\ 
$\alpha_8$ &$N(0, 1)$ & 2.15 (0.64, 3.67) & 1.78 (0.24, 3.39) & 2.58 (1.17, 3.94) \\ 
$\alpha_9$ &$N(0, 1)$ & -0.85 (-1.26, -0.41) & -0.01 (-0.49, 0.47) &  \\ 
$\sigma_{w, 09}$ &$N(4.83, 0.01)$ & 4.83 (4.81, 4.85) & 4.83 (4.81, 4.85) & 4.83 (4.81, 4.85) \\ 
\hline
\end{tabular}  
      \caption{Posterior estimates from the Bayesian model for the PHQ-9 outcome, complete-case and association models.}
 \label{tab:res-phq-post-est}
\end{table}

From table \ref{tab:res-phq-post-est}, the posterior mean of $e_{ate}$ is 0.08, which means that attending one HWC session can increase the PHQ-9 score by 0.08 on average. The 95\% posterior interval of $e_{ate}$ is (-0.01, 0.16) and covers 0, which indicates that the average effect of one HWC session on the PHQ-9 score is not statistically significant. The posterior estimates of $e_{ate}$ are small, which indicates the average effect of one HWC session on the PHQ-9 score is negligible. The PHQ-9 score ranges from 0 to 27, containing 5 categories: 0-4 as minimal depression, 5-9 as mild depression, 10-14 as moderate depression, 15-19 as moderately severe depression, 20-27 as severe depression. Hence, the PHQ-9 score may not be changed significantly even after many HWC sessions are attended, such as 10 HWC sessions.

The posterior estimates of $e_{ate}$ from the original Bayesian model for the PHQ-9 outcome are slightly larger than those from the complete-case and association model. By comparing the original Bayesian model and the association model, there may be slight unmeasured confounding. By comparing the original Bayesian model and the complete-case model, there tends to be more unmeasured confounding in missing data than in complete cases, because the coefficient of $U'$ is statistically insignificant in the complete-case model but statistically significant in the original Bayesian model. However, three models all indicate that the HWC intervention has no statistically significant effect on the PHQ-9 score and further mental health.

Sensitivity of the posterior estimates of $e_{ate}$ is tested against informativeness of the priors of $e_{ate}$, $U'$ and $\alpha_9$. Slightly more informative and less informative priors are considered in sensitivity analysis, with the prior means held fixed, in order to see how the posterior estimates of $e_{ate}$ would change with different informativeness from the priors to be tested. Zero prior means of $e_{ate}$, $U'$ and $\alpha_9$ would be suitable, as we do not have information about their means in reality. Table \ref{tab:results-phq-sa-prior} shows the results of prior sensitivity analysis. No suspicious multiple posterior solutions occur.

\begin{table}[htbp]
\small
   \begin{subtable}{1\textwidth}
   \centering
   \begin{tabular}{rrrr}
  \hline
Prior & $N(0, 0.5)$ & $N(0, 1)$ & $N(0, 10)$   \\ 
  \hline
$S_p$ &0.01&0.00&0.00  \\
$e_{ate}$ & 0.07 (-0.02, 0.16)&0.08 (-0.01, 0.16)&0.07 (-0.02, 0.16) \\
   \hline
   \end{tabular}
   \caption{Prior sensitivity measures and posterior mean (95\% interval) of $e_{ate}$ under priors of $e_{ate}$ with different informativeness. $S_p$ is the prior sensitivity measure.} 
   \end{subtable}
   
\vspace{1em}

    \begin{subtable}{1\textwidth}
   \centering
   \begin{tabular}{rrrrrr}
  \hline
Prior & $N(0, 1)$ & $N(0, 3)$ & $N(0, 5)$ & $N(0, 10)$ & $N(0, 100)$  \\ 
  \hline
$e_{ate}$ &0.02 (-0.05, 0.10)&0.08 (-0.01, 0.16)&0.14 (0.04, 0.25)&0.30 (0.13, 0.46)&- \\
$\alpha_9$ &-0.97 (-2.39, 0.97)&-0.85 (-1.26, -0.41)&-0.67 (-0.90, -0.43)&-0.49 (-0.63, -0.37)&-\\ 
   \hline
   \end{tabular}
   \caption{Posterior mean (95\% interval) of $e_{ate}$ and $\alpha_9$ under priors of $U'$ with different informativeness. ``-'' indicates the Bayesian model built in sensitivity analysis is bad. $\alpha_9$ is used as a reference for the strength of unmeasured confounding.} 
   \end{subtable}
 
 \vspace{1em}
 
       \begin{subtable}{1\textwidth}
   \centering
   \begin{tabular}{rrrr}
  \hline
Prior & $N(0, 0.5)$ & $N(0, 1)$ & $N(0, 10)$   \\ 
  \hline
$e_{ate}$ &0.07 (-0.02, 0.16) &0.08 (-0.01, 0.16)& 0.08 (-0.01, 0.17)\\
   \hline
   \end{tabular}
   \caption{Posterior mean (95\% interval) of $e_{ate}$ under priors of $\alpha_9$ with different informativeness.} 
   \end{subtable}
    \caption{Results of prior sensitivity analysis.}
 \label{tab:results-phq-sa-prior}
\end{table}

From table \ref{tab:results-phq-sa-prior}, the posterior estimates of $e_{ate}$ are insensitive to informativeness of the priors of $e_{ate}$ and $\alpha_9$, but they are sensitive to informativeness of the prior of $U'$. When the prior of $U'$ becomes less informative to an extent that the Bayesian model built in sensitivity analysis is good, the posterior estimates of $e_{ate}$ and $\alpha_9$ tend to be more positive. We are interested in $N(0, 5)$ or $N(0, 10)$ as the prior of $U'$, which leads to greater and statistically significant average treatment effect estimates. Due to model sensitivity, there is a possibility that the average effect of one HWC session on the PHQ-9 score is statistically positive. Under this possibility, the average increase in the PHQ-9 score by one HWC session may be very small, but there is a chance that the average increase in the PHQ-9 score by many HWC sessions may be moderate.

We also build a Bayesian instrumental variable model through two-stage least squares \autocite{angrist_two-stage_1995}, using equation \ref{eqn:results-phq-sa-iv}. The assignment $Z^{obs}$ is a valid instrument after conditioning on measured confounders. We still use sample standard deviations in the Bayesian instrumental variable model. The prior for $\alpha_0$ is $N(10, 1)$ and the prior for $\beta_0$ is $N(0, 3)$. Priors for the other parameters are $N(0, 1)$.
 \begin{eqnarray}
W^{09, obs} &\sim& N(\alpha_0 + \alpha_1~\text{Age$^{std}$} + \alpha_2~\text{Gender} + \alpha_3~\text{Ethnicity}  + \alpha_4~\text{Education level}  \nonumber \\
&& +~ \alpha_5~\text{Smoking status} + \alpha_6~\text{Alcohol frequency} + \alpha_7~\text{Baseline PHQ-9 score}  \nonumber \\
&& +~ \alpha_8~Z^{obs}, \sigma_{w, 09}) \\
Y^{phq, obs} &\sim& N(\beta_0 + \beta_1~\text{Age$^{std}$} + \beta_2~\text{Gender} + \beta_3~\text{Ethnicity}  + \beta_4~\text{Education level}  \nonumber \\
&& +~ \beta_5~\text{Smoking status} + \beta_6~\text{Alcohol frequency} + \beta_7~\text{Baseline PHQ-9 score}  \nonumber \\
&& +~ \beta_8~Z^{obs}, \sigma_{y, phq}) \\
e_{ate} &=& \frac{\beta_8}{\alpha_8}
\label{eqn:results-phq-sa-iv}
\end{eqnarray}

In addition, we build a Bayesian random intercept outcome model. The priors for unknown parameters including $U'$ in the random intercept outcome model are identical to those from the original Bayesian model for the PHQ-9 outcome correspondingly. Table \ref{tab:results-phq-sa-iv-ri} shows the posterior estimates of $e_{ate}$ from the instrumental variable model and the random intercept outcome model.
\begin{table}[htbp]
\small
   \centering
  \begin{tabular}{rrr}
  \hline
  & Instrumental variable model & Random intercept outcome model \\ 
  \hline
 $e_{ate}$ &0.04 (-0.03, 0.11) & 0.01 (-0.08, 0.10) \\
   \hline
   \end{tabular}
 \caption{Posterior mean (95\% interval) of $e_{ate}$ from the instrumental variable model and the random intercept outcome model.}
 \label{tab:results-phq-sa-iv-ri}
\end{table}

The average treatment effect estimates from the instrumental variable model and the random intercept outcome model are quite similar. They are slightly smaller than those from the original Bayesian model, but they are more statistically insignificant, which makes the original Bayesian model more convincing. Hence, it may be safe to conclude that there is no strong statistical evidence that the HWC intervention can increase the PHQ-9 score. However, due to model sensitivity, in reality the HWC intervention may increase the PHQ-9 score.

Then sensitivity of the posterior estimates of $e_{ate}$ is tested against variations in sample standard deviations of the exposure and the PHQ-9 outcome. Variations in sample standard deviations are represented by different locations of the priors of $\sigma_{w, 09}$ and $\sigma_{y, phq}$, where sample standard deviations are used as the prior means. Prior locations that are considered include point estimates of sample standard deviations and two boundary points from the 95\% bootstrap intervals of sample standard deviations from table \ref{tab:sd-expoout-boot}, possibly other values if necessary. Table \ref{tab:results-phq-sa-sd} shows the results of sensitivity analysis of sample standard deviations. 

\begin{table}[htbp]
\small
    \begin{subtable}{1\textwidth}
   \centering
   \begin{tabular}{rrrrrrr}
  \hline
Prior &$N(0.5, 0.01)$& $N(4.25, 0.01)$ & $N(4.83, 0.01)$ & $N(5.3, 0.01)$ & $N(10, 0.01)$\\ 
  \hline
$e_{ate}$&-& 0.10 (0.02, 0.19) &0.08 (-0.01, 0.16)&0.06 (-0.03, 0.15)& 0.01 (-0.07, 0.11)\\
   \hline
   \end{tabular}
   \caption{Posterior mean (95\% interval) of $e_{ate}$ under priors of $\sigma_{w, 09}$ with different locations.  ``-'' indicates the Bayesian model built in sensitivity analysis is bad.} 
   \end{subtable}
   
\vspace{1em}

    \begin{subtable}{1\textwidth}
   \centering
   \begin{tabular}{rrrrrr}
  \hline
Prior & $N(0.5, 0.01)$ & $N(3.84, 0.01)$ & $N(4.2, 0.01)$ & $N(4.76, 0.01)$ & $N(10, 0.01)$\\ 
  \hline
$e_{ate}$& -&0.08 (-0.01, 0.17)&0.08 (-0.01, 0.16)&0.08 (-0.02, 0.17)& 0.04 (-0.13, 0.22)\\
   \hline
   \end{tabular}
   \caption{Posterior mean (95\% interval) of $e_{ate}$ under priors of $\sigma_{y, phq}$ with different locations. ``-'' indicates the Bayesian model built in sensitivity analysis is bad.} 
   \end{subtable}
 \caption{Results of sensitivity analysis of sample standard deviations.}
 \label{tab:results-phq-sa-sd}
\end{table}

The posterior estimates of $e_{ate}$ are insensitive to variations in the sample standard deviation of the PHQ-9 outcome within a suitable range of this sample standard deviation, but they are more sensitive to variations in the sample standard deviation of the exposure. When the prior location of the sample standard deviation of the exposure decreases within the 95\% bootstrap interval of the sample standard deviation of the exposure, the posterior estimates of $e_{ate}$ tend to slightly increase and possibly become statistically significantly positive. This also indicates that there is a possibility that the average effect of the HWC intervention on the PHQ-9 score is statistically significant but may be very small. 

In conclusion, there is no clear evidence that the HWC intervention has a beneficial or harmful effect on mental health, when the PHQ-9 score is used as an indicator for mental health. On completion of the HWC intervention, on average the PHQ-9 score will be improved by about 1.2 (-0.15, 2.4) but more evidence supports the conclusion that the change in the PHQ-9 score is not statistically significant. However, due to model sensitivity, in reality there may be a statistically significant increase in the PHQ-9 score, which may be very small or moderate depending on how large the average effect of one HWC session would become and how many HWC sessions would be attended. Hence, due to model sensitivity, there is a risk that the HWC intervention may harm mental health.

\subsection{Comparison of unmeasured confounding among four outcomes}

\begin{figure}[htbp]
     \centering
     \includegraphics[width=1\textwidth]{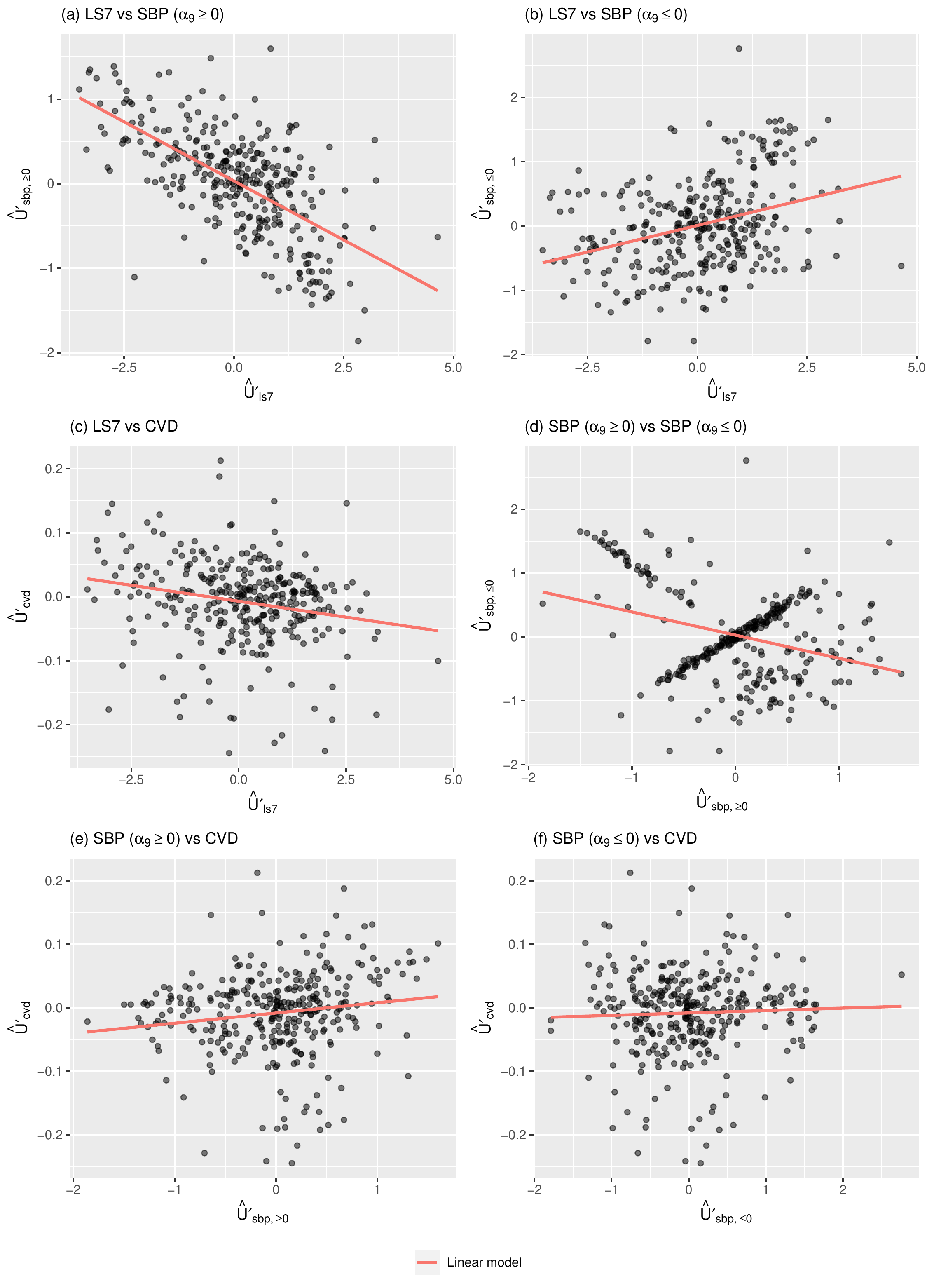}
        \caption{Scatter plots of posterior means of $U'$ between each two from three outcomes that measure cardiovascular health. Three outcomes are the LS7 outcome, the SBP outcome and the CVD risk outcome. The SBP outcome has two restricted Bayesian models, and $\geq 0$ is used to indicate the model with $\alpha_9 \geq 0$ while $\leq 0$ is used to indicate the model with $\alpha_9 \leq 0$. }
        \label{fig:results-comparison-cardiovascular}
\end{figure}

\begin{figure}[htbp]
     \centering
     \includegraphics[width=1\textwidth]{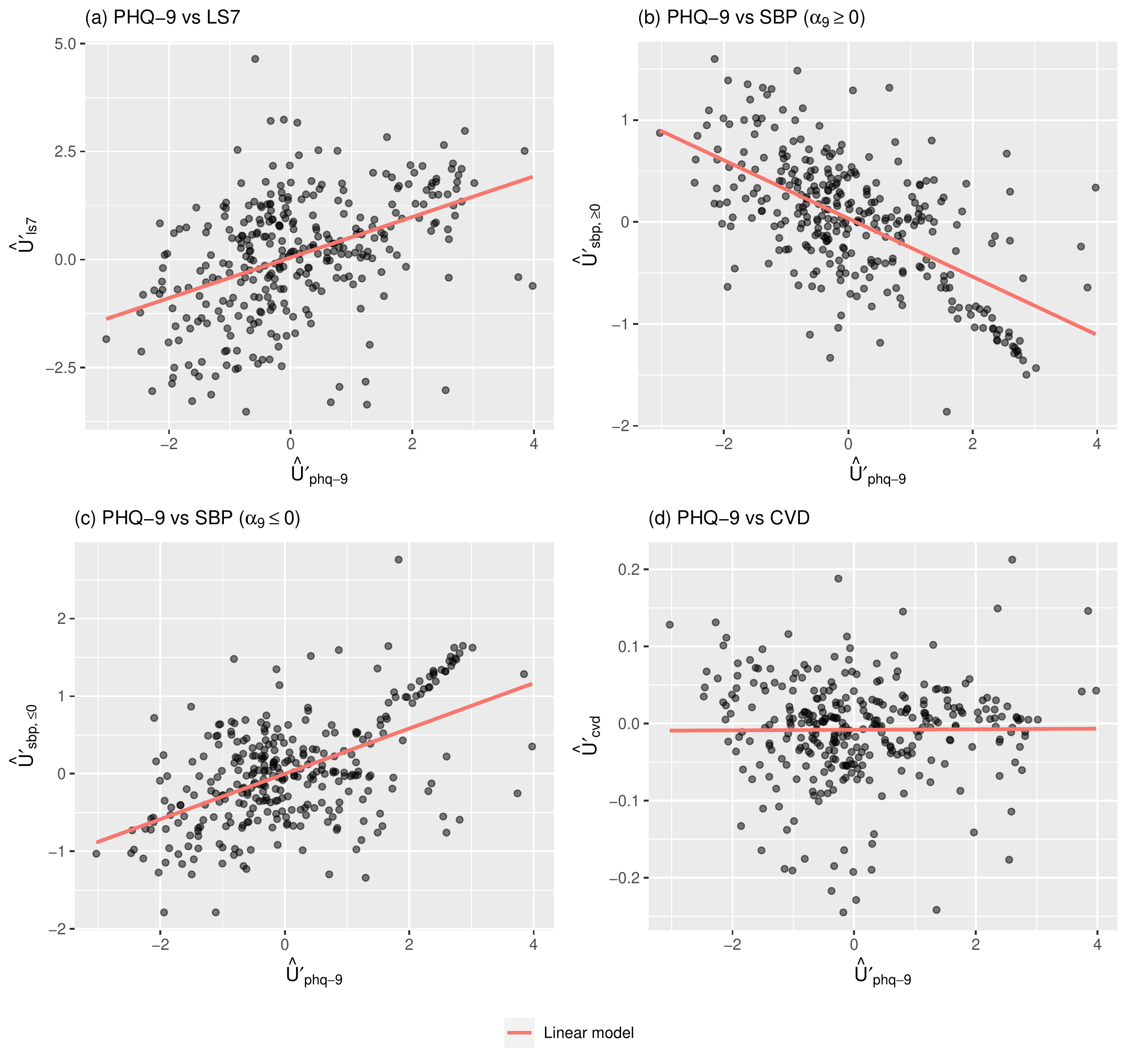}
        \caption{Scatter plots of posterior means of $U'$ between the PHQ-9 outcome and each from three outcomes that measure cardiovascular health. Three outcomes that measure cardiovascular health are the LS7 outcome, the SBP outcome and the CVD risk outcome. The SBP outcome has two restricted Bayesian models, and $\geq 0$ is used to indicate the model with $\alpha_9 \geq 0$ while $\leq 0$ is used to indicate the model with $\alpha_9 \leq 0$.}
        \label{fig:results-comparison-mental}
\end{figure}

Different outcomes usually have different confounders, given the same exposure. In the above analysis, four outcomes just happen to have measured confounders in common. In fact they can have different confounders, and their differences in confounders should be shown by unmeasured confounders used in Bayesian models when the same measured confounders are used. If the reparameterized unmeasured confounders from two outcomes are correlated, it is likely that the two outcomes have unmeasured confounders in common. Hence, we compare the estimated reparameterized unmeasured confounder among three outcomes that measure cardiovascular health, including the LS7 outcome, the SBP outcome and the CVD risk outcome, in order to understand how unmeasured confounding would be among the outcomes that measure the same thing. We also compare the estimated reparameterized unmeasured confounder between the PHQ-9 outcome that measures mental health and these three outcomes that measure cardiovascular health, in order to understand how unmeasured confounding would be among the outcomes that measure different things. Figure \ref{fig:results-comparison-cardiovascular} shows comparison of unmeasured confounding among three outcomes that measure cardiovascular health. Figure \ref{fig:results-comparison-mental} shows comparison of unmeasured confounding between the PHQ-9 outcome that measures mental health and three outcomes that measure cardiovascular health.

In figure \ref{fig:results-comparison-cardiovascular}, we find that the LS7 outcome and the SBP outcome seem to have similar unmeasured confounding. Two pairs, (1) the LS7 outcome and the CVD risk outcome, (2) the SBP outcome in the restricted Bayesian model with $\alpha_9 \geq 0$ and the CVD risk outcome, have slight similarity in unmeasured confounding. Two pairs, (1) the SBP outcome in the restricted Bayesian model with $\alpha_9 \geq 0$ and the SBP outcome in the restricted Bayesian model with $\alpha_9 \leq 0$, (2) the SBP outcome in the restricted Bayesian model with $\alpha_9 \leq 0$ and the CVD risk outcome, seem to have no similarity in unmeasured confounding. There is some evidence that the outcomes that measure the same thing can have  unmeasured confounders in common, but this cannot be concluded definitely because the SBP outcome has two Bayesian models and we cannot tell which model adjusts for unmeasured confounding better, and also because the CVD risk outcome is so small that unmeasured confounding may not be estimated completely. 

In figure \ref{fig:results-comparison-mental}, we find that the PHQ-9 outcome seems to have similar unmeasured confounding as the LS7 outcome and the SBP outcome, while it may have different unmeasured confounding from the CVD risk outcome. There is some evidence that the outcomes that measure different things may have similar or different unmeasured confounders. 

In conclusion, when we build Bayesian models for multiple outcomes together, it may be hard to determine the associations among these outcomes with regard to unmeasured confounding. Assumptions about unmeasured confounding in multiple outcomes should be carefully made.

\section{Analysis of three exposures and Life's Simple Seven total score\quad}
\label{sec:analysis-3}

Three exposures come from the second set of exposure, and we build Bayesian models for three exposures and the LS7 outcome, in order to understand how one HWC session would affect the LS7 outcome on average during three three-month periods over 9 months post randomization. 

Since three exposures are calculated from the HWC session attendance dates and the assessment dates, the initial way we impute missing exposures in a Bayesian model is to first construct separate distributions for the HWC session attendance dates and the assessment dates, impute all missing dates, and calculate the values for missing exposures from imputed dates. The HWC attendance intervals and the assessment intervals are used to construct distributions in order to impute missing HWC attendance dates and assessment dates. An HWC attendance interval is the duration between two consecutive HWC attendance dates, or the duration from the randomization date to the first HWC attendance date. An assessment interval is the duration between two consecutive assessment dates, or the duration from the randomization date to the first assessment date. Hence, there are 15 distributions for 15 HWC attendance intervals and 4 distributions for 4 assessment intervals. So many interval distributions are involved that a Bayesian model can easily have modelling issues such as high $\hat R$ values or low effective samples sizes. In addition, there are only 4 missing HWC attendance dates, but to impute them, we have to use 15 interval distributions, which results in model inefficiency and unnecessary flexibility that can produce multiple posterior solutions. Using the interval distributions may be able to protect observed HWC attendances alongside with missing attendances, but considering model validity, efficiency and identifiability, we choose to instead impute missing exposures by three exposures distributions themselves and exclude interval distributions from all Bayesian models.

We assume the compliance behavior is constant, as shown in assumption 2.1.8. The constant compliance behavior is defined by the first set of exposure rather than the second set of exposure. For the same person, three exposures in the second set of exposure may suggest different compliance behaviors in different periods, which makes it difficult to define an overall compliance behavior. Based on assumption 2.1.8, we do not consider the variability of the compliance behavior over different periods. On the other hand, due to no missingness in the first set of exposure, the compliance behavior in the treatment arm is fully observed.

We first assume three exposures and the LS7 outcome have unmeasured confounders in common. Distributions for the observed compliance behavior, three exposures, the LS7 outcome in a Bayesian model are shown in equation \ref{eqn:results-model-formula-ls7-3expo}. Missing data is not shown but uses the models for the corresponding observed variables. Bayesian models for three exposures and the LS7 outcome have already been reparameterized through the random intercept reparameterization. The reparameterized unmeasured confounder is denoted by $U'$. The symbol ``${\rm logit}^{-1}$'' is the logistic function. The symbol ``Age$^{std}$'' means the standardized age. Reference levels of categorical measured confounders are shown in table \ref{tab:descrip-stats-new}. For each categorical measured confounder, the coefficients of the levels except the reference level are presented in the format ``coefficient of the confounder, the level name''. $\beta_8$ is the average effect of the exposure $W^{03, obs}$, $\beta_9$ is the average effect of the exposure $W^{36, obs}$, and $\beta_{10}$ is the average effect of the exposure $W^{69, obs}$. We denote $\beta_8$, $\beta_9$, $\beta_{10}$ by $e_{ate, 03}$, $e_{ate, 36}$, $e_{ate, 69}$ when presenting results. 
\begin{eqnarray}
G^{obs} &\sim& Bern({\rm logit}^{-1}(\gamma_0 + \gamma_1~\text{Age$^{std}$} + \gamma_2~\text{Gender} + \gamma_3~\text{Ethnicity}  \nonumber \\
&& + ~\gamma_4~\text{Education level} + \gamma_5~\text{Smoking status} + \gamma_6~\text{Alcohol frequency} \nonumber \\
&& +~ \gamma_7~\text{Baseline LS7 total score})),  \nonumber \\
W^{03, obs} &\sim& 
    \begin{cases}
      0, & \text{if $Z^{obs}=0$, $G^{obs}=co$} \\
     N(\alpha_0^{03} + \alpha_1^{03}~\text{Age$^{std}$} + \alpha_2^{03}~\text{Gender} + \alpha_3^{03}~\text{Ethnicity} \\
     + ~\alpha_4^{03}~\text{Education level} + \alpha_5^{03}~\text{Smoking status} \\
     +~ \alpha_6^{03}~\text{Alcohol frequency} \\
     +~ \alpha_7^{03}~\text{Baseline LS7 total score}\\
     +~ \alpha_8^{03}~G^{obs} + \alpha_9^{03}~U', \sigma_{w, 03}), & \text{otherwise}  \nonumber 
    \end{cases} \\
 W^{36, obs} &\sim& 
    \begin{cases}
      0, & \text{if $Z^{obs}=0$, $G^{obs}=co$} \\
     N(\alpha_0^{36} + \alpha_1^{36}~\text{Age$^{std}$} + \alpha_2^{36}~\text{Gender} + \alpha_3^{36}~\text{Ethnicity} \\
     + ~\alpha_4^{36}~\text{Education level} + \alpha_5^{36}~\text{Smoking status} \\
     +~ \alpha_6^{36}~\text{Alcohol frequency} \\
     +~ \alpha_7^{36}~\text{Baseline LS7 total score}\\
     +~ \alpha_8^{36}~G^{obs} + \alpha_9^{36}~U', \sigma_{w, 36}), & \text{otherwise}  \nonumber 
    \end{cases} \\
W^{69, obs} &\sim& 
    \begin{cases}
      0, & \text{if $Z^{obs}=0$, $G^{obs}=co$} \\
     N(\alpha_0^{69} + \alpha_1^{69}~\text{Age$^{std}$} + \alpha_2^{69}~\text{Gender} + \alpha_3^{69}~\text{Ethnicity} \\
     + ~\alpha_4^{69}~\text{Education level} + \alpha_5^{69}~\text{Smoking status} \\
     +~ \alpha_6^{69}~\text{Alcohol frequency} \\
     +~ \alpha_7^{69}~\text{Baseline LS7 total score}\\
     +~ \alpha_8^{69}~G^{obs} + \alpha_9^{69}~U', \sigma_{w, 69}), & \text{otherwise}  \nonumber 
    \end{cases} \\
Y^{ls7, obs} &\sim& N(\beta_1~\text{Age$^{std}$} + \beta_2~\text{Gender} + \beta_3~\text{Ethnicity}  + \beta_4~\text{Education level}  \nonumber \\
&& +~ \beta_5~\text{Smoking status} + \beta_6~\text{Alcohol frequency} + \beta_7~\text{Baseline LS7 total score}  \nonumber \\
&& +~ \beta_8~W^{03, obs} + \beta_9~W^{36, obs} + \beta_{10}~W^{69, obs} + U', \sigma_{y, ls7}), \nonumber \\
U' &\sim& N(0, 3)
\label{eqn:results-model-formula-ls7-3expo}
\end{eqnarray}

The observed baseline LS7 total score with no missing values and the observed baseline LS7 total score with missing values use distributions from equation \ref{eqn:results-model-formula-ls7bl-missing}. We use sample standard deviations from table \ref{tab:sd-expoout-boot} as the prior means for standard deviations of the exposures and the LS7 outcome, $\sigma_{w, 03}$, $\sigma_{w, 36}$, $\sigma_{w, 69}$ and $\sigma_{y, ls7}$, and construct strongly informative priors for $\sigma_{w, 03}$, $\sigma_{w, 36}$, $\sigma_{w, 69}$ and $\sigma_{y, ls7}$ with a prior standard deviation of 0.01. 

\begin{figure}[htbp]
     \centering
     \includegraphics[width=1\textwidth]{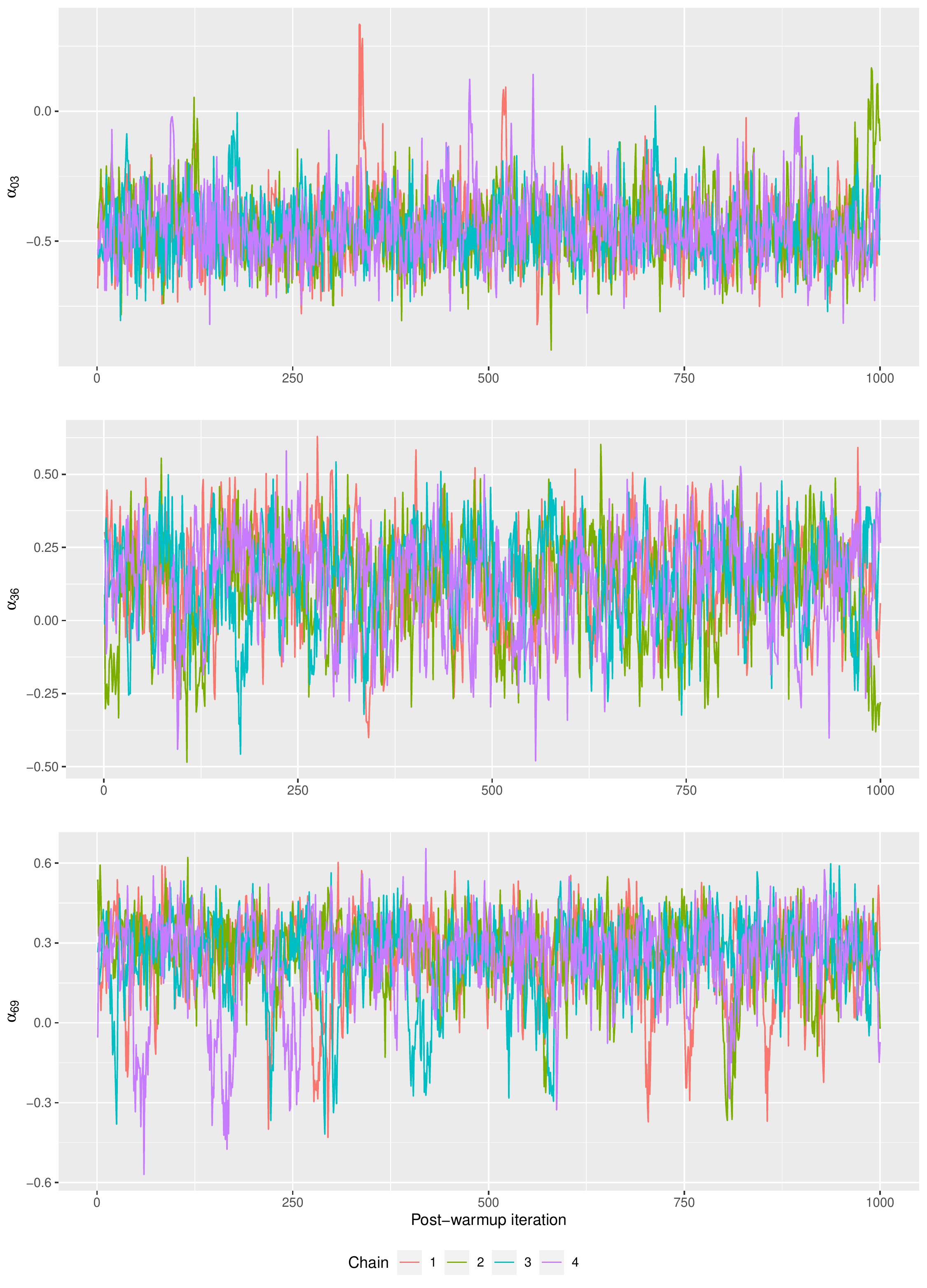}
        \caption{Samples for three coefficients of $U'$ in the exposure distributions over iteration. Three coefficients are $\alpha_9^{03}$, $\alpha_9^{36}$ and $\alpha_9^{69}$.}
        \label{fig:res-ls7-3w-ini}
\end{figure}

The initial Bayesian model for the LS7 outcome is not fitted well, because the coefficients of $U'$ in the exposure distributions including $\alpha_9^{03}$, $\alpha_9^{36}$ and $\alpha_9^{69}$ are not estimated well. Figure \ref{fig:res-ls7-3w-ini} show the samples for three coefficients of $U'$ in the exposure distributions over iteration. The samples of $\alpha_9^{03}$ shows several unusual transitions into the positive area that last for some iterations, which may indicate a potential bimodal posterior distribution. The samples of $\alpha_9^{69}$ seems to have a bimodal posterior distribution in this Bayesian model.

It is not helpful to make inference that we try both positive and negative signs of $\alpha_9^{03}$ and $\alpha_9^{69}$ and build separate Bayesian models, which leads to at least 4 posterior solutions. On the other hand, it may be because our assumption that three exposures have unmeasured confounders in common is wrong that the initial Bayesian model is not fitted well. Hence, we build a Bayesian model that adjusts for unmeasured confounding in the LS7 outcome and one exposure at a time. There will be three Bayesian models: (1) adjust for unmeasured confounding in $W^{03, obs}$ and $Y^{ls7, obs}$, exclude $U'$ from the distributions for $W^{36, obs}$ and $W^{69, obs}$ in equation \ref{eqn:results-model-formula-ls7-3expo}, (2) adjust for unmeasured confounding in $W^{36, obs}$ and $Y^{ls7, obs}$, exclude $U'$ from the distributions for $W^{03, obs}$ and $W^{69, obs}$ in equation \ref{eqn:results-model-formula-ls7-3expo}, (3) adjust for unmeasured confounding in $W^{69, obs}$ and $Y^{ls7, obs}$, exclude $U'$ from the distributions for $W^{03, obs}$ and $W^{36, obs}$ in equation \ref{eqn:results-model-formula-ls7-3expo}. Parameters remaining in each model still use symbols in equation \ref{eqn:results-model-formula-ls7-3expo}. Model (1) corresponds to an assumption that there are unmeasured confounders between $W^{03, obs}$ and $Y^{ls7, obs}$ and unmeasured confounding between the other two exposures and $Y^{ls7, obs}$, if any, is not considered. Model (2) corresponds to an assumption that there are unmeasured confounders between $W^{36, obs}$ and $Y^{ls7, obs}$ and unmeasured confounding between the other two exposures and $Y^{ls7, obs}$, if any, is not considered. Model (3) corresponds to an assumption that there are unmeasured confounders between $W^{69, obs}$ and $Y^{ls7, obs}$ and unmeasured confounding between the other two exposures and $Y^{ls7, obs}$, if any, is not considered.

\begin{figure}[htbp]
     \centering
     \includegraphics[width=1\textwidth]{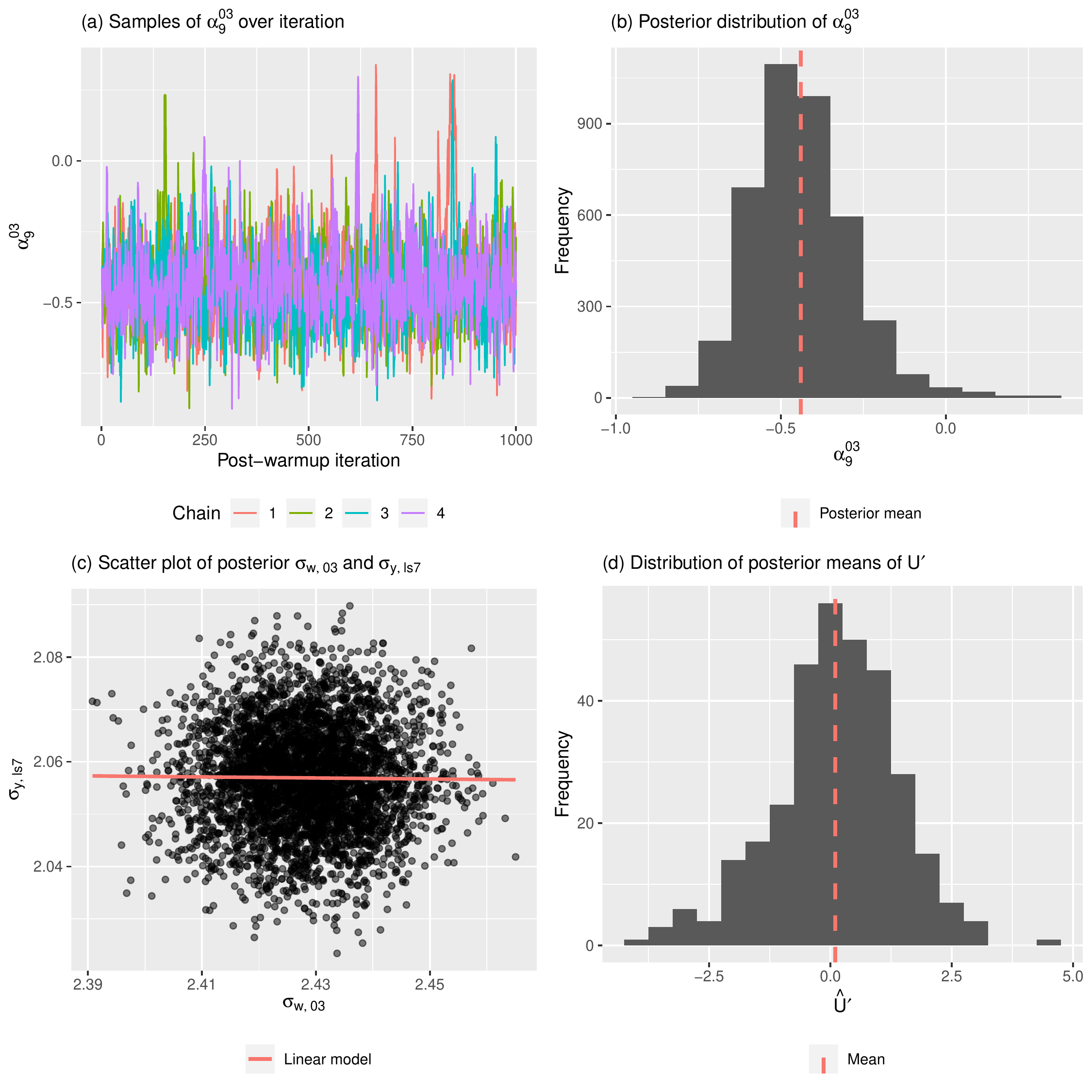}
        \caption{Identifiability diagnostics of Model (1). Figure (a) shows the Monte Carlo samples of $\alpha_9^{03}$ over iteration from 4 Markov chains. Monte Carlo transitions above zero may indicate a potential bimodal posterior distribution. Figure (b) shows the posterior distribution of $\alpha_9^{03}$ is proper. Figure (c) shows there is little posterior correlation between $\sigma_{w, 03}$ and $\sigma_{y, ls7}$ from a linear model between them. In Figure (d), the estimated $U'$ is denoted by $\hat U'$. We use posterior means of $U'$ as the point estimates of $U'$. Figure (d) shows that $U'$ is not estimated to be zero or constant.}
        \label{fig:res-ls7-diag-3w-u03}
\end{figure}

\begin{figure}[htbp]
     \centering
     \includegraphics[width=1\textwidth]{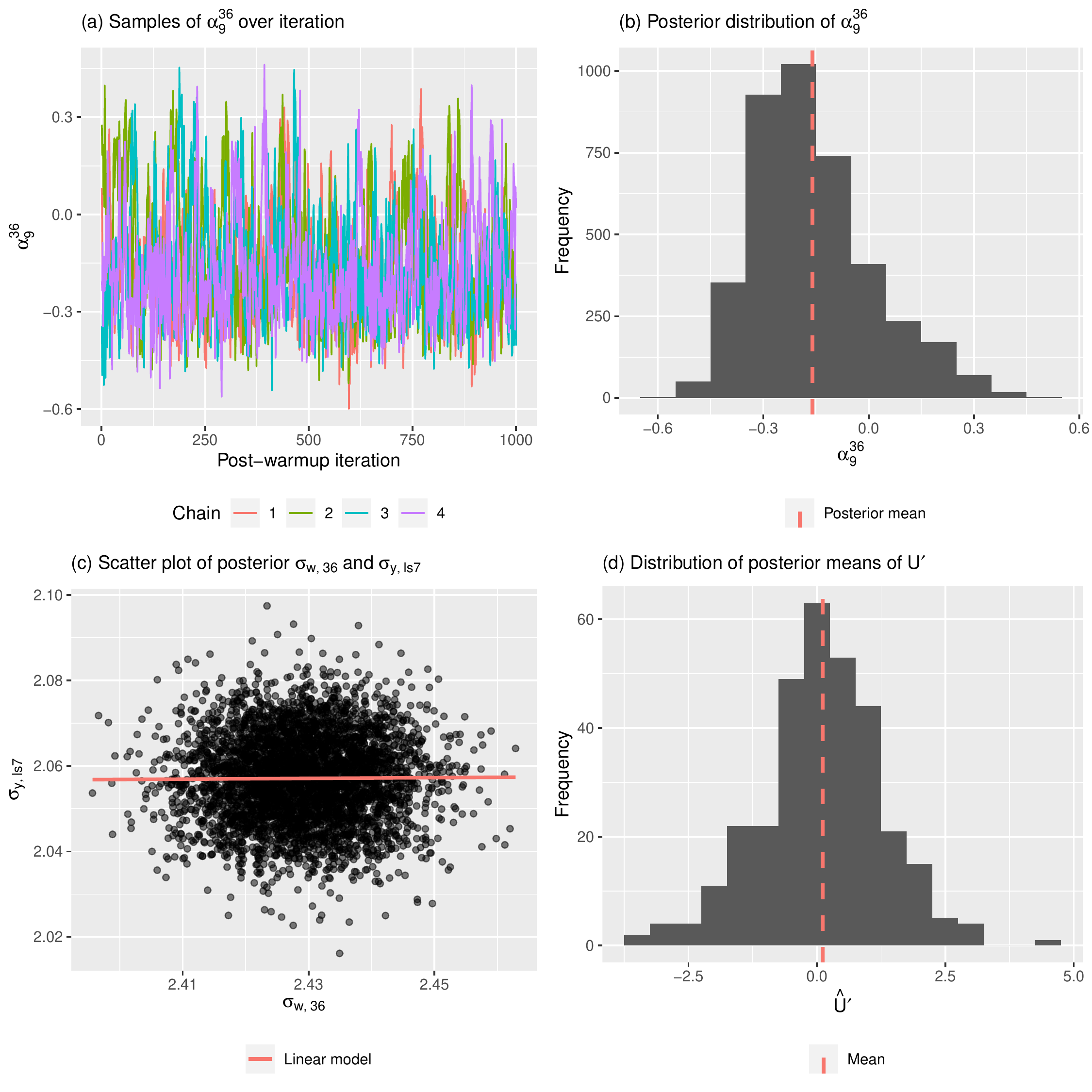}
        \caption{Identifiability diagnostics of Model (2). Figure (a) shows the Monte Carlo samples of $\alpha_9^{36}$ over iteration from 4 Markov chains. The samples do not show unusual behaviors. Figure (b) shows the posterior distribution of $\alpha_9^{36}$ is proper. Figure (c) shows there is little posterior correlation between $\sigma_{w, 36}$ and $\sigma_{y, ls7}$ from a linear model between them. In Figure (d), the estimated $U'$ is denoted by $\hat U'$. We use posterior means of $U'$ as the point estimates of $U'$. Figure (d) shows that $U'$ is not estimated to be zero or constant.}
        \label{fig:res-ls7-diag-3w-u36}
\end{figure}

\begin{figure}[htbp]
     \centering
     \includegraphics[width=1\textwidth]{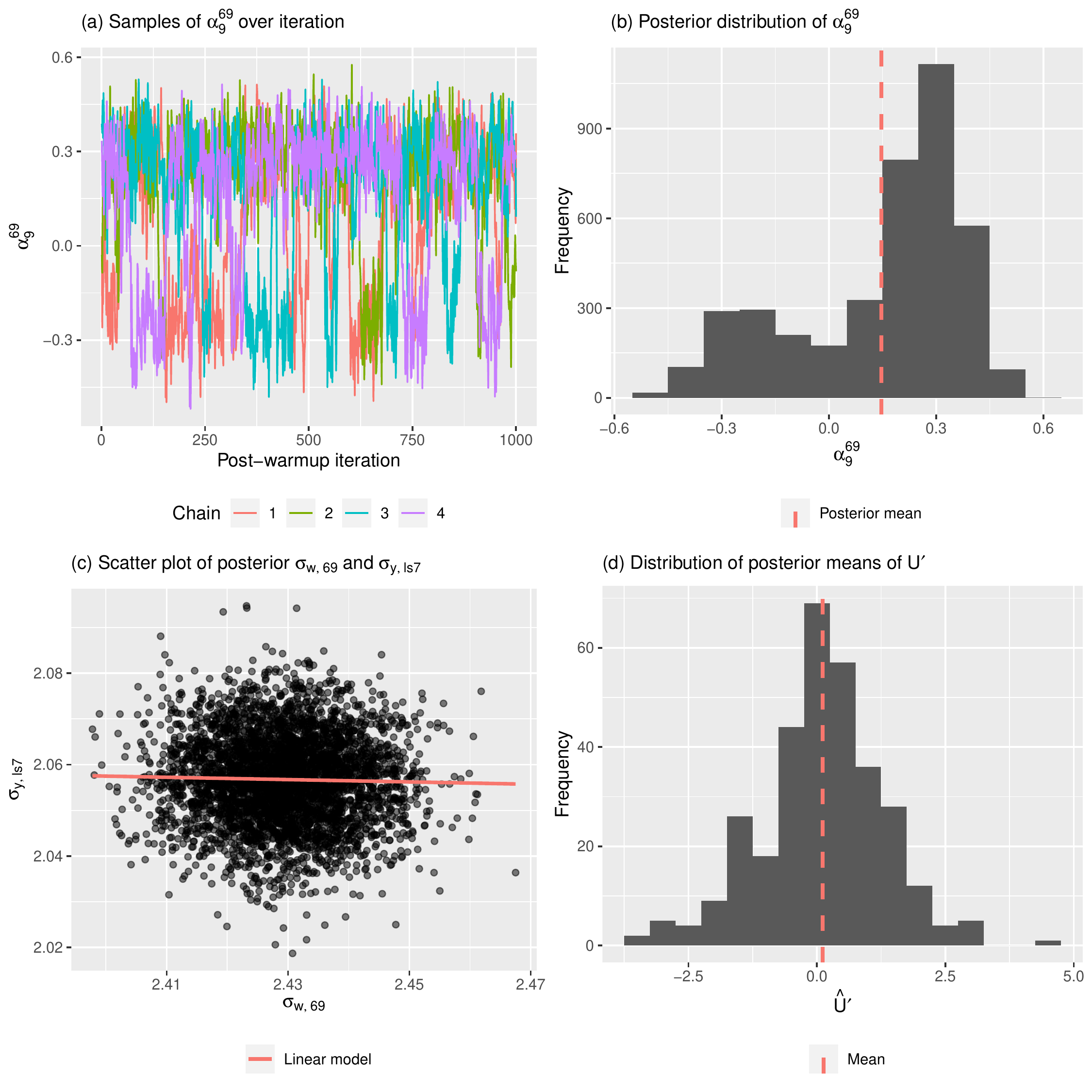}
        \caption{Identifiability diagnostics of Model (3). Figure (a) shows the Monte Carlo samples of $\alpha_9^{69}$ over iteration from 4 Markov chains. Figure (b) shows the posterior distribution of $\alpha_9^{69}$ is bimodal. Figure (c) shows there is little posterior correlation between $\sigma_{w, 69}$ and $\sigma_{y, ls7}$ from a linear model between them. In Figure (d), the estimated $U'$ is denoted by $\hat U'$. We use posterior means of $U'$ as the point estimates of $U'$. Figure (d) shows that $U'$ is not estimated to be zero or constant.}
        \label{fig:res-ls7-diag-3w-u69}
\end{figure}

Model (1) and Model (2) are fitted well, while Model (3) is not fitted well because $\alpha_9^{69}$ seems to have a bimodal posterior distribution. Diagnostic figures and comments for Model (1) are shown in figure \ref{fig:res-ls7-diag-3w-u03}. Diagnostic figures and comments for Model (2) are shown in figure \ref{fig:res-ls7-diag-3w-u36}. Diagnostic figures and comments for Model (3) are shown in figure \ref{fig:res-ls7-diag-3w-u69}. 

Then two restricted Bayesian models from Model (3) are well fitted, one from which restricts $\alpha_9^{69}$ to be non-negative and the other from which restricts $\alpha_9^{69}$ to be non-positive. Table \ref{tab:res-ls7-post-est-3w-4models} shows the posterior estimates of parameters related to $Y^{ls7, obs}$, three coefficients of $U'$ in the exposure distributions and the standard deviations of three exposures, from Model (1), Model (2) and two restricted Bayesian models from Model (3). The coefficient of $U'$ is denoted by $\alpha_9$ without any superscript, because each of the four Bayesian models only involves one coefficient of $U'$ and thus there will be no ambiguity.

\begin{table}
\small
\centering
 \begin{adjustwidth}{-2cm}{}
\begin{tabular}{lrrrrr}
  \hline
\multirowcell{2}[0pt][l]{Parameter} & \multirowcell{2}[0pt][r]{Prior} & \multicolumn{4}{c}{Posterior mean (95\% interval)}\\ 
  \cmidrule(l){3-6} 
 && \multirowcell{2}[0pt][r]{Model (1)} & \multirowcell{2}[0pt][r]{Model (2)} & Model (3): &Model (3): \\
& &  &  &$\alpha_9^{69} \geq 0$ &$\alpha_9^{69} \leq 0$ \\ 
  \hline
$\beta_1$ &$N(0, 1)$ & 0.18 (-0.28, 0.63) & 0.23 (-0.22, 0.70) & 0.23 (-0.21, 0.66) & 0.26 (-0.21, 0.72) \\ 
$\beta_2$, Male &$N(0, 1)$ & 0.24 (-0.59, 1.10) & 0.32 (-0.47, 1.14) & 0.27 (-0.56, 1.08) & 0.33 (-0.51, 1.15) \\ 
$\beta_3$, & &&& \\
\quad~ Maori &$N(0, 1)$  & 0.29 (-0.73, 1.30) & 0.34 (-0.64, 1.35) & 0.35 (-0.66, 1.32) & 0.33 (-0.71, 1.35) \\ 
\quad~ Pasifika &$N(0, 1)$  & 0.32 (-0.66, 1.31) & 0.33 (-0.65, 1.34) & 0.30 (-0.64, 1.27) & 0.44 (-0.56, 1.45) \\ 
\quad~ Asian &$N(0, 1)$ & 0.35 (-0.71, 1.41) & 0.46 (-0.66, 1.54) & 0.46 (-0.60, 1.54) & 0.42 (-0.66, 1.51) \\ 
$\beta_4$,  & &&& \\
\quad~ College &$N(0, 1)$& 0.54 (-0.41, 1.51) & 0.64 (-0.31, 1.59) & 0.65 (-0.28, 1.60) & 0.60 (-0.35, 1.56) \\ 
\quad~ University &$N(0, 1)$& 0.35 (-0.59, 1.26) & 0.45 (-0.51, 1.39) & 0.37 (-0.53, 1.31) & 0.42 (-0.51, 1.34) \\ 
$\beta_5$, Non-current &$N(0, 1)$ & -0.04 (-1.04, 0.97) & 0.08 (-0.98, 1.09) & 0.05 (-0.97, 1.06) & 0.05 (-0.99, 1.08) \\ 
$\beta_6$, More than &$N(0, 1)$ & 0.27 (-0.49, 1.04) & 0.29 (-0.52, 1.06) & 0.35 (-0.43, 1.08) & 0.30 (-0.48, 1.09) \\ 
$\beta_7$ &$N(0, 1)$ & -0.20 (-0.32, -0.08) & -0.21 (-0.33, -0.09) & -0.18 (-0.30, -0.07) & -0.21 (-0.33, -0.09) \\ 
$e_{ate, 03}$ &$N(0, 1)$ & 0.40 (0.12, 0.64) & -0.02 (-0.32, 0.33) & 0.24 (-0.02, 0.50) & -0.02 (-0.29, 0.25) \\ 
$e_{ate, 36}$ &$N(0, 1)$ & -0.18 (-0.53, 0.19) & 0.28 (-0.45, 0.80) & 0.23 (-0.15, 0.61) & -0.15 (-0.56, 0.27) \\ 
$e_{ate, 69}$ &$N(0, 1)$ & -0.10 (-0.48, 0.28) & 0.00 (-0.41, 0.41) & -0.53 (-0.99, 0.04) & 0.54 (-0.06, 1.04) \\ 
$\sigma_{y, ls7}$ &$N(2.06, 0.01)$ & 2.06 (2.04, 2.08) & 2.06 (2.04, 2.08) & 2.06 (2.04, 2.08) & 2.06 (2.04, 2.08) \\ 
$\alpha_9$ &$N(0, 1)$& -0.44 (-0.71, -0.11) & -0.16 (-0.43, 0.24) & 0.27 (0.05, 0.45) & -0.22 (-0.43, -0.01) \\ 
$\sigma_{w, 03}$ &$N(2.43, 0.01)$  & 2.43 (2.41, 2.45) & 2.43 (2.41, 2.45) & 2.43 (2.41, 2.45) & 2.43 (2.41, 2.45) \\ 
 $\sigma_{w, 36}$ &$N(1.8, 0.01)$ & 1.80 (1.78, 1.82) & 1.80 (1.78, 1.82) & 1.80 (1.78, 1.82) & 1.80 (1.78, 1.82) \\ 
 $\sigma_{w, 69}$ &$N(1.64, 0.01)$  & 1.64 (1.62, 1.66) & 1.64 (1.62, 1.66) & 1.64 (1.62, 1.66) & 1.64 (1.62, 1.66) \\
\hline
\end{tabular}  
 \end{adjustwidth}
 \caption{Posterior estimates from Model (1), Model (2) and two restricted Bayesian models from Model (3).}
\label{tab:res-ls7-post-est-3w-4models}
\end{table}

\begin{figure}[htbp]
     \centering
     \includegraphics[width=0.5\textwidth]{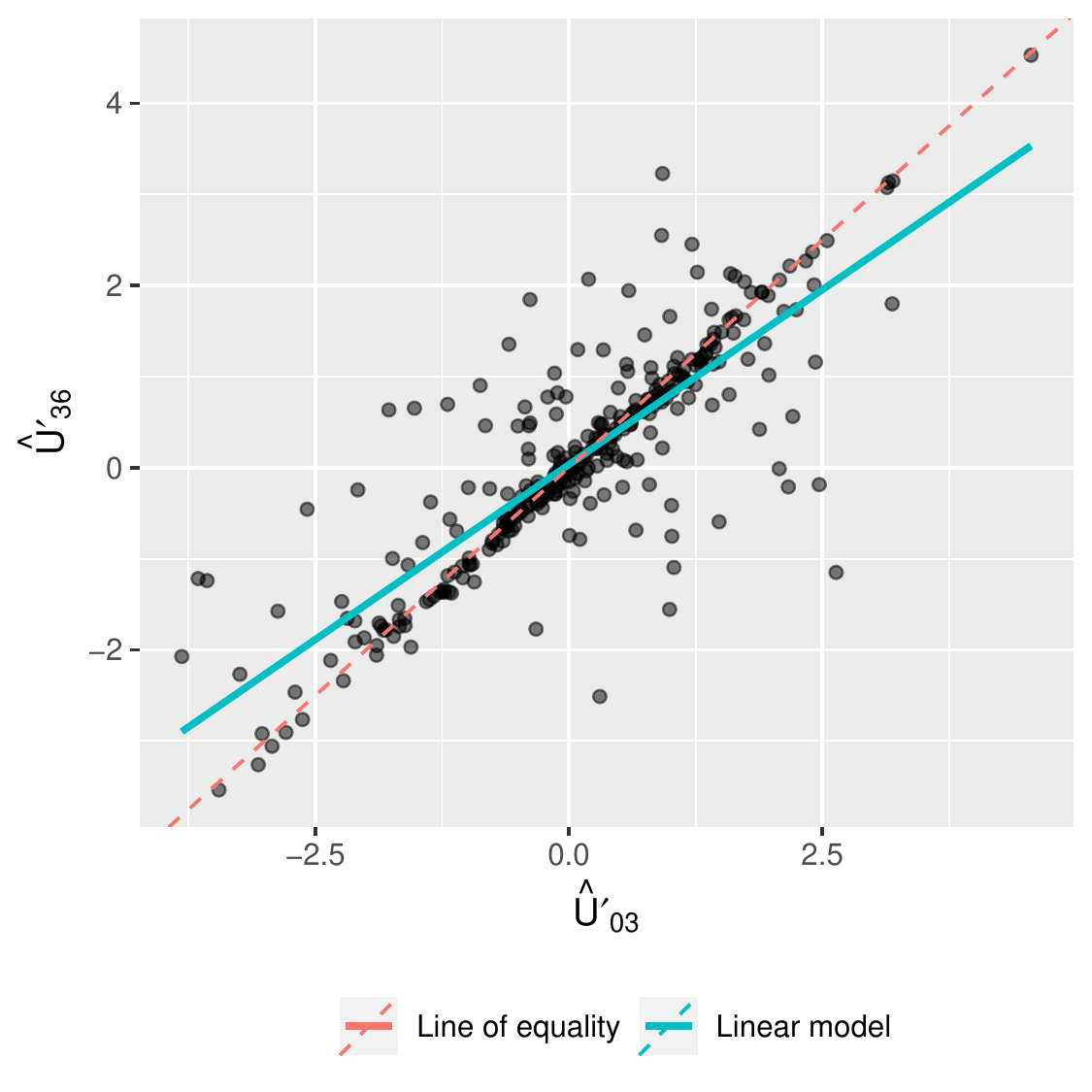}
        \caption{Scatter plot of posterior means of $U'$ from Model (1) and Model (2). $\hat U'_{03}$ is posterior means of $U'$ from Model (1). $\hat U'_{36}$ is posterior means of $U'$ from Model (2).}
        \label{fig:res-ls7-3w-diag-u2w}
\end{figure}

We find that the posterior means of $U'$ from Model (1) and Model (2) are correlated, as shown in figure \ref{fig:res-ls7-3w-diag-u2w}, which may indicate $W^{03, obs}$ and $W^{36, obs}$ have unmeasured confounders in common. Hence, we also build another Bayesian model that adjusts for unmeasured confounding in $W^{03, obs}$ and $W^{36, obs}$, where $U'$ is excluded from the distribution $W^{69, obs}$ in equation \ref{eqn:results-model-formula-ls7-3expo}.

A Bayesian association model and a Bayesian random intercept outcome model are also built. The association model uses the distribution for $G^{obs}$ in equation \ref{eqn:results-model-formula-ls7-3expo} and its distributions for three exposures and the LS7 outcome are shown in equation \ref{eqn:results-ls7-3w-sa-association}.
 
 \begin{eqnarray}
W^{03, obs} &\sim& 
    \begin{cases}
      0, & \text{if $Z^{obs}=0$, $G^{obs}=co$} \\
     N(\alpha_0^{03} + \alpha_1^{03}~\text{Age$^{std}$} + \alpha_2^{03}~\text{Gender} + \alpha_3^{03}~\text{Ethnicity} \\
     + ~\alpha_4^{03}~\text{Education level} + \alpha_5^{03}~\text{Smoking status} \\
     +~ \alpha_6^{03}~\text{Alcohol frequency} \\
     +~ \alpha_7^{03}~\text{Baseline LS7 total score}\\
     +~ \alpha_8^{03}~G^{obs}, \sigma_{w, 03}), & \text{otherwise}  \nonumber 
    \end{cases} \\
 W^{36, obs} &\sim& 
    \begin{cases}
      0, & \text{if $Z^{obs}=0$, $G^{obs}=co$} \\
     N(\alpha_0^{36} + \alpha_1^{36}~\text{Age$^{std}$} + \alpha_2^{36}~\text{Gender} + \alpha_3^{36}~\text{Ethnicity} \\
     + ~\alpha_4^{36}~\text{Education level} + \alpha_5^{36}~\text{Smoking status} \\
     +~ \alpha_6^{36}~\text{Alcohol frequency} \\
     +~ \alpha_7^{36}~\text{Baseline LS7 total score}\\
     +~ \alpha_8^{36}~G^{obs}, \sigma_{w, 36}), & \text{otherwise}  \nonumber 
    \end{cases} \\
W^{69, obs} &\sim& 
    \begin{cases}
      0, & \text{if $Z^{obs}=0$, $G^{obs}=co$} \\
     N(\alpha_0^{69} + \alpha_1^{69}~\text{Age$^{std}$} + \alpha_2^{69}~\text{Gender} + \alpha_3^{69}~\text{Ethnicity} \\
     + ~\alpha_4^{69}~\text{Education level} + \alpha_5^{69}~\text{Smoking status} \\
     +~ \alpha_6^{69}~\text{Alcohol frequency} \\
     +~ \alpha_7^{69}~\text{Baseline LS7 total score}\\
     +~ \alpha_8^{69}~G^{obs}, \sigma_{w, 69}), & \text{otherwise}  \nonumber 
    \end{cases} \\
Y^{ls7, obs} &\sim& N(\beta_0+\beta_1~\text{Age$^{std}$} + \beta_2~\text{Gender} + \beta_3~\text{Ethnicity}  + \beta_4~\text{Education level}  \nonumber \\
&& +~ \beta_5~\text{Smoking status} + \beta_6~\text{Alcohol frequency} + \beta_7~\text{Baseline LS7 total score}  \nonumber \\
&& +~ \beta_8~W^{03, obs} + \beta_9~W^{36, obs} + \beta_{10}~W^{69, obs}, \sigma_{y, ls7})
\label{eqn:results-ls7-3w-sa-association}
\end{eqnarray}

The priors for unknown parameters including $U'$ in the random intercept outcome model are identical to those from  Model (1), Model (2) and two restricted Bayesian models from Model (3) correspondingly. Table \ref{tab:results-ls7-3w-sa-ri} shows posterior estimates from the Bayesian model that adjusts for unmeasured confounding in $W^{03, obs}$ and $W^{36, obs}$, the association model and the random intercept outcome model.
\begin{table}[htbp]
\small
   \centering
  \begin{tabular}{lrrrr}
  \hline
 \multirowcell{2}[0pt][l]{Parameter} & \multirowcell{2}[0pt][r]{Prior} & \multicolumn{3}{c}{Posterior mean (95\% interval)}\\ 
  \cmidrule(l){3-5} 
 &&Bayesian model&Association model & Random intercept outcome model \\
  \hline
$\beta_0$ & $N(0, 3)$ &  & 3.56 (2.47, 4.73) &  \\ 
$\beta_1$ &$N(0, 1)$ & 0.27 (-0.20, 0.73) & 0.37 (0.09, 0.63) & 0.24 (-0.23, 0.71) \\ 
 $\beta_2$, Male &$N(0, 1)$  & 0.37 (-0.46, 1.21) & 0.03 (-0.49, 0.54) & 0.29 (-0.53, 1.14) \\ 
 $\beta_3$, & &&& \\
\quad~ Maori &$N(0, 1)$  & 0.37 (-0.63, 1.39) & -0.01 (-0.70, 0.67) & 0.33 (-0.67, 1.33) \\ 
 \quad~ Pasifika &$N(0, 1)$ & 0.36 (-0.66, 1.38) & 0.07 (-0.58, 0.71) & 0.36 (-0.64, 1.36) \\ 
\quad~ Asian &$N(0, 1)$ & 0.57 (-0.49, 1.61) & 0.31 (-0.43, 1.05) & 0.43 (-0.63, 1.50) \\ 
$\beta_4$,  & &&& \\
\quad~ College &$N(0, 1)$ & 0.69 (-0.27, 1.66) & 0.62 (0.00, 1.23) & 0.64 (-0.33, 1.57) \\ 
\quad~ University &$N(0, 1)$ & 0.50 (-0.45, 1.47) & 0.44 (-0.14, 1.08) & 0.40 (-0.54, 1.31) \\ 
$\beta_5$, Non-current &$N(0, 1)$& 0.11 (-0.90, 1.14) & -0.78 (-1.50, -0.06) & 0.06 (-0.95, 1.06) \\ 
$\beta_6$, More than &$N(0, 1)$& 0.29 (-0.51, 1.06) & -0.04 (-0.53, 0.47) & 0.32 (-0.44, 1.10) \\ 
$\beta_7$ &$N(0, 1)$ & -0.21 (-0.33, -0.09) & -0.54 (-0.67, -0.41) & -0.20 (-0.32, -0.08) \\ 
$e_{ate, 03}$ &$N(0, 1)$ & -0.25 (-0.56, 0.11) & 0.01 (-0.14, 0.16) & 0.10 (-0.14, 0.35) \\ 
$e_{ate, 36}$ &$N(0, 1)$ & 0.53 (-0.01, 0.94) & -0.03 (-0.27, 0.21) & 0.02 (-0.36, 0.40) \\ 
$e_{ate, 69}$ &$N(0, 1)$ & 0.06 (-0.34, 0.45) & 0.12 (-0.13, 0.37) & 0.04 (-0.37, 0.45) \\ 
$\sigma_{y, ls7}$ &$N(2.06, 0.01)$ & 2.06 (2.04, 2.08) & 2.06 (2.04, 2.08) & 2.06 (2.04, 2.08) \\ 
$\alpha_{9, 03}$ &$N(0, 1)$ & 0.21 (0.01, 0.41) &   &   \\ 
$\alpha_{9, 36}$ &$N(0, 1)$ & 0.00 (-1.98, 1.97) &  & \\ 
$\sigma_{w, 03}$ &$N(2.43, 0.01)$ & 2.43 (2.41, 2.45) & 2.43 (2.41, 2.45) & 2.43 (2.41, 2.45) \\ 
 $\sigma_{w, 36}$ &$N(1.8, 0.01)$  & 1.80 (1.78, 1.82) & 1.80 (1.78, 1.82) & 1.80 (1.78, 1.82) \\ 
 $\sigma_{w, 69}$ &$N(1.64, 0.01)$  & 1.64 (1.62, 1.66) & 1.64 (1.62, 1.66) & 1.64 (1.62, 1.66) \\ 
   \hline
   \end{tabular}
 \caption{Posterior estimates from the Bayesian model that adjusts for unmeasured confounding in $W^{03, obs}$ and $W^{36, obs}$, the association model and the random intercept outcome model.}
 \label{tab:results-ls7-3w-sa-ri}
\end{table}

We first focus on the posterior estimates of $e_{ate, 03}$. From Model (1) in table \ref{tab:res-ls7-post-est-3w-4models}, the posterior mean of $e_{ate, 03}$ is 0.40 and the 95\% posterior interval of $e_{ate, 03}$ is  (0.12, 0.64), which indicates a statistically significant positive average effect of one HWC session on the LS7 total score during 3 months post randomization. From Model (2) and Model (3) in table \ref{tab:res-ls7-post-est-3w-4models}, the average effect of one HWC session on the LS7 total score during 3 months post randomization is either statistically insignificantly positive or statistically insignificantly negative. From the Bayesian model that adjusts for unmeasured confounding in $W^{03, obs}$ and $W^{36, obs}$ in table \ref{tab:results-ls7-3w-sa-ri}, the average effect of one HWC session on the LS7 total score during 3 months post randomization is negative and statistically insignificant. From the association model and the random intercept outcome model in table \ref{tab:results-ls7-3w-sa-ri}, the average effect of one HWC session on the LS7 total score during 3 months post randomization is positive and statistically insignificant.

From figure \ref{fig:res-ls7-diag-3w-u03}, unusual Monte Carlo transitions above zero during sampling of $\alpha_9^{03}$ may indicate a potential bimodal posterior distribution. And the negative or insignificant posterior estimates of $e_{ate, 03}$ from the models except Model (1) are likely to be associated with the other posterior solution. We then build a Bayesian model that has the same structure as Model (1) but restricts $\alpha_9^{03}$ to be non-negative. The posterior estimates of $e_{ate, 03}$ from this new model are -0.25 (-0.56, 0.11), which align with the average treatment effect estimates from most models except Model (1). 

The conclusion that the average effect of one HWC session on the LS7 total score during 3 months post randomization is statistically significant positive, and another conclusion that the average effect of one HWC session on the LS7 total score during 3 months post randomization is statistically insignificant and may be negative, seem contradictory. We think the posterior estimates of $e_{ate, 03}$ are very sensitive to the ways for adjusting for unmeasured confounding, and in this case, adjusting for unmeasured confounding between $W^{03, obs}$ and $Y^{ls7, obs}$ fails to identify correct estimates of $e_{ate, 03}$. It seems appropriate to conclude that there is no clear evidence about how the average effect of one HWC session on the LS7 total score during 3 months post randomization is. If we tend to trust the association model and the random intercept outcome model, which produces unique estimates for each parameter, then there is evidence that the average effect of one HWC session on the LS7 total score during 3 months post randomization is positive, small and not statistically significant.

Then we focus on the posterior estimates of $e_{ate, 36}$. From two tables \ref{tab:res-ls7-post-est-3w-4models} and \ref{tab:results-ls7-3w-sa-ri}, the posterior means of $e_{ate, 36}$ are either positive or negative, but the 95\% posterior intervals of $e_{ate, 36}$ all indicate the average effect of one HWC session on the LS7 total score between 3 and 6 months post randomization is not statistically significant.  The posterior estimates of $e_{ate, 36}$ are also sensitive to the ways for adjusting for unmeasured confounding. We cannot conclude whether the average effect of one HWC session on the LS7 total score between 3 and 6 months post randomization is positive or not, but there may be evidence that the average effect of one HWC session on the LS7 total score between 3 and 6 months post randomization is not statistically significant. 

Next we focus on the posterior estimates of $e_{ate, 69}$. From Model (1) and Model (2) in table \ref{tab:res-ls7-post-est-3w-4models}, and from table \ref{tab:results-ls7-3w-sa-ri}, the posterior means of $e_{ate, 69}$ are either positive or negative, but the 95\% posterior intervals of $e_{ate, 69}$ all indicate the average effect of one HWC session on the LS7 total score between 6 and 9 months post randomization is not statistically significant. From Model (3) table \ref{tab:res-ls7-post-est-3w-4models}, one posterior solution indicates the average effect of one HWC session on the LS7 total score between 6 and 9 months post randomization is positive while the other posterior solution indicates the average effect of one HWC session on the LS7 total score between 6 and 9 months post randomization is negative, though the average effect of one HWC session on the LS7 total score between 6 and 9 months post randomization from either posterior solution is slightly statistically insignificant. These two posterior solutions also seem a little contradictory. We cannot conclude whether the average effect of one HWC session on the LS7 total score between 6 and 9 months post randomization is positive or not, but there may be evidence that the average effect of one HWC session on the LS7 total score between 6 and 9 months post randomization is not statistically significant. The posterior estimates of $e_{ate, 69}$ are also sensitive to the ways for adjusting for unmeasured confounding, and in this case, adjusting for unmeasured confounding between $W^{69, obs}$ and $Y^{ls7, obs}$ possibly fails to identify correct estimates of $e_{ate, 69}$. If we tend to trust the association model and the random intercept outcome model, then there is evidence that the average effect of one HWC session on the LS7 total score between 6 and 9 months post randomization is positive, small and not statistically significant.

To draw more clear inference, we need more reliable models. Model (1) and Model (3) will not be used further, because they seem to fail to identify correct estimates of three average treatment effects in three periods. The Bayesian model that adjusts for unmeasured confounding in $W^{03, obs}$ and $W^{36, obs}$ will not be used further, because it is prone to two contradictory posterior solutions related to $W^{03, obs}$. Model (2), the association model and the random intercept outcome model remain. Model (2) and the random intercept outcome model do successfully adjust for unmeasured confounding. Hence, they are likely to produce less biased average treatment effect estimates than the association model. However, Model (2) seems to favor greater estimation of $W^{36, obs}$ and unmeasured confounding seems insignificant, while the random intercept outcome model suffers a lack of thorough research.

We are interested in Model (2) and the random intercept outcome model, as they are still useful for inference. In analysis with one exposure from the first set of exposure, the random intercept outcome model is not the primary inferential model and only assists the primary inferential model that adjusts for unmeasured confounding in both the exposure and outcome distributions in making more clear inference. In the current analysis with three exposures from the second set of exposure, we try using the random intercept outcome model as an inferential model. Sensitivity of the posterior estimates of $e_{ate, 03}$, $e_{ate, 36}$, $e_{ate, 69}$ is tested against informativeness of the priors of $e_{ate, 03}$, $e_{ate, 36}$, $e_{ate, 69}$, $U'$ and $\alpha_9^{36}$ in the two models. The prior of $\alpha_9^{36}$ is only considered in Model (2), because the random intercept outcome model does not have $\alpha_9^{36}$. Slightly more informative and less informative priors are considered in sensitivity analysis, with the prior means fixed at zero and the prior standard deviations varying. Table \ref{tab:results-ls7-3w-sa-prior-model2} shows the results of prior sensitivity analysis for Model (2). Table \ref{tab:results-ls7-3w-sa-prior-ri} shows the results of prior sensitivity analysis for the random intercept outcome model.

\begin{table}[htbp]
\small
   \begin{subtable}{1\textwidth}
   \centering
   \begin{tabular}{rrrrrrr}
  \hline
 \multirowcell{2}[0pt][r]{Prior} & \multicolumn{2}{c}{$e_{ate, 03}$} & \multicolumn{2}{c}{$e_{ate, 36}$} & \multicolumn{2}{c}{$e_{ate, 69}$}  \\ 
  \cmidrule(l){2-3} \cmidrule(l){4-5} \cmidrule(l){6-7} 
&$S_p$ &Posterior&$S_p$&Posterior&$S_p$& Posterior \\ 
  \hline
$N(0, 0.5)$ & 0.11&-0.02 (-0.31, 0.33)&0.35&0.25 (-0.42, 0.74)&0.16&-0.01 (-0.40, 0.40) \\
$N(0, 1)$ &0.03&-0.02 (-0.32, 0.33)&0.10&0.28 (-0.45, 0.80)&0.04&0.00 (-0.41, 0.41) \\
$N(0, 10)$ &0.00&-0.02 (-0.31, 0.33)&0.00&0.31 (-0.49, 0.83)&0.00&-0.01 (-0.43, 0.39) \\
   \hline
   \end{tabular}
   \caption{Prior sensitivity measures and posterior mean (95\% interval) of $e_{ate, 03}$, $e_{ate, 36}$, $e_{ate, 69}$ under their own priors with different informativeness. $S_p$ is the prior sensitivity measure.``Posterior'' is the posterior mean (95\% interval). For each average treatment effect, $S_p$ and posterior estimates are found by only changing its own prior.} 
   \end{subtable}
   
\vspace{1em}

    \begin{subtable}{1\textwidth}
   \centering
   \begin{tabular}{rrrrr}
  \hline
Prior & $N(0, 1)$ & $N(0, 3)$ & $N(0, 5)$ & $N(0, 10)$   \\ 
  \hline
$e_{ate, 03}$ &0.08 (-0.09, 0.25)&-0.02 (-0.32, 0.33)&-0.37 (-0.70, -0.05)&-\\
$e_{ate, 36}$ &0.04 (-0.23, 0.31)&0.28 (-0.45, 0.80)&1.03 (0.55, 1.48)&-\\
$e_{ate, 69}$ &0.05 (-0.22, 0.33)&0.00 (-0.41, 0.41)&-0.10 (-0.58, 0.37)&- \\
$\alpha_{9, 36}$ & -0.11 (-0.63, 0.44)&-0.16 (-0.43, 0.24)&-0.31 (-0.44, -0.18)&-\\ 
   \hline
   \end{tabular}
   \caption{Posterior mean (95\% interval) of $e_{ate, 03}$, $e_{ate, 36}$, $e_{ate, 69}$ and $\alpha_{9, 36}$ under priors of $U'$ with different informativeness. ``-'' indicates the Bayesian model built in sensitivity analysis is bad. $\alpha_{9, 36}$ is used as a reference for the strength of unmeasured confounding.} 
   \end{subtable}
 
 \vspace{1em}
 
         \begin{subtable}{1\textwidth}
   \centering
   \begin{tabular}{rrrr}
  \hline
Prior & $N(0, 0.5)$ & $N(0, 1)$ & $N(0, 5)$   \\ 
  \hline
$e_{ate, 03}$ &-0.03 (-0.33, 0.33) &-0.02 (-0.32, 0.33)& -0.03 (-0.33, 0.33)\\
$e_{ate, 36}$ &0.29 (-0.44, 0.80) &0.28 (-0.45, 0.80)&0.31 (-0.46, 0.81) \\
$e_{ate, 69}$ &0.00 (-0.41, 0.44) &0.00 (-0.41, 0.41)& -0.01 (-0.39, 0.38)\\
   \hline
   \end{tabular}
   \caption{Posterior mean (95\% interval) of $e_{ate, 03}$, $e_{ate, 36}$, $e_{ate, 69}$ under priors of $\alpha_{9, 36}$ with different informativeness.} 
   \end{subtable}
 \caption{Results of prior sensitivity analysis for Model (2).}
 \label{tab:results-ls7-3w-sa-prior-model2}
\end{table}

\begin{table}[htbp]
\small
   \begin{subtable}{1\textwidth}
   \centering
   \begin{tabular}{rrrrrrr}
  \hline
 \multirowcell{2}[0pt][r]{Prior} & \multicolumn{2}{c}{$e_{ate, 03}$} & \multicolumn{2}{c}{$e_{ate, 36}$} & \multicolumn{2}{c}{$e_{ate, 69}$}  \\ 
  \cmidrule(l){2-3} \cmidrule(l){4-5} \cmidrule(l){6-7} 
&$S_p$ &Posterior&$S_p$&Posterior&$S_p$& Posterior \\ 
  \hline
$N(0, 0.5)$ &0.06 &0.09 (-0.15, 0.33)&0.13&0.03 (-0.33, 0.39)& 0.15&0.04 (-0.34, 0.42) \\
$N(0, 1)$ &0.02& 0.10 (-0.14, 0.35) &0.04&  0.02 (-0.36, 0.40)&0.04& 0.04 (-0.37, 0.45) \\
$N(0, 10)$ &0.00 &0.11 (-0.14, 0.35)&0.00&0.03 (-0.35, 0.42)&0.00&0.04 (-0.37, 0.45) \\
   \hline
   \end{tabular}
   \caption{Prior sensitivity measures and posterior mean (95\% interval) of $e_{ate, 03}$, $e_{ate, 36}$, $e_{ate, 69}$ under their own priors with different informativeness. $S_p$ is the prior sensitivity measure.``Posterior'' is the posterior mean (95\% interval). For each average treatment effect, $S_p$ and posterior estimates are found by only changing its own prior.} 
   \end{subtable}
   
\vspace{1em}

    \begin{subtable}{1\textwidth}
   \centering
   \begin{tabular}{rrrrr}
  \hline
Prior & $N(0, 1)$ & $N(0, 3)$ & $N(0, 5)$ & $N(0, 10)$   \\ 
  \hline
$e_{ate, 03}$ &0.09 (-0.07, 0.26)&0.10 (-0.14, 0.35) &0.10 (-0.25, 0.44)&-\\
$e_{ate, 36}$ &0.02 (-0.25, 0.27)&0.02 (-0.36, 0.40)&0.04 (-0.48, 0.54)&-\\
$e_{ate, 69}$ &0.05 (-0.23, 0.32)&0.04 (-0.37, 0.45)&0.04 (-0.52, 0.64)&-\\
   \hline
   \end{tabular}
   \caption{Posterior mean (95\% interval) of $e_{ate, 03}$, $e_{ate, 36}$, $e_{ate, 69}$ under priors of $U'$ with different informativeness. ``-'' indicates the Bayesian model built in sensitivity analysis is bad.} 
   \end{subtable}
 \caption{Results of prior sensitivity analysis for the random intercept outcome model.}
 \label{tab:results-ls7-3w-sa-prior-ri}
\end{table}

The posterior estimates of $e_{ate, 03}$, $e_{ate, 36}$, $e_{ate, 69}$ from Model (2) are sensitive to the informativeness of the prior of $U'$. When the prior of $U'$ becomes less informative to an extent that a Bayesian model in sensitivity analysis is fitted well, the posterior mean of $e_{ate, 03}$ and $e_{ate, 69}$ decreases from positive values to negative values, while the positive posterior mean of $e_{ate, 36}$ increases. For $e_{ate, 03}$ and $e_{ate, 69}$, we may only conclude their posterior estimates are not statistically significant, but there is no clear evidence that their posterior means are positive or negative. For $e_{ate, 36}$, there is evidence that its posterior mean is positive but not statistically significant, but the magnitude of its posterior mean seems to change much with the prior of $U'$, between 0.04 and 1.03. Hence, the posterior mean of $e_{ate, 36}$ from Model (2) may not be reliable.

Overall, the posterior estimates of $e_{ate, 03}$, $e_{ate, 36}$, $e_{ate, 69}$ from the random intercept outcome model are insensitive to the informativeness of the priors of $e_{ate, 03}$, $e_{ate, 36}$, $e_{ate, 69}$ and $U'$. The random intercept outcome model indicates the posterior means of $e_{ate, 03}$, $e_{ate, 36}$, $e_{ate, 69}$ are positive and the three average treatment effects are not statistically significant. In addition, the posterior mean of $e_{ate, 03}$ is larger than the posterior means of $e_{ate, 36}$ and $e_{ate, 69}$, which indicates the period during 3 months post randomization may have the relatively bigger treatment effect compared to the other two periods between 3 and 9 months post randomization.

Then sensitivity of the posterior estimates of $e_{ate, 03}$, $e_{ate, 36}$, $e_{ate, 69}$ is tested against variations in sample standard deviations of the exposures and the LS7 outcome. Variations in sample standard deviations are represented by different locations of the priors of $\sigma_{w, 03}$, $\sigma_{w, 36}$, $\sigma_{w, 69}$ and $\sigma_{y, ls7}$, where sample standard deviations are used as the prior means. Prior locations that are considered include point estimates of sample standard deviations and two boundary points from the 95\% bootstrap intervals of sample standard deviations from table \ref{tab:sd-expoout-boot}, possibly other values if necessary. Table \ref{tab:results-ls7-3w-sa-sd-model2} shows the results of sensitivity analysis of sample standard deviations for Model (2). Table \ref{tab:results-ls7-3w-sa-sd-ri} shows the results of sensitivity analysis of sample standard deviations for the random intercept outcome model.

\begin{table}[htbp]
\small
    \begin{subtable}{1\textwidth}
   \centering
   \begin{tabular}{rrrrrr}
  \hline
Prior & $N(1, 0.01)$ & $N(2.16, 0.01)$ & $N(2.43, 0.01)$ & $N(2.66, 0.01)$ & $N(10, 0.01)$\\ 
  \hline
$e_{ate, 03}$ &-0.02 (-0.33, 0.37)&-0.02 (-0.32, 0.35)&-0.02 (-0.32, 0.33)&-0.03 (-0.32, 0.34)&-0.04 (-0.31, 0.29)\\
$e_{ate, 36}$ &0.28 (-0.49, 0.80)&0.29 (-0.48, 0.81)& 0.28 (-0.45, 0.80)&0.31 (-0.44, 0.80)&0.32 (-0.40, 0.83)\\
$e_{ate, 69}$ &0.00 (-0.40, 0.40)&0.00 (-0.41, 0.39)&0.00 (-0.41, 0.41)&-0.01 (-0.39, 0.40)& 0.00 (-0.40, 0.41)\\
   \hline
   \end{tabular}
   \caption{Posterior mean (95\% interval) of $e_{ate, 03}$, $e_{ate, 36}$, $e_{ate, 69}$ under priors of $\sigma_{w, 03}$ with different locations.} 
   \end{subtable}
   
\vspace{1em}

   \begin{subtable}{1\textwidth}
   \centering
   \begin{tabular}{rrrrr}
  \hline
Prior & $N(1.6, 0.01)$ & $N(1.8, 0.01)$ & $N(1.99, 0.01)$ & $N(10, 0.01)$\\ 
  \hline
$e_{ate, 03}$ &-&-0.02 (-0.32, 0.33)&0.02 (-0.27, 0.34)&0.10 (-0.13, 0.34)\\
$e_{ate, 36}$ &-& 0.28 (-0.45, 0.80)&0.19 (-0.43, 0.69)&0.00 (-0.26, 0.27)\\
$e_{ate, 69}$ &-&0.00 (-0.41, 0.41)&0.02 (-0.40, 0.43)&0.06 (-0.33, 0.46) \\
   \hline
   \end{tabular}
   \caption{Posterior mean (95\% interval) of $e_{ate, 03}$, $e_{ate, 36}$, $e_{ate, 69}$ under priors of $\sigma_{w, 36}$ with different locations. ``-'' indicates the Bayesian model built in sensitivity analysis is bad.} 
   \end{subtable}
   
\vspace{1em}

   \begin{subtable}{1\textwidth}
   \centering
   \begin{tabular}{rrrrrr}
  \hline
Prior & $N(0.5, 0.01)$ & $N(1.5, 0.01)$ & $N(1.64, 0.01)$ & $N(1.78, 0.01)$ & $N(10, 0.01)$\\ 
  \hline
$e_{ate, 03}$ &-0.03 (-0.32, 0.34)&-0.02 (-0.31, 0.37)&-0.02 (-0.32, 0.33)&-0.02 (-0.32, 0.35)&-\\
$e_{ate, 36}$ &0.31 (-0.46, 0.83)&0.28 (-0.46, 0.80)& 0.28 (-0.45, 0.80)&0.29 (-0.46, 0.80)&-\\
$e_{ate, 69}$ &-0.01 (-0.44, 0.43)&0.00 (-0.42, 0.43)&0.00 (-0.41, 0.41)&-0.01 (-0.41, 0.39)&- \\
   \hline
   \end{tabular}
   \caption{Posterior mean (95\% interval) of $e_{ate, 03}$, $e_{ate, 36}$, $e_{ate, 69}$ under priors of $\sigma_{w, 69}$ with different locations. ``-'' indicates the Bayesian model built in sensitivity analysis is bad.} 
   \end{subtable}
   
\vspace{1em}

    \begin{subtable}{1\textwidth}
   \centering
   \begin{tabular}{rrrrrr}
  \hline
Prior & $N(0.5, 0.01)$ & $N(1.95, 0.01)$ & $N(2.06, 0.01)$ & $N(2.26, 0.01)$ & $N(10, 0.01)$\\ 
  \hline
$e_{ate, 03}$ &-&-0.04 (-0.32, 0.32)&-0.02 (-0.32, 0.33)&-0.01 (-0.31, 0.35)& 0.10 (-0.51, 0.71)\\
$e_{ate, 36}$ &-&0.32 (-0.37, 0.79)& 0.28 (-0.45, 0.80)&0.27 (-0.45, 0.80)& 0.05 (-0.85, 0.93)\\
$e_{ate, 69}$ &-&-0.01 (-0.40, 0.39)&0.00 (-0.41, 0.41)&0.00 (-0.39, 0.43)&0.02 (-0.94, 1.00) \\
   \hline
   \end{tabular}
   \caption{Posterior mean (95\% interval) of $e_{ate, 03}$, $e_{ate, 36}$, $e_{ate, 69}$ under priors of $\sigma_{y, ls7}$ with different locations. ``-'' indicates the Bayesian model built in sensitivity analysis is bad.} 
   \end{subtable}
 \caption{Results of sensitivity analysis of sample standard deviations for Model (2).}
 \label{tab:results-ls7-3w-sa-sd-model2}
\end{table}

\begin{table}[htbp]
\small
    \begin{subtable}{1\textwidth}
   \centering
   \begin{tabular}{rrrrrr}
  \hline
Prior & $N(1, 0.01)$ & $N(2.16, 0.01)$ & $N(2.43, 0.01)$ & $N(2.66, 0.01)$ & $N(10, 0.01)$\\ 
  \hline
$e_{ate, 03}$ &0.12 (-0.13, 0.38)&0.12 (-0.14, 0.36)& 0.10 (-0.14, 0.35)&0.11 (-0.13, 0.35)& 0.09 (-0.14, 0.32)\\
$e_{ate, 36}$ &0.00 (-0.39, 0.37)&0.00 (-0.38, 0.37)&0.02 (-0.36, 0.40)&0.01 (-0.37, 0.39)&0.02 (-0.36, 0.40)\\
$e_{ate, 69}$ &0.04 (-0.35, 0.44)&0.04 (-0.36, 0.44)& 0.04 (-0.37, 0.45)&0.03 (-0.38, 0.43)&0.05 (-0.34, 0.44) \\
   \hline
   \end{tabular}
   \caption{Posterior mean (95\% interval) of $e_{ate, 03}$, $e_{ate, 36}$, $e_{ate, 69}$ under priors of $\sigma_{w, 03}$ with different locations.} 
   \end{subtable}
   
\vspace{1em}

   \begin{subtable}{1\textwidth}
   \centering
   \begin{tabular}{rrrrrr}
  \hline
Prior & $N(0.5, 0.01)$ & $N(1.6, 0.01)$ & $N(1.8, 0.01)$ & $N(1.99, 0.01)$ & $N(10, 0.01)$\\ 
  \hline
$e_{ate, 03}$ &0.11 (-0.14, 0.37)&0.11 (-0.13, 0.35)& 0.10 (-0.14, 0.35)&0.12 (-0.13, 0.36)&0.12 (-0.10, 0.35)\\
$e_{ate, 36}$ &0.01 (-0.40, 0.41)&0.01 (-0.37, 0.41)&0.02 (-0.36, 0.40)&0.00 (-0.38, 0.39)&-0.01 (-0.27, 0.23)\\
$e_{ate, 69}$ &0.04 (-0.37, 0.44)&0.04 (-0.37, 0.45)& 0.04 (-0.37, 0.45)&0.04 (-0.37, 0.45)& 0.04 (-0.33, 0.44)\\
   \hline
   \end{tabular}
   \caption{Posterior mean (95\% interval) of $e_{ate, 03}$, $e_{ate, 36}$, $e_{ate, 69}$ under priors of $\sigma_{w, 36}$ with different locations.} 
   \end{subtable}
   
\vspace{1em}

   \begin{subtable}{1\textwidth}
   \centering
   \begin{tabular}{rrrrrr}
  \hline
Prior & $N(0.5, 0.01)$ & $N(1.5, 0.01)$ & $N(1.64 0.01)$ & $N(1.78, 0.01)$ & $N(10, 0.01)$\\ 
  \hline
$e_{ate, 03}$ &0.11 (-0.14, 0.36)&0.11 (-0.13, 0.36)& 0.10 (-0.14, 0.35)&0.11 (-0.13, 0.36)&0.13 (-0.11, 0.37)\\
$e_{ate, 36}$ &0.00 (-0.39, 0.39)&0.00 (-0.39, 0.38)&0.02 (-0.36, 0.40)&0.01 (-0.38, 0.39)&0.02 (-0.34, 0.38)\\
$e_{ate, 69}$ &0.04 (-0.40, 0.49)&0.05 (-0.38, 0.47)& 0.04 (-0.37, 0.45)&0.03 (-0.36, 0.43)&-0.01 (-0.29, 0.26) \\
   \hline
   \end{tabular}
   \caption{Posterior mean (95\% interval) of $e_{ate, 03}$, $e_{ate, 36}$, $e_{ate, 69}$ under priors of $\sigma_{w, 69}$ with different locations.} 
   \end{subtable}
   
\vspace{1em}

    \begin{subtable}{1\textwidth}
   \centering
   \begin{tabular}{rrrrrr}
  \hline
Prior & $N(0.5, 0.01)$ & $N(1.95, 0.01)$ & $N(2.06, 0.01)$ & $N(2.26, 0.01)$ & $N(10, 0.01)$\\ 
  \hline
$e_{ate, 03}$ &0.11 (-0.09, 0.32)&0.12 (-0.12, 0.35)& 0.10 (-0.14, 0.35)&0.12 (-0.13, 0.37)&0.11 (-0.48, 0.70)\\
$e_{ate, 36}$ &0.00 (-0.32, 0.31)&0.00 (-0.38, 0.36)&0.02 (-0.36, 0.40)&0.01 (-0.38, 0.39)&0.01 (-0.87, 0.85)\\
$e_{ate, 69}$ &0.04 (-0.31, 0.40)&0.04 (-0.36, 0.44)& 0.04 (-0.37, 0.45)&0.03 (-0.38, 0.46)&0.03 (-0.91, 0.98) \\
   \hline
   \end{tabular}
   \caption{Posterior mean (95\% interval) of $e_{ate, 03}$, $e_{ate, 36}$, $e_{ate, 69}$ under priors of $\sigma_{y, ls7}$ with different locations. ``-'' indicates the Bayesian model built in sensitivity analysis is bad.} 
   \end{subtable}
 \caption{Results of sensitivity analysis of sample standard deviations for the random intercept outcome model.}
 \label{tab:results-ls7-3w-sa-sd-ri}
\end{table}

\begin{figure}[htbp]
     \centering
     \includegraphics[width=\textwidth, height = \textheight]{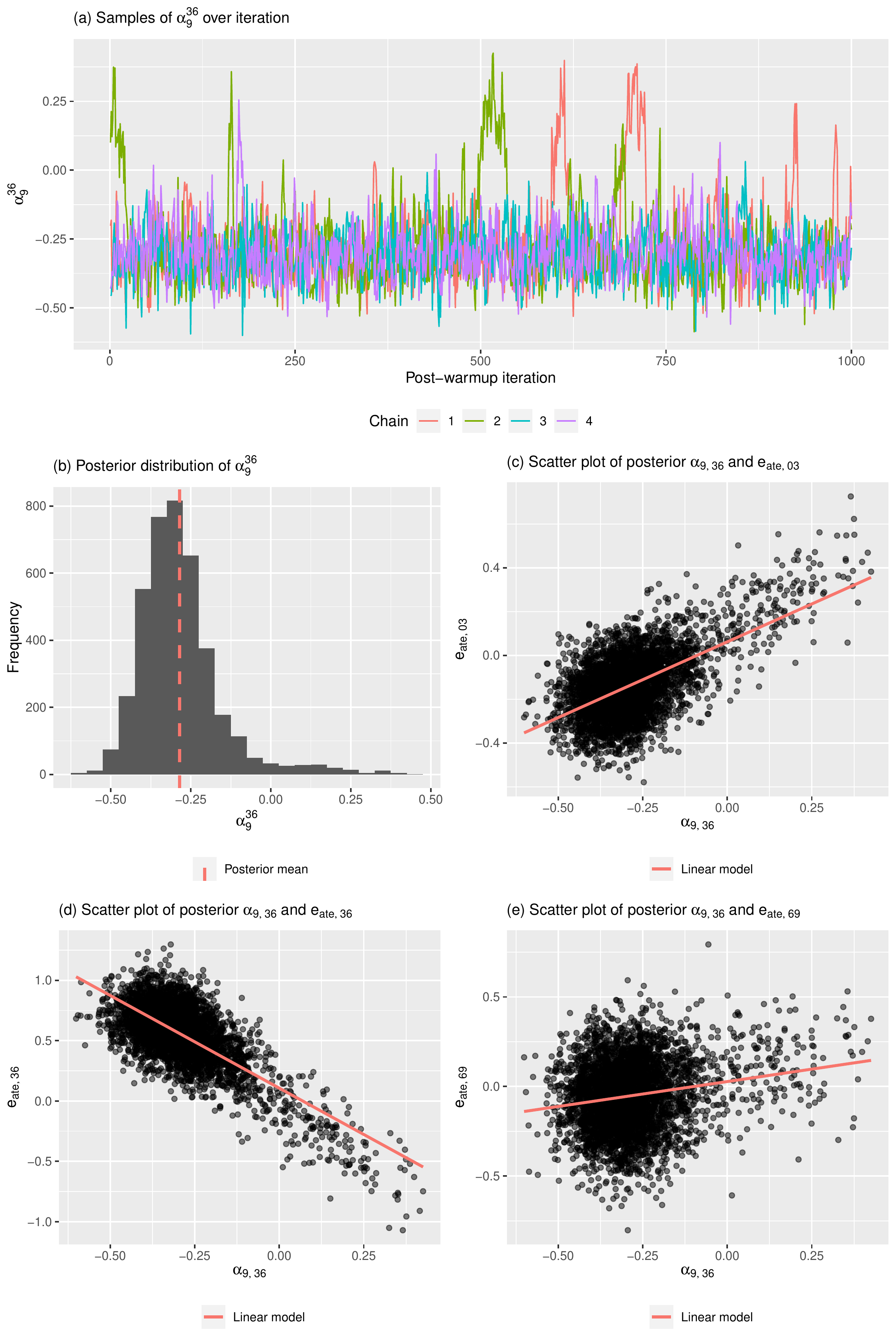}
        \caption{Diagnostics of Model (2) when the prior of $\sigma_{w, 36}$ is $N(1.6, 0.01)$ and the other priors are unchanged. Figure (a) shows the Monte Carlo samples of $\alpha_{9, 36}$ over iteration from 4 Markov chains. Unusual transitions between the negative and positive areas occur, which indicates a potential bimodal posterior distribution. Figure (b) shows the posterior distribution of $\alpha_{9, 36}$ with a long tail in the positive area. Figures (c)-(e) show the association between $\alpha_{9, 36}$ and $e_{ate, 03}$, $e_{ate, 36}$, $e_{ate, 69}$, which indicates the posterior distributions of $e_{ate, 03}$ and $e_{ate, 36}$ may be bimodal.}
        \label{fig:-ls7-3w-diag-u36-sa-prior}
\end{figure}

From table \ref{tab:results-ls7-3w-sa-sd-model2}, the posterior estimates of $e_{ate, 03}$, $e_{ate, 36}$, $e_{ate, 69}$ are sensitive to variations in sample standard deviations of the exposures and the LS7 outcome. The posterior means of $e_{ate, 03}$ and $e_{ate, 69}$ can vary between positive and negative values. When the prior of $\sigma_{w, 36}$ is changed to $N(1.6, 0.01)$, the posterior distribution of $\alpha_{9, 36}$ is potentially bimodal, because the chains can transit from the negative area to the positive area and stay in the positive area for several iterations. Figure \ref{fig:-ls7-3w-diag-u36-sa-prior} shows the samples over iteration and the posterior distribution of $\alpha_{9, 36}$, and the association between $\alpha_{9, 36}$ and three average treatment effects, $e_{ate, 03}$, $e_{ate, 36}$, $e_{ate, 69}$.

From figure \ref{fig:-ls7-3w-diag-u36-sa-prior}, the posterior distribution of $\alpha_{9, 36}$ is potentially bimodal, and if we want to use Model (2) to draw inference, we have to build another two restricted Bayesian models where $\alpha_{9, 36}$ is restricted to be non-negative or non-positive. We do not need to build these two restricted Bayesian models, because we can infer multiple posterior solutions from the associations between $\alpha_{9, 36}$ and $e_{ate, 03}$, $e_{ate, 36}$, $e_{ate, 69}$ through figure \ref{fig:-ls7-3w-diag-u36-sa-prior}. When the samples of $\alpha_{9, 36}$ are negative, the samples of $e_{ate, 03}$ center in the negative area and the samples of $e_{ate, 36}$ center in the positive area. When the samples of $\alpha_{9, 36}$ are positive, the samples of $e_{ate, 03}$ center in the positive area and the samples of $e_{ate, 36}$ center in the negative area. These results indicate there are likely to be two posterior solutions, one from which has negative $\alpha_{9, 36}$, negative $e_{ate, 03}$ and positive $e_{ate, 36}$, and the other from which has positive $\alpha_{9, 36}$, positive $e_{ate, 03}$ and negative $e_{ate, 36}$. Two likely posterior solutions can explain why the posterior means of $e_{ate, 03}$ can vary between positive and negative values in sensitivity analysis very obviously. In addition, two likely posterior solutions seem contradictory, because we cannot conclude whether the posterior means of $e_{ate, 03}$, $e_{ate, 36}$ and $e_{ate, 69}$ are positive or not. Hence, we think Model (2) is useless for further inference.

From table \ref{tab:results-ls7-3w-sa-sd-ri}, the posterior estimates of $e_{ate, 03}$, $e_{ate, 36}$, $e_{ate, 69}$ are relatively insensitive to variations in sample standard deviations of the exposures and the LS7 outcome. When the prior of $\sigma_{w, 69}$ becomes $N(10, 0.01)$ and relatively uninformative, the posterior mean of $e_{ate, 69}$ becomes negative. Compared to the posterior mean of $e_{ate, 69}$ under other priors of $\sigma_{w, 69}$, this negative posterior mean is very small and seems to only reflect changes with the prior of $\sigma_{w, 69}$ rather than changes in estimation of $e_{ate, 69}$. We think the random intercept outcome model is relatively robust and useful for further inference.

A complete-case model is then built for the random intercept outcome model. It uses the same distributions and the priors for parameters in common as the random intercept outcome model. Table \ref{tab:results-ls7-3w-complete-ri} shows the posterior estimates of $e_{ate, 03}$, $e_{ate, 36}$, $e_{ate, 69}$ from this complete-case model. The posterior estimates of $e_{ate, 03}$, $e_{ate, 36}$ and $e_{ate, 69}$ are relatively insensitive to missingness. The posterior mean of $e_{ate, 03}$ from the complete-case model increases by about 0.05 and the posterior mean of $e_{ate, 36}$ decreases to negative by about 0.05. These changes in the posterior means are very small and seem normal. We think missing data does not affect the average treatment effect estimates much.

\begin{table}[htbp]
\small
   \centering
  \begin{tabular}{rrr}
  \hline
$e_{ate, 03}$  &$e_{ate, 36}$&$e_{ate, 69}$ \\ 
  \hline
0.15 (-0.12, 0.42) &-0.03 (-0.49, 0.42)&0.04 (-0.41, 0.49)\\
   \hline
   \end{tabular}
 \caption{Posterior mean (95\% interval) of $e_{ate, 03}$, $e_{ate, 36}$, $e_{ate, 69}$ from the complete-case model for the random intercept outcome model.}
 \label{tab:results-ls7-3w-complete-ri}
\end{table}

Our methods that adjust for unmeasured confounding in both the exposure and outcome distributions fail in this analysis, because multiple posterior solutions are contradictory. Adjusting for unmeasured confounding provides a potential to obtain less biased or unbiased average treatment effect estimates, but possibly due to residual confounding, Bayesian models suffer great flexibility and may become useless for inference. Although the random intercept outcome model has not been sufficiently developed, we make conclusions from the random intercept outcome model, because of its estimation robustness in this analysis and ability to adjust for unmeasured confounding shown in simulation study.

In conclusion, there is evidence that the average effect of one HWC session on the LS7 outcome is positive, during 3 months post randomization, or between 3 and 6 months post randomization, or between 6 and 9 months post randomization. During 3 months post randomization, attending one HWC session reduces the LS7 outcome by 0.10 (-0.14, 0.35) on average. Between 3 and 6 months post randomization, attending one HWC session reduces the LS7 outcome by 0.02 (-0.36, 0.40) on average. Between 6 and 9 months post randomization, attending one HWC session reduces the LS7 outcome by 0.04 (-0.37, 0.45) on average. The average effect of one HWC session on the LS7 outcome is slightly larger during 3 months post randomization than in the other two periods. However, the average effect of one HWC session on the LS7 outcome in each of these three periods is relatively small and not statistically significant. Compared to the analysis with one exposure on the LS7 outcome, the average effect of one HWC session on the LS7 outcome during 9 months post randomization is slightly larger than that during each of three periods mentioned above and slightly more statistically significant. Hence, there is evidence that the HWC intervention has a larger beneficial effect on cardiovascular health during 3 months post randomization than the other two subsequent periods, but the cumulative average effect of one HWC session on cardiovascular health during 9 months post randomization is slightly more statistically significant than the single average effect of one HWC session on cardiovascular health in each of three three-month periods over 9 months post randomization. These conclusions indicate that the duration of the HWC intervention may be shortened to 3 months because the average effect of one HWC session on cardiovascular health in two three-month periods between 3 and 9 months post randomization is negligible, but maintaining the original design of the HWC intervention may lead to a more statistically significant average effect of one HWC session on cardiovascular health.

\chapter{Discussions}
\label{sec:dis}

Our causal framework with one or multiple multivalued exposures in the presence of unmeasured confounding is general. It starts from causal relationships among counterfactual variables and can be easily extended to more complex versions that deal with more complex causal relationships. The way we adjust for unmeasured confounding is basically to impute unmeasured confounders as completely missing variables. Model non-identifiability is a major problem in estimation of unmeasured confounders. Some types of model non-identifiability have been identified, but we cannot guarantee there will be no other type of model non-identifiability. Pearl \autocite*{pearl_parameter_2001} proposed mathematical definitions of parameter identification and model identification, but we find it difficult to directly prove a Bayesian model is identified by definition. Hence, our approach to make a Bayesian model identified is to eliminate all possible types of model-identifiability. We also recognize that direct criteria to test if a model is identified may be better and they need development. If counterexamples to our methods are found, we suggest first exploring if there are new model non-identifiability conditions, since we believe that our methods have generalizability to different causal questions. In order to prevent standard deviations from being estimated to zero, we propose using sample standard deviations in Bayesian models, which avoids estimation of standard deviations by Bayesian models themselves. We choose a Bayesian approach to implement our causal framework and estimation methods of unmeasured confounders, where the frequentist method to find sample standard deviations may be incorporated, but also we expect all these methods applies to frequentist models. However, no further investigation on application in frequentist models, because imputing unmeasured confounders in frequentist models is challenging. We suspect imputation through chained equations such as the \texttt{mice} package \autocite{buuren_mice_2011} may be viable.

Our causal framework and estimation methods of unmeasured confounders show usefulness in simulation study and analysis of the PreventS trial data. When Bayesian models are fitted well, our methods can effectively estimate unmeasured confounding. However, the biggest drawback of our methods is that, they fail when there are contradictory multiple posterior solutions from a Bayesian model. Multiple posterior solutions can occur due to many reasons including residual confounding and model flexibility. When multiple posterior solutions are not contradictory, there is a chance to draw inference from all of them, otherwise Bayesian models are useless. When a Bayesian model produces multiple posterior solutions, we can instead use the random intercept outcome model that only adjusts for unmeasured confounding in the outcome distribution and exclude the exposure distribution(s) in modelling. The random intercept outcome model is expected to produce a single posterior solution. We find it useful from simulation study and further apply it in data analysis. However, we do not focus on developing the random intercept outcome model from the beginning. It is born because the exposure distribution(s) in compliers in the control arm and possibly other people cannot be used to estimate unmeasured confounders and thus it becomes natural to only adjust for unmeasured confounding in the outcome distribution. The theoretical background of the random intercept outcome model seems simple: just use the structural equation of the outcome in the presence of unmeasured confounders, reparameterize this structural equation and develop the complete-data model. Our findings can be easily applied to the random intercept outcome model by simply excluding all the exposures. However, we think further understanding on the modelling performance of the random intercept outcome model is necessary. On the other hand, the random intercept outcome model can be regarded as a mixed model. Mixed models have been used for a long time and there may have been research about using mixed models to adjust for unmeasured confounding. However, we have not yet carried out any literature search about using mixed models to adjust for unmeasured confounding. We think the development of the random intercept outcome model in our thesis is insufficient and possibly more work will be done.

In our causal frameworks, we clearly separate the control intervention and the treatment of research interest. In our thesis, we only have considered a fixed control intervention, which is always zero for all participants. Multivalued control interventions have not been considered. The definition of the control intervention in a two-arm randomized trial is of vital importance in estimation of the average treatment effect(s), because (1) in order to define the average treatment effect(s), we need to define the effect of the control intervention first and then define the average treatment effect(s) compared to the control intervention effect, and (2) the control intervention may need a different structural equation or distribution from the treatment in Bayesian models. In other words, we need to first consider whether the control intervention can represent the natural zero level of the treatment before doing data analysis.

Non-linearity of the treatment effect on each individual is not considered. We assume the treatment effect is constant for all the individuals, which means that the treatment effect on each individual is constant over time and the treatment effect can vary among individuals because we want to estimate an average treatment effect. Non-linearity of the treatment effect means that the treatment effect on each individual is not constant with regards to some factors, such as time; for example, in the PreventS trial, the treatment effect may become smaller at 9 months post randomization than at 3 months post randomization. Without considering non-linearity of the treatment effect, our methods and data analysis are simplified yet still valid, since we can regard the individual treatment effect as an average of the treatment effect that varies with some factors for one individual, but more interesting results may be obtained if we consider non-linearity of the treatment effect.

Our analyses are based on Normal distributions or the normality assumption.
This is because (1) Normal distributions are simpler and faster to fit in Bayesian modelling, (2) the mean and the variance from a Normal distribution are relatively independent, which guarantees validity of using sample standard deviations without considering the relationship between the mean and the variance. However, when using other distributions where the means and the variances are associated, such as Poisson distributions, our methods can still apply, but use of sample standard deviations should be adjusted to incorporate the relationship between the mean and the variance. More complicated variances may need additional modelling and diagnostic procedures. On the other hand, possibly with other distributions except Normal distributions, the variance may not be easily estimated to be zero by Bayesian models. In addition, with Normal distributions, we obtain average treatment effect estimates on the risk difference scale. The risk ratio scale has not been studied but we suspect our causal framework and estimation methods of unmeasured confounders can be extended to the risk ratio scale.

There are some other limitations in our methods. The structural equations we develop from one causal framework are mainly based on additive relationships among variables. Multiplicative relationships and complex interaction among variables have not been considered. How misspecification on distributions of the exposure(s) and the outcome will affect estimation of the average treatment effects has not been investigated. Study on the properties of estimators from Bayesian models built through our methods has not been started. How to incorporate more complex censoring in our causal framework has not been studied. Causal inference covers a very large scope of methodologies and applications. Hence, the literature review may be fuller and thus more helpful.

\chapter{Conclusions}
\label{sec:con}

When we want to understand the average effect of one multivalued exposure on one outcome in the presence of unmeasured confounding, Bayesian models that use our causal framework and estimation methods of unmeasured confounders can produce less biased average treatment effect estimates and have the potential to deal with more complex causal relationships between the exposures and the outcome. When estimation methods of unmeasured confounders fail, the random intercept outcome models can be a good surrogate causal model. Hence, our methods have some generalizability.

Through our methods, there is evidence that the Health Wellness Coaching intervention is beneficial to cardiovascular health at 9 months post randomization by slightly increasing the Life’s Simple Seven total score and effectively reducing systolic blood pressure. There is evidence that the Health Wellness Coaching intervention has no effect on the 5-year cardiovascular disease risk score recalculated with PREDICT, possibly because changes in this 5-year cardiovascular disease risk score are too small to detect the effect of the Health Wellness Coaching intervention. 

There is evidence that based on the Life’s Simple Seven total score, the Health Wellness Coaching intervention has a larger beneficial effect on cardiovascular health during 3 months post randomization than between 3 and 6 months post randomization and than between 6 and 9 months post randomization, but its beneficial effect on cardiovascular health is slightly more statistically significant during 9 months post randomization than in each of three consecutive 3-month periods over 9 months post randomization.

On the other hand, there is no clear evidence that the Health Wellness Coaching intervention benefits or harms mental health based on the Patient Health Questionnaire-9 score, but due to sensitivity of Bayesian models, there is a risk that the Health Wellness Coaching intervention may harm mental health.



\printbibliography

\addcontentsline{toc}{chapter}{Bibliography}


\end{document}